\documentstyle[amssymb]{book}

\input{tcilatex}

\begin{document}

\title{Bose-Einstein Condensation of Trapped Atomic Gases}
\author{Ph.W. Courteille$^{\ast }$, V.S. Bagnato$^{\ast }$, V.I. Yukalov$^{\ast
,\ast \ast }\bigskip $ \\
$^{\ast }${\em Instituto de F\'{i}sica de S\~{a}o Carlos, Universidade de
S\~{a}o Paulo,}\\
{\em \ P.O. Box 369, S\~{a}o Carlos, SP 13560-250, Brazil\bigskip }\\
$^{\ast \ast }${\em Bogolubov Laboratory of Theoretical Physics,}\\
{\em \ Joint Institute for Nuclear Research, Dubna 141980, Russia\bigskip }\\
e-mail: \ yukalov@thsun1.jinr.ru}
\maketitle

\begin{abstract}
This article reviews recent investigations on the phenomenon of
Bose-Einstein condensation of dilute gases. Since the experimental
observation of quantum degeneracy in atomic gases, the research activity in
the field of coherent matter-waves literally exploded. The present topical
review aims to give an introduction into the thermodynamics of Bose-Einstein
condensation, a general overview over experimental techniques and
investigations, and a theoretical foundation for the description of bosonic
many-body quantum systems.
\end{abstract}

\tableofcontents

\chapter{Introduction}

\setcounter{page}{1}In classical physics, light is a wave and matter
consists of particles. At the beginning of the twentieth century, new
experiments like the discovery of the photo effect shattered the common view
of life. Those observations could only be explained by the assumption that
light consists of quantized energy packets, similar to particles. The
feature that light sometimes appears as a wave and sometimes as a particle
seemed incompatible. This duality of light was understood within the
framework of the newly developed {\it quantum theory} which benefitted from
important contributions from scientists including Max Planck, Niels Bohr,
Werner Heisenberg and Albert Einstein. Together with Einstein's {\it theory
of relativity} the quantum theory today constitutes the fundamental pillar
of modern physics. Louis de Broglie applied the duality principle also to
material particles. According to him, very cold particles should under
certain conditions behave like waves whose wavelengths increase as their
velocity drops. The particle is delocalized over a distance corresponding to
the de Broglie wavelength. These features were soon discovered
experimentally and are today even used commercially, {\em e.g.} in electron
microscopes.

The {\it laser} was discovered in 1956. In a laser, light particles are
forced to oscillate synchronously, {\em i.e.} coherently. By analogy, we may
now raise the question if a similar phenomenon can occur for material
particles, and if it should in principle be possible to construct an {\it %
atom laser}. Such a device would emit coherent matter-waves just like the
laser emits coherent light. When a gas is cooled down to very low
temperatures, the individual atomic de Broglie waves become very long and,
if the gas is dense enough, eventually overlap. If the gas consists of a
single species of bosonic particles all being in the same quantum state, the
de Broglie waves of the individual particles constructively interfere and
build up a huge coherent matter-wave. The matter-wave is described by a
single quantum mechanical wavefunction exhibiting long range order and
having a single phase. If this wavefunction is formed in a trap, all the
atoms pile up in its ground state. The transition from a gas of individual
atoms to the mesoscopic quantum degenerate many-body state occurs as a phase
transition and is named after Bose and Einstein who calculated the effect as
early as 1924 \cite{Bose24,Einstein24} {\it Bose-Einstein condensation}
(BEC).

The vast interest in Bose-Einstein condensation arises partly from the fact
that this phenomenon touches several physical disciplines thus creating a
link between them: In thermodynamics BEC occurs as a phase transition from
gas to a new state of matter, quantummechanics view BEC as a matter-wave
coherence arising from overlapping de Broglie waves of the atoms and draw an
analogy between conventional and ''atom lasers'', quantumstatistics explain
BEC as more than one atom sharing a phase space cell, in the quantum theory
of atomic traps many atoms condense to the ground state of the trap, in
quantum field theory BEC is closely related to the phenomenon of spontaneous
breaking of the gauge symmetry.

The experimental verification of Bose-Einstein condensation has been a long
cherished dream in physics. On one hand, several phenomena have been related
to BEC in the past, {\em e.g.} the phenomenon of superfluidity in liquid
helium and the superconductivity. On the other hand, those strongly
interacting systems\ are not pure enough to clearly identify the role of the
Bose-condensation. A few years ago, however, Bose-Einstein condensation in
weakly interacting confined atomic gases was achieved in experiments \cite
{AndersonMH95}\nocite{Davis95b,Bradley95}-\cite{Han98}. The observation of
Bose-Einstein condensation has now been confirmed by more than twenty groups
worldwide and triggered an enormous amount of theoretical and experimental
work on the characterization of Bose-condensed gases. While the early work
focussed on the equilibrium thermodynamics of condensates close to the phase
transition, very soon the dynamical response of the condensate wavefunction
to perturbations was subject of thorough investigations. Subsequently, the
general attention turned to the study of the superfluid characteristics of
BECs, phenomena of quantum transport and the interaction of BECs with light.
Meanwhile, exotic states like multiple species condensates \cite
{Myatt97,Stenger98} and vortices \cite{Matthews99b,Madison99} have been
created, Feshbach collision resonances have been found \cite{Inouye98}\nocite
{Courteille98b}-\cite{Vuletic99} various kinds of atom lasers have been
constructed \cite{Mewes97}\nocite{AndersonBP98,Bloch99,Hagley99}-\cite
{Martin99} BEC interferometers have been realized \cite{Hall98b},
diffraction experiments have been carried out with BECs \cite{Kozuma99},
nonlinear matter-wave interactions \cite{Deng99} and matter-wave
amplification \cite{Inouye99}\nocite{Kozuma99b}-\cite{Inouye99b} have been
observed.

One of the most exciting features is the possibility to construct atom
lasers. The technical advances made in the past few years in controlling and
manipulating matter-waves have raised a new field called atom optics. Nearly
all optical elements which are used to manipulate light beams have found
their atom-optical counterpart within the past ten years, including mirrors,
lenses, waveguides, acousto-optical modulators, and so on. The occurrence of
large-scale coherent quantum objects like BECs and atom lasers will
definitively lead to a modernization of the fields of atomic interferometry,
holography, lithography and microscopy. Collisions between atoms add a rich
variety of phenomena to the field of coherent matter-wave optics where they
play a role similar to the role played by atom-photon interactions in
quantum and nonlinear optics. Since the experimental observation of
matter-wave four-wave mixing \cite{Deng99} the field of nonlinear
matter-wave optics \cite{Lenz93} is evolving at very high speed.

The characteristics (shape, stability, quantum depletion,...) and the
dynamics (superfluidity, nonlinear excitations,...) of BECs are largely
governed by interactions between the atoms. The importance of atomic
collisions for BEC turns them into interesting subject for studies.
Low-energy scattering phenomena, like the recently found Feshbach collision
resonances \cite{Inouye98,Courteille98b} may be used to coherently couple a
bound state of two atoms to the unbound continuum \cite{Timmermans99b}. This
is particularly interesting for the development of techniques capable of
producing ultracold molecules right inside a trap (ultracold chemical
engineering), or even to produce molecular BECs.

Finally, the field of atomic quantum optics is being launched with many
interesting theoretical predictions and ideas. Atomic quantum optics could
be defined as the matter-wave counterpart of quantum optics with light
fields. In analogy, one might expect the possibility of building up
''nonclassical'' quantum correlations, {\em e.g.} Schr\"{o}dinger cat like
quantum states in a truly mesoscopic quantum system (expanded BEC
wavefunctions may easily range up to millimeter sizes) \cite{Ruostekoski98}.
Those states have been studied in various quantum optical systems. But even
more important is the possibility of coherently coupling the optical,
motional, and internal degrees of freedom and therefore the entanglement of
the related modes. In such systems, quantum optics of laser modes (Cavity
QED) and matter-wave optics will merge. There are already several ideas
about the implementation of mutual coherent quantum control between optical
and matter-wave modes \cite{Goldstein99}, and an ultracold version of the
Correlated Atomic Recoil Laser (CARL), an atomic analogue of the
Free-Electron Laser (FEL), may play the role of an interface between optical
and matter-wave fields \cite{Heurich99}.

\bigskip

This topical review is organized as follows. The introduction into the basic
notions of the thermodynamics of Bose-Einstein condensation (Chapter~\ref
{SecBasicNotions}) is brief, since many excellent papers and textbooks have
been published on this subject. Chapter~\ref{SecMakingProbing} reviews
experimental approaches to BEC and points out the essential techniques to
achieve and probe condensates. These have been covered by several review
articles as well, so that we just give a short overview. Ever since the
first achievement of BEC in a dilute gas, the experimental progress has been
very fast. Almost every month a new milestone-experiment is published and\
any attempt of writing a review is hence outdated at the time of
publication. Nevertheless, we believe that a review of the recent
experiments is helpful to show the state of the art in BEC manipulation and
to point out the challenges which still lie ahead. We devoted Chapters~\ref
{SecExperiments}, \ref{SecAtomOptics} and \ref{SecCollisionResonances} to
this subject.

There are two good reviews discussing the physics of trapped Bose-Einstein
condensates \cite{Parkins98,Dalfovo99}. The theoretical part of our review
differs from the latter in several aspects. First of all, we thoroughly
investigate those principal notions, whose discussion is rarely met in
literature but which are crucial to answer such basic questions as: What is
Bose-Einstein condensation actually and what are the mathematically correct
criteria for this phenomenon? What is the relation of Bose condensation to
coherence and to gauge symmetry breaking? What is the true meaning of the
famous Gross-Pitaevskii equation? Is it possible to produce non-groundstate
condensates?

Trapped atoms compose a nonlinear nonuniform system, whose description is
essentially more complicated than that of uniform systems. We explain in
more details than usually done mathematical techniques helping to treat such
nonlinear and nonuniform problems. This especially concerns those original
methods that have been developed recently and which cannot be found in other
reviews.

Our paper contains some fresh topics that have not been reviewed earlier.
Among these are: the stratification of condensate components moving with
respect to each other; resonance formation of topological coherent modes and
critical effects that can arise during this resonance process; non-adiabatic
dynamics of trapped atoms and their escape from a trap.

Finally, in Chapter~17, the problems of describing and measuring
Bose-Einstein condensate in quantum liquids, such as superfluid helium, are
discussed. This makes it possible to better understand analogies and
differences between liquids and gases.

\chapter{Basic Notions}

\label{SecBasicNotions}The canonical approach to statistical mechanics
starts with Boltzmann's probabilistic analysis of the velocity distribution
of an ideal gas. For a gas composed of particles of mass $m$ at temperature $%
T$, the velocity distribution is given by the well-known {\it %
Maxwell-Boltzmann} (MB) law \cite{Huang87} 
\begin{equation}
g({\bf v})=\left( \frac{m}{\sqrt{2\pi k_{B}T}}\right) ^{3}\exp \left( -\frac{%
m{\bf v}^{2}}{2k_{B}T}\right) ,  \label{EqMaxwellBoltzmann}
\end{equation}
where $k_{B}$ is the Boltzmann constant. The Maxwell-Boltzmann law was first
experimentally tested by Otto Stern in 1920 using a primitive atomic beam
and a simple time-of-flight technique based on a velocity-selective rotating
drum. With the advent of laser spectroscopy, the MB law and its limitations
could be tested with highly improved precision. This law describes well the
behavior of weakly interacting atoms at high temperatures. Deviations from
it are insignificant until quantum mechanical effects assert themselves, and
this does not occur until the temperature becomes so low that the atomic de
Broglie wavelength becomes comparable to the mean distance between
particles. For a gas in equilibrium the characteristic wavelength is 
\begin{equation}
\lambda _{dB}=\sqrt{\frac{2\pi \hbar ^{2}}{mk_{B}T}},
\label{EqThermalDeBroglie}
\end{equation}
where $\hbar =h/2\pi $ is Planck's constant. For a general system with
density $n$, the mean distance between particles is $n^{-1/3}$. Quantum
effects are expected to show up for $n^{-1/3}\sim \lambda _{dB}(T)$, so that
the boundary to this regime is defined by 
\begin{equation}
k_{B}T(n)=\frac{2\pi \hbar ^{2}}{m}n^{2/3}.  \label{EqStep1}
\end{equation}
For example, an atomic gas at $900~$K and $n\sim 10^{16}~$cm$^{-3}$ is
safely within the classical regime, since $n^{-1/3}\sim 10^{6}~$cm$~\gg
\lambda _{dB}=10^{-9}~$cm. To witness quantum effects one needs atoms at low
temperature and relatively high density. For most gases, lowering the
temperature or increasing density promotes the system to liquidity before
the quantum regime is reached. Well-known exceptions are spin-polarized
hydrogen (H$\uparrow $) which does not get liquid at all and helium which
exhibits effects of quantum degeneracy in the liquid phase, although those
effects are rather complex due to strong interparticle forces.

All particles of the quantum world are either {\it bosons} with integer spin
or {\it fermions} with half-integer spin. Fermions do not share a quantum
state, because they must follow Pauli's exclusion principle. They obey a
quantum statistical distribution called {\it Fermi-Dirac} distribution. In
contrast, bosons enjoy sharing a quantum state and even encourage other
bosons to join them in a process called bosonic stimulation. Bosons follow a
quantum statistical distribution called {\it Bose-Einstein} distribution
(BE). In this article, we will mainly focus on the Bose-Einstein
distribution. The basic difference between MB statistics and BE statistics
is that the former applies to identical particles that nevertheless are
distinguishable from one another, while the latter describes identical
indistinguishable particles. For Bose-Einstein statistics, one can derive 
\cite{Kittel76}\ the Bose-Einstein distributed occupation number for a
non-degenerate quantum state at energy $\varepsilon $ when the system is
held at temperature $T$, 
\begin{equation}
f(\varepsilon )=\frac{1}{{\normalsize e}^{\beta (\varepsilon -\mu )}-1},
\label{EqBoseEinstein}
\end{equation}
where we used the short-hand notation $\beta \equiv 1/k_{B}T$. The chemical
potential $\mu $ is an important parameter of the system, which helps
normalizing the distribution $f(\varepsilon )$ to the total number of
particles, 
\begin{equation}
N=\sum\nolimits_{\varepsilon }f(\varepsilon ).  \label{EqTotalNumber}
\end{equation}
Similarly, the total energy of the system is given by 
\begin{equation}
E=\sum\nolimits_{\varepsilon }\varepsilon f(\varepsilon ).
\label{EqTotalEnergy}
\end{equation}

A very remarkable effect occurs in a bosonic gas at a certain characteristic
temperature: below this temperature a substantial fraction of the total
number of particles occupies the lowest energy state, while each of the
remaining states is occupied by a negligeable number of particles. Above the
transition temperature the macroscopic observables of the gas, like
pressure, heat capacity, {\em etc.}, receive contributions from all states
with a certain statistical weight but without favouring the lowest energy
state. Below the transition temperature, the observables are altered by the
macroscopic occupation of the ground state, which results in dramatic
changes in the thermodynamic properties. The phase transition is named after
Shandrasekar Bose \cite{Bose24} and Albert Einstein \cite{Einstein24} {\it %
Bose-Einstein Condensation} (BEC).

\section{Bose-Einstein Condensation of Ideal Gas}

One of the keys to understand BEC is the behaviour of the chemical potential 
$\mu $ at very low temperatures. The chemical potential is responsible for
the stabilization of the large number of atoms in the ground state $N_{0}$.
A system of a large number $N$ of noninteracting bosons condenses to the
ground-state as the temperature goes to zero, $N_{0}\rightarrow N$. The
Bose-Einstein distribution function (\ref{EqBoseEinstein}) gives the ground
state population, $\varepsilon =0$, in the zero-temperature limit, $%
N=\lim_{T\rightarrow 0}\left( e^{-\beta \mu }-1\right) ^{-1}=-1/\beta \mu $,
or in terms of the fugacity $Z=e^{\beta \mu }$, 
\begin{equation}
Z\sim 1-1/N.  \label{EqFugacity}
\end{equation}
Note that the chemical potential in a bosonic system must always be lower
than the ground-state energy, in order to guarantee non-negative occupancy $%
f(\varepsilon )$ of any state. $Z\sim 1$ denotes macroscopic occupation of
the ground state. We define the critical temperature for Bose-Einstein
condensation via the occupancy of the ground state. Above this temperature
the occupancy of the ground state is not macroscopic, below this point it is.

For a noninteracting Bose gas with $N$ particles of mass $m$ confined in a
hard-wall box of volume $V=L^{3}$ the critical temperature for BEC can be
calculated by equation (\ref{EqStep1}). The boundary conditions require that
the momenta satisfy $p_{j}=2\pi \hbar l_{j}/L$, where $j=x,y$ or $z$ and $%
l_{j}$ are integers. Each state is labeled by a set of three integers $%
(l_{x},l_{y},l_{z})$. In the thermodynamic limit, the sum over all quantum
states may be converted to an integral over a continuum of states, 
\begin{equation}
\sum\nolimits_{{\bf p}}\stackrel{N\rightarrow \infty }{\longrightarrow }%
\frac{V}{h^{3}}\int d^{3}{\bf p.}  \label{EqThermodynamicLimit}
\end{equation}
For a free gas with energy $\varepsilon =p^{2}/2m$, we can derive the
density of states $\rho (\varepsilon )$ from the normalization of the phase
space, 
\begin{eqnarray}
1 &=&h^{-3}\int \int d^{3}{\bf r}d^{3}{\bf p}  \label{EqDensityOfStatesBox}
\\
&=&2\pi \sqrt{2m}^{3}\frac{V}{h^{3}}\int_{0}^{\infty }\sqrt{\varepsilon }%
d\varepsilon \equiv \int_{0}^{\infty }\rho (\varepsilon )d\varepsilon . 
\nonumber
\end{eqnarray}
The density of states basically depends on the geometry of our system. For a
homogeneous system we find $\rho (\varepsilon )=2\pi \sqrt{2m}^{3}V/h^{3}%
\sqrt{\varepsilon }$, but we can easily extend this result to inhomogeneous
systems (Section~2.2). We should, however, keep in mind that the
density-of-states approach is an approximation which is not valid for
experiments with limited numbers of atoms (Section~2.5). Using the
occupation number $f(\varepsilon )$ for the Bose-Einstein distribution~(\ref
{EqBoseEinstein}), in the thermodynamic limit, we calculate the total number
of particles, 
\begin{eqnarray}
N &=&N_{0}+h^{-3}\int \int f(\varepsilon ({\bf r},{\bf p}))d^{3}{\bf r}d^{3}%
{\bf p}  \label{EqTotalNumberIntegral} \\
&=&N_{0}+\int_{0}^{\infty }f(\varepsilon )\rho (\varepsilon )d\varepsilon
=N_{0}+2\pi \sqrt{2m}^{3}\frac{V}{h^{3}}\int_{0}^{\infty }\frac{\varepsilon
^{1/2}d\varepsilon }{{\normalsize e}^{\beta (\varepsilon -\mu )}-1}, 
\nonumber
\end{eqnarray}
where the ground state population $N_{0}$ is explicitly retained. In the
process of converting the sum into an integral (\ref{EqThermodynamicLimit})
the density of states goes to zero approaching the ground state. This error
is corrected by adding a contribution $N_{0}$ to the integral. At this
point, we introduce the Bose-function that will help to simplify the
notation by 
\begin{equation}
g_{\eta }(z)=\sum\nolimits_{t=1}^{\infty }\frac{z^{t}}{t^{\eta }}
\label{EqBoseFunction}
\end{equation}
and its integral representation 
\begin{equation}
g_{\eta }(z)=\frac{zr^{\eta }}{\Gamma (\eta )}\int_{0}^{\infty }\frac{%
x^{\eta -1}dx}{{\normalsize e}^{rx}-z},  \label{EqBoseIntegral}
\end{equation}
where $\Gamma (\eta )$ denotes the Gamma function. With this definition,
equation (\ref{EqTotalNumberIntegral}) reads 
\begin{equation}
N=N_{0}+\frac{V}{\lambda _{dB}^{3}(T)}g_{3/2}({\normalsize e}^{\beta \mu }).
\label{EqTotalNumberBox}
\end{equation}
We can use equation~(\ref{EqTotalNumberBox}) to calculate the critical
temperature $T_{c}^{0}$, defined through $N_{0}\rightarrow 0$ and $\mu
\rightarrow 0$. Above the phase transition, $T>T_{c}^{0}$, the population is
distributed over all the states, each state being weakly occupied. Below $%
T_{c}^{0}$ the chemical potential is ''pinned'' at $\mu =0$ and the number
of particles occupying the excited states is 
\[
N_{therm}=\frac{V}{\lambda _{dB}(T)^{3/2}}g_{3/2}({\normalsize 1})=N\left( 
\frac{T}{T_{c}^{0}}\right) ^{3/2}, 
\]
with $g_{3/2}(1)=2.612$. Since $N_{0}+N_{therm}=N$, the number of particles
in the ground state becomes 
\begin{equation}
\frac{N_{0}}{N}=1-\left( \frac{T}{T_{c}^{0}}\right) ^{3/2},
\label{EqCondensedFractionBox}
\end{equation}
which is the fraction of the atomic cloud being condensed in the ground
state. The abrupt occurrence of a finite occupation in a single quantum
state at $T_{c}^{0}$ indicates a spontaneous change in the system and a
thermodynamic phase transition. We will come back to this in Section~2.7.

\section{Thermodynamics of Ideal Confined Bose-Gas}

If the atoms are confined in a spatially varying potential, the critical
temperature $T_{c}^{0}$ can be significantly altered. The critical
temperature depends on the general shape and on the steepness of the
potential. We consider $N$ particles of an ideal Bose-gas distributed over
various quantum states of an arbitrary potential. The occupation number $%
f(\varepsilon )$ of particles in an energy level $\varepsilon $ is still
given by (\ref{EqBoseEinstein}), the ground state energy is set to zero. In
the thermodynamic limit, the relation between the chemical potential and the
total number of particles is given by generalization of equation~(\ref
{EqTotalNumberIntegral}), with the appropriate density of states $\rho
(\varepsilon )$. The density of state for an arbitrary confining potential $%
U({\bf r})$ can be found by a generalization of the calculation for the free
gas. The volume in phase space between the surfaces of energy $\varepsilon $
and $\varepsilon +d\varepsilon $ is proportional to the number of states in
that energy interval. However, the external potential limits the space
available to the gas. The density of states is calculated in analogy to
equation~(\ref{EqDensityOfStatesBox}) and yields \cite{Groot50,Bagnato87} 
\begin{equation}
\rho (\varepsilon )=2\pi \sqrt{2m}^{3}\frac{1}{h^{3}}\int_{V^{\ast
}(\varepsilon )}\sqrt{\varepsilon -U({\bf r})}d^{3}{\bf r,}
\label{EqDensityOfStatesConfined}
\end{equation}
where $V^{\ast }(\varepsilon )$ is the available space for particles with
energy $\varepsilon $. We assume a generic power-law potential confining an
ideal Bose-gas in $\alpha $ dimensions, 
\begin{equation}
U({\bf r})=\sum\nolimits_{i=1}^{\alpha }\left| \frac{x_{i}}{a_{i}}\right|
^{t_{i}},  \label{EqPowerLawPotential}
\end{equation}
and define a parameter describing the confinement power of the potential, 
\begin{equation}
\eta =\frac{\alpha }{2}+\sum\nolimits_{i=1}^{\alpha }\frac{1}{t_{i}}.
\label{EqConfinementPower}
\end{equation}
Although the temperature is the basic thermodynamical state variable, the
system needs to be characterized by other variables. Heat is not a state
variable, because the amount of heat required to raise the temperature of
the system depends on the way the heat is transferred. The heat capacity
quantifies the ability of the system to retain energy. In conventional
systems, the heat capacity is typically either given at constant volume or
at constant pressure. With this specification heat capacities are extensive
state variables. When crossing a phase transition, the temperature
dependence of the heat capacity measures the degree of changes in the system
above and below the critical temperature and provides valuable informations
about the general type of phase transitions.

The total energy of the system is given by: 
\begin{equation}
E(T)=\int_{0}^{\infty }\varepsilon f(\varepsilon )\rho (\varepsilon
)d\varepsilon .  \label{EqTotalEnergyIntegral}
\end{equation}
For a confined gas, volume and temperature are connected, and the concept of
pressure is somewhat vague. In this case we cannot refer to heat capacity at
constant volume or constant pressure. However, we may define the heat
capacity at a fixed number of particles, 
\begin{equation}
C(T)=\frac{\partial E(T)}{\partial T}.  \label{EqHeatCapacityDefinition}
\end{equation}
Keeping the implicit temperature dependencies of the thermodynamic variables
in mind, we can evaluate~(\ref{EqHeatCapacityDefinition}): 
\begin{equation}
C(T)=\beta \int_{0}^{\infty }\varepsilon f(\varepsilon )^{2}\rho
(\varepsilon )\left[ \mu ^{\prime }(T)+\frac{\varepsilon -\mu }{T}\right] 
{\normalsize e}^{\beta (\varepsilon -\mu )}d\varepsilon ,
\label{EqHeatCapacityEvaluated}
\end{equation}
where the derivative of the chemical potential from above $T\rightarrow
T_{c}^{0}$ is 
\begin{equation}
\mu ^{\prime }(T_{c}^{+})=-\frac{1}{T}\frac{\int_{0}^{\infty }\varepsilon
f(\varepsilon )^{2}\rho (\varepsilon ){\normalsize e}^{\beta \varepsilon
}d\varepsilon }{\int_{0}^{\infty }f(\varepsilon )^{2}\rho (\varepsilon )%
{\normalsize e}^{\beta \varepsilon }d\varepsilon }.
\label{EqChemicalPotentialDeriv}
\end{equation}

It is especially interesting to compare the discontinuity of the heat
capacity and of its derivative $\partial C(T)/\partial T$ for various
potential power laws and dimensions of confinement, since this may clarify
the nature of the phase transition. The thermodynamic quantities take a
particularly simple form for power law potentials. The calculations are
analogous to those carried out for homogeneous Bose-gases (last section),
and we restrict ourselves here to giving the general results for the
thermodynamic quantities \cite{Bagnato87,Yan99}, {\em e.g.} internal energy $%
E$ and heat capacity $C$: 
\begin{eqnarray}
\frac{N_{0}}{N} &=&1-\left( \frac{T}{T_{c}}\right) ^{\eta }\frac{g_{\eta }(Z)%
}{g_{\eta }(1)},  \label{EqIdealThermodynamics} \\
\frac{E}{Nk_{B}T} &=&\eta \frac{g_{\eta +1}(Z)}{g_{\eta }(Z)},  \nonumber \\
\frac{C_{T>T_{c}}}{Nk_{B}} &=&\eta (\eta +1)\frac{g_{\eta +1}(Z)}{g_{\eta
}(Z)}-\eta ^{2}\frac{g_{\eta }(Z)}{g_{\eta -1}(Z)}\text{ \ \ \ \ , \ \ \ \ }%
\frac{C_{T<T_{c}}}{Nk_{B}}=\eta (\eta +1)\frac{g_{\eta +1}(1)}{g_{\eta }(1)},
\nonumber \\
\frac{\Delta C_{T_{c}}}{Nk_{B}} &=&\frac{C_{T_{c}^{-}}-C_{T_{c}^{+}}}{Nk_{B}}%
=\xi ^{2}\frac{g_{\eta }(1)}{g_{\eta -1}(1)}.  \nonumber
\end{eqnarray}
The Bose-functions at zero chemical potential are just the familiar Rieman
zeta functions, $g_{\eta }(1)=\zeta (\eta )$. The expression for the
critical temperature for $N$ particles confined in a generic power-law
potential in $\alpha $ dimensions reads 
\begin{equation}
T_{c}^{0}=k_{B}^{-1}\left[ \frac{h^{\alpha }}{(2\pi m)^{\alpha /2}}\frac{N}{%
2^{\alpha }}\frac{1}{g_{\eta }(1)\prod_{i=1}^{\alpha }a_{i}\Gamma
(t_{i}^{-1}+1)}\right] ^{1/\eta }.  \label{EqCriticalTemperatureConfined}
\end{equation}
To evaluate the temperature dependence of the thermodynamic variables, we
calculate the fugacity $Z(T)=e^{\beta \mu }$ from the second equation of (%
\ref{EqIdealThermodynamics}) \cite{Huang87}. For $T>T_{c}$ we derive the
fugacity as the root of $g_{\eta }(Z)=g_{\eta }(1)(T_{c}/T)^{\eta }$ and for 
$T<T_{c}$ the fugacity is simply $Z=1$.

Let us give two examples for three-dimensional confinement, $\alpha =3$. The
homogeneous 3D box potential inside a volume $V$ is obtained from the
power-law formula by setting $t_{i}\longrightarrow \infty $ so that $\eta
\equiv 3/2$. Evaluating the density of states (\ref
{EqDensityOfStatesConfined}), we find $N=N_{0}+g_{3/2}(Z)V/\lambda _{dB}^{3}$%
.

For an anisotropic harmonic potential, $U({\bf r})=\frac{m}{2}\omega
_{x}^{2}x^{2}+\frac{m}{2}\omega _{y}^{2}y^{2}+\frac{m}{2}\omega
_{z}^{2}z^{2} $, we have $\eta \equiv 3$. We introduce the geometrically
averaged secular frequency $\omega _{trap}\equiv \left( \omega _{x}\omega
_{y}\omega _{z}\right) ^{1/3}$, and the size of the ground state $a_{trap}=%
\sqrt{\hbar /m\omega _{trap}}$. Evaluating the density of states (\ref
{EqDensityOfStatesConfined}), we find $N=N_{0}+\left( k_{B}T/\hbar \omega
_{trap}\right) ^{3}g_{3}(Z)$. The values for confinement power, critical
temperatures, heat capacity and its discontinuity at the phase transition
for several potential configurations are shown in Figure~\ref
{FigBoseCondensation} and summarized in Table~\ref{TabCriticalTemperature}.

Table~\ref{TabCriticalTemperature} shows that steeper potential wells ({\em %
i.e.} smaller $a,b,$ and $c$) give higher values for $T_{c}^{0}$. The
critical temperature also depends on the confinement power of the potential 
\begin{equation}
\eta =-\frac{T_{c}^{0}}{N}\left( \frac{dN_{0}}{dT}\right) _{T=T_{c}^{0}}.
\label{EqConfinementPowerDer}
\end{equation}
Larger values of the confinement power result in higher $T_{c}^{0}$. A
strongly confining potential can lead to quantum degeneration at much higher
critical temperatures and greatly facilitate experimental efforts to achieve
BEC. At a given temperature, a strongly confining potential reduces the
minimum number of trapped particles required for condensation.

It is also interesting to note, that from the values presented in Table~\ref
{TabCriticalTemperature}, the changes in heat capacity at the phase
transition are larger for any power-law potential than for a rigid wall
container. This is due to the fact that increasing the energy of the gas
requires to work against the confining potential.

\section{Low-Dimensional Systems}

The trapping potentials can technically be designed to be very anisotropic,
reaching almost two-dimensional pancake-shaped or one-dimensional
needle-shaped configurations. The thermodynamics for such systems can easily
be formulated as limiting cases of the general formulae presented in
Section~2.2 \cite{Yan99,Bagnato91}. We will first discuss a Bose-gas
confined in a one-dimensional power law potential, $U(x)=\left( \frac{x}{a}%
\right) ^{t}$. In this case, the confinement power~(\ref{EqConfinementPower}%
) reads $\eta =1/t+1/2$. From the general formula for the critical
temperature~(\ref{EqCriticalTemperatureConfined}) we get 
\begin{equation}
k_{B}T_{c}^{1D}=\left( \frac{h}{\left( 2\pi m\right) ^{1/2}}\frac{N}{2a}%
\frac{1}{\Gamma (\eta +1/2)g_{\eta }({\normalsize 1})}\right) ^{1/\eta }.
\label{EqCriticalTemperature1D}
\end{equation}
According to the properties of the zeta function, $g_{\eta }(1)=\varsigma
(\eta )$ is finite only if $t<2$. Therefore, the one-dimensional confined
gas will exhibit BEC only if the potential power is less than $2$, {\em i.e.}
only if the external potential is more confining than a parabolic potential.

For a two dimensional power-law potential which is symmetric in both
directions, $U(x,y)=\left( \frac{x}{a}\right) ^{t}+\left( \frac{y}{a}\right)
^{t}$, equation~(\ref{EqConfinementPower}) reads $\eta =2/t+1$ and the
critical temperature is 
\begin{equation}
k_{B}T_{c}^{2D}=\left( \frac{h^{2}}{2\pi m}\frac{N}{4a^{2}}\frac{1}{\Gamma
(\eta /2+1/2)^{2}g_{\eta }({\normalsize 1})}\right) ^{1/\eta }.
\label{EqCriticalTemperature2D}
\end{equation}
Unlike in the $1D$-case, $g_{\eta }(1)$ remains finite for all positive
values of $t$. Consequently, for a confined $2D$ system, BEC can in
principle occur, except for homogeneous systems where $t\rightarrow \infty $.

\section{Semiclassical Density Distribution}

An effect of the inhomogeneous confining potential is a spatial compression
of the cloud during cooling and crossing $T_{c}^{0}$. The behavior of this
spatial compression can be and has been used as a signature for the
occurrence of BEC (Section~3.2.1). In the following, we will calculate the
temperature dependence of the spatial density profile near the critical
point. We start with \cite{Bagnato96} 
\begin{equation}
n({\bf r})=\left\{ 
\begin{array}{ccc}
\sum_{\varepsilon =0}^{\infty }f(\varepsilon )\left| \psi _{\varepsilon }(%
{\bf r})\right| ^{2} & \text{for} & T>T_{c}^{0} \\ 
&  &  \\ 
N_{0}\left| \psi _{0}({\bf r})\right| ^{2}+\sum_{\varepsilon _{1}}^{\infty
}f(\varepsilon )\left| \psi _{\varepsilon }({\bf r})\right| ^{2} & \text{for}
& T<T_{c}^{0}
\end{array}
\right\} .  \label{EqSemiclassicalDensitySum}
\end{equation}
The temperature dependent occupation numbers $f(\varepsilon )$ and the
wavefunctions $\psi _{\varepsilon }({\bf r})$ for all states have to be
known. Furthermore, we must know $\mu (T)$, which is an important parameter
for determining the occupation number $f(\varepsilon )$ for $T>T_{c}^{0}$.
Therefore, equation~(\ref{EqSemiclassicalDensitySum}) is quite difficult to
evaluate analytically. There is, however, a different way to do this. The
number of particles occupying a given phase space cell is 
\begin{equation}
dN=h^{-3}f(\varepsilon )d^{3}{\bf p}d^{3}{\bf r,}
\label{EqSemiclassicalPhaseSpaceCell}
\end{equation}
where $f(\varepsilon )$ is 
\begin{equation}
f(\varepsilon ({\bf r,p}))=\frac{1}{{\normalsize e}^{\beta \left( {\bf p}%
^{2}/2m+U({\bf r})-\mu \right) }-1}.  \label{EqSemiclassicalDistribution}
\end{equation}
The total density of the normal fraction in position space is found by
integrating over momentum space 
\begin{equation}
n_{th}({\bf r})=h^{-3}\int f(\varepsilon )d^{3}{\bf p=}\lambda
_{dB}^{-3}g_{3/2}\left[ {\normalsize e}^{\beta \left( \mu -U({\bf r})\right)
}\right] ,  \label{EqSemiclassicalDensity}
\end{equation}
where we make use of the integral representation of the Bose-function. This
formula holds for any trapping potential. If we may now for simplicity
assume a harmonic oscillator, $U({\bf r})=\frac{m}{2}\omega _{x}^{2}x^{2}+%
\frac{m}{2}\omega _{y}^{2}y^{2}+\frac{m}{2}\omega _{z}^{2}z^{2}$, we can
similarly calculate the momentum distribution by integrating over position
space 
\begin{equation}
\tilde{n}_{th}({\bf p})=h^{-3}\int f(\varepsilon )d^{3}{\bf r=}\hbar
^{-3}\lambda _{dB}^{-3}a_{trap}^{6}g_{3/2}[{\normalsize e}^{\beta \left( \mu
-{\bf p}^{2}/2m\right) }],  \label{EqSemiclassicalMomentum}
\end{equation}
where $a_{trap}$ is the size of the ground state of the harmonic trap. Of
course, by integrating the distributions (\ref{EqSemiclassicalDensity}) and (%
\ref{EqSemiclassicalMomentum}) we recover the normalization (\ref
{EqTotalNumberIntegral}): 
\begin{equation}
N=\int \tilde{n}_{th}({\bf p})d^{3}{\bf p=}\int n_{th}({\bf r})d^{3}{\bf r}
\label{EqSemiclassicalNumber}
\end{equation}

When evaluating~(\ref{EqSemiclassicalDensity}) using a semiclassical
approach, we left out the ground-state contribution, which is in fact
negligeable above the phase transition $T_{c}^{0}$. Below $T_{c}^{0}$, the
contribution of the ground-state to the density (\ref{EqSemiclassicalDensity}%
) is given by $N_{0}\left| \psi _{0}\right| ^{2}$, where $\psi _{0}$
describes the ground state of the trap. If we assume a harmonic oscillator
potential $U({\bf r})=\frac{m}{2}\omega ^{2}{\bf r}^{2}$, we expect a
Gaussian distribution for the ground state density

\begin{equation}
n({\bf r})=\left\{ 
\begin{array}{ccc}
\lambda _{dB}^{-3}(T)g_{3/2}\left[ Z(T){\normalsize e}^{-\beta \frac{m}{2}%
\omega ^{2}{\bf r}^{2}}\right] & \text{for} & T\geq T_{c}^{0} \\ 
&  &  \\ 
\sqrt{8Ng_{3}(1)}\lambda _{dB}^{-3}(T_{c}^{0})\left[ 1-\left(
T/T_{c}^{0}\right) ^{3}\right] {\normalsize e}^{-\frac{m}{\hbar }\omega {\bf %
r}^{2}}+\lambda _{dB}^{-3}(T)g_{3/2}\left[ {\normalsize e}^{-\beta \frac{m}{2%
}\omega ^{2}{\bf r}^{2}}\right] & \text{for} & T<T_{c}^{0}
\end{array}
\right\} .  \label{EqSemiclassicalDensityIntegral}
\end{equation}

To obtain the evolution of $n({\bf r})$ while the system is cooled down
across the phase transition, it is necessary to know the fugacity $Z$ as a
function of temperature. We can either numerically solve the second equation
of (\ref{EqIdealThermodynamics}) separately above and below the phase
transition, or we can approximate the fugacity by a series as described
below. Above $T_{c}^{0}$ the equation~(\ref{EqTotalNumber}) can be written 
\[
N=\sum_{\varepsilon }\frac{Z{\bf e}^{-\beta \varepsilon }}{1-Z{\bf e}%
^{-\beta \varepsilon }}=\sum_{\varepsilon }\sum_{j=1}^{\infty }Z^{j}{\bf e}%
^{-j\beta \varepsilon }. 
\]
The sum can be transformed into an integral in the continuum-of-states
approximation: 
\begin{equation}
N=\sum\nolimits_{j=1}^{\infty }D_{j}Z^{j},  \label{EqTotalNumberSeries}
\end{equation}
where $D_{j}=\int_{0}^{\infty }\rho (\varepsilon ){\bf e}^{-j\beta
\varepsilon }d\varepsilon $. This series relates the number of particles
with the fugacity, where the coefficients $D_{j}$ carry all information
about the external potential. The series can be inverted yielding values for
the fugacity $Z$. For the harmonic oscillator, we obtain the explicit
expression 
\begin{equation}
Z=1.200\left( \tfrac{T_{c}^{0}}{T}\right) ^{3}-0.180\left( \tfrac{T_{c}^{0}}{%
T}\right) ^{6}-0.010\left( \tfrac{T_{c}^{0}}{T}\right) ^{9}+\text{higher
orders.}  \label{EqFugacityExpansion}
\end{equation}
We can now evaluate (\ref{EqSemiclassicalDensityIntegral}) at any
temperature. For simplicity, we consider the density at ${\bf r}=0$. As a
function of temperature, the peak density $n(0)$ exhibits a sudden jump,
proportional to $N^{1/2}$, at the critical temperature (Fig.~\ref
{FigPeakDensity}). This behavior is used as an experimental indication for
the occurrence of BEC \cite{AndersonMH95}. Larger total particle numbers $N$
make the identification easier. For small numbers, the density jump
decreases and may eventually be washed out by fluctuations in the thermal
distribution.

Intuitively, one expects Bose condensation to set on when the mean distance
between the particles is approximately $\lambda _{dB}$. Indeed, the density
distribution~(\ref{EqSemiclassicalDensityIntegral}) takes a value of $%
n_{c}=\lambda _{dB}^{-3}g_{3/2}(1)$ at the critical point irrespective of
the nature of the confining potential. The main effects of inhomogeneous
trapping is to concentrate the density at a smaller region of space and to
facilitate the formation of BEC in this region. The quantity $n(0)\lambda
_{dB}^{3}=g_{3/2}(Z)$ is often called {\it phase space density} of the gas.

\section{Finite Number of Particles}

The condensates experimentally produced in alkali gases consisted of
relatively small atom numbers between $1000$ to $10^{7}$, so that the
validity of the thermodynamic approximation and the use of the
density-of-states approach has been questioned \cite{Grossmann95}.
Furthermore, the decision whether to use the {\it grand canonical}, the {\it %
canonical} or the {\it microcanonical} ensemble for calculating the
thermodynamic quantities noticeably influences the results. Herzog and
Olshanii \cite{Herzog97} have shown that for small atom numbers on the order
of $100$ the canonical and grand canonical statistics lead to predictions on
the condensed fraction that differ by up to $10\%$. On the other hand, they
give the same results if the particle numbers are large. Which canonical
statistics is more appropriate is not a trivial question and depends on the
experimental setup and in particular on the time scale of the measurements.
If we look at the sample for short times, the number of condensed atoms will
be fixed, and we can assume a canonical ensemble. For longer times, however,
the atom number may be an equilibrium parameter depending on the contact of
the sample with a reservoir, and the grand canonical statistics is better
suited.

Assuming grand canonical ensembles, we will now discuss the impact of finite
atom numbers on the properties of a Bose-gas at the condensation threshold
and, in particular, on the transition temperature and the heat capacity \cite
{Grossmann95}\nocite{Kirsten96,Ketterle96b,Herzog97}-\cite{Napolitano97}. To
illustrate this point, we numerically calculate the heat capacity of a
Bose-gas confined in a three-dimensional isotropic harmonic trap. The energy
eigenvalues $\varepsilon _{m}$ are: 
\begin{equation}
\varepsilon _{m}=m\hbar \omega .  \label{EqOscillatorQuantization}
\end{equation}
For a three-dimensional trap, we must take the degeneracy $\gamma _{m}$ for
the levels into account, 
\begin{equation}
\gamma _{m}=\tfrac{1}{2}(m+1)(m+2).  \label{EqDegeneracy}
\end{equation}
We numerically integrate the expression for the number of atoms 
\begin{equation}
N=\sum\nolimits_{m=0}^{\infty }\gamma _{m}f(\varepsilon _{m}),
\label{EqTotalNumberFinite}
\end{equation}
in order to extract the chemical potential $\mu (T)$ from the occupation
number (\ref{EqBoseEinstein}). We start using a certain limited number of
levels $m$ and subsequently add more until the result converges. Knowing $%
\mu (T)$, we can easily estimate the other thermodynamical quantities. The
total energy of the system is 
\begin{equation}
E(T)=\sum\nolimits_{m=0}^{\infty }\gamma _{m}f(\varepsilon _{m})\varepsilon
_{m},  \label{EqTotalEnergyFinite}
\end{equation}
and the heat capacity is derived from its definition~(\ref
{EqHeatCapacityDefinition}) in analogy to the continuum-of-states formula~(%
\ref{EqHeatCapacityEvaluated}) \cite{Napolitano97}, 
\begin{equation}
C(T)=\beta \sum_{m=0}^{\infty }\frac{\gamma _{m}\varepsilon _{m}{\normalsize %
e}^{\beta (\varepsilon _{m}-\mu )}}{\left[ {\normalsize e}^{\beta
(\varepsilon _{m}-\mu )}-1\right] ^{2}}\left( \beta (\varepsilon _{m}-\mu )+%
\frac{\partial \mu }{\partial T}\right) ,  \label{EqHeatCapacityFinite}
\end{equation}
where 
\begin{equation}
\frac{\partial \mu }{\partial T}=-\frac{1}{T}\frac{\sum_{m=0}^{\infty
}\gamma _{m}(\varepsilon _{m}-\mu ){\normalsize e}^{\beta (\varepsilon
_{m}-\mu )}f(\varepsilon _{m})^{2}}{\sum_{m=0}^{\infty }\gamma _{m}%
{\normalsize e}^{\beta (\varepsilon _{m}-\mu )}f(\varepsilon _{m})^{2}}.
\label{EqChemicalPotentialDerivFinite}
\end{equation}
Figure~\ref{FigHeatCapacity} shows the results of the numerical calculations
of the heat capacity for different values of $N$. The critical temperature $%
T_{c}^{0}$ is defined at this discontinuity. If we define the critical
temperature $T_{c}$ to coincide with the maximum heat capacity (where $%
\partial C/\partial T=0$), we find $T_{c}/T_{c}^{0}=0.813,0.898$ and $0.946$
for $N=100,10^{3}$ and $10^{4}$, respectively. The lowering of the critical
temperature for decreasing numbers of particles is due to the fact that
smaller systems have larger available effective volume. In the thermodynamic
limit $(N\rightarrow \infty )$ the discontinuity appears very clearly. As $N$
decreases, $C(T)$ gets smoother at the transition and the discontinuity
disappears. Strictly speaking, the finite system does not undergo a phase
transition. On the other hand, the deviation of the behavior of a large
finite number system from the thermodynamic limit is reasonably small to
justify talking about phase transition.

Grossmann and Holthaus \cite{Grossmann95} derived analytic expressions for
grand canonical ensembles and harmonic traps with $\tilde{\omega}=\left(
\omega _{x}\omega _{y}\omega _{z}\right) ^{1/3}$ and $\bar{\omega}=\frac{1}{3%
}\left( \omega _{x}+\omega _{y}+\omega _{z}\right) $. For the condensed
fraction and the critical temperature they found: 
\begin{equation}
\frac{N_{0}}{N}=g_{2}(Z)\approx 1-\left( \frac{T}{T_{c}^{0}}\right) ^{3}%
\frac{g_{3}(Z)}{g_{3}(1)}-\frac{3\tilde{\omega}}{2\bar{\omega}N^{1/3}}\left( 
\frac{T}{T_{c}^{0}}\right) ^{2}\frac{g_{2}(Z)}{g_{3}(1)^{2/3}},
\label{EqCondensedFractionFinite}
\end{equation}
\begin{equation}
T_{c}\approx T_{c}^{0}\left( 1-\frac{\tilde{\omega}}{2\bar{\omega}N^{1/3}}%
\frac{g_{2}(1)}{g_{3}(1)^{2/3}}\right) .  \label{EqCriticalTemperatureFinite}
\end{equation}

\section{Atomic Interactions in Nonideal Confined Bose-Gas}

Until now, we only considered non interacting ideal gases. The thermodynamic
behavior of such systems is solely governed by statistics or, at low
temperatures, by quantum statistics. Real systems are always affected by
particle interactions. Often particle interactions are so dominant that they
blur the quantum effects. Interactions cause {\it quantum depletion} of the
condensate phase even at zero temperature. In the case of superfluid$\ ^{4}$%
He only a small fraction, typically around $10\%$, is in the ground state.
However, far from being only a nuisance, atomic interactions enrich the
multitude of physical phenomena. They give rise to nonlinear behavior of the
de Broglie matter-wave exploited in nonlinear atom optics (Section~5.4), and
the strength of the interactions can even be tuned close to so-called {\it %
Feshbach collision resonances} (Section~6.1).

The grand canonical many-body Hamiltonian of a trapped Bose-gas, interacting
through the local $s$-wave collision potential, in second quantization reads 
\begin{equation}
\hat{H}=\int d^{3}{\bf r}\hat{\psi}^{\dagger }({\bf r})\left[ -\frac{\hbar
^{2}}{2m}\triangle +U_{trap}({\bf r})-\mu +\frac{g}{2}\left| \hat{\psi}({\bf %
r})\right| ^{2}\right] \hat{\psi}({\bf r}),  \label{EqMeanfieldHamiltonian}
\end{equation}
where $\hat{\psi}({\bf r})$ denotes the bosonic field atomic annihilation
operator and satisfies the Heisenberg equation of motion. The interaction
strength $g=4\pi \hbar ^{2}a/m$ only depends on a single atomic parameter,
the {\it scattering length} $a$. A common approximation is the {\it %
Bogolubov prescription}, where the field operators describing the condensate
and thermal phase can be decomposed into a complex function $\psi _{0}({\bf r%
})\equiv <\hat{\psi}({\bf r})>$ called {\it condensate wavefunction} which
can be chosen as the {\it order parameter} of the system and into a small
perturbation $\delta \hat{\psi}({\bf r})\equiv \hat{\psi}({\bf r})-\psi _{0}(%
{\bf r})$ describing the{\it \ thermal excitations}. At zero temperature, we
can neglect the thermal excitations \cite{Baym96} and our system is
completely described by a single wavefunction $\psi _{0}({\bf r},t)$ that
follows the {\it Gross-Pitaevskii equation}, 
\begin{equation}
\left[ \frac{-\hbar ^{2}}{2m}\Delta +U_{trap}({\bf r})+g\left| \psi _{0}(%
{\bf r},t)\right| ^{2}\right] \psi _{0}({\bf r},t)=i\hbar \frac{\partial }{%
\partial t}\psi _{0}({\bf r},t).  \label{EqGrossPitaevski}
\end{equation}

\subsection{Semiclassical Approximation}

If we additionally assume $\hbar \omega \ll k_{B}T$, we can apply the {\it %
semiclassical WKB approximation}, {\em i.e.} the atomic motion does not have
to be quantized and the trap has a continuous energy spectrum. We can then
replace coordinate and momentum operators by their expectation values and,
with the abbreviations ${\cal L}({\bf r},{\bf p})\equiv {\bf p}%
^{2}/2m+U_{trap}({\bf r})-\mu +2gn({\bf r})$ and $n({\bf r})=n_{0}({\bf r}%
)+n_{th}({\bf r})$, we get a set of two semiclassical {\it Bogolubov
equations} 
\begin{eqnarray}
{\cal L}({\bf r},{\bf p})u({\bf r},{\bf p})-gn_{0}({\bf r},{\bf p})v({\bf r},%
{\bf p}) &=&\varepsilon ({\bf r},{\bf p})u({\bf r},{\bf p}),
\label{EqBogoliubovEquations} \\
{\cal L}({\bf r},{\bf p})v({\bf r},{\bf p})-gn_{0}({\bf r},{\bf p})u({\bf r},%
{\bf p}) &=&-\varepsilon ({\bf r},{\bf p})v({\bf r},{\bf p}),  \nonumber
\end{eqnarray}
where the phonon creation amplitude $u({\bf r},{\bf p})$ and the phonon
annihilation amplitude $v({\bf r},{\bf p})$ obey the normalization condition 
$u({\bf r},{\bf p})^{2}-v({\bf r},{\bf p})^{2}=1$. They relate the particle
distribution function $F({\bf r},{\bf p})$ and the quasiparticle
distribution function $f(\varepsilon )=\left( e^{\beta \varepsilon ({\bf r},%
{\bf p})}-1\right) ^{-1}$ by 
\begin{equation}
F({\bf r},{\bf p})=\left( \left| u({\bf r},{\bf p})\right| ^{2}+\left| v(%
{\bf r},{\bf p})\right| ^{2}\right) f(\varepsilon ).
\label{EqBogoliubovDistribution}
\end{equation}
The spatial distribution of the thermal density is calculated from 
\begin{equation}
n_{th}({\bf r})=h^{-3}\int F({\bf r},{\bf p})d^{3}{\bf p}
\label{EqDensityDistributionLowT}
\end{equation}
and analogously for the momentum distribution. The last equation represents
a generalization of equation~(\ref{EqSemiclassicalDensity}) for interacting
particles at all excitation energies. The Bogolubov equations~(\ref
{EqBogoliubovEquations}) yield a simple expression for the excitation
spectrum 
\begin{equation}
\varepsilon ({\bf r},{\bf p})=\sqrt{{\cal L}({\bf r},{\bf p}%
)^{2}-g^{2}n_{0}^{2}({\bf r})}.  \label{EqBogoliubov}
\end{equation}
They can be solved numerically \cite{Giorgini97} or be approximated
analytically. All thermodynamic quantities can be derived from the
distribution functions and the spectrum. For example the entropy reads $%
S=k_{B}h^{-3}\int \left( \beta \varepsilon f(\varepsilon )-\ln \left(
1-e^{-\beta \varepsilon }\right) \right) d^{3}{\bf r}d^{3}{\bf p}$, the heat
capacity is $C=T\left( \partial S/\partial T\right) _{N}$, and the total
energy follows from $C=\left( \partial E/\partial T\right) _{N}$. For
homogeneous systems, where the wavefunctions are plane waves, the energy
spectrum takes the well-known local-density form of the {\it Bogolubov
dispersion relation}, 
\begin{equation}
\varepsilon ({\bf r},{\bf p})=\sqrt{\frac{{\bf p}^{2}}{2m}\left( \frac{{\bf p%
}^{2}}{2m}+2gn_{0}({\bf r})\right) },  \label{EqBogoliubovDispersion}
\end{equation}
which has been used to calculate particle-like excitations, ${\bf p}%
^{2}/2m\gg gn({\bf r})$, and phonon-like excitations, ${\bf p}^{2}/2m\ll gn(%
{\bf r})$ (Sections~4.3.1 and 5.4.2).

\bigskip

Several interesting results can be obtained by restricting the analysis to
energies that are much larger than the chemical potential, $\varepsilon (%
{\bf r},{\bf p})\gg \mu $. From the Bogolubov equations, we then derive a
particularly simple {\it Hartree-Fock type spectrum} 
\begin{equation}
\varepsilon ({\bf r},{\bf p})=\frac{{\bf p}^{2}}{2m}+U({\bf r})+2gn({\bf r}).
\label{EqHartreeFockDispersion}
\end{equation}

As a rough approximation, above $T_{c}$, we can neglect the influence of the
interactions on the density distribution, plug the semiclassical expression~(%
\ref{EqSemiclassicalDensity}) into the Hamiltonian and recalculate the
thermodynamic potentials with the effective potential $U_{eff}({\bf r})=U(%
{\bf r})+g\lambda _{dB}^{-3}g_{3/2}\left[ e^{\beta \left( \mu -U({\bf r}%
)\right) }\right] $ \cite{Bagnato87}. In the case of a harmonic potential,
the condensate fraction is 
\begin{equation}
\frac{N_{0}}{N}=1-\left( \frac{T}{T_{c}^{0}}\right) ^{3}-4\frac{a}{\lambda
_{dB}}\left( \frac{T}{T_{c}^{0}}\right) ^{7/2},
\label{EqCondensedFractionInteractions}
\end{equation}
where the $T_{c}^{0}$ is the critical temperature in the ideal gas limit
(Table~\ref{TabCriticalTemperature}). The critical temperature modified by
interactions is estimated from (\ref{EqCondensedFractionInteractions}) by
setting $N_{0}=0$. For positive scattering lengths, the phase transition
occurs at lower temperatures. This can be understood intuitively, because
the repulsive particle interaction associated with positive scattering
lengths counteracts the density compression required for initiating the
condensation process.

Giorgini {\em et al. }\cite{Giorgini97} numerically integrated the
semiclassical Bogolubov equations and derived the density distributions and
the main thermodynamic quantities for atoms trapped in harmonic potentials.
Among other results, they found that repulsive interactions strongly enhance
the thermal depletion of the condensate. They also confirmed the decrease of
the transition temperature and noticed a smoothing of the temperature
dependance of the heat capacity at the phase transition due to collisions.
For ideal gases, we set $g\rightarrow 0$ in the Bogolubov equations~(\ref
{EqBogoliubovEquations}) and recover the results of the previous sections.
The excitation spectrum is simply, 
\begin{equation}
\varepsilon ({\bf r},{\bf p})=\frac{{\bf p}^{2}}{2m}+U({\bf r})-\mu .
\label{EqIdealDispersion}
\end{equation}

\subsection{Attractive Interactions}

The atomic interaction potential decides on the value of the scattering
length $a$: A repulsive potential corresponds to a positive $a$. For a
purely attractive potential that supports no bound state $a$ is negative,
and for an attractive potential that supports bound states $a$ can be either
positive or negative depending on the proximity of the last bound state to
the dissociation limit.

A negative scattering length may, at first, seem desirable, because it rises
the BEC threshold temperature according to equation~(\ref
{EqCondensedFractionInteractions}). However, attractive interactions raise
other problems. The interaction energy of a Bose-Einstein condensate is
given by $4\pi \hbar ^{2}an/m$ and, if the scattering length is negative,
decreases with increasing density $n$. The condensate attempts to lower its
interaction energy by increasing its density until it collapses \cite
{Stoof94} by inelastic two-body spin exchange processes or three-body
recombination processes. This holds for homogeneous condensates. However,
when confined in a trap, the zero-point energy exerts a kinetic pressure
which balances the destabilizing influence of the interactions, so that {\it %
small condensates} are expected to be stable. Calculations for spherical
traps predict \cite{Ruprecht95} $N_{min}\approx 0.575a_{trap}/\left|
a\right| $, where $a_{trap}=\sqrt{\hbar /m\omega _{trap}}$.

\section{Classification of Phase Transitions}

{\em Ehrenfest Classification.} --- Because of the huge variety of phase
transition phenomena, a general classification is not easy. The first
attempt was undertaken in 1933 by Ehrenfest. He proposed the following
classification scheme founded on the thermodynamic properties of the phases.
A phase transition is of $n^{th}$ order if the $n^{th}$ derivative with
temperature of at least one of the state variables, {\em e.g.} chemical
potential $\mu (T)$, internal energy $E(T)$ or entropy $S(T)$, is
discontinuous at the transition point whereas all lower derivatives are
continuous \cite{Fisher64}. As an example: the liquid-gas phase transition
is of first order, because $\partial \mu /\partial T$ is discontinuous.

In order to characterize the Bose-Einstein phase transition, we investigated
the temperature dependence of the heat capacity in Section~2.2 (Table~\ref
{TabCriticalTemperature} and Figure~\ref{FigBoseCondensation}). We saw that,
depending on the type of the confining potential, the occurrence of a
thermodynamic phase transition can be quite remarkable through a
discontinuity of the heat capacity $C(T)=\partial E/\partial T$ at the
critical temperature $T_{c}^{0}$. For a generic power law potential, the
discontinuity depends on the confinement power $\eta $. If the confinement
power is $\eta \leq 2$, for example for a homogeneous system (3D-box), the
discontinuity of $C(T)$ disappears, but $\partial C/\partial T$ remains
discontinuous. However, in any case the chemical potential exhibits an
abrupt change of its temperature dependence at $T_{c}^{0}$, {\em i.e.} $%
\partial \mu /\partial T$ is discontinuous. This aspect is very similar to
liquid-gas phase transitions. Therefore, adopting Ehrenfest's classification
scheme, Bose-Einstein condensation of an ideal gas takes place as a {\it %
first order phase transition} regardless of the shape of the confining
potential.

The Bose-Einstein condensation of a homogeneous system is often called a 
{\it condensation in momentum space}, because the phases do not separate 
\cite{London38}. It is important to note, that the phase separation is not
an essential feature for BEC and does even occur in a homogeneous system
under the influence of gravity \cite{Huang87}, because the dense condensate
has a negative buoyancy inside the normal fraction. In harmonically trapped
gases, the condensed and thermal fractions spatially separate to a large
extent, since the condensate nucleates at the center of the thermal cloud,
where the density is highest. The process must then be considered a
condensation in phase space.

We have seen in the Sections~2.4 and 2.6, that atomic interactions and
finite ensemble sizes smooth out the discontinuities. As a result, in
Ehrenfest's classification scheme first order transitions become{\it \
second or higher order transitions} \cite{Huang87}. The different
classification suggests a fundamental change in the quality of the
transition due to interactions. We should, however, keep in mind that the
reason for the occurrence of the BEC phase transition is the symmetry of the
bosonic single-particle wavefunction, and that forces between the particles
rather tend to blur the quantum statistical nature of the process. In the
case of the strongly interacting liquid $^{4}$He, the heat capacity changes
smoothly and exhibits a $\lambda $-shaped peak at the critical point.

\bigskip

{\em Ginzburg-Landau Classification.} --- Landau emphasized the role of
symmetry in thermodynamics by introducing the notion of the {\it order
parameter} \cite{Landau37}, which he defined as a very general macroscopic
measure for the amount of symmetry in a system. Symmetry considerations play
an important role at phase transitions, and many types of phase transitions
change the symmetry of the system. Typically the phase with the higher
temperature is more symmetric. The order parameter is zero for this phase
and non-zero for the less symmetric phase. An order parameter can also be
defined, if the symmetry apparently does not change as it is the case for
the phase transition from liquid to gas: Both phases are isotropic. Order
parameters may be very different in nature, depending on the specific system
and type of force driving the phase transition, {\em e.g.} they may be $c$%
-numbers, vectors or even many-body quantum fields. In the case of
Bose-condensation, the condensate density is often taken as the order
parameter. When crossing the phase transition from high to low temperatures,
the system can spontaneously adopt a symmetry that its Hamiltonian does not
have, {\em i.e.} the {\it symmetry is broken}, and the order parameter takes
a value different from zero. For example: the transition from liquid to
solid breaks the translational symmetry.

Landau labeled a transition {\it first order}, if there is a discontinuous
change in the order parameter\ and {\it continuous transition} or {\it %
critical phenomenon}, if the order parameter goes smoothly to zero at $T_{c}$%
. Applying the Landau criterion to Bose-gases, we find that BEC is a
second-order phase transition, because the temperature dependence of the
order parameter is continuous at the critical point regardless of whether
the gas interacts or not (Figure~\ref{FigBoseCondensation}). The Landau
classification seems therefore more appropriate for the Bose-Einstein phase
transition.

\chapter{\bf Making and Probing Bose-Einstein Condensates}

\label{SecMakingProbing}The first hint, that Bose-Einstein condensation was
more than just a theoretical fantasy came from London \cite{London38} who
connected the newly found phenomenon of superfluidity in $^{4}$He to BEC.
However, the interpretation of the $\lambda $ point in terms of BEC was not
obvious because strong particle interaction blur the quantumstatistics, and
the thermodynamic quantities exhibit divergences at $T_{c}$ rather than
discontinuities as expected from ideal gas BECs. The occurrence of BEC is
inferred from its influence on the bulk properties of the system. These
uncertainties motivated intense search in other systems. In 1954, Schafroth
pointed out that electron pairs can be viewed as composite Bosons and might
Bose-condense at low temperatures \cite{Schafroth54}. In 1957, Bardeen,
Cooper and Schrieffer developed the microscopic theory of superconductivity 
\cite{Bardeen57}, a phenomenon that has been related to Bose-condensation of
electron- or {\it Cooper-pairs} by other researchers including Blutt,
Schaffrot, Fr\"{o}hlich and Bogolubov.

Motivated by the need to test the concept of composite-particle or
quasi-particle condensation in weakly interacting systems, in 1962, Blatt 
{\em et al.} proposed the investigation of BEC in {\it exciton} gases \cite
{Blatt62}. Excitons are bound electron-hole pairs that can form a weakly
interacting gas in certain nonmetallic crystals. They are interesting
because their small mass permits BEC at high temperatures, their density can
be controlled over a wide range by modifying the optical excitation level,
and they are destructible. Excitons were discovered in 1968 and the first
evidence of Bose-Einstein condensation of biexciton-molecules in CuCl
crystal dates from 1979 \cite{Chase79}. One year later, the influence of
Bose-Einstein statistics on orthoexcitons ($S=1$) was observed by Hulin {\em %
et al.} in CuO$_{2}$, and finally the condensation of paraexcitons ($S=0$)
in $2~\mu $m thick stressed CuO$_{2}$ films by Lin {\em et al.} \cite{Lin93}%
. They achieved BEC at densities above $10^{19}~$cm$^{-3}$ and transition
temperatures close to $T_{c}=50~$K.

Hecht \cite{Hecht59} suggested in 1959, followed by Stwalley and Nosanow 
\cite{Stwalley76} in 1976, that spin polarized atomic hydrogen would be a
suitable candidate for BEC. In 1978 Greytak and Kleppner at the MIT started
intensive efforts to form BECs in dilute hydrogen gases. In the nineties,
advances in cooling atoms by laser light led to really low temperatures, and
the invention of traps for neutral atoms allowed their confinement and the
compression of their density. This initiated efforts to try to realize BEC
in alkalis, which have an electronic level scheme that lends itself to
optical cooling. Later, it turned out that the phase space density in
optical traps is limited by optical rescattering effects. As a solution to
this problem people started to trap the atoms by their magnetic dipole
moment and to replace optical cooling by evaporative cooling. This was the
final step towards BEC in alkali gases. The hydrogen experiment, that
initially stimulated the alkali experiments, now taking advantage from their
success has led to BEC, as well.

Today, hybrid optical plus evaporative cooling in alkali-metals has
increased the phase space density by a factor of $10^{15}$. BEC has been
achieved in rubidium, lithium, sodium and also in hydrogen. In the present
chapter, we will outline the experimental progress that led to BEC in the
alkalis by discussing the various techniques employed.

\section{Techniques for Cooling and Trapping}

Light interacts in two different ways with the mechanical degrees of freedom
of atoms \cite{Gordon80}. One way is through the force 
\begin{equation}
{\bf F}=-\nabla _{r}\left( {\bf d\cdot E}\right) ,  \label{EqLightForce}
\end{equation}
where ${\bf d}$ denotes the atomic dipole moment and ${\bf E}$ the electric
field of the light. The {\it dipole force} arises from the interaction of
the light with the dipole, which the light induces in the atom \cite
{Dalibard85}. It can be understood as stimulated scattering of photons
between the modes of the light field by the atoms. The force acts in the
direction of the intensity gradient. It is a conservative spatially varying
force and therefore interesting for realizing trapping potentials for atoms 
\cite{Chu86}. A light field with intensity $I$, detuned from a resonance $%
\omega _{0}$ by $\Delta =\omega -\omega _{0}$, gives rise to the dipole
force 
\begin{equation}
{\bf F}=d~\left( \nabla _{r}E_{0}\right) ~\frac{2\Omega \Delta }{\Gamma ^{2}}%
~\frac{\sigma (\Delta )}{\sigma _{0}},  \label{EqDipoleForce}
\end{equation}
where the absorption profile is described by the optical cross-section $%
\sigma $, 
\begin{equation}
\sigma (\Delta )=\frac{\sigma _{0}\Gamma ^{2}}{4\Delta ^{2}+2\Omega
^{2}+\Gamma ^{2}},  \label{EqDetunedOptCrosssection}
\end{equation}
and $\sigma _{0}=3\lambda ^{2}/2\pi $ is the resonant optical cross-section
on a transition whose Clebsch-Gordon factor is equal to one. Furthermore,
the Rabi frequency is introduced by $\Omega \equiv {\bf d\cdot E}/\hbar =%
\sqrt{\sigma _{0}\Gamma I/\hbar \omega }$. The force~(\ref{EqDipoleForce})
can be expressed as the gradient of the conservative trapping potential 
\begin{equation}
U=\frac{\hbar \Delta }{2}~\ln \left( 1+\frac{2\Omega ^{2}}{4\Delta
^{2}+\Gamma ^{2}}\right) .  \label{EqDipolePotential}
\end{equation}

The second force is called {\it radiation pressure} and comes from
spontaneous scattering of photons at an atomic resonance. It was first
observed as early as 1933 by Frisch \cite{Frisch33}. The absorption of a
photon from the light field (wavevector ${\bf k}$) imparts a recoil momentum 
${\bf p}=\hbar {\bf k}$\ to the atom. The subsequent spontaneous emission is
isotropic in the time-average, so that in the average over many emission
processes no momentum is transferred to the atom. The radiation pressure is
dissipative and has been used in optical cooling schemes \cite
{Kastler50,Haensch75}.

The radiation pressure is velocity dependent. The velocity dependence comes
from the Doppler effect, which links the external degrees of freedom of the
atom (its motion) to the internal ones (detuning between light and atomic
resonance frequency): The frequency $\omega $ of a light field is increased
or decreased in the inertial system of the atom, {\em i.e.} relative to the
atomic resonance frequency, depending on whether the atom moves towards or
away from the light propagation direction. In a red-detuned light field
photons are only absorbed by counterpropagating atoms, while copropagating
atoms are out of resonance. One can therefore use the radiation pressure to
manipulate the velocity of the atoms and if need be decelerate them. Often
the reduction in kinetic energy is accompanied by a reduction in kinetic
energy spread. Those cases are called {\it optical cooling}. Radiation
pressure has been used to decelerate atomic beams in Zeeman-slowers \cite
{Phillips82} and chirped-frequency slowers \cite{Ertmer84}.

The radiation pressure force of a light field on a two-level atom (velocity $%
{\bf v}$, linewidth $\Gamma $)\ averaged over many absorption-spontaneous
emission cycles, is \cite{Stenholm86} 
\begin{equation}
{\bf F}=\hbar {\bf k}~\frac{I}{\hbar \omega }~\sigma (\Delta -{\bf k\cdot v}%
).
\end{equation}
The cooling force is proportional to the laser intensity, as long as the
transition is not saturated, $I/I_{s}=2\Omega ^{2}/\Gamma ^{2}<1$. The
smallest temperature that two-level atoms can attain by Doppler cooling is
limited by diffusion of the momentum in phase space due to the stochastic
process of spontaneous emission. Cooling and diffusion heating are leveled
when the atom has the kinetic energy ${\bf p}^{2}/2m=\hbar \Gamma /2$.

\subsection{Magneto-Optical Traps}

A frequently used optical cooling scheme for low temperatures consists of
irradiating the atoms with three orthogonal pairs of counterpropagating
red-detuned laser beams. Radiation pressure slows down the atoms without
confining them, and the atoms move like in a viscous medium, the so called 
{\it optical molasses} \cite{Chu85}. Surprisingly, the temperatures measured
in optical molasses were well below the Doppler limit. The responsible
cooling mechanisms have been identified to be based on optical pumping
between the Zeeman sublevels induced by polarization gradients \cite
{Dalibard89,Ungar89}. These polarization gradients are also responsible for
the low temperatures found in {\it Magneto-Optical Traps} (MOT). The MOT was
invented by Dalibard and first realized by Raab {\em et al.} \cite{Raab87}
and is presently the most commonly used trap for atoms. It consists of a
magnetic field gradient produced by a quadrupole field and three pairs of
circularly polarized, counterpropagating optical beams, detuned red from an
atomic transition and intercepting at right angles in the position of the
magnetic field zero. The MOT exploits the position-dependent Zeeman shifts
of the electronic levels when the atom moves in the radially increasing
magnetic field. The use of circularly polarized, slightly red-detuned light, 
$\Delta \approx \Gamma $, results in a spatially dependent transition
probability whose net effect aims at producing a restoring force that pushes
the atoms towards the origin. The force exerted by one of the laser beams
(wavevector ${\bf k}=k{\bf \hat{e}}_{z}$) acts primarily on atoms with
velocity ${\bf v}=v{\bf \hat{e}}_{z}$, 
\begin{equation}
F_{z}=\hbar k~\frac{I}{\hbar \omega }~\sigma (\Delta -kv_{z}{\bf -}\hbar
^{-1}\mu ~z~{\bf \partial }_{z}B).  \label{EqMOTForce}
\end{equation}
Analogous expressions hold for all other beams. For small displacements and
velocities the total force can be linearized, 
\begin{equation}
m\ddot{z}=F_{z}+F_{-z}=-\alpha \dot{z}-\varkappa z,  \label{EqLinearizedMOT}
\end{equation}
where $\alpha $ and $\varkappa $ denote the friction and the spring constant
respectively, and describe a damped motion inside a harmonic potential, 
\begin{equation}
\varkappa =\frac{16\hbar k\Gamma ~\Omega ^{2}\Delta ~\left( {\bf \partial }%
_{z}\omega _{Zeeman}\right) }{\left( 4\Delta ^{2}+12\Omega ^{2}+\Gamma
^{2}\right) ^{2}}\text{ \ \ \ \ \ , \ \ \ \ \ }\alpha =\varkappa \frac{k}{%
{\bf \partial }_{z}\omega _{Zeeman}}.  \label{EqMOTSpringFriction}
\end{equation}

The dissipative character of the MOT makes it a very powerful and versatile
tool: At the same time, the MOT traps up to $10^{9}$ atoms from the
environment, cools them down to very low temperatures and then confines them
in a potential at densities approaching $10^{10}~$cm$^{-3}$. However, at
such high density the atomic cloud gets so optically thick, that photons are
scattered several times before they find their way out. This phenomenon is
termed {\it radiation trapping}. The atomic repulsion induced by the photons
at each absorption and emission blows up the cloud size \cite{Sesko91}. To
overcome this radiation trapping, Ketterle {\em et al.} \cite{Ketterle93}
proposed to keep the colder atoms that are close to the origin of the trap
in a dark electronic state in order to prevent them from scattering light.
His scheme, presently known as dark MOT or {\it dark Spontaneous Force
Optical Trap} (dark SPOT), takes advantage of the large hyperfine splitting
of the $^{2}S_{1/2}$ ground state, which allows to selectively excite and
pump both hyperfine substates.

Typically, dark MOTs capture up to $5\cdot 10^{9}$ atoms at densities
approaching $n=10^{11}~$cm$^{-3}$ and temperatures below $100$~$\mu $K. The
phase space density for such clouds is $\rho =n\lambda _{dB}^{3}<10^{-6}$,
which is still more than six orders of magnitudes away from BEC. Since it
seems impossible to reach BEC in MOTs, alternative trapping schemes have
been developed, the most promising of which are {\it dipole force traps}
using laser light and {\it magnetic traps} operating without light beams at
all.

\subsection{Far-off Resonance Dipole Traps}

For large detunings, the potential depth (or light-shift) estimated from
equation~(\ref{EqDipolePotential}) reads: 
\begin{equation}
U=\frac{\hbar \Omega ^{2}}{4\Delta },  \label{EqFORTPotential}
\end{equation}
while the spontaneous light scattering rate $\gamma _{s}$ is proportional to 
$I/\Delta ^{2}$. Heating of the atoms due to spontaneous scattering of
photons can therefore be avoided by using intense, far-detuned laser beams.
Such dipole force trapping potentials are called {\it Far-off Resonance Traps%
} (FORT) and can be engineered with various geometries. For example, one-,
two- or three-dimensional configurations of red-detuned standing light waves
give rise to arrays of potential valleys in the intensity antinodes called 
{\it optical lattices} \cite{Hemmerich93,Grynberg93}. The simplest optical
dipole trap (and the first that has been realized \cite{Chu86}) consists of
a tightly focussed red-detuned laser beam, that confines the atoms at its
waist. Atoms trapped in a crossed dipole beam trap have even been
evaporatively cooled \cite{Adams95}. Alternatively, one can use blue-detuned
FORTs, where the atoms are confined in local minima of the intensity profile
and suffer less from spontaneous light scattering \cite{Davidson95}.
Finally, dipole beams can be used in conjunction with other trapping
techniques (a blue-detuned FORT was used to repel atoms from the center of a
magnetic quadrupole trap \cite{Davis95b}) (Section~3.2.1), to manipulate
Bose-Einstein condensates \cite{Andrews97b} (Section~4.3.1) and even to trap
them \cite{Stamper-Kurn98} (Section~4.2.2).

The optical approach offers the advantage of high spatial definition and
temporal control, {\em e.g.} the trap depth and location can easily be
manipulated and modulated. The trap can be turned on and off very fast
compared to magnetic traps and offers the advantage of being insensitive to
magnetic fields, {\em i.e.} all magnetic substates can be trapped.
Furthermore, optical subrecoil cooling schemes that do not work for
magnetically trapped atoms, may be implemented in optical traps. {\it %
Velocity Selective Coherent Population Trapping} (VSCPT) led to extremely
low temperatures in the nano-Kelvin regime \cite{Lawall95} and {\it Raman
cooling} led to very high phase-space densities \cite{Lee96}. Several groups
attempt to attain BEC using all-optical methods, and it seems possible to
cross the phase transition in the near future \cite{Depue99}.

\subsection{Magnetic Traps}

Magnetic traps hold the atoms by their magnetic dipole moment ${\bf \mu =}%
\mu _{B}g_{F}{\bf F}$, where $\mu _{B}$ denotes the Bohr magneton and $g_{F}$
is the Land\'{e} $g$-factor for the total atomic spin ${\bf F}={\bf J}+{\bf I%
}=\left( {\bf L}+{\bf S}\right) +{\bf I}$. The symbol ${\bf I}$ denotes the
nuclear spin, ${\bf S}$ is the electron spin, and ${\bf L}$ is the electron
angular momentum. The Land\'{e} factor can be calculated from 
\begin{eqnarray}
g_{F} &\approx &g_{J}\frac{F(F+1)+J(J+1)-I(I+1)}{2F(F+1)},
\label{EqLandeFactor} \\
g_{J} &=&1+\frac{J(J+1)+S(S+1)-L(L+1)}{2J(J+1)}.  \nonumber
\end{eqnarray}
If the spin adiabatically follows the magnetic field, the force that the
magnetic field gradient exerts on an atom is 
\begin{eqnarray}
{\bf F}_{magn} &=&-\bigtriangledown _{{\bf r}}U,  \label{EqMagneticPotential}
\\
U &=&-{\bf \mu \cdot B}=-g_{F}m_{F}\mu _{B}B.  \nonumber
\end{eqnarray}
Thus, depending on the atom's magnetic sublevel $m_{F}$ (positive or
negative), it is attracted towards or repulsed from a magnetic field
extremum. Unfortunately, static magnetic field maxima are not possible, so
that only {\it low-field seekers} can be trapped in magnetic field minima.
These have the disadvantage of not being the energetically lowest state. The
spontaneous decay rate out of those metastable Zeeman states is only $\sim
10^{-10}~$s$^{-1}$, but spin-changing collisions can induce decay and trap
losses. Even in the lowest atomic state the trapped alkali gases are
metastable: In {\it three-body recombination} processes two atoms can form
an energetically more favorable dimer if a third atom is nearby to take away
the excess energy. Since the event rate for three-body collisions scales
with the third power of the density, this process becomes dominant at high
densities \cite{Soeding99}, {\em e.g.} for Bose-Einstein condensates.

Dynamic traps based on time-dependent magnetic fields have been proposed for
both low- and strong-field seekers by Lovelace {\em et al.} \cite{Lovelace85}%
. The first demonstration of a static magnetic trap for neutral atoms \cite
{Migdall85} used an anti-Helmholtz coil configuration to produce an axially
symmetric quadrupole field. The magnetic field geometry of a spherical
quadrupole trap has a linear spatial dependence and provides the tightest
confinement for atoms.

The assumption that the spin adiabatically follows the magnetic field,
unfortunately does not hold in all cases. Especially, near zero magnetic
field the Larmor frequency $g_{F}m_{F}\mu _{B}B/\hbar $ may get smaller than
the rate of change of magnetic field amplitude experienced by the atom
through its motion with velocity $v$, {\em i.e.} $v\partial _{r}B/B$. At
this location, missing a quantization axis, the atoms are free to reorient
themselves arbitrarily and undergo so called {\it Majorana spin flip
transitions} to untrapped magnetic sublevels. This is exactly the case for
a\ quadrupole trap. In this trap, atoms are lost due to Majorana spin flips
as they pass near the trap center due to the sudden change of the magnetic
field. The colder the atoms are, the more time they spend near the center
''hole'', and the situation is even worse.

The {\it Time-Orbiting-Potential} (TOP) trap was designed to suppress the
Majorana loss by adding a rotating transverse bias field $B_{bias}$ to the
quadrupole trap (gradient $\partial _{r}B_{quad}$) \cite{Petrich95}. The
bias field shifts the ''hole'' away from the region where atoms are trapped
to a distance $r_{d}=B_{bias}/\partial _{r}B_{quad}$ from the center. The
hole rotates in a so called {\it death circle} around the harmonic trap,
fast enough for the atoms to only respond to the time-averaged potential. An
alternative approach are {\it Ioffe-Pritchard} (IP) type magnetic field
configurations. In the magnetic field minimum the field amplitude does not
vanish, so that Majorana spin-flip transitions do not occur. The original IP
trap consists of a combination between a quadrupole waveguide and a magnetic
bottle \cite{Pritchard83,Bergeman87}. Variations of IP traps include the
cloverleaf trap \cite{Mewes96}, the baseball trap \cite{Myatt97,Monroe93},
the Ioffe bar trap \cite{Ernst97}, and the Quadrupole Ioffe Configuration
(QUIC) trap \cite{Esslinger98}.

Laser cooling in magnetic traps has the advantage over evaporative cooling
(next section), that the heat dissipation does not rely on the irreversible
removal of hot atoms and does not require high densities and collision
rates. On the other hand, Doppler-cooling in magnetic traps is possible \cite
{Helmerson92}, but it meets its natural limit at temperatures in the
milli-Kelvin range. Raman-cooling and VSCPT are not compatible with magnetic
traps. However, other cooling schemes are possible or will probably be
developed in the future. One example is the idea of {\it gravitational
Sisyphus cooling} tested by Newbury {\em et al.} \cite{Newbury95}.

\subsection{Evaporative Cooling}

As we have seen earlier, laser cooling becomes ineffective when the density
of the gas is high. We need another dissipative mechanism to cool atoms
confined in magnetic traps. A new method called {\it evaporative cooling}
has been proposed by Hess \cite{Hess86} for spin-polarized hydrogen (H$%
\uparrow $) and has been observed by Masuhara {\em et al.} \cite{Masuhara88}%
. It was later utilized for the alkali-metals \cite
{Adams95,Petrich95,Davis95}. A detailed review has been published by
Ketterle and van Druten \cite{Ketterle96}.

Evaporation always occurs when energetic particles leave a system with
finite binding energy, taking away more than their share in mean energy per
particle. We consider here the case of a magnetic trapping potential with a
finite extension, {\em i.e.} the potential has an edge or a spout through
which hot atoms with enough kinetic energy to reach that region can leave
the trap.\ In the ideal case, this will lead to a complete truncation of the
hot tail of the equilibrium Maxwell-Boltzmann velocity distribution. If the
remaining system finds back to thermal equilibrium, it will do so at a lower
temperature. The redistribution of kinetic energy among the atoms that
ultimately leads to rethermalization, happens through elastic collisions. It
takes about three collisions per atom to rethermalize a cloud \cite
{Monroe93,Wu96}. The rate for elastic collisions between trapped atoms is 
\begin{equation}
\gamma _{coll}=n_{0}\sigma _{el}\bar{v}\sqrt{2}\sim \rho ^{3}N^{2/3},
\label{EqCollisionRate}
\end{equation}
where $n_{0}$ is the peak density, $\sigma _{el}$ is the elastic scattering
collision cross-section, and $\bar{v}$ is the average relative velocity
between two atoms \cite{Ketterle96}. Obviously, the evaporation process
slows down when the cloud gets colder, unless the potential edge is lowered
so that the hottest atoms of the colder cloud are evaporated. By
continuously lowering the potential edge while the atomic cloud is
rethermalizing (this procedure is called {\it forced evaporation}), very low
temperatures in the nano-Kelvin regime can be reached and the phase space
density can be increased by six orders of magnitude up to the threshold of
Bose-Einstein condensation. This is of course only possible by sacrificing
many hot atoms. Even in a well optimized evaporation ramp, only $1\%$ may
reach the condensate stage after about $500$~collisions per atom.

Two aspects should be pointed out concerning the optimization of the
evaporation path, {\em i.e.} the down-ramping of the truncation edge. The
first aspect is, that inelastic collisions with atoms from the background
vapour limit the trap lifetime. Therefore, the evaporation needs to be fast,
which requires high elastic collision rates or a good vacuum. There is a
trade-off between an efficiently slow evaporative cooling on one hand and
avoiding the accumulation of trap losses on the other hand. The second
aspect is, that the dimensionality of the evaporation edge determines the
efficiency of evaporative cooling. The first demonstration of evaporation in
H$\uparrow $ ejected hot atoms over a saddle point. The saddle point
constitutes a small region of space away from the trap origin, and only
atoms that have enough kinetic energy in one direction $E_{z}>U_{edge}$ can
leave the trap. The evaporation is then called 1-dimensional (1D). It is
true, that ergodic mixing due to anharmonicities in the trapping potential
will sooner or later drive all the atoms through this region, but this
effect becomes less pronounced when the cloud cools down, since the atoms
settle down in the harmonic (and therefore separable) region at the bottom
of the trapping potential. This fact has inhibited efficient evaporation in H%
$\uparrow $ below $120~\mu $K \cite{Fried98}.

A second evaporation technique has been demonstrated in TOP traps \cite
{Petrich95}. As explained in Section~3.1.3, it is a feature of TOP traps to
be surrounded by a {\it death-circle} that ejects the atoms passing through.
This death-circle can act as a 2D evaporation surface if its radius $r_{d}$
is large enough. Under the influence of gravity the dimensionality is even
reduced to 1D \cite{Ketterle96}. For small atomic clouds, typically less
than $10^{8}$ atoms, a TOP trap only needs a moderate death circle radius.
But a small death circle is an obstacle for the creation of Bose condensates
with large numbers of atoms. For efficient evaporation, it is necessary for
the death circle radius $r_{d}$ to be greater than $3$ to $4$ times the {\em %
rms} radius $r_{0}$ of the trapped atom cloud. On the other hand, it is also
necessary that the elastic collision rate be very large. This is usually
achieved in part with an adiabatic compression of the magnetic trapping
potential, in which the trap frequency increases according to $\omega
_{r}\sim \partial _{r}B_{quad}/B_{bias}^{1/2}$. Thus, we wish to increase $%
\partial _{r}B_{quad}$ and decrease $B_{bias}$. However, this also shrinks
the ratio $r_{d}/r_{0}\sim B_{bias}^{3/4}/B_{quad}{}^{1/2}$. For large atom
numbers the initial radius $r_{0}$ tends to be large and the ratio $%
r_{d}/r_{0}$ small unless the magnetic field strengths are very large. Thus,
only modest compression can be achieved before the death circle loss sets
in, and the elastic collision rate must already be large enough for
efficient evaporation at this point. This means that for large clouds it is
not possible to depend on a large compression of the density, and that the
initial densities in the trap must not be too far from those required for
efficient evaporation. This is achieved by efficient optical cooling and
compression, and efficient transfer of the atoms into the TOP \cite{Han98}.

The most successful evaporation technique implemented so far is based on
radiative coupling of trapped and untrapped states in an energy-selective
way and is termed {\it radiofrequency} ({\em rf}) evaporation. It originates
from an idea proposed by Pritchard and coworkers \cite{Pritchard89}, who
already had some experience with rf-spectroscopy on magnetically trapped
neutral atoms \cite{Martin88}. The spatially dependent Zeeman-splitting is
an intrinsic feature of magnetic traps. Irradiation of a radiofrequency wave
with a given frequency couples the trapped and untrapped Zeeman-substates at
a well-defined distance from the trap origin. This gives rise to a 3D
evaporation surface, where crossing atoms can undergo Landau-Zener
transitions and be expelled from the trap. The technical advantages of this
scheme are substantial: The magnetic trapping potential does not need to be
manipulated, and the potential edge can easily be controlled by the
radiofrequency. If the evaporation is forced by continuously lowering the
radiofrequency and if the evaporation path is suitably chosen, the density
will increase and therefore the collision rate. The rethermalization will
speed up and initiate run-away evaporation. Rf-evaporation was first
demonstrated by Ketterle and coworkers \cite{Davis95}.

Another cooling mechanism based on collisions is {\it sympathetic cooling}.
The technique originally used in ion traps was later applied to neutral
atoms confined in magnetic traps. The idea consists of bringing the gas into
thermal contact with a cold buffer gas. In some cases, the buffer gas can be
optically or evaporatively cooled. Buffer gas loading of conservative traps, 
{\em e.g.} magnetic traps \cite{Doyle95}, is an alternative to the transfer
from MOTs. Sympathetic cooling has been used in magnetic traps to create
double condensates \cite{Myatt97} (Section~4.2.1) and to cool fermions down
to the regime of quantum degeneracy \cite{DeMarco99} (Section~3.2.3).

\section{Realization of Bose-Einstein Condensation}

In early 1995, several research groups were very close to the long pursued
goal. Several improvements of the optical trap led to large atom numbers
transferred to the quadrupole magnetic trap, evaporative cooling had been
observed. The first observation of Bose-Einstein condensation in a dilute
gas was made at JILA \cite{AndersonMH95} in rubidium. It was followed by
Rice \cite{Bradley95} in lithium and MIT in sodium \cite{Davis95b}, and
later by Han {\em et al.} \cite{Han98} and Hau {\em et al.} \cite{Hau98},
and has now been confirmed by more than twenty groups worldwide. This
section will briefly relate the first three experiments. A remarkable
achievement is the condensation of atomic hydrogen \cite{Fried98}. Other
candidates for BEC are thoroughly investigated, like cesium \cite
{Depue99,Guery-Odelin98,Kokkelmans98}, potassium \cite{Prevedelli99}, helium 
\cite{Shlyapnikov94} and neon \cite{Beijerinck00}.

\subsection{BEC in Alkalis}

The JILA experiment led by Cornell and Wieman worked with a Rubidium vapour
cell dark MOT. Operating at $10^{-11}~$torr it took $300~$s to load $10^{7}$
atoms. In order to optimize the loading efficiency into a magnetic trap, the
MOT gradient and laser frequency detuning were adjusted and a short pulse of
circularly polarized laser light pumped the atoms in presence of a weak
homogeneous magnetic field defining the quantization axis into the fully
stretched $F=2,m_{F}=2$ Zeeman state. Then, all laser light was switched off
and the quadrupole TOP trap quickly switched on. The effective time-averaged
potential $\mu _{B}B$ was pancake-shaped with secular frequencies close to $%
\omega _{z}=\sqrt{8}\omega _{r}=2\pi \times 120~$Hz. The TOP trap rotating
frequency $\omega _{TOP}=2\pi \times 7.5~$kHz, was chosen to satisfy $\hbar
\omega _{z}\ll \hbar \omega _{TOP}\ll \mu _{B}B$. The potential was
adiabatically compressed by ramping up the quadrupole field gradient and
then reducing the rotating bias field amplitude. This enhanced the collision
rate to about three per second, which was (in view of the $70~$s magnetic
trap lifetime) enough to initiate run-away evaporative cooling. At this
stage, the cloud consisted of $4\times 10^{6}$ atoms at $90~\mu $K
temperature and $2\times 10^{10}~$cm$^{-3}$ density. Death circle induced
Majorana spin flips and {\em rf}-induced Landau-Zener transitions both acted
as an edge to the potential and contributed to evaporative cooling. The {\em %
rf}-scalpel was ramped down thus skimming off the hot atoms from the
continuously rethermalizing cloud. The phase transition was crossed with the 
{\em rf}-frequency at $3.6~$MHz. With a $5~$G rotating bias field amplitude,
this {\em rf}-frequency made the trapping potential about $800~$kHz deep.
Finally, after an equilibration time of two seconds, the released and
ballistically expanded condensate was probed after $40~$ms time-of-flight
with a circularly polarized laser beam tuned to a cycling transition
(Section~3.3.1). Figure~\ref{FigAbsorptionPictures} shows typical absorption
pictures taken with that method. The signatures of BEC are 1) a bimodal
density distribution with a sharp increase in peak density in the center of
the cloud, 2) a critical dependence on the final {\em rf}-ramp frequency,
and 3) an anisotropic shape of the central condensed feature. As discussed
later (Section~4.1), only the BEC wavepacket remembers the shape of the trap
that confined it before being released.

Of course, the condensed atom number and quality of the results in general
have been largely improved in subsequent experiments at JILA and in other
groups. {\em E.g.} the loading of the optical trap from the background
vapour has been replaced by a Zeeman-slower, a double MOT or an axicon trap
configuration which permitted faster loading rates at a lower vapour
pressure. Other beam and magnetic trap configurations have been used, and
other imaging systems have been developed. Today, atom numbers higher than $%
10^{6}$ can routinely be condensed in TOP traps. However, the essential
features of the method described above has been the same for all alkali BEC
experiments.

The Rice group led by Hulet opted for lithium \cite{Bradley95}. Lithium has
a slightly negative scattering length $a=-27.3a_{B}$, for which only small
condensates are expected to be stable (Section~2.6.2). The Rice group used a
magnetic trap configuration made of permanent magnets in an arrangement that
produces a harmonic potential with a magnetic field minimum offset by $823~$%
G. This has the advantage of a simple experimental setup, but at the price
of flexibility. The magnetic trap is directly loaded from a Zeeman-slower. A 
$10^{-12}~$torr background pressure corresponding to $10~$mn magnetic trap
lifetime allows $1000$ collisions per lifetime with an elastic scattering
cross-section of $\sigma =5\times 10^{-13}~$cm$^{2}$. After typically $5~$mn
evaporation, a sample of $10^{5}$ atoms reaches temperatures close to $300$%
~nK. Because the magnetic field cannot be switched off, {\it in-situ}
imaging of the dense condensed cloud is necessary. Near-resonant imaging of
the optically thick cloud introduces lensing effects, which make the
interpretation of the recorded images difficult \cite{Bradley97} and first
led to erroneous claims about the numbers of condensed atoms \cite{Bradley95}%
. Later, the use of phase-contrast imaging which exploits the birefringence
of the atoms in the strong magnetic field offset{\em \ }(Section~3.3.2)
resolved this problem and resulted in the observation of limited atom
numbers of about $1400$ atoms. Subsequent experiments monitored the dynamics
of collective collapse of lithium BECs as soon as a critical atomnumber is
surpassed \cite{Sackett99}.

The MIT group led by Ketterle used sodium. Instead of using a vapour cell,
they loaded their MOT from a Zeeman slower. Apart from this, their approach
was similar to the JILA experiment, except for the method used for
suppressing the Majorana spin flips. In their first experiment they used a
far-off resonant optical beam, $\lambda =514~$nm and $I=2\times 10^{5}~$W/cm$%
^{2}$, causing $7~$MHz light shift to repel the atoms from the center of
their quadrupole trap and thus plug its hole \cite{Davis95b}. They condensed 
$N=5\times 10^{5}$ atoms at $T=2~\mu $K and densities around $n=4\times
10^{14}~$cm$^{-3}$. In a modified setup, the MIT group replaced the
quadrupole trap by a cloverleaf trap, which has a Ioffe-Pritchard type
potential \cite{Mewes96} and is free of the spin-flip problem
(Section~3.1.3). In this trap, Mewes {\em et al.} produced condensates of $%
N=5\times 10^{6}$ atoms. In contrast to the pancake-shaped fixed aspect
ratio of TOP traps, cloverleaf traps are cigar-shaped and have a large
tunable aspect ratio, which proved useful for a range of subsequent
experiments.

\subsection{BEC in Hydrogen}

Hydrogen is a very interesting element to study BEC, because its small
scattering length, $a_{s}\approx 1.23a_{B}$, makes it an almost ideal gas.
This has the advantage of small three-body losses allowing dense
condensates. The small mass results in a high critical temperature. Its
simple level structure allows precise calculations of the interaction
potentials based on first principles, which may, in this way, be tested by
experiments. The narrow $1.3~$Hz wide $1S-2S$ (Lyman-$\alpha $) transition
at $121.56~$nm might be a good candidate for frequency standards in atomic
fountains \cite{Niering00}. Narrow lasers with $0.6~$Hz emission bandwidth,
that would be able to exploit this narrow reference line have already been
constructed in other wavelength regions \cite{Young99}.

In 1978 Greytak and Kleppner started intensive efforts to form BECs in
dilute hydrogen gases. Twenty years later they finally reached their goal 
\cite{Fried98}. The sequences of this difficult experiment recapitulate the
advances in the historical development towards higher densities and lower
temperatures. In this experiment, large numbers of hydrogen molecules were
dissociated in a cryogenic discharge, spinpolarized and trapped in a
magnetic Ioffe-Pritchard potential and confined in a cell with $120~$mK cold
liquid $^{4}$He coated walls. The atoms thermalized by collisions with the
walls until they settled down in the $500~$mK deep trap and thermally
disconnected from the wall. Because cryogenic cooling is limited to $40~$mK,
a new technique had to be invented, namely evaporative cooling \cite{Hess86}
over a saddle point of the potential. The evaporation could be forced by
lowering the trapping potential down to $1.1~$mK. However, the evaporation
surface is one-dimensional (Section~3.1.4) and becomes increasingly
inefficient at temperatures below $120~\mu $K, because rethermalizing
collisions are rare due to the small scattering length. The problem was
solved by applying the technique of radiofrequency evaporation, which had
been developed for the alkali atoms and yields three-dimensional evaporation
surfaces. Unfortunately, radiofrequency evaporation requires low-field
seeking atoms which have the disadvantage of being in excited spin states
and therefore metastable with respect to dipolar spinflips.

The phase transition was finally crossed at $50~\mu $K temperature and
atomic densities of about $5\times 10^{15}~$cm$^{-3}$. The number of
condensed atoms was $10^{9}$, corresponding to a limited condensed fraction
of below $10\%$. At higher condensed fractions and densities the losses due
to dipolar heating overrule the gain from evaporative cooling of the thermal
cloud \cite{Hijmans93}, which is rather poor because of the small scattering
length. The needle-shaped condensed cloud had $15~\mu $m radial and $5~$mm
axial extension.

The atomic sample was probed by ultra-high resolution two-photon Raman
spectroscopy on the forbidden $1S-2S$ transition. The fluorescence photons
could be observed by Stark-quenching the $2S$ level via the rapidly decaying 
$2P$ level. The spectrum shown in Figure~\ref{FigHydrogenSpectrum} consists
of a Doppler-sensitive and a Doppler-free part which result from photons
being absorbed from the same laser beam\ or from counterpropagating beams,
respectively. The Doppler-free narrow peak is red-shifted by cold collisions
by an amount $\Delta \nu _{1S-2S}=-3.8~$Hz$~$cm$^{3}\times n$, and its width
of a few kilo-Hertz is mainly due to transit time broadening of the atoms
flying through the small $50~\mu $m waist of the laser beam. The density
dependence of the pressure shift is useful for studying the density
distribution of the cloud. The Doppler-sensitive broad peak is blue-shifted
by $6.7~$MHz, {\em i.e.} twice the recoil-energy. It represents an {\it %
in-situ} measurement of the atomic momentum distribution and is in this
respect complementary to imaging techniques that either measure the spatial
distribution in-situ (Section~3.3.2) or the momentum distribution after a 
{\it time-of-flight} (Section~3.3.1).

The condensate leaves its imprints on the two-photon spectrum. The
Doppler-free peak has a shoulder, that is red-shifted by $-0.9~$MHz and
originates from a region in the trap where the density is significantly
higher. This is a signature of BEC. The intrinsic width of the shoulder is
determined by the position-momentum uncertainty, but is overruled by the
broadening due to the very inhomogeneous density distribution of the BEC.
The broad Doppler-sensitive peak develops a narrow structure that is
qualitatively similar to the one of the Doppler-free peak.

\subsection{Fermions}

Atoms are either bosons or fermions depending on whether their spin is
integer or half-integer. At high phase-space densities the atoms have to
sort out how they will organize their coexistence. Bosons encourage each
other to occupy the same phase-space cell, in contrast to the reluctant
fermions which follow the Pauli exclusion principle. The different behavior
is described by different quantum statistics which settle how the
phase-space ({\em i.e.} the available energy levels) has to be populated by
the atoms. The Bose-Einstein distribution holds for bosons, the Fermi-Dirac
distribution for fermions and both asymptotically approach the Boltzmann
distribution at high temperatures. We have seen that bosons undergo a
phase-transition and all condense in the ground state as the temperature is
lowered. On the other hand, the fermions must organize their phase-space so
that their energy levels are organized like a ladder. This has observable
consequences at low temperatures. (1) The internal energy of a fermionic gas
is a little higher than expected according to classical statistics, because
the atoms drive each other out of the lowest energy levels. (2) In a
confining trap, the cloud resists compression, its spatial density
distribution is shaped by atoms pushing each other out of the trap center,
where the potential energy is lower. (3) The collision rate is strongly
suppressed, because the density cannot increase beyond a certain limit. This
last point makes ultracold Fermi-gases interesting for frequency standard
applications, because of the absence of pressure shifts.

It is, of course, very interesting to experimentally confirm the impact of
quantum statistics on a cold Fermi-gas. There are various fermions among the
alkali isotopes (see Table~\ref{TabBoseSpecies}). DeMarco and Jin \cite
{DeMarco99} carried out experiments on potassium. They loaded a magnetic
cloverleaf trap from a MOT with $150~\mu $K hot atoms and initiated
evaporative cooling. The problem with evaporative cooling of fermions is
that at ultralow temperatures $s$-wave collisions between the spin-polarized
identical fermions dominate, and those are forbidden by Pauli's exclusion
law. Because $s$-wave collisions between distinguishable particles are
possible, DeMarco and Jin could circumvent this problem by dividing the
potassium cloud into two different internal energy states and simultaneously
cooling them. The two energy states used were the $|F,m_{F}\rangle
=|9/2,7/2\rangle $ and the $|9/2,9/2\rangle $ Zeeman sublevels of the ground
state. Inside the magnetic trapping field the Zeeman states are split. In
order to maintain a $50\%$ ratio during forced evaporation, the evaporation
edge of both Zeeman states had to be controlled and ramped down separately
and simultaneously by two microwave frequencies tuned between each of the
Zeeman states and an untrapped level of the $F=7/2$ hyperfine state.

DeMarco and Jin cooled a two-component Fermi-gas of$\ 7\times 10^{5}$
potassium atoms down to $300~$nK, which corresponded to $60\%$ of the atoms
populating energy levels below the Fermi-temperature $T_{F}=\hbar
k_{B}^{-1}(6\omega _{z}\omega _{r}^{2}N)^{1/3}$. Then they selectively
removed the $|7/2\rangle $ atoms, took time-of-flight absorption pictures
and analyzed the momentum distribution. The onset of the Fermi-statistical
regime was observed as a barrier to evaporative cooling at temperatures
lower than $0.5~T_{F}$ and also left its imprint on the momentum
distribution. The shape of the distribution deviated from the Gaussian
expected for classical gases, and the analysis of the second moment of the
distribution (which is independent of any assumption concerning the exact
statistical distribution) showed a higher total energy than classically
expected.

A possible next step could be the attempt of inducing Cooper-pairing of
fermions. Cooper-pairs are bosons and it might be possible to cool them down
to a BEC phase transition. This effect is known from superconductivity,
where Bose-condensed Cooper-paired electrons move through a metal without
friction. The superfluidity of fermionic $^{3}$He is also explained by
Cooper-pairing. In dilute gases, however, this is most likely a very
difficult task, due to the lack of efficient cooling mechanism at such low
temperatures.

Table~\ref{TabBoseSpecies} lists the scattering lengths and transition
frequencies for some bosonic and fermionic alkalis. The scattering lengths
are either measured or calculated. Many calculations on scattering lengths
and interaction potentials were carried out by Verhaar, Julienne, Greene,
Stwalley and their respective coworkers.

\section{Imaging Techniques}

All methods of extracting information from a condensate used up to now are
based on its interaction with light. However, the methods differ in the kind
of information they yield. A detailed discussion can be found in reference~ 
\cite{Ketterle98}. We will restrict ourselves here to tracing the essential
points.

We can either use frequency-domain high-resolution Doppler-sensitive
Raman-spectroscopy \cite{Fried98} or just take pictures of the spatial
density distribution of the atomic sample. Pictures are taken by irradiating
the light of a probe beam and either imaging the fluorescence of the atoms
or the imprint of the atomic cloud on the intensity distribution of the
probe beam. The fluorescence method does not reproduce the density
distribution of optically dense clouds and is afflicted with a small light
collection angle. The imprint method can be absorptive or dispersive.
Absorptive imaging destroys the sample and only works for optically thin
clouds. It is therefore commonly used in {\it Time-of-Flight} (TOF) schemes
on ballistically expanded clouds. Dispersive imaging usually preserves the
sample. It works for thick clouds and is commonly used {\it in-situ} on
trapped samples with far-off resonant probe light. Several methods of
dispersive imaging have been applied, notably dark-ground imaging,
phase-contrast imaging and polarization-contrast imaging. It is worth
noticing, that in order to get specific informations, one may manipulate the
sample before imaging it. {\em E.g.} one can convert momentum into
coordinate information by ballistic expansion or one can let a condensate
interfere with a reference condensate and extract information about the
matter-wave phase distribution.

\subsection{Time-of-Flight Imaging}

Time-of-flight (TOF) absorption images are taken after non-adiabatic removal
of the trapping potential within a time much shorter than the trap
oscillation period, $t_{switch}\ll \omega _{trap}^{-1}$, which is typically
on the order of a milli-second. After a time long enough to allow the
initial velocity distribution of the trapped BEC to convert into a spatial
distribution of the expanded cloud that is large enough to neglect the
initial spatial distribution, the cloud is irradiated with a weak resonant
probe beam. The shadow imprinted by the cloud into the probe beam is imaged
onto a CCD camera. Several reasons make TOF imaging well-suited for probing
BECs. (1)~The total number of atoms can be derived from the missing photons
in the probe beam, (2)~the velocity distribution directly reflects the
temperature, (3)~the expanded cloud is optically thin enough, not to
saturate the probe beam shadow even if the probe frequency is on resonance,
(4)~the expanded cloud is large enough to be easily resolved with standard
imaging systems and even to reveal structural details. {\em E.g.} the shapes
and aspect ratios of the condensed and thermal parts are different and
permit their visual separation. On the other hand, it is important for TOF
imaging to guarantee the sudden and free expansion of the cloud, which is a
technically demanding task.

The density of the thermal part $n_{th}({\bf r})$ of an ideal trapped
Bose-gas has been estimated in the semiclassical limit (Section~2.4) by
equation~(\ref{EqSemiclassicalDensity}), while its momentum distribution $%
\tilde{n}_{th}({\bf p})$ follows equation~(\ref{EqSemiclassicalMomentum}).
Switching off the trap suddenly removes all potential energy. While the gas
ballistically expands, the self-interaction energy is adiabatically
converted into kinetic energy \cite{Holland97}. If we wait sufficiently
long, $t\gg \omega _{trap}^{-1}$, the expanded spatial density distribution $%
n_{TOF}({\bf r},t)$ just traces the original momentum distribution, 
\begin{equation}
n_{TOF}({\bf r},t)=m^{3}t^{-3}~\tilde{n}_{th}(m{\bf r}/t)=\left( \lambda
_{dB}\omega _{trap}t\right) ^{-3}~g_{3/2}(e^{\beta (\mu -m{\bf r}%
^{2}/2k_{B}Tt^{2})}).  \label{EqSemiclassicalTOF}
\end{equation}
In the far wings, $m{\bf r}^{2}/2t^{2}\gg \mu $, the density of the thermal
cloud approaches a Gaussian distribution, 
\begin{equation}
n_{TOF}({\bf r},t)\approx \frac{N}{\left( 2\pi \right) ^{3/2}r_{rms}^{3}}%
~e^{-{\bf r}^{2}/2r_{rms}^{2}},  \label{EqSemiclassicalTOFGaussian}
\end{equation}
where we used $k_{B}T\approx m{\bf r}_{rms}^{2}/t^{2}$ and $N=\left(
k_{B}T/\hbar \omega _{trap}\right) ^{3}~\exp \left( \mu /k_{B}T\right) $.
The quantity $\left( 2\pi \right) ^{3/2}r_{rms}^{3}$ is sometimes called 
{\it effective volume}. If we assume thermal equilibrium between the thermal
and condensed fraction, the temperature can easily be extracted from the far
wings of the Gaussian (\ref{EqSemiclassicalTOFGaussian}), which are beyond
the spatial extension of the BEC wavefunction. The generalization of the
distribution (\ref{EqSemiclassicalTOFGaussian}) to nonisotropic potentials
is straightforward. The ideal-gas approximation is valid, because the
thermal fraction is very dilute compared to the condensed fraction.

While the thermal atomic cloud constitutes a statistical ensemble, the
condensed part is described by a single complex wavefunction $\psi _{0}({\bf %
r})$ that is a solution of the Gross-Pitaevskii equation~(\ref
{EqGrossPitaevski}). Let us assume a harmonic trapping potential $U_{trap}(%
{\bf r})$. In the absence of interactions, the ground state wavefunction is
a Gaussian, and it stays a Gaussian during ballistic expansion although it
changes its size and aspect ratio. In most experiments, however, the
interactions are strong, $g\left| \psi _{0}({\bf r},t)\right| ^{2}\gg \hbar
\omega _{trap}$. In this so-called {\it Thomas-Fermi limit}, we can neglect
the kinetic energy term in equation~(\ref{EqGrossPitaevski}). The solution
is a parabolically shaped density distribution, which preserves its shape
during ballistic expansion \cite{Castin96} (Section~4.1). The chemical
potential $\mu $ can be estimated in the Thomas-Fermi limit from the width
of the expanded condensate $\mu \approx m{\bf r}_{rms}^{2}/t^{2}$.

In summary, absorptive TOF imaging permits the unambiguous measurement of $N$%
, $N_{c}$, $T$, and $\mu $ and the derivation of $T_{c}$ from $%
N_{c}=g_{3}(1)~\left( k_{B}T_{c}/\hbar \omega _{trap}\right) ^{3}$. However,
we still have to establish the relationship between the expanded cloud
density~(\ref{EqSemiclassicalTOFGaussian}) and the absorption profile that
the cloud imprints into the probe beam.

\bigskip

{\em Optical Density.} --- The influence of the atoms on the probe beam is
described by the complex refraction index $n_{ref}$, which can be linked to
the optical cross-section $\sigma (\Delta )$ and easily be calculated from
the optical Bloch equations for a two-level atom in the rotating-wave
approximation \cite{Loudon83}: 
\begin{equation}
n_{ref}\approx 1+\frac{n({\bf r})\lambda \sigma (\Delta )}{4\pi }\left( i-%
\frac{2\Delta }{\Gamma }\right) .  \label{EqRefractionIndex}
\end{equation}
If the atomic cloud is small and dilute, the amplitude of the probe beam is
locally attenuated and phase shifted: 
\begin{equation}
E=E_{0}~{\normalsize \exp }\left[ \tfrac{2\pi i}{\lambda }\tint (n_{ref}-1)dz%
\right] \equiv E_{0}~{\normalsize e}^{-\alpha /2}{\normalsize e}^{i\delta }.
\label{EqAbsorptionDispersion}
\end{equation}
The absorption coefficient or {\it optical density} $\alpha $ is the product
of the column density $\tint n({\bf r})dz$ and the optical cross-section. $z$
is chosen to be the imaging direction. Close to resonance, the interaction
is well described by the Lambert-Beer law, $I=I_{0}e^{-\alpha }$,\ where 
\begin{equation}
\alpha (x,y)=\sigma _{0}\tint n({\bf r})dz.  \label{EqLambertBeer}
\end{equation}
The CCD camera actually records the intensity distribution of the probe beam 
$I(x,y)$ that went through the atomic cloud. Inhomogeneities in the
intensity distribution can be compensated by taking a reference picture
without atoms and calculating $I(x,y)/I_{0}$.

\subsection{In-Situ Imaging}

TOF imaging is a one-shot measurement and destroys the sample. As mentioned
earlier, absorptive in-situ imaging is accompanied by the problem of large
local phase shifts of the probing beam due to the optical thickness of the
cloud. Measurements have been carried out in this regime \cite{Hau98}, but
they have only yielded reliable values at the surface of the trapped cloud,
where the density is small. The problem can be circumvented by dispersive
imaging, which additionally possesses several other advantages.

If we increase the detuning, the {\it dispersion coefficient} $\delta $
defined in equation~(\ref{EqAbsorptionDispersion}) decreases, but the
absorption $\alpha $ decreases faster, because $\delta =\alpha \Delta
/\Gamma $. Very far from resonance, the heating due to absorption and
spontaneous emission is insignificant and the BEC will not be destroyed. The
probe beam also phase-shifts the matter-wave and should have uniform
intensity distribution in order to prevent the formation of matter-wave
phase gradients. The local phase shift in the probe beam cross-section can
be turned into an intensity profile using the method of {\it dark ground
imaging} or {\it phase contrast imaging }\cite{Ketterle98}.

Andrews {\em et al.} have used dispersive, non-destructive imaging
techniques \cite{Andrews96}, that allows taking up to 20 images of the
trapped BEC without significantly perturbing it. This permits to watch the
condensate dynamics, {\em e.g.} the response to trap perturbations, on-line.
Bradley {\em et al.} \cite{Bradley97} took phase-contrast images which
exploit the birefringence of the atoms in the strong magnetic field offset
of their trap. Confined in a trap, the thermal and condensed fractions of a
cloud are not well separated. Therefore, in-situ imaging is not well suited
for determining the thermodynamic quantities at the phase transition.
However, at low temperatures, where most atoms are condensed, $N_{c}$ and $%
\mu $ can be measured from the size and shape of the observed cloud. A
technical disadvantage of in-situ imaging is the need of a very high
resolution imaging system allowing to map the tiny sample. The size of some
small structures in the condensate wavefunction, {\em e.g.} vortices and
solitons, may even be beyond the diffraction limit.

\section{Measurements on Condensate Equilibrium Thermodynamics}

The measurement of the temperature dependence of thermodynamical quantities
at the phase transition reveals important information about deviations from
the ideal gas behavior due to particle interaction, finite size effects and
spatial confinement (Chapter~2). Dilute gases are almost ideal systems, the
nonideal features are quite small. Therefore they present a better
opportunity to study the thermodynamics of Bose phase transitions than other
systems, {\em e.g.} $^{4}$He where the condensed fraction $N_{c}/N$ is
difficult to measure and the critical temperature $T_{c}$ is difficult to
calculate.

The time-of-flight absorption pictures yield a number of informations. One
can extract the total number of atoms $N$ from the missing photons in the
probe beam shadow. Two independent two-dimensional Gaussians are fit to the
thermal and the condensed part, thus allowing the determination of $N_{c}$.
The temperature\ $T$ of the sample can be estimated by fitting Gaussians to
the thermal cloud (outer wings of the total TOF density distribution)
assuming thermal equilibrium between thermal and condensed clouds and a
Maxwell-Boltzmann velocity distribution for the thermal cloud. The scaling
temperature $T_{c}$\ is chosen to be the critical temperature (\ref
{EqCriticalTemperatureConfined}) for an ideal harmonically trapped Bose-gas
in the thermodynamic limit (Table~\ref{TabCriticalTemperature}). Various
experiments \cite{Han98,Ensher96} verified that the condensed fraction $%
N_{c}/N$ reproduces well the theoretical dependence (\ref
{EqIdealThermodynamics}) expected for ideal Bose-gases in the thermodynamic
limit (Table~\ref{TabCriticalTemperature}). This means, that the gas is so
ideal, that any nonideal gas feature is difficult to be quantitatively
measured against the experimental shot-to-shot noise and calibration
uncertainties.

However, the measurement of other thermodynamic quantities like the energy
and the heat capacity \cite{DeMarco99,Ensher96} displayed significant
deviations from ideal gas behavior and emergence of interaction effects. The
same ballistic expansion data set used for determining the condensed
fraction can be fit by an arbitrary {\it model-independent} density
distribution,$\ \tilde{n}({\bf k})$ if we only make sure, that its zeroth
moment is normalized to the atom number $N=\int_{V}\tilde{n}({\bf k})d^{3}%
{\bf k}$, where $k=mr/\hbar t_{TOF}$. Depending on the chosen distribution,
the quality of the fit may be better or worse, but in any case, the second
moment gives the kinetic energy, $U=\int_{V}E_{kin}\tilde{n}({\bf k})d^{3}%
{\bf k}$, where $E_{kin}=\hbar ^{2}k^{2}/2m$. For trapped ideal gases the
virial theorem ensures $E=E_{kin}+E_{pot}=2E_{kin}$, however for real gases
the repulsive meanfield energy adds up, $E=E_{kin}+E_{pot}+E_{self}$. The
sudden switch-off of the trapping potential nonadiabatically removes the
potential energy $E_{pot}$. The kinetic and the self energy of the BEC are
converted into pure kinetic energy during the ballistic expansion. This
energy, $E_{kin}+E_{self}$, (sometimes called {\it release energy}) is
measured by TOF measurements. We can expect the temperature dependence of
the measured release energy to correspond to the dependence of the total
energy. The heat capacity derived from both quantities should give the same
results. Despite the experimental noise, the release energy measurement data
clearly show a change of slope at the phase transition. Figure~\ref
{FigReleaseEnergy} shows measurements of the kinetic energy after $10$~ms
ballistic expansion.

The measurement was strictly model-independent and contained no assumptions
on the quantum statistical nature of the particles or on particle
interactions. It would be interesting to compare the results to theories in
various limits, {\em i.e.} the ideal-gas in the thermodynamic limit or by
taking into account finite number effects \cite{Grossmann95} and particle
interactions. However, the accuracy of the experiment does not permit
quantitative conclusions. It only shows, that the effect of mean-field
repulsion is to increase the energy and to reduce the discontinuity of the
temperature-derivative of the energy at the phase transition.

\chapter{Experiments on Condensate Dynamics}

\label{SecExperiments}The experimental achievement of Bose-Einstein
condensation in dilute gases triggered a wealth of theoretical and
experimental work on the characteristics and dynamics of Bose-condensed
gases. The early work focussed on the equilibrium thermodynamics of
condensates (discussed in Section~3.4) and their dynamic response to
perturbations, especially when they are near the critical point. Since then,
breathtaking progress has advanced the field of investigations: Exotic
states like vortices, solitons and multispecies condensates have been
created, collision resonances have been found, experiments on the
interaction of BEC with light have been carried out, and various kinds of
atom lasers have been constructed. Chapters~4 to 6 are devoted to giving a
brief overview and r\'{e}sum\'{e} of recent experiments involving
Bose-Einstein condensates. In the present chapter, we focus on experiments
on the dynamics of condensates and of binary mixtures of condensates.

\section{Wavepacket Dynamics}

The dynamics of Bose-Einstein condensates is generally studied by observing
the modification of the shape of the condensate wavepacket in response to
time-dependent variations of the trapping potential. The simplest
time-dependence imaginable is the sudden removal of the trapping potential.
Indeed, the first experiment performed on BECs was the study of free
expansion \cite{AndersonMH95} (Section~3.3.1). Contrary to the thermal phase
which expands isotropically independently of the shape of the trap, the
shape of the expanded condensate reflects the trap geometry. The condensed
ground state is described by a single macroscopic quantum mechanical
wavepacket and expands predominantly into those dimensions originally
constrained by the trap. {\em E.g.} the BEC aspect ratio reverses during
free expansion \cite{Ernst97,Castin96}. Additionally, as the condensate is
much denser than the thermal cloud, the condensate self-interaction is much
stronger. This repels the atoms and enforces the dynamical evolution
described above. Figure~\ref{FigAspectRatio} shows the evolution of the
aspect ratio of a released sodium condensate.

The time-dependence of the trapping potential may also be an oscillatory or
pulsed small perturbation, {\em e.g.} a tiny modification of the magnetic
trapping fields, a pulsed local anisotropy induced by the dipole force of
far-detuned laser beams, a manipulation of the condensate self-interaction
or density oscillations induced by light coupling to internal or motional
degrees of freedom. We will return to these techniques in Section~4.3.

\section{Multicomponent Condensates}

\subsection{Double Species Condensates in $^{87}$Rb}

Mixtures of Bose-condensates in different internal states are expected to
suffer from relaxation due to spin exchange processes during mixed
collisions. Indeed, a collision may scatter atoms into untrapped states or
else grant sufficient kinetic energy to eject the atoms from the trap. In
the case of rubidium, the fortuitous coincidence between the singlet and
triplet ground state scattering lengths reduces the collisional losses for
any mixture of spin states. In particular, it allows the coexistence of BECs
in the low-field seeking states $|F,m_{F}\rangle =|2,2\rangle $ and $%
|1,-1\rangle $\ \cite{Burke97}. Such double-species condensates have been
observed in experiment \cite{Myatt97}. When the magnetic trap is loaded with
atoms in both hyperfine states, the $|1,-1\rangle $ cloud being less
severely confined, extends to larger radii and thus experiences larger
magnetic trapping fields. During forced radiofrequency evaporation hot $%
|1,-1\rangle $ atoms see a lower potential edge than $|2,2\rangle $ atoms
and are evaporated faster. If the rate of rethermalizing elastic
interspecies collisions is large enough, the $|2,2\rangle $ cloud is cooled
sympathetically and simultaneously with the $|1,-1\rangle $ cloud. Under the
influence of gravity $|1,-1\rangle $ atoms sag further down into the
trapping field than $|2,2\rangle $ atoms, but the displacement is typically
smaller than the size of the condensates. Separate imaging of the two BEC
components therefore requires appropriate hyperfine-pumping and probing
cycles in order to discriminate the two hyperfine states \cite{Myatt97}.

An alternative method to create two-component BECs has been demonstrated by
Matthews {\em et al.} \cite{Matthews98}. They irradiated a $|1,-1\rangle $
BEC with a microwave-radiofrequency two-photon radiation pulse tuned to the $%
|2,1\rangle $ state, $\omega _{micro}/2\pi \approx 6.8$~GHz and $\omega
_{radio}/2\pi \approx 2$~MHz. With $600$~Hz\ Rabi-frequency, which is much
faster than the secular frequencies of the (uncompressed) trap, $\omega
_{trap}/2\pi \approx 100$~Hz, they were able to suddenly transfer nearly $%
100\%$ of the lower state population to the excited $|2,1\rangle $ state. By
transferring only part of the population, they could suddenly spatially mix
the two BEC species and study the complicated nonlinear dynamics of spatial
reorganization and component separation \cite{Hall98}. The influence of
gravity on the TOP trap makes it possible to control the relative vertical
displacement of the $|1,-1\rangle $ and $|2,1\rangle $ states by judiciously
choosing the values for the trapping field strengths and the TOP rotating
frequency \cite{Hall98c} and thus allows to realize a considerable overlap
region of the two BEC species. The reproducibility of the experimental
conditions were good enough to trace the dynamical evolution with
destructive time-of-flight imaging by repeating the whole procedure of
creating and manipulating the sample. The dynamics is essentially governed
by the ratios of scattering lengths\ between the different components, which
have been determined to be $a_{|1,-1\rangle }:a_{|1,-1\rangle |2,1\rangle
}:a_{|2,1\rangle }=1.03:1:0.97$ \cite{Matthews98,Hall98}. Because $%
a_{|1,-1\rangle |2,1\rangle }/\sqrt{a_{|1,-1\rangle }a_{|2,1\rangle }}%
\approx 1.0005>1$, the components weakly repel each other. The $|1,-1\rangle 
$ cloud has a slightly positive buoyancy with respect to the $|2,1\rangle $
cloud.

The coherent coupling of two distinguishable BECs occupying the same region
of space has proven a useful tool for several important experiments: The
observation of {\it compression oscillations} induced by the sudden transfer
between the hyperfine states (Section~4.3.1), the creation of {\it vortices}
in double condensate systems (Section~4.3.4) and the realization of an {\it %
internal state BEC interferometer} (Section~5.3.1).

\subsection{Spinor Condensates in $^{23}$Na}

In order to create a spatial overlap of condensates in different internal
states, it is necessary for the atoms to be confined in a common trap and
not to suffer from spin relaxation processes. Pure dipole-force optical
traps, {\em e.g.} red-detuned far-off resonance traps (FORT) are not
sensitive to the Zeeman state of the atoms. Therefore, they can trap high
field seeking states which cannot be confined in magnetic traps, or even
mixtures of atoms being in all Zeeman substates of a hyperfine level.
Furthermore, they avoid (spatially inhomogeneous) Zeeman shifts. While
attempts to directly produce BECs in optical traps have not been successful
yet, $|1,-1\rangle $ sodium condensates have been loaded from a magnetic
trap into a FORT \cite{Stamper-Kurn98}. Due to the low kinetic energy of
BECs, the optical trap can be made very shallow, so that low power
far-detuned lasers may be used. In their experiment, Stamper-Kurn {\em et al.%
} \cite{Stamper-Kurn98} needed only a few milli-Watts of laser power at $985$%
~nm wavelength focussed down to $6$~$\mu $m waist creating a few
micro-Kelvin deep potential to reach trapping lifetimes on the order of $10$%
~s. On the other hand, dipole traps are an order of magnitude steeper than
magnetic traps, $\tilde{\omega}_{trap}/2\pi \approx 40$--$400$~Hz.
Consequently, very high densities between $n=3\times 10^{14}$ and $3\times
10^{15}$~cm$^{-3}$, mainly limited by three-body recombination, can be
reached.

In a subsequent experiment, Stenger {\em et al.} \cite{Stenger98} lifted the
Zeeman degeneracy by application of a weak magnetic field and coupled the $%
|1,-1\rangle $ state to the other Zeeman states $|1,0\rangle $ and $%
|1,1\rangle $ by irradiation of resonant radiofrequency. The population
could be completely or partially transferred between the states. The
resulting three component BEC quantum field is described by a {\it spinor}.
Several interesting features of the dynamics of spinor BECs have already
been observed, {\em e.g.} the formation of {\it spin domains}, the
miscibility of the $|1,1\rangle $ and $|1,-1\rangle $ and the immiscibility
of the $|1,\pm 1\rangle $ and $|1,0\rangle $, the antiferromagnetic behavior
of the spin-dependent atomic interaction \cite{Stenger98}, the metastability
of spin domains against very small ($0.1~$nK) energy barriers \cite
{Miesner99}, and quantum tunneling across spin domains \cite{Stamper-Kurn99}
(Section~5.3.2). A review of the experiments on spinor condensates can be
found in \cite{Stamper-Kurn00}. Spinor BECs are also interesting candidates
for studies of nonlinear four-wave mixing processes \cite{Goldstein99,Law98}
(Section~5.4.5).

\section{Collective Excitations}

\subsection{Elementary Excitations}

Elementary excitations (also called {\it quasiparticles} or {\it normal modes%
}) of the Bose-Einstein condensate are solutions of the linearized
Gross-Pitaevskii equation. The coherent excitation of many quasiparticles
leads to collective oscillations or density modulations (also called {\it %
sound}) of the trapped atomic cloud. Technically, the excitations are
generated in response to small time-dependent perturbations of the trapping
potential. There are various procedures and consequently various types of
excitations. The first experiments have been performed by modulating the
trapping potential \cite{Jin96,Mewes96b} and resulted in the observation of 
{\it center-of-mass oscillations} (also called {\it sloshing modes}) and 
{\it shape oscillations}. The frequencies of those excitations agreed well
with theoretical calculations \cite{Singh96}\nocite{Edwards96}- \cite
{Stringari96}. The oscillations were damped by interactions between the
collective mode and thermal excitations. The temperature dependence of the
damping has been experimentally studied by Jin {\em et al.} \cite{Jin97}.

At JILA, microwave-radiofrequency double-resonance transitions have been
used to suddenly transfer a $^{87}$Rb condensate from the $|1,-1\rangle $
into the $|2,2\rangle $ internal atomic state. This state has the same
magnetic moment, but a slightly different scattering length, $%
a_{|1,-1\rangle }\neq a_{|2,1\rangle }$, and therefore a different
mean-field energy $gn_{0}=4\pi \hbar ^{2}a_{|2,1\rangle }n_{0}/m$, where $%
n_{0}$ is the peak density of the condensed cloud. The wavefunction
immediately starts to adjust its shape to fit into the modified environment
and commences oscillating. These damped {\it compression oscillations} have
been recorded (Fig.~\ref{FigCompressionOscillations}) and by comparison to a
theoretical model permitted the determination of the ratio of scattering
lengths of the two states \cite{Matthews98} (Section~4.2.1).

The range of excitations accessible by modulating the magnetic trapping
fields is limited. Sophisticated engineering of the perturbation is required
to excite more complicated excitations like higher multipolar order {\it %
surface oscillations} \cite{Onofrio99}. Focussed far-off resonant laser
beams are a useful tool to push around the atoms inside a BEC. They have
also been used to observe the {\it propagation of sound pulses }by
generating a short small local density perturbation and tracing its way
through the condensate \cite{Andrews97b}. The temporal behavior of
excitations is best studied by taking a rapid sequence of nondestructive
images (Section~3.3.2).

The nature of the excitations profoundly depends on their de Broglie
wavelengths $k^{-1}=(\hbar /2m\omega _{k})^{1/2}$ compared to three
characteristic lengths. Those are the mean free path for quasi-particles $%
l_{mfp}\approx (n_{th}\sigma _{el})^{-1}$, the size of the ground state of
the trapping potential $a_{trap}=(\hbar /m\omega _{trap})^{1/2}$ and the
healing length 
\begin{equation}
\xi =1/\sqrt{8\pi an_{0}}.  \label{EqHealingLength}
\end{equation}
Here $a=53a_{B}$ is the sodium $s$-wave scattering length for the $%
F=1,m_{F}=-1$ state, the cross-section for elastic collisions is $\sigma
_{el}\approx 8\pi a^{2}$. Typical experimental values for a sodium
condensate are $\hbar \omega _{trap}\approx h~100$~Hz and $gn_{0}\approx h~7$%
~kHz, the characteristic lengths are roughly on the order of $l_{mfp}\approx
100$~$\mu $m, $a_{trap}\approx 2$~$\mu $m, and $\xi \approx 0.2$~$\mu $m.
The mean free path marks the boundary between the hydrodynamic regime, $%
k^{-1}\gg l_{mfp}$, and the collisionless regime (in the sense of
quasiparticle collisions), $k^{-1}\ll l_{mfp}$. The trap size delimits the
regime of discrete collective modes, $k^{-1}\gtrsim a_{trap}$, from the
regime of pulsed localized excitations, $k^{-1}<a_{trap}$. The healing
length, finally, sets the boundary between the regimes of phonon-like
excitations, $k^{-1}\gg \xi $, and particle-like excitations, $k^{-1}\ll \xi 
$.

The trap manipulation methods mentioned so far are restricted to the
collisionless regimes of discrete collective modes and pulsed localized
excitations. The hydrodynamic regime of low energy excitations has barely
been approached in experiments \cite{Stamper-Kurn98b}, and the opposite
regime of high energy localized excitations, $k^{-1}\lll a_{trap}$, has been
investigated employing the newly developed method of {\it Bragg spectroscopy}
\cite{Stenger99}, \cite{Stamper-Kurn99b}. This technique allows to probe
excitations over a very large range of frequencies. It is particularly well
suited for high frequencies and therefore provides a useful tool to probe
the boundary between the regimes of phonon and particle-like excitations. We
will return to this in Section~5.4.2. The hierarchy of length scales formed
by the various regimes of collective excitations and the experimental method
to produce them are summarized in Table~\ref{TabExcitationRegimes}.

\subsection{Non-Circulating Topological Modes}

Topological modes are stationary solutions of the Gross-Pitaevskii equation~(%
\ref{EqGrossPitaevski}), that are not the ground-state. There have been some
proposals on the creation of such non-groundstate Bose-condensates \cite
{Yukalov97}, and an antisymmetric dipole topological mode has recently been
created in experiment. Williams {\em et al.} \cite{Williams99b,Matthews99}
used a coherently coupled double-species condensate. The total order
parameter for such a system is a two-dimensional spinor $\Psi ({\bf r}%
,t)\equiv (\psi _{|1,-1\rangle },\psi _{|2,1\rangle })$. We may also view
the spinor field as a spatially varying Bloch-vector describing the local
internal coherence and inversion of the two-level atoms that form the BEC 
\cite{Matthews99}, 
\begin{equation}
\vec{\Psi}({\bf r},t)=\left( 
\begin{array}{c}
\psi _{|1,-1\rangle }\psi _{|2,1\rangle }^{\ast } \\ 
\psi _{|1,-1\rangle }^{\ast }\psi _{|2,1\rangle } \\ 
|\psi _{|2,1\rangle }|^{2}-|\psi _{|1,-1\rangle }|^{2}
\end{array}
\right) .  \label{EqSpinorBlochvector}
\end{equation}
The gravitation, whose direction is assumed to coincide with the symmetry
axis of the TOP trap, modifies the magnetic trapping potential and displaces
the trapped $|1,-1\rangle $\ and $|2,1\rangle $\ clouds vertically from one
another by a tunable amount. When coupling the states with a
microwave-radiofrequency two-photon radiation, the axial displacement makes
the generalized Rabi-frequency position-dependent, $G(z)\equiv \sqrt{\Omega
^{2}+\Delta (z)^{2}}$. The effect of the inhomogeneous coupling strength is
a non-uniform precession and nutation speed of the local Bloch-vector and
consequently a spatial modulation of the inversion, which comes down to
generating a differential torque on each of the single-component
wavefunctions. The torque corresponds to a matter-wave phase shift and
ultimately reaches a point, where the matter-wave phase twist is $2\pi $
across the condensate along the $z$-axis. This state is the dipole
topological mode. Applying the torque furthermore untwists the phase winding
until the condensate finds back to its original state.

In the JILA experiments \cite{Matthews99,Williams99}, the time-evolution of
the local inversion (third component of the Bloch-vector~(\ref
{EqSpinorBlochvector})) and the total space-integrated inversion were
monitored non-destructively and on-line by taking sequences of
phase-contrast pictures. The probe beam was tuned between the hyperfine
states, so that the $|2,1\rangle $\ atoms stepped out as rising and the $%
|1,-1\rangle $\ atoms as deepening from the background. The Rabi-flopping of
the total space-integrated inversion exhibited a behavior reminiscent to
quantum-collapse and -revival well-known to occur in Jaynes-Cummings type
systems studied in quantum optics. The epitome of a Jaynes-Cummings system
is a two-level atom coherently coupled to a single-mode light field. The
driven atom is able to momentarily bury its coherence by transferring it to
quantum correlations of the light field. Tracing over the degrees of freedom
of the light field, the atomic coherence appears to momentarily collapse and
revive at a later time. The coupled system consisting of the BEC matter-wave
and the internal atomic degrees of freedom behaves similarly. If all atomic
dipoles oscillate in phase (groundstate BEC), the space-integrated inversion
exhibits strong Rabi-oscillations (Figure~\ref{FigRabiTwist}). If the
matter-wave field is strongly modulated (twisted condensate, higher
topological mode), the atomic dipoles oscillate at different phases, and the
Rabi-oscillations of the space-integrated inversion cancel out. Under the
influence of the inhomogeneous coupling strength $G(z)$ the system gradually
changes its topology and thus causes the collapse and revival of the
Rabi-oscillations. A quantitative model can be found in ref.~\cite
{Williams99}.

Matthews {\em et al.} \cite{Matthews99} extended the experiment later by
adjusting the condensate shape for long axial extension, exploiting the
tricky TOP-gravitation interplay. The radiation twists the system more and
more, successively cranking up to four windings into the BEC. Past some
point, the radiation untwists the system again until it finds its way back
to the original form.

\subsection{Superfluid Flow}

Superfluid liquids or gases are distinguished by their ability to support
dissipationless flow, {\em i.e.} flow that is exempt from viscous damping.
The phenomenon of superfluidity is a well-known property of liquid $^{4}$He,
but the relationship between superfluidity and Bose-Einstein condensation in
this strongly interacting system is not trivial. The situation is much
simpler in weakly-interacting Bose-gases, where the superfluid phase is
nearly identical with the condensed fraction, and the normal fluid phase
with the thermal fraction. The availability of dilute gas Bose-condensates
offers the unique opportunity to study the superfluid-condensate
interdependence. The early experiments on the dynamical behavior of
condensates at very low temperatures already provided indirect signatures of
their superfluid nature, because the hydrodynamic theory of superfluidity
describes well the collective excitations, as we have pointed out in
Section~4.3.1. Furthermore, the observation of matter-wave interference
(Section~5.2.2) is an indication for superfluidity, since quantum coherence
is a characteristic of superfluids.

Several experiments provided direct evidence for the superfluid nature of
condensates. Raman {\em et al.} \cite{Raman99} performed a calorimetric
measurement of the dissipation induced by stirring a condensate with a far
blue-detuned laser beam. The thermal fraction of the atomic cloud was
measured before and after stirring by standard time-of-flight imaging and
fitting a bimodal density distribution to the condensed and thermal phases
of the cloud (Section~3.3.1). They observed a {\it critical value for the
stirring velocity} $v_{c}$: For lower velocities, the perturbation was found
to be dissipationfree, at higher velocities phonons were excited and the
cloud was heated. In a subsequent experiment, the density fluctuations
induced by stirring were observed {\it on-line} and {\it in-situ} \cite
{Onofrio00}. When the stirring velocity was inferior to the critical
velocity, the density was quasi stationary at any instant of time thus
indicating superfluid flow. However, when the stirring velocity exceeded $%
v_{c}$, the stirring beam dragged the atoms piling them up in front of it.
The resulting pressure gradients led to turbulent flow around the
perturbation and dissipation. Taking the asymmetry of the instantaneous
density distribution resulting from the bow and the stern wave of the moving
laser beam as a measure for the amount of dissipation Onofrio {\em et al.}
determined a critical velocity that agreed with the result of the
calorimetric measurement.

The critical velocity $v_{c}$ found in the stirring experiment was about ten
times smaller than the local speed of sound which is inversely proportional
to the superfluid healing length (\ref{EqHealingLength}), 
\begin{equation}
c_{s}=\hbar /\sqrt{2}m\xi .  \label{EqSoundSpeed}
\end{equation}
In fact, while the onset of dissipation is accelerated by turbulence around
the {\it macroscopic} object traversing the superfluid, the local speed of
sound (\ref{EqSoundSpeed}) is derived for a {\it microscopic} object.
Chikkatur {\em et al.} \cite{Chikkatur00} studied the motion of impurity
atoms through a condensate. For that purpose, they produced an impurity BEC
with well defined initial velocity out of the original BEC by inducing Raman
transition from the trapped $|F=1,m_{F}=-1\rangle $ state to the untrapped $%
|F=1,$ $m_{F}=0\rangle $. The initial velocity was set by arranging the
Raman laser beams (polarization, encompassed irradiation angle, relative
detuning) to satisfy the Bragg condition (Section~5.3.3) and the selection
rules for the Raman transition. The impurity, not being constrained by the
trap, traversed the BEC before the trap was switched off, a Stern-Gerlach
magnetic field gradient was pulsed to spatially separate the atoms being in
different hyperfine states and the atoms were probed by time-of-flight
imaging. When the initial velocity was well above a critical value that
coincided with the local speed of sound, ultracold $s$-wave collisions
between the impurity atoms and the stationary condensate distributed the
momentum of the collision partners evenly and, in the TOF images (Fig.~\ref
{FigImpurityScattering}), gave rise to a circular halo centered around the
center-of-mass of the collision partners. However, when the initial velocity
was reduced, the collision rate between the impurity and the stationary
condensate was suppressed and the trajectory was more superfluid.

\subsection{Vortices}

Important manifestations of superfluidity are associated with rotational
phenomena. An example is the occurrence of {\it scissors modes} \cite
{Guery-Odelin99} that are excited when an angular momentum is suddenly
applied to an anisotropic BEC. Scissors modes have been generated by
Marag\'{o} {\em et al.} \cite{Marago99}. They produced a BEC in an
anisotropic trapping potential and then suddenly reoriented the inclination
of the symmetry axis of the trap. The response of the BEC wavefunction is a
pure oscillation of its tilt generated by irrotational superfluid flow. The
excitation spectrum reflects the strong reduction of the moment of inertia
for the superfluid.

The most stringent manifestation of superfluidity, however, is the
occurrence of quantized and persistent currents, called {\it vortices}.
Vortices are stationary solutions (or {\it topological modes}) of the
Gross-Pitaevskii equation~(\ref{EqGrossPitaevski}) that, under the
assumption of a cylindrically symmetric system, additionally satisfy the
condition $\psi ({\bf r})=\psi (r,z)e^{i\kappa \varphi }$, where $\kappa $
is an integer. In a vortex, the superfluid current is driven by the phase
gradient, ${\bf v}=\hbar /m\triangledown \varphi $. The superfluid flow
around a close path must by quantized to make the wavefunction
single-valued, $\oint {\bf v}d{\bf r}=2\pi \hbar /m$, {\em i.e.} the phase
winds up to multiples of $2\pi $. Also, the flow must be persistent,
because\ its winding number can only be changed discontinuously by
overriding an energy barrier, which requires energy from thermal
excitations. The normal component of a gas can have, of course, circular
flow, as well. However, the disorderly microscopic motion of every
individual particle causes a viscous drag that precludes the persistence of
the flow in the absence of a driving torque. This is in contrast to
superfluid flow which persists even without an externally imposed rotation.
Questions about the stability, the formation and the topology of vortices
have been addressed in recent experiments \cite{Matthews99b,Madison99}, \cite
{Madison00}\nocite{Chevy00}-\cite{AndersonBP00}.

\bigskip

{\em Stability.} --- In a topologically singly-connected trap ({\em e.g.}
harmonic potential), vortices are not the lowest energy eigenstate, and they
must decay into the ground state. If the mean-field interaction energy of
the condensate is weak compared to the kinetic energy, $gn_{0}/\hbar \omega
_{z}\ll 1$, the healing length is larger than the size of the condensate, $%
\xi \gg a_{trap}$, and the vortex rapidly decays dissipating the excess
energy to thermal excitations. Such Bose-condensates cannot be considered
superfluid. If the mean-field interaction is strong, the vortex
spontaneously breaks the azimuthal symmetry, dislocates from the center and
spirals out of the condensate \cite{Rokhsar97}. However, the decay time may
be pretty long, and under certain conditions a vortex might be trapped off
center. On the other hand, a vortex can be the ground-state in
multiply-connected traps ({\em e.g.} torus-shaped potentials). Such a
potential can be realized as the time-average of a harmonic potential with a
small rotating anisotropy \cite{Madison99}. Another scheme uses harmonically
trapped double-condensates where a ground-state BEC is located at the trap
center. An excited internal state BEC can form a vortex in a circular orbit
around the ground-state BEC \cite{Matthews99b}. If the condensates repel
each other, the vortex core is pinned by the ground state BEC, so that the
vortex is very stable.

\bigskip

{\em Formation.} --- The ideas on how to create vortices can be divided into
two classes. Some propose to imprint an angular momentum into the atomic
cloud by rotating the (anisotropic) trap during the process of forced
evaporation. This can be done with rotating magnetic fields or by {\it %
stirring} the atomic cloud with a laser beam. When crossing the phase
transition, a vortex state should nucleate within the stirring path. Others
propose to imprint a circular $2\pi $ {\it phase gradient} into a previously
created condensate. These procedures must drive the local density in the
center of the trap to zero and then rely on dissipative relaxation of the
BEC into the vortex state. Dobrek {\em et al.} suggested to exploit the
inhomogeneous Stark-shift which a far-detuned optical beam with an
appropriately designed intensity profile generates in a BEC \cite{Dobrek99}.
An alternative method based on the phase imprinting idea but avoiding the
need of relaxation processes has been suggested by Williams {\em et al.} 
\cite{Williams99b}. In their configuration the phase gradient is created via
adiabatic Raman-transfer between two internal states of the condensate atoms
within a small rotating area of space. Thus a coherent process is used to
directly build and shape the vortex wavefunction.

The first experimental evidence for vortices was reported by Matthews {\em %
et al.} \cite{Matthews99b}. They produced and recorded vortices in a coupled
double-species condensate system using a method based on the phase
imprinting idea. The method consisted in dynamically converting atoms from a
nonrotating $|1,-1\rangle $ ground-state BEC to $|2,1\rangle $ atoms (or
vice versa) having a torus-shaped topology by time-dependent and spatially
inhomogeneous adiabatic population transfer.

In the experiment, Matthews {\em et al.} \cite{Matthews99b} produced a
standard $|1,-1\rangle $ ground-state BEC with a size of typically $%
r_{rms}=54$~$\mu $m in an isotropic harmonic TOP trap with $\omega
_{trap}/2\pi =7.8$~Hz secular frequency. They coherently coupled the two
hyperfine states using two-photon microwave radiation tuned $\Delta
_{rf}/2\pi =94$~Hz below (or above) the resonance and adjusted the radiation
power to produce $\Omega _{rf}/2\pi =35$~Hz Rabi frequency\ thus causing the
Bloch vector of the two-level system to precess with the generalized Rabi
frequency $G_{rf}=(\Omega _{rf}{}^{2}+\Delta _{rf}^{2})^{1/2}=2\pi \times
100 $~Hz. Spatial and temporal control over the conversion rate between the
hyperfine states was achieved by additionally focussing a moveable laser
beam ($P=10~$nW, $w_{0}=180$~$\mu $m) onto the cloud and rotating it with
frequency $\omega _{rot}$ at a distance $r_{rot}=75$~$\mu $m around the
symmetry axis of the trap. The laser was detuned $\Delta _{l}/2\pi =800$%
~MHz\ blue from the $^{87}$Rb $D2$ line thus giving rise to an\
inhomogeneous time-dependent AC Stark-shift $\Omega _{l}({\bf r}%
,t)^{2}/4\Delta _{l}$, where $\Omega _{l}$ is the Rabi frequency on the $D2$
transition. While ground-state atoms located at the center of the trap did
not sense the modulation of the Stark-shift, atoms located at distance $%
r_{rot}$ from the trap center were subject to the full modulation depth and
experienced the microwave radiation on two modulation sidebands located at $%
\Delta _{rf}\pm \omega _{rot}$. In order to fulfill the resonance condition
for one of the sidebands and drive transitions from the ground-state to the $%
|2,1\rangle $ state, the rotation frequency was adjusted to the precession
frequency, $|\omega _{rot}|=G_{rf}$. This is equivalent to ensuring, that
the phase delay of the precessing two-level Bloch vector, $G_{rf}t$, was
equal to the azimuthal matter-wave phase variation of the newly created $%
|2,1\rangle $ atoms, $\omega _{rot}t$, along the rotation path, and that the
matter-wave phase was single-valued around a closed loop. The matter-wave
phase gradient caused circular flow and formed a visible vortex after about $%
70$~ms. The direction of the vortex rotation could be arbitrarily chosen
through the sign of the detuning $\Delta _{rf}$. Vortices could be formed
either in the $|1,-1\rangle $\ state around a central $|2,1\rangle $ BEC or
vice versa. The small positive buoyancy of the\ state $|1,-1\rangle $ with
respect to $|2,1\rangle $ made the first option more stable.

While in harmonically trapped single-species condensates the diameter of the
vortex core is on the order of the healing length $2\xi $ and too small to
be seen by {\it in-situ} spatial imaging, in the double-species
configuration the diameter of the vortex core is much larger, because it is
determined by the size of the central core BEC. The core BEC can partially
or completely be removed with resonant light pressure and the vortex be
studied as a function of the core size and the filling material. To see the
vortex, Matthews {\em et al.} took a non-destructive image of the density
distribution of the upper $|2,1\rangle $ state. Then, on the same sample,
they applied a resonant two-photon radiofrequency $\pi /2$ pulse which mixed
the vortex with the core BEC. The resulting ring-shaped matter-wave
interference pattern reveals the phase profile of the vortex. This Ramsey
type interference technique will be detailed in Section~5.3.1. Finally, a
second radiofrequency $\pi /2$ pulse completely inverted the population of
the states $|2,1\rangle $ and $|1,-1\rangle $ and permitted recording the
density distribution of the nonrotating ground state (Fig.~\ref{FigVortex}).

\bigskip

{\em Vortex precession.} --- A radial force acting on a vortex results in
its azimuthal displacement and precession around the symmetry axis. The
effect is known as Magnus effect \cite{JacksonB00} and is due to pressure
imbalances at the vortex surface. A radial force arises naturally when the
core is displaced from the center, because local pressure gradients tend to
force it outwards to lower density regions. Anderson {\em et al.} \cite
{AndersonBP00}\ observed a roughly $1$~Hz slow precession of the vortex core
by a succession of non-destructive images for various sizes of the core BEC.
A radial motion of the vortex as expected for dissipative interaction with a
thermal cloud was not observed. Instead, the vortex core decreased its size
due to gradually reduced repulsion of the slowly decaying core BEC.

\bigskip

{\em Vortex lattices.} --- Superfluid $^{4}$He in a rotating bucket
spontaneously develops patterns of symmetrically arranged vortices. Similar
phenomena can be expected when a dilute gas Bose-condensate is forced to
rotate. Butts {\em et al.} \cite{Butts99} have calculated the vortex
patterns that will form as a response to forcing a BEC to rotate with a
predefined frequency $\Omega $. The energy in the corotating frame gets an
additional contribution from the centrifugal term $U_{rot}({\bf r})=U_{trap}(%
{\bf r})-\Omega L_{z}$, where $L_{z}=\hbar Nl_{z}$, and $l_{z}=i\left(
y\partial _{x}-x\partial _{y}\right) $ is the single-particle angular
momentum. If the rotation is slow, the energy $\Omega L_{z}$ is too small to
force the condensate wavefunction to rotate. If the rotation frequency is
higher than a critical value $\Omega _{c}$, the time-averaged potential, $%
U_{rot}({\bf r})\ $eventually develops a local maximum in the center (torus
shaped potential). For non-interacting gases, the critical frequency
coincides with the radial secular frequency, $\Omega _{c}=\omega _{r}$, and
the radial restoring force of the trap does not balance the centrifugal
force anymore, so that the atoms escape from the trap. However, for
superfluid gases the critical frequency is reduced, $\Omega _{c}<\omega _{r}$%
. Between the rotation frequencies $\Omega =\Omega _{c}$ and $\Omega =\omega
_{r}$, the lowest energy state in the torus shaped potential is a vortex
filament around the center. At even higher rotation frequencies, one might
expect single vortices with a higher winding number (more than $2\pi $ phase
winding for a single path around one vortex). However, single multiple-order
vortices in harmonic traps are always very unstable. Instead, vortex
lattices \cite{Butts99} are formed. For a given trapping potential and
mean-field interaction, the symmetry of the lattice and the number of
vortices depend on the rotation frequency $\Omega $. Counterintuitively, the
single-particle angular momentum $l_{z}$ is {\it not quantized}. Upon
varying $\Omega $, forbidden ranges of $l_{z}$ alternate with allowed bands.
The discontinuous transition from one vortex pattern to another is a
first-order phase transition and spontaneously breaks the previous azimuthal
symmetry to form another one. An upper limit for the rotation speed is set
by the balance of the centrifugal force and the radial restoring force of
the trap at $\Omega =\omega _{r}$.

These vortex patterns have been observed in a recent experiment, that
employed the stirring method of rotating the trap \cite{Madison99}. Madison 
{\em et al.} produced a cigar-shaped $^{87}$Rb condensate with $N=10^{5}$
atoms in a cloverleaf trap with $\omega _{z}=2\pi \times 12$~Hz, $\omega
_{r}=2\pi \times 220$~Hz. The ratio of mean-field interaction to kinetic
energy was $gn_{0}/\hbar \omega _{z}=a_{trap}^{2}/2\xi ^{2}>100$. Along the
symmetry axis but slightly displaced from the center, they focussed a far
red-detuned dipole-force laser beam in order to create a weak anisotropy in
the trapping potential. During forced evaporation and while crossing the BEC
phase transition, this optical ''spoon'' is rotated around the symmetry
axis. Beyond a certain critical rotation frequency, $\Omega _{c}\approx 2\pi
\times 150$~Hz, they observed the formation of a central vortex. At higher
frequencies, they could image vortex lattices with up to eleven vortices
symmetrically arranged in the transversal plane (Fig.~\ref{FigVortexLattice}%
). At stirring frequencies approaching the radial secular frequency of the
trap $\omega _{r}$, the BEC wavefunction got more and more turbulent and
finally vanished altogether. After removing the optical spoon, the lifetime
of a single vortex was measured to be approximately one second (the lifetime
of the condensate being much longer). The vortex decayed to the ground-state
of the unperturbed harmonic potential most likely by spiraling out of the
center. Vortex patterns were found to decay by successively losing one
vortex at a time.

Madison {\em et al.} probed the density distribution of the vortices by
absorption imaging. The diameter of the dark core of a vortex in the
unperturbed trap (no stirring spoon) is set by the healing length and
measures about $2\xi \approx 0.4$~${\rm \mu }$m. This size is too small for
optical imaging. However, after a $30$~ms period of ballistic expansion the
core diameter reached $15$~$\mu $m\ and could be probed easily.

\bigskip

{\em Angular momentum.} --- Zambelli {\em et al.} \cite{Zambelli98} have
suggested a method for measuring the flow around a vortex via the excitation
of quadrupolar surface modes. In an axisymmetric trap, the transversal
quadrupole modes can be linearly decomposed into two counterrotating modes
with angular momentum $\pm 2\hbar $ and frequencies $\omega _{\pm }=\pm
\omega $. In a rotating BEC, the degeneracy of the frequencies is lifted by
an amount that corresponds to the rotational energy of a single atom, 
\begin{equation}
\omega _{+}-\omega _{-}=\frac{2L_{z}}{mr^{2}},  \label{EqSagnacRotation}
\end{equation}
where $r$ is the average radius of the orbit. This behavior is known as
Sagnac effect. Chevy {\em et al.} \cite{Chevy00} performed an experiment
where they first stirred the BEC and then excited the quadrupolar surface
modes similarly to the earlier experiment of Onofrio {\em et al.} \cite
{Onofrio99}. They observed the quadrupolar oscillation in time-of-flight
measurements and noticed a continuous rotation of the principal axis of the
quadrupolar mode if vortices had been excited. Also, being in the
Thomas-Fermi limit, they could infer $r$ from TOF images and calculate the
angular momentum $L_{z}$ of the rotating BEC from equation~(\ref
{EqSagnacRotation}) as a function of the chosen stirring velocity. They
found $L_{z}=0$ below the critical velocity. At the stirring velocity $%
\Omega _{c}$, the angular momentum suddenly jumped to $L_{z}=\hbar $ and
gradually increased (in fractions of $\hbar $) up to $L_{z}\approx 3\hbar $
as the stirring was further accelerated. At stirring velocities approaching
the radial trap frequency $\omega _{r}$, the vortex pattern got turbulent
and $L_{z}$ diminished again and finally vanished.

\subsection{Matter-Wave Solitons}

Solitons are localized non-singular solutions of any nonlinear wave equation
satisfying $|\psi ({\bf r},t)|=|\psi ({\bf r}-{\bf v}t)|$. Solitons are well
known to occur in nonlinear optical media, {\em e.g.} in optical fibers when
the dispersion is counterbalanced by self-phase modulation so that they
propagate without spreading. The Gross-Pitaevskii equation is another
example of a nonlinear wave equation that can exhibit soliton-like
solutions. Correspondingly, so called {\it dark solitons} or {\it kink-wise
states}, {\em i.e.} states with dynamically stable density minima, are
expected in condensates with repulsive interactions. They have been
predicted for one-dimensional BECs \cite{Morgan97}\nocite
{Reinhardt97,JacksonAD98}-\cite{Muryshev99} and may occur in higher
dimensions, as well. In contrast to truly topologically stabilized defect
states like vortices, dark solitons are pseudodefects, whose decay may be
very slow although they are topologically trivial. Due to the greater
motional freedom of their wavefunctions they may be untwisted by complex
deformations \cite{Busch99}. Matter-wave soliton-like states have first been
observed in superfluid $^{3}$He-B \cite{Backhaus98}. In dilute gases, their
size is expected to be on the order of the healing length which typically
corresponds to a few hundred nanometers.

Dum {\em et al.} \cite{Dum98} proposed to engineer dark solitons in
Bose-condensates using adiabatic Raman-transfer, and many other schemes have
been suggested. Burger {\em et al.} \cite{Burger99} and the NIST group at
Gaithersburg \cite{Denschlag00} recently successfully created and observed
solitons. Both groups employed a method based on the application of an
inhomogeneous matter-wave phase-shift. They created and magnetically
confined a rubidium {\em resp.} sodium condensate and irradiated half of the
condensate with a far-off resonance laser beam pulse (detuning $\Delta $,
Rabi frequency $\Omega $, duration $\tau \ll \hbar /gn_{0}$) thus advancing
the phase of this half condensate by $\varphi =\Omega ^{2}\tau /4\Delta $.
When the phase-shift was adjusted to be on the order of $\pi $, a steep
phase gradient developed at the boundary plane driving the density
distribution in the condensate to adjust itself until a density minimum
formed along the plane. The density distribution of the condensate was
mapped by time-of-flight imaging at various delays after application of the
phase shifting laser pulses (Fig.~\ref{FigSoliton}). Denschlag {\em et al.}
additionally used an interferometric technique based on Bragg diffraction
(Section~5.3.3) to monitor the phase distribution of the condensate. The
observed density kink corresponds to the node of a topological dipole mode
and can also be interpreted as a one-dimensional dark soliton on a finite
background, where the kink and the background move synchronously \cite
{Williams99b,Morgan97,Muryshev99}. These states represent standing
matter-waves for which the trap serves as a cavity.

The steep phase gradient at the nodal plane exerts a force that tries to
enhance the gap, while the repulsive interaction works to fill it. At zero
temperature, this balance guarantees the dynamical stability of the soliton.
While a perfectly dark soliton should be stationary, the experiments \cite
{Burger99,Denschlag00} exhibited a propagation of the density kink
perpendicularly to the nodal plane. This feature is a result of the finite
contrast of the kink. However, the propagation velocity must always be
inferior to the local speed of sound, 
\begin{equation}
v_{sol}=c_{s}\sqrt{\frac{n_{sol}}{n}},  \label{EqSolitonVelocity}
\end{equation}
where $n$ is the condensate peak density and $n_{sol}$ the density at the
bottom of the dark soliton \cite{Reinhardt97,JacksonAD98}. Figure~\ref
{FigSoliton} shows that the soliton develops a curvature as it propagates.
The reason for this is the decrease of the local speed of sound, $c_{s}=%
\sqrt{gn_{0}/m}$, at the edge of the BEC where the density gets smaller. A
second reason is that the density in the dip $n_{sol}$ tends to zero towards
the edge. In the presence of a thermal cloud, dissipation reduces the
contrast of the density kink and accelerates the soliton until it reaches
the speed of sound $c_{s}$ and finally vanishes.

\chapter{\bf Atom Optics with Bose-Einstein Condensates}

\label{SecAtomOptics}In the past decade, various methods and schemes of
laser cooling and trapping of atoms became powerful tools in atom physics
and quantum optics. In achieving always lower temperatures and extreme
densities, the whole field moved to the boundaries of the new regime, where
coherent matter-wave interactions become dominant. This development
culminated in the experimental achievement of Bose-Einstein condensation.
The atoms confined on microscopic or macroscopic scales at high phase space
density are governed by collective and quantum statistical effects. This
opens up new perspectives for many-body studies in regimes, where standard
approximations cease to be valid. It is, for example, particularly
interesting to explore atomic two--body interactions, which may play a role
in coherent matter-wave optics similar to the role played by atom-photon
interactions in quantum and nonlinear optics. At the same time, the field of
atom optics developed rapidly with the demonstration of atom optical
elements like atom mirrors, atom lenses and beamsplitters. These two fields
are now being combined and form the basis of the new emerging field of
coherent atom optics.

The {\it atom laser} is the matter-wave analogue of the photon laser. It is
a coherent atom source ''pumped'' by an ultracold and dense ensemble, which
is stimulated to feed one mode of the atomic de Broglie field. A
Bose-condensate trapped in the ground state of a confining potential and fed
from an evaporatively cooled thermal cloud may already be understood as a
rudimentary stationary pulsed atom laser. The trap plays the role of the
laser cavity. However, while lasers can oscillate in any cavity mode, BECs
generally condense in the ground state of the trap. Similar to the invention
of the laser in the early sixties with all its fascinating scientific
applications, coherent atom sources will open new areas of fundamental
physics and applications, in part still unforeseen. These may include atom
interferometry, atom lithography, atom microscopy, atom holography, atom
sensoring or nanostructuring. Many applications demand dense, bright and
coherent sources of atoms in order to exhaust their capabilities. In this
respect, atom lasers are much superior to thermal atomic beams. While a
typical thermal beam has about $10^{-12}$\ atoms per mode, a Bose-condensed
mode contains $\gg 1$ atoms. The recent demonstrations of the experimental
feasibility of Bose-Einstein condensation \cite{AndersonMH95}\nocite
{Davis95b}-\cite{Han98}, \cite{Bradley97,Courteille98} boosted theoretical
and experimental work and accelerated the development of this whole field.

As an example, atomic holography may become practical with the availability
of spatially coherent matter-wave sources. Microfabricated holograms may
have typical dimensions of a few $100$~$\mu $m and minimum feature sizes on
the order of $1$~$\mu $m. Exploiting the repulsive self interaction, one may
let a BEC expand, pass it through a transmission hologram computed
judiciously to produce the desired diffraction pattern, and refocus it.
Small chromatic aberration is due to small velocity spread. The resolution
is limited on one hand by the atomic de Broglie wavelength, on the other
hand by the size of the smallest structures of the hologram \cite
{Goldstein99,Soroko97,Morinaga96} which can be made as small as $10$~nm.
Another important quantity, the resolving power, is limited by the number of
holes in the hologram and the velocity spread of the atoms. While the
reduction of the velocity spread by spatial filtering of an incoherent
atomic beam is only possible at the cost of huge loss in intensity, the
velocity spread of coherent matter-waves is at its quantum limits.

The present chapter reviews recent experimental work on coherent atom
optics. In order to place our topic into the right context, we start with a
very brief overview of conventional atom optics. We discuss the impact of
the advent of BEC on the field of atom optics in the Sections~5.2 and 5.3,
on the basis of recent realizations of atom lasers and interferometers.
Section~5.4 is devoted to experiments in nonlinear atom optics, Section~5.5
relates the recent demonstration of a coherent matter-wave amplifier and
gives a brief outlook on the evolution of the field of quantum optics with
atoms.

\section{Conventional Atom Optics}

The far-reaching analogy of light waves and atomic beams is a result of the
particle-wave duality and thus of the quantum nature of both light and
matter. It motivated de Broglie in 1924 to assign a wavelength to material
particles that depends on the particle's momentum: 
\begin{equation}
\lambda _{p}=\frac{\hbar }{p}.  \label{EqDeBroglie}
\end{equation}
Contrary to photons \cite{Lamb95}, there is no doubt about the fact that
atoms are (also) particles. Whether an atom rather behaves like a particle
or a wave depends on the specific experimental situation. In
interferometers, atoms interfere with themselves if their de Broglie
wavelength is coherently split and recombined. Atoms are capable of
interfering with one another if their de Broglie wavelength is larger than
their distance. This requires high densities and very low temperatures (at
least in some dimensions). In fact, what matters is not the small kinetic
energy of the atoms, but a small velocity spread, {\em i.e.} a high
phase-space density. At phase-space densities so high that the atomic de
Broglie waves get into contact, quantum statistical effects start to
influence the atomic dynamics, {\em i.e.} Bosons behave differently from
Fermions.

Analogously to the distinction between classical optics and laser optics, we
may divide the field of atom optics into conventional single-atom optics
with atoms that are not mutually coherent and atom optics with
Bose-condensed atoms. In conventional atom interferometers, one takes
advantage of the interference of every atom with itself, and most atom
optical devices do not rely on the mutual coherence of the atoms. On the
other hand, nonlinear interactions between the atoms make the dynamics of
coherent matter-waves interacting with atom optical devices much more
complex than single-atom optics. Atom optics with condensates offers the
advantage of large de Broglie wave amplitudes and ultra-long de Broglie
wavelengths. In fact, the coherence length of a BEC is equal to its physical
size. This has obviously an important impact on the sensitivity and
resolution of atom optical devices, as we shall soon see.

We will not go into details about conventional (single-particle) atom optics
here, since there are many excellent topical reviews \cite{Mlynek92}\nocite
{Adams94,Pillet94}-\cite{Prokhorov94}. However, for the sake of
completeness, we list below the most important atom optical devices that
have been developed and used in experiments.

\subsection{Atom Optical Devices}

In analogy to the manipulation of light beams by optical elements,
atom-optical components have been developed for manipulating atomic
matter-waves. The basic equipment of an optics lab consists of cavities,
lenses, refractive, dispersive and birefringent media, mirrors,
beamsplitters, transmission and reflection gratings, fibers, acousto-optic
modulators \cite{Mlynek92,Foot94}. The matter-wave counterparts of all those
elements have been realized today. Most elements exploit the interaction of
the mechanical degrees of freedom of the atoms with light. Note that the
atom optical devices only manipulate the atomic field density and the first
order coherence.

{\it Atomic beams} have a long history of applications in ultra-high
precision experiments, {\em e.g.} in atomic clocks \cite{Ramsey49}. Since
the development of laser cooling techniques, atomic beam slowing and cooling
has proven a powerful source for many applications \cite{Phillips82,Ertmer84}%
. Today, atomic beams are often used to load magnetic, optical, and
magneto-optical traps for atoms.

{\it Traps} for atoms are to some extent analogous to optical cavities for
light. In second quantization the radiation field inside a cavity is
described by harmonic oscillators, just like the motion of atoms confined in
a harmonic trap \cite{Blockley92}. At very low temperatures, $k_{B}T<\hbar
\omega _{trap}$, the effects of quantized motion can be directly observed 
\cite{Grynberg93,Wineland98}.

{\it Lenses} for atomic waves may be realized by exploiting the radiation
forces of laser beams, or if the atoms are moving within a waveguide, by
arranging for spatial or temporal variations of the fields \cite{Timp92} 
\nocite{Landragin96}-\cite{Ruprecht94}. {\it Mirrors} for atoms can be made
by a far blue-detuned evanescent wave emerging from the surface of a glass
substrate \cite{Balykin88}. Aspect and his group let atoms bounce in the
gravitational field on a curved and (for matter-waves) achromatic mirror
more then twenty times \cite{Henkel96}. This is a rudiment for a
gravitational cavity for atoms, where many atomic bosons could occupy one
cavity mode in analogy to optical resonators. Another option for atomic
mirrors is a microfabricated magnetic surface that repels the atoms
approaching the strongly inhomogeneous magnetic surface field \cite{Drndic99}%
. Already, falling Bose-Einstein condensates have been reflected from a far
blue-detuned sheet of light \cite{Bongs99}. {\it Gratings} are
microfabricated \cite{Pritchard99} or based on standing light waves. They
are at the heart of atom interferometers and already permitted the
development of high precision applications and experiments (atomic
gyroscopes, measurement of the gravitational acceleration $g$). {\it %
Waveguides} are the atom-optical counterpart of fibers. Forces that guide
the atoms can be exerted by electric or magnetic fields (single wire \cite
{Denschlag99}, quadrupolar waveguides \cite{Hinds98}), or by light beams via
the dipole interaction. Possible geometries are evanescent wave hollow
fibers \cite{Renn95,Ito96} or blue-detuned hollow-core laser beams \cite
{Schiffer98}. Recently, Bose-Einstein condensates have been transferred to
such hollow-core laser beam waveguides \cite{Bongs00}. Inhomogeneous
magnetic fields ({\em e.g.} magnetic trapping fields) act as Stern-Gerlach
filters and can be thought of as {\it matter-wave polarizers}.

De Broglie wave {\it frequency shifters} are the matter-wave analogue of
acousto-optical modulators (AOM). They have been implemented in conventional
atom interferometers \cite{Bernet96} and more recently in coherent
matter-wave optics \cite{Kozuma99}. The matter-wave experiment will be
discussed in Section~5.3.3. Finally, we want to mention atom holography \cite
{Morinaga96}, atom lithography \cite{Timp92} and atom microscopy \cite
{Doak99} as examples for the successful application of matter-wave optics.
It is very likely that the availability of coherent matter-waves will have a
strong impact on these fields, too.

\subsection{Atom Interferometers}

Atom interferometers split and recombine a single atom or an atomic ensemble
in time or in space (or both). If the temporal or spatial evolution is
coherent, we observe interference phenomena. In many experiments one
attempts to produce large splittings of the atomic de Broglie wave, but even
a motionless single particle can act as an interferometer and produce Rabi-
or even Ramsey fringes \cite{Huesmann99}. Recoil effects in the interaction
of atoms with light become important when the atoms are so cold that the
atomic momentum verges on the wavevector of the photons, ${\bf p}\approx
\hbar {\bf k}$. Each absorbed photon adds a quantized amount of momentum, $%
\hbar {\bf k}$, to the motion. The application of light-induced $\pi /2$
pulses to the atoms splits the de Broglie wave and entangles the internal
and motional degrees of freedom. Variations of this idea led to the
development of Ramsey-Bord\'{e} interferometers and atomic fountains \cite
{Weiss94}.

\section{Atom Laser}

Probably the most striking feature of Bose-condensed atoms is their mutual
(first-order) coherence spectacularly demonstrated by Andrews {\em et al.} 
\cite{Andrews97}. In quantum optics, the epitome of a coherent light source
is the laser, and we may ask, whether there is a matter-wave analogon, and
what the relationship between such an {\it atom laser} and a Bose-condensate
would be. In fact, we may already consider a BEC to be a rudimentary
stationary atom laser pulse, the trapping potential taking over the role of
the cavity. The atom laser, in the sense of a coherent atomic wave emitting
device, must satisfy a few more requirements. Generally, we ask for a
continuously working output coupler for a coherent atomic beam and an
irreversible pump process that refills the atom-lasing medium. Many theories
on atom lasers or bosers have been developed \cite{Wiseman95}\nocite
{Borde95,Spreeuw95,Holland96,Guzman96,Moy97,Wiseman97,Kneer98}-\cite
{Hutchinson99}, and we will not explain them here. Instead, along the lines
set by the analogy between optical and conventional atom optical devices, we
will briefly describe the experimental progress that has been made on the
way towards an atom laser that deserves this name.

\subsection{Bosonic Stimulation and Evaporation}

The gain mechanism for optical lasers can be understood as photons in a
laser mode stimulating atoms to emit more photons into the same laser mode.
The atom laser works similarly. The atoms trapped in a potential constitute
a thermal reservoir. Binary collisions redistribute the atoms over the
energy states. If a state already contains a population of $N$ atoms, the
Bose quantum statistics encourages an atom involved in a collision process
to join this state. The bosonic enhancement factor is proportional to $N+1$.
Bose-condensation is necessarily a result of bosonic stimulation. However,
the dynamics and the time scale of the formation process were
controversially discussed, until recent experiments performed at the MIT
clearly demonstrated that BECs form at finite times and develop long-range
order.

In order to directly observe bosonic stimulation, Miesner {\em et al.} \cite
{Miesner98} evaporatively cooled $2\times 10^{7}$ magnetically trapped atoms
close to the condensation threshold{\em \ }at $1.5~\mu $K. The final
temperature was set by the final {\em rf} frequency of the evaporation ramp.
Then they suddenly decreased the {\em rf} frequency by $200$~kHz, thus
initiating a fast truncation of the hot tail of the energy distribution. The
quick subsequent relaxation produced an oversaturated ''thermal'' cloud, and
the nucleation process and exponential growth of the BEC within the thermal
cloud was observed time-resolved by non-destructive dispersive imaging of
the atomic cloud (see Section~3.3.2).

Figure~\ref{FigStimulationCurves} shows the growth of the condensate atom
numbers towards equilibrium starting with various condensed atom numbers at
the time of the fast {\em rf} truncation. If no condensate was present, the
growth started slowly and increased exponentially until thermal equilibrium
was reached. The exponential acceleration of the growth is a clear
indication of bosonic stimulation and is in contrast to pure thermal
relaxation, which slows down exponentially. For the experimental conditions
(the trap secular frequencies were $\omega _{r}=2\pi \times 83$~Hz and $%
\omega _{z}=2\pi \times 18$~Hz) the formation of BEC took about $40$~ms,
while elastic collisions happened on the time scale of $2$~ms. The large
collision rate ensured that during the process of forced radio-frequency
evaporation (Section~3.1.4) the atomic sample is always held in thermal
equilibrium so that, even while crossing the phase transition to BEC, the
condensed fraction of atoms reflects the instantaneous temperature rather
than the dynamics of condensate formation.

Other experiments have confirmed the role of bosonic stimulation and
matter-wave amplification. They will be discussed in Section~5.4.4 and 5.5.2.

\subsection{Coherence and Interference}

First-order coherence and long-range order are necessary prerequisites for
the assignment of a single global phase to a condensate. The coherence
properties of BECs and the possibility to measure a condensate's phase have
in the past been questioned. The phase of a BEC is certainly not observable
by itself, but only the {\it relative phase} of two condensates. In
superconductors, phase-differences between the order parameters of coupled
systems are measured through Josephson-oscillations. For dilute gases, the
first-order coherence and long-range order of the condensate wavefunction
have been demonstrated in a remarkable experiment by observing matter-wave\
interference fringes generated by two overlapping condensates \cite
{Andrews97}.

Andrews {\em et al.} \cite{Andrews97} produced a cigar-shaped BEC made of $%
5\times 10^{6}$ atoms and no apparent thermal cloud in a cloverleaf trap
with secular frequencies $\omega _{r}\approx 140\cdot \omega _{z}\approx
2\pi \times 243$~Hz. They subsequently cut it into two parts distributed
along the weak axis with a $12$~$\mu $m thin laser light sheet (Fig.~\ref
{FigInterferenceScheme}). The laser light was blue-detuned by $75$~nm below
the $D2$ resonance, so that heating due to Rayleigh scattering could be
neglected. The two parts of the condensate were then released from the trap
by suddenly removing all magnetic fields and laser beams. During free
expansion, the condensates progressively overlapped and formed interference
fringes. After $40$~ms time-of-flight, the interference patterns were probed
by absorption imaging (Section~3.3.1). Standard absorption techniques only
sense the integrated column density and blur the images of the slightly
curved interference patterns (Fig.~\ref{FigInterferenceFringes}) \cite
{Wallis98}. Andrews {\em et al.} solved this problem by only probing atoms
within a $100$~$\mu $m thin slice orthogonal to the imaging direction. This
was achieved by selectively pumping the atoms within this slice to the $F=2$
hyperfine level of the groundstate which in turn is resonant to the probing
transition.

Two condensates interpenetrating at a velocity $v$ exhibit interference
fringes with a periodicity that corresponds to the relative de Broglie
wavelength $\lambda =h/mv$, where $v=d/t$. Here, $t$ is the time-of-flight
and $d$ is the initial separation of the BECs assumed to be ideal point
sources, but the finite extension makes only small modifications. The
interference patterns observed in experiment \cite{Andrews97} depended on
the initial separation of the condensates, on the time-of-flight and on the
way they were released from the trap (pulsed or cw). The interference fringe
contrast was found to be between $50\%$ and $100\%$. The interference
fringes periodicity was on the order of a few micrometers, which corresponds
to atoms having a kinetic energy much lower than the mean-field energy and
the zero-point energy of the harmonic trap. The reason for this is the large
anisotropy of the trapping potential: The released condensates expand
predominantly in radial direction, but are very slow in axial direction.

The ability of freely expanding condensates to interfere proves that there
are no random local phase-shifts during ballistic expansion, and that the
BECs preserve their long range order. The homogeneity of the intrinsic phase
of trapped BECs has recently been confirmed in other experiments \cite
{Stenger99,Hagley99b,Bloch00} (Section~5.4.2) based on the technique of
Bragg diffraction (Section~5.3.3). Simsarian {\em et al.} \cite{Simsarian00}
measured the evolution of the local phase of released condensates and found
that, under the influence of mean-field repulsion, the phase develops a
non-uniform profile during the ballistic expansion.

In the original interference experiment \cite{Andrews97}, the magnetic
trapping fields created in conjunction with the laser light sheet a {\it %
double-well potential}. However, the potential well was so large, that it
prevented tunneling between the condensates and decoupled their dynamics.
Different atom numbers in the condensates, imperfections in the exact
symmetry of the two traps and technical noise caused the condensate phases
to evolve independently and asynchronously. However, it might be possible in
future experiments ({\em e.g.} by employing a very narrow light sheet) to
allow for quantum tunneling and, ultimately, to observe Josephson
oscillation between two condensates (Section~5.3.2).

The degree of coherence ({\em i.e.} the amount of fluctuations in the field
amplitudes) is measured by the first-order correlation function. Similar to
optical double-slit experiments, the observed matter-wave interference only
indicates first-order coherence of the interfering beams. However,
signatures for higher-order short range coherence of condensates have been
found in other experiments: The second-order correlation function, which is
a measure of the amount of fluctuations in the field intensities, has been
estimated from measurements of the release energy of BECs \cite{Ketterle97}.
The third-order correlation function revealed itself by comparison of the
three-body recombination rates of condensed and thermal clouds \cite{Burt97}.

\subsection{Output Coupling}

\label{SecOutputCoupling}Coherent output coupling from Bose-condensates out
of magnetic traps is generally achieved by radiatively coupling trapped $%
|-\rangle $ and untrapped $|+\rangle $ Zeeman-states. A condensate of $N$
atoms driven with Rabi frequency $\Omega $ evolves with time $\tau $ into a
superposition $\left[ b_{-}|-\rangle +b_{+}|+\rangle \right]
^{N}=\sum_{n=0}^{N}\sqrt{\binom{N}{n}}~b_{-}^{N-n}b_{+}^{n}|N-n,n\rangle $,
with $b_{-}=\cos \Omega \tau /2$ and $b_{+}=\sin \Omega \tau /2$. The total
wavefunction describes an entanglement between trapped and untrapped states, 
$|N-n,n\rangle \equiv \sqrt{\binom{N}{n}~}|-\rangle ^{N-n}|+\rangle ^{N}$,
which is analog to the coherent splitting of a photon Fock state by an
optical beamsplitter. The inhomogeneous trapping potential acts like a
Stern-Gerlach filter and ejects the untrapped atoms. We have, however, seen
in the previous section, that the release process preserves the intrinsic
coherence of the released BEC, which propagates according to a single-mode
wave equation. The coupling between trapped and untrapped condensates
therefore remains truly coherent.

Coherent output coupling of parts of condensates out of magnetic traps has
been realized in several different ways. {\it Radiofrequency radiation} was
used for pulsed \cite{Mewes97,Martin99} and continuous \cite{Bloch99} output
coupling. Laser beams in {\it Raman configuration} have been used to create
a quasi-continuous, well-collimated coherent atomic beam \cite{Hagley99},
and a {\it mode-locked system} has been demonstrated \cite{AndersonBP98}. In
this section, we will briefly discuss the experiment of Mewes {\em et al.} 
\cite{Mewes97}.

The first output coupling experiment was performed by Mewes {\em et al.} in
a sodium condensate with $5\times 10^{6}$ atoms and no discernible thermal
fraction confined in a cigar-shaped cloverleaf trap with secular frequencies 
$\omega _{r}\approx 20~\omega _{z}\approx 2\pi \times 400$~Hz. The magnetic
trapping field had a bias of $B_{0}=1.1$~G, which removed the degeneracy of
the trapped $m_{F}=-1$ and untrapped $m_{F}=0,1$\ Zeeman states within the
lower hyperfine multiplet $F=1$. Mewes {\em et al.} coupled them via
radiofrequency radiation. With time, the system evolved into a coherent
superposition of Zeeman states $\left[ b_{-1}|-1\rangle +b_{0}|0\rangle
+b_{1}|1\rangle \right] ^{N}$, where $b_{-1}=\cos ^{2}\Omega \tau /2$ and $%
b_{0}=i\sqrt{2}\sin \Omega \tau $ and $b_{1}=-\sin ^{2}\Omega \tau /2$.
Atoms in the $m_{F}=1$ state were quickly repelled from the trap center by
the magnetic field, while atoms in the $m_{F}=0$ state were slowly
accelerated by gravity. The spatial dependence of the Zeeman shift in the
magnetic trap inhomogeneously broadened the radiofrequency resonance and
made the output coupling efficiency spatially dependent. This problem was
solved by either sweeping the radiofrequency through the resonance or by
applying pulses so short, that the Fourier broadening dominated the
inhomogeneous broadening. Repetitive application of $5$~$\mu $s long pulses
gave rise to the absorption images shown in Figure~\ref{FigOutputCoupling}.
By controlling the amplitude of the radiofrequency, the output coupling
could be adjusted between $0\%$ and $100\%$. In a subsequent experiment, it
was verified that the output coupling preserves the coherence by observing
interference fringes between outcoupled pulses \cite{Andrews97}. This also
shows that this output coupler may be understood as the analogue of a pulsed
mode-locked laser.

It is also important to consider collisions between the output coupled atoms
and the atoms remaining in the condensate. Those collisions represent losses
for the output mode and may even lead to bosonically fed momentum sidemodes
(Section~5.4.4). A low condensate density is advantageous for reducing the
atomic scattering. On the other hand, the BEC gets superfluid at high
densities thus allowing the dissipationless motion of the output coupled
atoms through the BEC. We remind here of the study of the motion of impurity
atoms through a condensate by Chikkatur {\em et al.} \cite{Chikkatur00}.

Within the small region of space occupied by the trapped condensate, the
magnetic field is harmonic to first order except for a tiny deformation at
the bottom side due to gravity. For the experiment described above, the
deformation corresponded to about $10~$mG magnetic field variation. Precise
tuning of the radiofrequency to this value results in a {\it spout} through
which slow atoms may continuously escape, thus generating continuous and
precisely localized output coupling. Of course this method requires very
stable magnetic fields. This method has been used by \cite{Bloch99}, to
create a quasi-continuous atom laser beam.

\bigskip

We have seen, that atom lasers can be built including all features that make
up an optical laser. We can generate coherent matter-waves taking atoms from
a thermal reservoir by irreversible bosonically stimulated scattering, and
we can couple (quasi-)continuous coherent atomic beams out of a single mode
of the trap. However, the mode only contains a finite number of condensed
atoms. In order to realize a true cw atom laser, an incoherent pump
mechanism that would continuously refill the BEC being depleted by output
coupling still remains to be developed.

\section{Atom Interferometry}

The most obvious use of an atom laser is within an atom interferometer. We
already saw in Section~5.2.2 that we get matter-wave interference by just
splitting and recombining a Bose-condensate.\ Andrews' experiment \cite
{Andrews97} thus realizes an external degree of freedom coherent matter-wave
interferometer, where the atoms in the interferometer arms are distinguished
by their being at different locations. It is also possible to build an
interferometer based on splitting the BEC in momentum space as we will see
in Section~5.3.3 \cite{Kozuma99}. Alternatively, we may consider BEC atom
interferometers, where the interfering components are in different internal
states (Zeeman-states \cite{Stenger98}, hyperfine states \cite{Myatt97},
dressed states \cite{Lee99}). We will briefly discuss an experimental
implementation of an internal state BEC interferometer in the following
section.

\subsection{Double Species Interferometer and Phase Measurements}

The possibility to coherently couple two-species Bose-condensates, {\em i.e.}
two BECs that are distinguishable by their internal degrees of freedom,
suggests their application on an internal-state time-domain atom
interferometer \cite{Huesmann99}. The phases of the two internal states $%
|\pm \rangle $ evolve according to their respective chemical potential, $%
\varphi _{|\pm \rangle }(t)=\mu _{|\pm \rangle }t$. The phases are not
observable, but their difference $\Delta \varphi (t)$ can be measured by
Ramsey interferometry.\ The idea of a Ramsey interferometer is the
following: First, a coherent superposition of the internal states is created
by coupling the two internal states for a short time. The two-level Bloch
vector then starts to precess according to the difference in the chemical
potentials. After a while, the internal states are mixed again, and the
Bloch vector is projected onto the internal state energy axis. The
population distribution between the internal states depends on the
accumulated precession angle. Thus, the Ramsey method of separated
oscillatory fields converts the phase measurement into a measurement of
populations, which can easily be carried out experimentally.

Hall {\em et al.} \cite{Hall98b} started with a single $|1,-1\rangle $ BEC
having a well-defined global phase. A first two-photon
microwave-radiofrequency pulse prepared a coherent superposition of $%
|1,-1\rangle $ and $|2,1\rangle $ BECs. With a resonant $\pi /2$ pulse, they
got $50\%$ population in both levels. The converted and the remaining atoms
were not immediately in the stationary ground-states of their respective
trapping potentials, because they had to adjust the spatial shape of their
condensate wavefunction to the modified conditions. The smaller partial atom
numbers, a slight change of the scattering length and of the trapping
potentials for the atoms turned into $|2,1\rangle $ altered the chemical
potentials for both states.\ So, it took some time for the two BECs to relax
to their respective ground states. During a time $T$, the two-level Bloch
vector freely precessed, and the BECs accumulated a differential phase
proportional to the difference in their chemical potentials. A second $\pi
/2 $ pulse now remixed the components. Finally, the populations in $%
|1,-1\rangle $ and $|2,1\rangle $ were separately probed via time-of-flight
imaging.

The TOP trap offers the possibility to precisely tune the relative
displacement of the two clouds. The interpenetration can be made
considerable and typically amounts to $20\%$.{\em \ }The overlap region
constitutes the interfering portion of the BEC interferometer, and its size
determines the fringe visibility. A simple model describes the fringe
contrast as a function of the local densities $n_{|F,-m_{F}\rangle }$. The
final population of the $|2,1\rangle $ state after completion of the Ramsey
sequence reads \cite{Hall98b}: 
\begin{equation}
n_{|2,1\rangle }^{f}({\bf r})=\tfrac{1}{2}n_{|1,-1\rangle }({\bf r})+\tfrac{1%
}{2}n_{|2,1\rangle }({\bf r})+\sqrt{n_{|1,-1\rangle }({\bf r})n_{|2,1\rangle
}({\bf r})}~\cos \left[ \int_{0}^{T}\Delta \mu ({\bf r},t)dt\right] .
\label{EqRamseyFringes}
\end{equation}
Ramsey fringes were recorded \cite{Hall98b} by repeating the whole sequence
of BEC creation, Ramsey interferometry and destructive imaging with
different free precession times $T$. In equation~(\ref{EqRamseyFringes}), we
assumed an inhomogeneous and time-dependent evolution of the {\it local}
relative phase at a rate proportional to the{\it \ local} difference in
chemical potentials $\Delta \mu ({\bf r},t)$. This assumption accounts for
the complicated transient relaxation of the two partial condensates into
their respective ground-states. The transients should lead to phase
diffusion in the spatial average and engender strong decoherence. However,
in experiment \cite{Hall98b} the fringes turned out to be surprisingly clear
and reproducible, thus indicating lower phase diffusion than naively
expected. The relaxation typically lasted $45$~ms, but even after $T=100$~ms
the double BEC system remembered the initial phases and could interfere.
Furthermore, $\Delta \mu ({\bf r},t)$ depends on the numbers of atoms in the
upper and lower BECs and therefore on the total condensed atom numbers. The
high fringe visibility indicates a very good experimental reproducibility.

\bigskip

Note that phase and atomnumber are non-commuting observables. Measuring the
difference in atomnumbers of two coupled Bose-condensates destroys the
relative coherence and decouples the BECs. {\it Internal} coherence of a BEC
means predictable phase between any two atoms. Atomnumber measurements yield
BEC number states, but of course this does not diminish the{\it \ inherent}
coherence of the BEC.

\subsection{Quantum Transport and Josephson Tunneling}

When two superconductors are brought into contact, a {\em dc} voltage that
is applied to a tunnel junction between the superconductors generates an
oscillating current proportional to the electric potential difference. This
phenomenon, called Josephson effect \cite{Josephson62}, is a general feature
of coupled macroscopic quantum systems and can be observed {\em e.g.} with
gaseous Bose-condensates confined in a double-well potential. Here, the
oscillating quantum current is proportional to the difference in chemical
potentials of the BECs.

Anderson {\em et al.} \cite{AndersonBP98} have directly observed another
manifestation of Josephson tunneling. They loaded a Bose-condensate into a
vertical standing light wave. Accelerated by gravity, the BEC tunneled from
antinode to antinode. Since the tunneling process was coherent, the partial
BECs quasi-trapped in the antinodes were phase-locked and interfered. This
feature is in close analogy to mode-locked lasers, so that the tunnel array
can also be considered a {\it mode-locked atom laser}.

Another example for coherent matter-wave tunneling is the experiment by
Stamper-Kurn {\em et al.} \cite{Stamper-Kurn99} on spinor BECs trapped in a
focussed far-detuned laser beam (Section~4.2.2). In a weak magnetic bias
field, the BEC was transferred into a superposition of the spinor components 
$|F,m_{F}\rangle =|1,1\rangle $ and $|1,0\rangle $, which were then
separated with a Stern-Gerlach type magnetic field gradient and formed spin
domains. Then the magnetic field gradient was reversed, thus generating a
force in the opposite direction. Since the Zeeman components are immiscible,
the domains repel each other. The energy barrier is higher than the chemical
potential of the domains. This means that the domains are metastable against
decay into their respective equilibrium positions. The experiment \cite
{Stamper-Kurn99} observed quantum tunneling of the spinor components through
each other and measured the tunneling rate.

\subsection{Bragg Diffraction}

We now turn our attention again to the external degrees of freedom of the
Bose-condensed atoms and take a closer look at the interaction of their
center-of-mass motion with light. We consider two laser beams with
wavenumbers $k_{\omega }=\omega /c$ and $k_{\omega +\Delta \omega }$
enclosing an angle $\vartheta $ (Fig.~\ref{FigBraggDiffractionScheme}). The
light field amplitude is described by 
\begin{equation}
E({\bf r,t)}=E_{0}\left[ \cos ({\bf k_{\omega }\cdot r}-\omega t)+\cos ({\bf %
k_{\omega +\Delta \omega }\cdot r-(}\omega +\Delta \omega )t)\right] \approx
E_{0}\cos ({\bf k_{\omega }\cdot r-\omega }t)~\cos (\tfrac{1}{2}{\bf q\cdot
r-}\tfrac{1}{2}\Delta \omega t),  \label{EqSuperposingField}
\end{equation}
where we defined ${\bf q}\equiv \hbar {\bf k}_{\omega }-\hbar {\bf k}%
_{\omega +\Delta \omega }$. The time-average over an oscillation period
yields the light intensity 
\begin{equation}
I({\bf r,t)}=I_{0}\left[ 1+\cos ({\bf q\cdot r-}\Delta \omega t)\right] ,
\label{EqSuperposingIntensity}
\end{equation}
which describes a one-dimensional light grating moving in the direction of $%
{\bf q}$ with a velocity that depends on $\Delta \omega $. A useful
approximation for the momentum transfer at small $\Delta \omega $ is: 
\begin{equation}
q\approx 2\hbar k_{\omega }\sin \vartheta /2.  \label{EqBraggAngle}
\end{equation}

The interaction process can be understood in two ways. In position space it
may be interpreted as {\it Bragg scattering}, {\em i.e.} stimulated Rayleigh
scattering of the atomic de Broglie wave at the optical grating induced by
the standing wave and subsequent interference of the phase-modulated de
Broglie sidemodes. Alternatively, it can be interpreted in momentum space as 
{\it Compton scattering}, {\em i.e.} stimulated Raman scattering between two
different motional states of the atoms. The recently observed Recoil-Induced
Resonances (RIR) \cite{Courtois94} are another manifestation of the same
process.

\bigskip

{\em Compton Picture.} --- In the Compton picture, the atoms being in the
standing wave light field may absorb photons from any of the two laser modes
and be stimulated to reemit the photons into the modes. Let us assume that
an atom with momentum ${\bf p}_{i}$\ first absorbs a photon of frequency $%
\omega $ from the laser beam ${\bf k}_{\omega }$ and is then stimulated by
the laser beam ${\bf k}_{\omega +\Delta \omega }$ to\ emit a photon of
frequency $\omega +\Delta \omega $ and to acquire the final momentum ${\bf p}%
_{f}$ (Fig.~\ref{FigBraggDiffractionScheme}). Because this is a two-photon
process, its amplitude is proportional to the square of the light field
amplitude, and thus to the light intensity: 
\begin{equation}
H_{Compton}({\bf r},t)\sim I({\bf r,t)}~\sum\nolimits_{{\bf p}_{i}}|{\bf p}%
_{f}\rangle \langle {\bf p}_{i}|+c.c.  \label{EqBraggHamiltonian}
\end{equation}
The momentum ${\bf q=p}_{f}-{\bf p}_{i}$ and the energy $\hbar \Delta \omega
=p_{f}^{2}/2m-p_{i}^{2}/2m$ are transferred to the atom, so that the atom
must follow the Bragg condition: 
\begin{equation}
\hbar \Delta \omega =\frac{q^{2}}{2m}+\frac{{\bf p}_{i}{\bf \cdot q}}{m}.
\label{EqBraggCondition}
\end{equation}
The Bragg condition~(\ref{EqBraggCondition}) can also be fulfilled by
higher-order Raman scattering processes, as we can see by substituting $%
q\rightarrow nq$, where $2n$ is the number of absorbed and reemitted
photons. This general case is depicted in Figure~\ref
{FigBraggDiffractionScheme} for ${\bf p}_{i}=0$.

\bigskip

{\em Bragg Picture.} --- In order to explain the scattering process in the
Bragg picture, we choose our reference frame so that ${\bf k}_{\omega }=-%
{\bf k}_{\omega +\Delta \omega }$ in equation~(\ref{EqSuperposingField}). In
this moving frame, the standing wave amplitude can be written as $%
E(z,t)=2E_{0}\cos k_{\omega }z\sin \omega t$. The (single-photon) Rabi
frequency $\Omega $ generated by a single travelling wave laser beam has
been introduced in equation~(\ref{EqDipolePotential}). For large red-detuned
laser frequencies, $\left| \Delta \right| \gg \Omega $, the standing wave
creates a light shift modulation described by 
\begin{equation}
U(z)=U_{0}\cos ^{2}k_{\omega }z,  \label{EqStandingWave}
\end{equation}
where $U_{0}=\hbar \Omega ^{2}/\Delta $ according to equation~(\ref
{EqFORTPotential}). Consequently, the condensate matter-wave develops a
spatial phase modulation according to: 
\begin{equation}
\psi (z,t)=\psi _{0}(z)\exp \left[ i\hbar ^{-1}U(z)t\right] =\psi
_{0}(z)\sum\nolimits_{n}\Im _{n}^{2}(U_{0}t/2\hbar )\exp \left( 2\pi
ink_{\omega }z\right) .  \label{EqBraggSidemodes}
\end{equation}
The condensate wavefunction evolves into a superposition of sidemodes, which
are just the diffraction orders of the Bragg scattering and whose strengths
are given by the Bessel functions $\Im _{n}$. The diffraction efficiency
increases with laser intensity and with time.

The above description neglects the atomic motion during the interaction with
the standing wave. This {\it thin grating approximation} can be satisfied in
experiment by irradiating the standing wave only for very short times. The
time scale is set by the oscillation period of the atoms in the optical
potential valleys generated by the standing wave via the dipole force
interaction. At the locations of the antinodes, we may harmonically
approximate the potential and introduce the secular frequencies $\omega
_{opt}$ by $\frac{m}{2}\omega _{opt}^{2}=k_{\omega }^{2}U_{0}$. The thin
grating approximation holds for laser pulse durations $\tau \ll 2\pi /\omega
_{opt}$. For longer pulse durations, in the {\it thick grating limit}, the
atoms perform on average several oscillations in the optical potential
during the interaction time. This causes periodic focussing (decollimation)
and defocussing (collimation), which manifests itself in a oscillating Bragg
diffraction efficiency.

Bragg diffraction of Bose-condensates was first experimentally observed by
Kozuma {\em et al.} \cite{Kozuma99} in the thin grating limit. They briefly
irradiated a standing wave into a trapped Bose-condensate, then released the
BEC from the trap and recorded the momentum distribution with standard
time-of-flight imaging. They observed a splitting of the condensate
wavefunction into the Bragg diffracted modes. The efficiency of the Bragg
diffraction could be made as high as $100\%$. By variation of the relative
detuning $\Delta \omega $, the diffraction orders could be selected.
Subsequent experiments also investigated the thick grating limit \cite
{Ovchinnikov99}, by applying the standing wave pulse to released condensates
and arranging for large secular frequencies $\omega _{opt}$.

The BEC Bragg scattering technique described above displays many
similarities with acousto-optical modulators (AOMs), which are commonly used
in laser optics. However, while AOMs deflect photons passing through the
interaction zone, the matter-wave Bragg scattering described here is a
time-domain process. This diffraction method constitutes an important atom
optical device that will certainly prove a powerful tool in many
applications. It has already been used to excite phonons in a controlled way
(Section~5.4.2) and to study the intrinsic phase of a condensate \cite
{Stenger99,Hagley99b}. In the reference~\cite{Hagley99b}, small condensate
replica sequentially generated from a large BEC by coherent Bragg
diffraction interfered with each other and yielded information about
intrinsic phase variations of the BEC. In reference~\cite{Deng99b}, the
Bragg diffraction scheme has been extended to demonstrate a time domain
matter-wave analogue of the Talbot effect. And in reference~\cite
{Simsarian00}, a Bragg diffraction interferometer has been used to map the
autocorrelation function of a BEC and to image its phase evolving in time.

\section{Nonlinear Atom Optics}

In classical nonlinear optics the interaction between matter ({\em e.g.}
dilute gases) and light is described by Maxwell's equations: 
\begin{eqnarray}
{\bf P}({\bf r},t) &=&\chi ({\bf E}){\bf E}({\bf r},t)=\chi ^{\left(
1\right) }\cdot {\bf E}+\chi ^{\left( 3\right) }:{\bf EEE}+...
\label{EqMaxwellEquation} \\
\square {\bf E}({\bf r},t) &=&\frac{4\pi }{c}\stackrel{..}{{\bf P}}({\bf r}%
,t).  \nonumber
\end{eqnarray}
The electromagnetic field ${\bf E}$ creates a macroscopic polarization ${\bf %
P}$, which in turn acts back on the field via $\square {\bf E}$. Higher
order processes like self-focussing, second harmonic generation, four-wave
mixing, {\em etc.} are described by the nonlinear susceptibility $\chi
^{\left( 3\right) }$. These processes require the presence of a nonlinear
medium, because the polarizability of the vacuum itself is pretty small. For
visible wavelength the photon-photon scattering cross-section is well
approximated by $45^{-2}~(973/5)~(\alpha ^{4}/\pi )~(\hbar ^{8}\omega
^{6}/m_{e}^{8}c^{14})$, which is only on the order of $10^{-63}$~cm$^{2}$ 
\cite{Karplus50} and very difficult to reach even with high intensity
lasers. In contrast to this, the scattering cross-section for shapeless
two-body collisions in ultracold sodium gases is on the order of $2\times
10^{-12}$~cm$^{2}$, so that two-body collisions are frequent processes at
currently available densities and temperatures.

\subsection{Self-Defocussing}

Two-body collisions play a role in coherent matter-wave optics which is very
similar to that of the nonlinear susceptibility in quantum optics. Within
the mean-field theory, the groundstate wavefunction of the condensate is
described by the {\it nonlinear Schr\"{o}dinger equation}: 
\begin{equation}
\left[ \frac{-\hbar ^{2}}{2m}\Delta +U_{trap}({\bf r})+g\left| \psi ({\bf r}%
,t)\right| ^{2}\right] \psi ({\bf r},t)=\mu \psi ({\bf r},t),
\label{EqNonlinearSchroedinger}
\end{equation}
where $g\equiv 4\pi \hbar ^{2}a/m$. The nonlinear term describes the
condensate self-interaction and is analogous to the third order contribution
to the polarization in the nonlinear Maxwell equations~(\ref
{EqMaxwellEquation}). If the atomic interaction is repulsive, the nonlinear
term causes the condensate to expand as far as the trapping potential
permits. This behavior is analogous to the nonlinear optical
self-defocussing in local Kerr media with instantaneous response. For large
condensates, the self-interaction can be so overwhelming, that the kinetic
energy may be neglected (at least in the center of the trap, where the
density is highest). This approximation defines the so-called {\it %
Thomas-Fermi limit}.

\subsection{Dispersion}

The nonlinear mean-field interaction in a weakly interacting condensate is
at the origin of the phenomenon of dispersion, {\em i.e.} the de Broglie
wavelength of a single atom with a given momentum ${\bf p}$ inside the
condensate depends on the local density. For homogeneous condensates, the
dispersion relation~(\ref{EqBogoliubovDispersion}) can easily be derived
from the semiclassical Bogolubov equations (Section~2.6). In the
Thomas-Fermi limit, the region inside the condensate has a nearly
homogeneous density, $n({\bf r})\approx n_{0}$, so that the Bogolubov
dispersion describes the excitation spectrum quite well. For low excitation
energies, $\varepsilon _{rec}({\bf p})\equiv {\bf p}^{2}/2m\ll gn_{0}$, the
spectrum is {\it phonon-like} (quasi-particle-like): 
\begin{equation}
\varepsilon _{phon}({\bf p})=c_{s}{\bf p}\text{ \ \ \ \ \ \ \ \ where \ \ \
\ \ \ \ }c_{s}=\sqrt{gn_{0}/m}.  \label{EqPhononSpectrum}
\end{equation}
The excitation energy then depends linearly on the momentum, and density
perturbations travel without spreading inside the condensate at the speed $%
c_{s}$ of the Bogolubov $0^{th}$ sound. In contrast, for high energy
excitations, ${\bf p}^{2}/2m\gg gn_{0}$, the spectrum is particlelike: 
\begin{equation}
\varepsilon _{part}({\bf p})={\bf p}^{2}/2m+gn_{0}.
\label{EqParticleSpectrum}
\end{equation}

Phonon-like collective excitations have been driven very soon after the
achievement of Bose-Einstein condensation using trap modulation methods
(Section~4.3.1). The excitation energies were quite low, {\em i.e.} in the
same order of magnitude as the trap secular frequencies, $\varepsilon
_{phon}({\bf p})\approx \hbar \omega _{trap}$. The de Broglie wavelength of
the phonons is then comparable to the condensate size, so that the phonon
spectrum is influenced by boundary conditions. It is interesting to tap
other regimes of excitation energies that are free from this limitation. The
newly developed Bragg diffraction technique can be employed to {\it %
optically imprint} high energy phonons and even particle-like excitations
into the condensate \cite{Stamper-Kurn99b} and thus to investigate the
boundary between these two regimes. Bragg diffraction has been observed
earlier with a two laser beam standing wave arrangement as splitting of the
matter-wave in momentum space \cite{Kozuma99} (Section~5.3.3). The energy
transfer $\varepsilon _{rec}({\bf p})$ could be tuned by adjusting the angle
between the two laser beams according to equation~(\ref{EqBraggCondition})
(Fig.~\ref{FigBraggDiffractionScheme}). We have also seen, that the
efficiency of Bragg scattering atoms into the first diffraction order
depends on the fulfillment of the Bragg condition~(\ref{EqBraggCondition}), 
{\em i.e.} Bragg scattering is velocity-selective. One can therefore measure
the number of deflected atoms versus the relative detuning of the lasers
that drive the Raman transition, record the recoil-induced resonances and
call this procedure {\it Bragg spectroscopy} \cite{Stenger99}. The spectrum
closely reflects the momentum distribution of the atoms. Since the
mean-field interaction causes a finite momentum spread of the condensate
wavefunction, the shift and broadening of the RIR reveal detailed
information about the condensate self-interaction (Fig.~\ref
{FigBraggDiffractionScheme}).

For a real condensate, there are several contributions to the width of the
momentum distribution: 1)\ The finite size of the trapped condensate limits
the width of the momentum distribution according to Heisenberg's uncertainty
relation \cite{Baym96}. 2) The inhomogeneous density distribution of the
trapped condensate shifts and smears out the momentum distribution in
equation~(\ref{EqParticleSpectrum}). Since this is an inhomogeneous
broadening, it adds to the other linewidth as a quadrature sum. 3) The
finite length of the Bragg scattering pulse produces a broadening\ analogous
to the time-of-flight broadening in atomic beam spectroscopy, which is
inversely proportional to the pulse length. 4) Acoustic noise may
Doppler-broaden the linewidth of the frequency difference of the lasers and
reduce the resolving power of the Bragg spectroscopy scheme. The shifts and
broadenings of the recoil-induced resonances have been calculated for a
realistic condensate density distribution and verified in two experiments,
one carried out in the particle regime \cite{Stenger99} and one in the
phonon regime \cite{Stamper-Kurn99b}.

Stenger {\em et al.} \cite{Stenger99} performed the particle regime Bragg
scattering experiment, $\varepsilon _{rec}({\bf p})\gg gn_{0}$, with
counterpropagating laser beams. For this case, the recoil shift for sodium
condensates at fulfilled Bragg condition~(\ref{EqBraggCondition}) was $%
\varepsilon _{rec}({\bf p})=h\times 100$~kHz, which was much larger than the
mean-field energy at typical condensate densities, $gn_{0}=g\times 5\times
10^{14}$~cm$^{-3}=hg\times 7.3$~kHz. The experiment could closely reproduce
the expected shift and broadening of the RIR shown in Figure~\ref
{FigBraggSpectroscopy} ({\bf a}).

Stamper-Kurn {\em et al.} \cite{Stamper-Kurn99b} carried out the phonon
regime Bragg scattering experiment with laser beams enclosing an angle of $%
14^{o}$. In this case, the recoil shift at fulfilled Bragg condition~(\ref
{EqBraggCondition}) was $\varepsilon _{rec}({\bf p})=h\times 1.54$~kHz,
which was now smaller than the mean-field energy at typical densities. The
results of this experiment were found in good agreement with calculations of
the shift and strength of the RIR shown in the Figures~\ref
{FigBraggSpectroscopy} ({\bf b}) and ({\bf c}). In order to understand the
density dependence of the RIR, we first have a look at the spectrum of light
scattered from a homogeneous dilute gas of atoms. If the gas is
non-degenerate, the spectrum mirrors the velocity distribution of the atoms.
In the presence of condensed atoms, photon recoil events that take atoms to
an already occupied state are enhanced by Bose-stimulation, and if the atoms
do not interact, according to Javanainen {\em et al.} \cite{Javanainen95},
the spectrum $S({\bf p},\varepsilon )$ exhibits two characteristic peaks at $%
\varepsilon =\pm \varepsilon _{rec}({\bf p})$. Later, Graham {\em et al.} 
\cite{Graham96} extended the calculations by taking into account two-body
collisions and found the characteristic peaks at an energy $\varepsilon =\pm
\varepsilon _{Bog}({\bf p})$ given by the Bogolubov dispersion relation~(\ref
{EqBogoliubovDispersion}): 
\begin{equation}
S({\bf p,}\varepsilon )=\frac{\varepsilon }{\varepsilon _{Bog}({\bf p})}%
~\delta (\varepsilon _{Bog}({\bf p})-\varepsilon ).
\label{EqStructureFactor}
\end{equation}
The experiments of Stamper-Kurn and Stenger \cite{Stenger99,Stamper-Kurn99b}
measured exactly these spectra. However, instead of looking at scattered
photons, they analyzed the shifts, widths and strengths of the
recoil-induced Bragg resonances. They measured, in particular, the line
strength, $S({\bf p})=\int S({\bf p,}\varepsilon )d\varepsilon $, and the
shift from the free particle resonance, $\Delta \varepsilon \equiv
\varepsilon _{Bog}({\bf p})-\varepsilon _{rec}({\bf p})$, as a function of
the mean-field energy. In order to compare with the experiment, the formula~(%
\ref{EqStructureFactor}) needs to be slightly modified to take into account
the inhomogeneity of the trapped condensate. Figures~\ref
{FigBraggSpectroscopy} ({\bf b}) and ({\bf c}) show the shift and strength
of the RIR as a function of the condensate density. At low densities, when
the excitations are particle-like, the line shift tends to zero, $%
\varepsilon _{Bog}({\bf p})\approx \varepsilon _{part}({\bf p})\rightarrow
\varepsilon _{rec}({\bf p})$, and the line strength tends to its maximum
value, $S({\bf p})\rightarrow 1$. At high densities, the excitations are
phonon-like, $\varepsilon _{Bog}({\bf p})\approx \varepsilon _{phon}({\bf p}%
)>\varepsilon _{rec}({\bf p})$, and the RIR is shifted towards higher
energies, while the line strength rapidly decreases. The relative weakness
of phonon-like excitations is due to the presence of correlated pair
excitations. The direct comparison of the two regimes of excitations thus
reveals important information about correlation effects \cite
{Stamper-Kurn99b}.

It is interesting to note, that the spectrum is equivalent to the structure
factor, which is itself the Fourier transform of the density correlation
function of the condensate quantum field. The structure factor plays a
similar role in the theory of many-body Schr\"{o}dinger fields as the
familiar $Q$-function in quantum optics. The correlations are probed by
scattering quasiparticles back and forth: 
\begin{equation}
S({\bf p})\sim \langle g|\widehat{a}_{{\bf p}}\widehat{a}_{{\bf p}}^{+}+%
\widehat{a}_{-{\bf p}}^{+}\widehat{a}_{-{\bf p}}+\widehat{a}_{-{\bf p}}^{+}%
\widehat{a}_{{\bf p}}^{+}+\widehat{a}_{{\bf p}}\widehat{a}_{-{\bf p}%
}|g\rangle ,  \label{EqStructureFactorDefinition}
\end{equation}
where $\widehat{a}_{{\bf p}}$ stands for the annihilation of a phonon with
wavevector ${\bf p}$.

\subsection{Second Harmonic Generation}

The elementary excitations ({\em i.e.} small oscillations around the
many-body ground state) discussed in Section~4.3.1 are well described by a
linearized Gross-Pitaevskii equation. In contrast, large amplitude
oscillations are sensitive to anharmonicities induced by the nonlinear
mean-field interaction. Nonlinear effects may result in frequency shifts of
the normal modes and mode coupling.

For mode coupling, the anisotropy of the trapping potential plays an
important role. Dalfovo {\em et al. }\cite{Dalfovo97} calculated the
excitation frequencies for the normal modes of cylindrically symmetric traps
($\omega _{r},\omega _{z}$). The modes are usually labelled with the
projection of the angular momentum onto the symmetry axis $m$. The lowest
lying modes are the breathing mode (high-lying $m=0$), the radial
compression oscillation with axial sloshing (low-lying $m=0$), and the
quadrupolar radial shape oscillation ($m=2$). The oscillations depend
differently on variations of the trap geometry. For example, at the aspect
ratio $\omega _{z}/\omega _{r}=\frac{1}{6}\sqrt{77+5\sqrt{145}}$, the
excitation frequencies are shifted such that $\omega _{high}(m=0)=2\omega
_{low}(m=0)$. Thus, through active control of the trap aspect ratio, it is
possible to arrange for degeneracies of the modes where the anharmonic
mixing diverges. Under such conditions, frequency doubling effects should
occur analogous to {\it Second Harmonic Generation} (SHG) in quantum optics.

Second harmonic generation has recently been observed in the collective
dynamics of a Bose-Einstein condensate by Hechenblaikner {\em et al.} \cite
{Hechenblaikner00}. They modified the potential of their TOP trap by an
additional magnetic field oscillating along the symmetry axis with twice the
frequency of the rotating bias field. In the time-average, this trap has a
variable aspect ratio which can be set by the amplitude of the additional
field. Similar to earlier experiments \cite{Jin96,Mewes96b}, the
hydrodynamic mode was excited by sinusoidal modulation of the rotating bias
field amplitude. The response of the condensate wavefunction, {\em i.e.} the
shape oscillation, was observed by standard time-of-flight imaging. When the
aspect ratio of the trap was set to the degeneracy condition, the condensate
responded nonlinearly by oscillating with twice the driving frequency.

\bigskip

In contrast to light, the material de Broglie wave also depends on the
particle's mass. Therefore, modifying the mass and keeping the momentum
fixed modifies the de Broglie wavelength. Two free atoms can be coherently
coupled to a molecular bound state. The coupling may be realized through a
Feshbach resonance \cite{Timmermans99} (Section~6.1) or by exciting a Raman
transition with laser beams \cite{Javanainen99} (Section~6.3). This process
may also be understood as Second Harmonic Generation.

\subsection{Four-Wave Mixing and Phase Conjugation}

\label{SecFourWaveMixing}The idea of phase conjugation with coherent
matter-waves has been proposed by Goldstein {\em et al.} \cite{Goldstein96}.
The authors proposed dropping a condensate onto a {\em cw} standing light
wave which was tilted by the Bragg angle from the horizontal plane. When
falling through the standing wave, a first-order Bragg diffracted BEC would
be generated. This wavepacket would four-wave mix with the zero-order
diffracted BEC and the falling BEC to create a phase conjugate BEC. Just
recently, {\it Four-Wave Mixing} (4WM) has been experimentally demonstrated.
Slightly different from the proposal \cite{Goldstein96}, Deng {\em et al.} 
\cite{Deng99} produced three condensates out of one right inside the trap
using the method of Bragg scattering described in Section~5.3.3. The
scattering process produced the three condensate parts in the same region of
space, but with different momenta. The initially overlapping condensates
carried out {\it half collisions} that nonlinearly mixed the de Broglie
waves before they flew apart.

The temporal evolution of four-wave mixing BEC wavepackets has been
numerically investigated by Trippenbach {\em et al.} \cite{Trippenbach98}.
They considered three BEC wavepackets with the initial atomnumbers $%
N_{j}^{0} $\ and wavefunctions, $\psi _{0}({\bf r-r}_{j})$, $j=1,2,3$, each
one being the solution of a Gross-Pitaevskii equation (GPE) with a potential
centered around ${\bf r}_{j}$. The initial locations ${\bf r}_{j}$ and the
initial momenta ${\bf p}_{j}$ were chosen to let the three wavepackets
perform {\it full collisions}. The evolution of the total wavefunction $\psi
_{tot}({\bf r},t)$ starting from the initial state $\psi _{tot}({\bf r}%
,0)=\sum_{j=1}^{3}N_{j}^{0}\psi _{0}({\bf r-r}_{j})\exp (\frac{i}{\hbar }%
{\bf p}_{j}\cdot {\bf r})$ was monitored by solving the time-dependent GPE~(%
\ref{EqGrossPitaevski}). The wavepackets mix due to the nonlinear mean-field
interaction term in the GPE giving birth to new wavepackets $\psi _{4}\sim
g\psi _{j}^{+}\psi _{m}\psi _{n}\exp \frac{i}{\hbar }{\bf p}_{4}\cdot {\bf r}
$ with momenta ${\bf p}_{4}=-{\bf p}_{j}+{\bf p}_{m}+{\bf p}_{n}$. Mixing
configurations like $\psi _{j}^{+}\psi _{j}\psi _{j}$ and $\psi _{j}^{+}\psi
_{j}\psi _{m}$\ do not produce wavepackets with new momenta ${\bf p}_{4}\neq 
{\bf p}_{j},{\bf p}_{m},{\bf p}_{n}$, but describe self-phase modulation
(Section~5.4.1) and cross-phase modulation, respectively. Only terms that
combine atoms from all three wavepackets can produce new momenta. Further
restrictions on the possible mixing configurations $j,m,n=1,2,3$ arise from
particle number, momentum and energy conservation laws: 
\begin{eqnarray}
N_{4}
&=&-N_{j}+N_{j}^{0}=N_{m}-N_{m}^{0}=N_{n}-N_{n}^{0}=\tsum\nolimits_{\kappa
=1}^{3}\left( N_{\kappa }^{0}-N_{\kappa }\right)  \label{EqConservation} \\
{\bf p}_{4} &=&-{\bf p}_{j}+{\bf p}_{m}+{\bf p}_{n}  \nonumber \\
p_{4}^{2} &=&-p_{j}^{2}+p_{m}^{2}+p_{n}^{2}~.  \nonumber
\end{eqnarray}

In order to generate three BEC wavepackets with different momenta, Deng {\em %
et al.} \cite{Deng99} applied two short Bragg diffraction sequences in rapid
succession. The geometry of the standing wave laser beams is shown in Figure~%
\ref{FigFourWaveMixing} ({\bf a})\ in the laboratory frame. The first
standing wave was generated by lasers ${\bf k}_{1}$ and ${\bf k}_{2}$
detuned from one another, so that the Bragg condition~(\ref{EqBraggCondition}%
) was satisfied and the momentum ${\bf p}_{2}=\hbar {\bf k}_{1}-\hbar {\bf k}%
_{2}$ was imparted to the diffracted atoms. The second standing wave was
formed by the lasers ${\bf k}_{1}$ and ${\bf k}_{3}=-{\bf k}_{1}$ and
transferred the momentum ${\bf p}_{3}=2\hbar {\bf k}_{1}$ to the atoms. The
durations and intensities of the standing waves were adjusted to distribute
the atoms in more or less equal parts into the three momentum states ${\bf p}%
_{1}=0,$ ${\bf p}_{2}$ and ${\bf p}_{3}$. A fourth momentum state ${\bf p}%
_{4}$ was generated by four-wave mixing.

The conservation laws only permit processes that can be viewed as degenerate
4WM in an appropriate reference frame. Figure~\ref{FigFourWaveMixing} ({\bf b%
})\ shows the process in a moving frame defined by ${\bf p}_{1}\equiv -{\bf p%
}_{3}$. Two atoms from $\psi _{1}$ and $\psi _{3}$ are bosonically scattered
by an atom from $\psi _{2}$ into the wavepackets $\psi _{2}$ and $\psi _{4}$%
. Each of the wavepackets $\psi _{1}$ and $\psi _{3}$ sacrifices $N_{4}$
atoms to create the new wavepacket $\psi _{4}$ and to increase the
wavepacket $\psi _{2}$. The redistribution is coherent. Figure~\ref
{FigFourWaveMixing} ({\bf c})\ shows the process in a moving frame defined
by ${\bf p}_{1}\equiv -{\bf p}_{2}$. Energy conservation only allows the
terms satisfying $p_{4}=p_{3}$. These terms are $\psi _{1}^{+}\psi _{2}\psi
_{3}$\ and $\psi _{2}^{+}\psi _{1}\psi _{3}$. In this frame, the process may
be interpreted as Bragg scattering of wavepacket $\psi _{3}$ by the
matter-wave grating formed by $\psi _{1}$ and $\psi _{2}$. The wavepacket $%
\psi _{4}$ is just the first-order Bragg diffracted wavepacket. In contrast
to Bragg diffraction at an optical grating (Section~5.3.3), Bragg
diffraction at a matter-wave grating relies on nonlinear mixing by two-body
collisions. The amount of redistributed atoms therefore depends on
parameters like\ the atomic interaction strength, the condensate size, and
the collision time, {\em i.e.} the time that the wavepackets spend together
before they separate. Time-of-flight images of the total condensate
wavefunction after 4WM are shown in Figure~\ref{FigFourWaves}.

The occurrence of four-wave mixing was foreseeable in view of the
equivalence between the nonlinear coupling strength $g$ in the
Gross-Pitaevskii equation~(\ref{EqNonlinearSchroedinger}) and the
higher-order susceptibility $\chi ^{(3)}$ in nonlinear optics, which is
known to produce such phenomena. But despite the similarities with the
optical counterpart, four-wave mixing with matter-wave is fundamentally
different. Particle numbers must be conserved and the energy-momentum
dispersion relation is different from the one that holds for massless
photons. Furthermore, while photons generally require the presence of a
nonlinear medium to undergo higher-order processes, the atomic matter-waves
mix via binary collisions.

\subsection{Spin Mixing}

In the four-wave mixing scheme discussed above, the nonlinearly interacting
condensates are distinguished by their different center-of-mass momenta.
Another possibility is to nonlinearly mix overlapping BECs in different
internal states, {\em e.g.} Zeeman substates. The experimental feasibility
of confining spinor condensates of sodium atoms distributed over all $F=1$
hyperfine states in the same trap has triggered extensive theoretical work 
\cite{Goldstein99,Law98}. Spin-exchange interactions constantly mix the
different spin components and drive complex nonlinear spin population
dynamics. For example, two $m_{F}=0$ atoms may collide and change their
internal state to one $m_{F}=-1$ and one $m_{F}=+1$ atom. A recent
experiment has demonstrated, how a condensate (initially in the $m_{F}=0$
state) evolves into a mixture of populations of all three hyperfine states
and subsequently forms spin domains \cite{Stenger98}. However, the
observation of nonlinear spin mixing is a challenge still lying ahead.

\subsection{Dielectric Properties of Bose-Einstein Condensates}

\qquad In the preceding sections, we discussed several matter-wave effects
with Bose-Einstein condensates that were due to their intrinsic
collision-induced nonlinearity. However, regardless of this {\it atom
optical nonlinearity}, Bose-condensed gases can also behave as highly {\it %
dielectric media for light} and be useful objects for studies in nonlinear
quantum optics.

Under normal circumstances, the refractive index of a gas can only be
increased at the detriment of transmission. However, in a gas of
laser-driven $\Lambda $-shaped atomic three-level systems, quantum
interference can occur cancelling out the absorption and leaving transparent
the otherwise opaque medium. The phenomenon is termed {\it %
Electromagnetically Induced Transparency} (EIT). In this system, when both
lasers are tuned to resonance, the excited state is not\ populated and can
be adiabatically eliminated. Upon tuning one of the lasers, a {\it dark
resonance} can be observed whose width is power-broadened by the laser
intensities, if Doppler broadening, broadening by laser phase fluctuations
or by collisions between atoms can be neglected \cite{Siemers92}. Close to
the dark resonance, the dispersion ({\em i.e.} the frequency dependence of
the refractive index) is very large. It depends on the width of the dark
resonance. The group velocity for a propagating light pulse is $%
v_{g}=c\left( n(\omega _{probe})+\omega _{probe}\frac{dn}{d\omega _{probe}}%
\right) ^{-1}$, where $n(\omega _{probe})$ is the refractive index at the
probe beam frequency $\omega _{probe}$. The propagation velocity is slowed
down if the dispersion is large \cite{Kasapi95}.

Taking advantage of their sodium BEC apparatus, Hau {\em et al.} \cite{Hau99}
produced a dense ($n\sim 8\times 10^{13}$~cm$^{-3}$) gas of ultracold ($%
T\sim 400~$nK) atoms in an oblong magnetic trap and probed the cloud {\it %
in-situ} and {\it time-resolved}. They shone along the long axis of the
cloud a short pulse of circularly polarized probe light resonantly tuned
between the levels $|F^{\prime }=2,m_{F}^{\prime }=-2\rangle $ and $%
|F=1,m_{F}=-1\rangle $. The transmitted pulse is detected with a
photomultiplier. In the presence of a low-intensity linearly polarized laser
beam irradiated perpendicular to the probe beam and tuned between the levels 
$|F^{\prime }=2,m_{F}^{\prime }=-2\rangle $ and $|F=2,m_{F}=-2\rangle $
which dressed the atomic cloud and kept the probe light from being absorbed,
the light pulse was delayed. If BECs were used, the delay corresponded to a
speed of light on the order of only $17~$m/s. This corresponds to an
unprecedentedly large nonlinear refractive index. Inouye {\em et al.} \cite
{Inouye00}\ later reported light group velocities of $1~$m/s in the context
of their experiment on the amplification of light and atoms in a BEC
(Section~5.5.2). It is worth mentioning, that the effect does not require
quantum degeneracy, but rather high density and low temperature, and a
similar reduction of the speed of light has subsequently been observed in
hot gases \cite{Kash99}. The group velocity reduction scales with the gas
density and inversely with probe beam intensity. At low temperature, one can
afford lower probe beam intensity without being dominated by the Doppler
effect.

Such strong nonlinearities may prove useful for a variety of applications in
nonlinear quantum optics. An interesting proposal \cite{Leonhardt00} points
out, that strongly dielectric moving media may exhibit detectable
relativistic effects of light when the speed of light gets comparable to the
local speed of sound or the flow of mass. In particular, a vortex flow
imprints a long-ranging topological phase shift on incident light that can
be understood in terms of an optical Aharonov-Bohm effect. This may prove
useful for the detection of quantum vortices in BECs (Section~4.3.4). At
short ranges, vortices should behave similar to gravitational black holes
and deviate light towards the vortex singularity. Beyond an\ ''optical
Schwarzschild radius'', the light is trapped by the vortex.

\section{Coherent Coupling of Optical Fields and Matter-Waves}

\subsection{Superradiant Rayleigh Scattering}

An early example for the influence of the dynamic coupling between optical
fields and matter-waves on the center-of-mass motion of the material system
is the {\it Free Electron Laser} (FEL). In this device, a combination of
periodic magnetic and optical fields causes a spatial density modulation of
a relativistic electron beam. This modulation generates an oscillating
current which amplifies the optical field and increases the density
modulation again, thus initiating a runaway amplification process. In an
appropriate reference frame, the fundamental mechanism that coherently
scatters photons into the optical field can be understood as cooperative
Compton scattering or Bragg scattering of the particles ({\em i.e.}
electrons) at a moving standing light wave. This point of view together with
the experimental observation of recoil-induced resonances in atomic gases 
\cite{Courtois94} triggered a few years ago the idea of an atomic analogue
to the FEL: the {\it Collective Atomic Recoil Laser} (CARL) \cite
{Bonifacio94}. In the CARL, photons are coherently redistributed between the
modes of a moving standing light wave by mediation of the atomic
center-of-mass motion. Cooperative Compton scattering leads to collective
atomic recoil and self-bunching of the matter-wave which results in
exponential gain. The recent observation of Bragg scattering in
Bose-condensates brought up the question whether BECs could serve the
purpose of an ultra-cold version of the CARL \cite{Heurich99}. The
superradiant Rayleigh scattering of laser light by a BEC seen by Inouye {\em %
et al.} \cite{Inouye99} already shows several features peculiar to CARL. The
long coherence time of BECs strongly correlates successive Rayleigh
scattering events via long-lived quasiparticle excitations. The positive
feedback of these excitations on the laser light results in exponential gain
and directional bundling of the scattered light.

When an incoming photon with wave vector ${\bf k}_{0}$ is scattered by a
condensed atom into the mode ${\bf k}_{s}$, with $k_{s}=k_{0}$, this atom
receives the recoil momentum ${\bf q}=\hbar {\bf k}_{s}-\hbar {\bf k}_{0}$\
and, while it propagates with a speed of a few centimeters per second
through the condensate, it interferes with the other atoms of the BEC to
form a matter-wave grating. The grating, which is long-lived compared to the
scattering rate, now stimulates subsequent photons from the incoming laser
beam to scatter into the same direction ${\bf k}_{s}$ and for its part picks
up the recoiled atoms. The process is self-amplifying, {\em i.e.} the number
of photons in ${\bf k}_{s}$ grows exponentially in time. The scheme can also
be interpreted the other way round as scattering of atoms into the BEC
momentum sidemode ${\bf q}$ stimulated by spontaneously scattered photons
and bosonically enhanced by the numbers of atoms already being in the
sidemode. The inversion that produces the exponential gain is readily
understood in a {\it dressed atom} picture. The resting BEC and the
irradiated laser light form together an excited state that decays into
recoiling atoms and scattered photons. The photons quickly leave the BEC
which maintains the inversion and permits, in principle, the complete
transfer of the BEC into the momentum sidemode. The spontaneity of the
scattering process ensures the irreversibility of the gain process.

For their experiment, Inouye {\em et al.} produced a cigar-shaped sodium
condensate with spatial extensions $z_{rms}=200$~$\mu $m and $r_{rms}=20$~$%
\mu $m confined in a cloverleaf trap. The BEC was irradiated from a radial
direction with a single linearly polarized laser light pulse tuned $\Delta
=-1.7$~GHz below the $D2$ line. The variable laser intensity, $I=1..100$%
~mW/cm, and duration, $\tau =10..800$~$\mu $s, permitted the adjustment of
the single-atom far-off resonance Rayleigh scattering rate to $R\approx
I/\hbar \omega \times \sigma _{0}\Omega ^{2}/4\Delta ^{2}=45..4500$~s$^{-1}$%
. After the application of the laser pulse, the magnetic trap was switched
off and a time-of-flight picture was taken after $20..50$~ms of free
expansion (Fig.~\ref{FigRaleighWaves}). Additionally, the scattered light
could be recorded either spatially resolved with a CCD camera or
time-resolved with a photomultiplier. In the following, we will discuss some
of the observations made in this experiment.

\bigskip

{\em Rayleigh Scattering.} --- The total gain depends on the size of the
condensate, {\em i.e.} the distance over which single-path gain can happen.
Mode competition quenches the scattering in all but the maximum gain
directions \cite{Moore99}. Non-spherical BECs therefore yield highly
anisotropic Rayleigh scattering. The competing process of Raman scattering
into different Zeeman sublevels is not bosonically stimulated. For
cigar-shaped BECs the gain path is longest along the symmetry axis, which
results in so-called {\it end-fire modes}. Scattering recoils the atoms and
has to stop when all the atoms are transferred to higher momentum sidemodes.
Consequently, Inouye {\em et al.} observed highly directional fluorescence
light bursts along the symmetry axis, whose durations were shortened as the
irradiated laser intensity was increased.

Since the end-fire modes enclose a $90^{o}$ angle with the incoming laser
beam and the frequency of the light does not change during Rayleigh
scattering, the scattered matter-wave gets a $45^{o}$ momentum kick. The
time-of-flight images in Figure~\ref{FigRaleighWaves} show the momentum
distribution of the condensate after irradiation of a single laser pulse
with various durations. For longer pulse durations, repeated Rayleigh
scattering at the higher momentum sidemodes gives rise to additional peaks.

\bigskip

{\em Superradiance.} --- The process is equivalent to Dicke superradiance,
where the overlapping radiation fields of a dense sample of excited atomic
dipoles stimulate each other to synchronously emit light, thus leaving the
sample in a coherent superposition state. The total emission time is reduced
to short fluorescence bursts. While in classical superradiance the sample of
two-level systems evolves into an oscillating coherence of internal
electronic states, in the MIT experiment, we have a coherent oscillation of
translational sidemodes. Superradiance does not require quantum degeneracy,
but the dipoles must have a long coherence time. Doppler broadening
accelerates relaxation. In the MIT experiment, where the coherence is stored
in the motional degrees of freedom, having long coherence time is equivalent
to having a large coherence length. BECs have a large coherence length that
corresponds to their size, while for thermal clouds the coherence length is
just its thermal de Broglie wavelength. This explains why Inouye {\em et al.}
could not observe superradiance using thermal clouds.

The superradiance was found to be very sensitive to the polarization of the
incoming laser light. Since the atoms were polarized in axial direction by
the magnetic field of the cloverleaf trap, photons polarized in the same
direction were absorbed and spontaneously reemitted according to the
(torus-shaped) dipole radiation pattern for $\pi $\ radiation, {\em i.e.}
not in axial direction. On the other hand, if the laser beam was polarized
perpendicular to the long BEC axis, the (bow-tie-shaped) dipole radiation
pattern for $\sigma ^{\pm }$\ radiation supported superradiance.

\subsection{Matter-Wave and Light Amplification}

The superradiance experiment of Inouye {\em et al.} realizes a {\it %
matter-wave amplifier} along the lines described by Law {\em et al.} \cite
{Law98b} and Moore {\em et al.} \cite{Moore99b}. The momentum sidemodes
which they observed may be regarded as amplified vacuum fluctuations.
However, the proof that the amplification process is coherent, {\em i.e.}
that the original matter-wave has a well-defined phase relation to the
amplified matter-wave was still lacking. This proof has recently been
provided by two experiments by Kozuma {\em et al.} \cite{Kozuma99b} and at
the MIT \cite{Inouye99b,Inouye00}.

In extension of the superradiance experiment, the MIT group seeded the
matter-wave amplifier with a very small condensate ($\sim 0.1\%$ of the
total condensate) thus substituting the quantum fluctuations in their role
of input wavepacket. The seed condensate was provided by a matter-wave Bragg
diffraction pulse (Section~5.3.3). It interfered with the main condensate to
form a matter-wave grating which was then amplified by a subsequent Rayleigh
scattering pulse. The gain in atom number for the seed mode could be set
between $10$ and $100$ by controlling the intensity and duration of the
Rayleigh pulse. Inouye {\em et al.} also demonstrated the coherence of the
amplification process by setting up a Ramsey type {\it active atom
interferometer} scheme whose one arm consisted of the amplified seed
condensate and the other arm of a reference condensate created from the
original condensate by Bragg diffraction. The observation of interference
proved the coherence of the amplification process.

\bigskip

Kozuma {\em et al.} chose a similar approach. They produced an elongated
rubidium condensate in a cloverleaf trap and, in contrast to the MIT group,
irradiated the superradiance and Bragg diffraction pulses into the long axis
of the condensate after releasing it from the trap. They reduced the
superradiant gain of their system so much that spontaneous quantum
fluctuations were not amplified, produced a seed condensate wavepacket by
Bragg diffraction ($\sim 6.5\%$) and showed that this was amplified to up to 
$66\%$ of the total BEC by a Rayleigh scattering pulse. They could also
demonstrate interference between the amplified and the original BEC
wavepackets in a Mach-Zehnder type atom interferometric setup \cite
{Hagley99b}. In\ a traditional Mach-Zehnder atom interferometer, a
wavepacket is first split with a $\pi /2$ interaction pulse, thus recoiling
half of the atoms and leaving the other half unaffected. A subsequent $\pi $
pulse reverses the momentum, so that the wavepackets move towards each
other. A final $\pi /2$ pulse recombines the components and produces
interference, provided every interaction was really coherent. Kozuma {\em et
al.} used Bragg diffraction interaction pulses in their Mach-Zehnder
interferometer with an essential modification: The first $\pi /2$ pulse
consisted of a combination of a Bragg pulse which produced the seed
condensate and a superradiance pulse which amplified it to a size
corresponding to half the BEC. The observation of interference thus proved
that the first composite $\pi /2$ pulse maintained the coherence, that the
long-range order was preserved for the amplified BEC and that it was
phase-locked to the seed BEC. An important drawback for matter-wave
amplifiers and atom lasers (Section~5.2.3) is the limited reservoir of
atoms. The amplification imperatively stops when all the atoms of the BEC
have been transferred into the amplified momentum sidemode.

The atom optical devices listed in Section~5.1.1 are all {\it passive devices%
}. In contrast, the phase-coherent matter-wave amplifiers discussed above 
{\it actively} stimulate the atoms to scatter into the amplified mode. It is
worth pointing out the analogy between this scattering process and four-wave
mixing. While matter-wave 4WM, which may be viewed as bosonically enhanced
redistribution of atoms between momentum sidemodes mediated by the
mean-field, involves four atoms (two in the input and two in the output
channel) and quantum optical 4WM, which may be viewed as coherent
redistribution of photons between light modes, involves four photons, the
process underlying the superradiant Rayleigh scattering takes place between
two atoms and two photons. In all three cases, bosonic stimulation plays a
key role.

\bigskip

We have seen in the superradiance experiment, that the Rayleigh scattered
light is stimulated into the end-fire modes. The process is self-amplifying
and can be used as a {\it light amplifier} for optical seed pulses. In a
subsequent experiment, Inouye {\em et al.} \cite{Inouye00} demonstrated the
amplification of light pulses. The occurrence of Rabi oscillations in the
temporal behavior of the gain showed that the gain process was coherent.

\subsection{Quantum Optics with Atoms}

The intrinsic coherence of Schr\"{o}dinger fields implies the possibility of
''exotic'' quantum correlations. Laser light is, normally, best described by
a coherent or {\it Glauber state}. But other quantum states of light are
possible, {\em i.e.} squeezed states, Schr\"{o}dinger cat states, states
with sub-Poissonian photon distribution, {\em e.g.} pure number or {\it Fock
states}, and even single photon states. All these states have been observed
in ultrahigh finesse micromasers. A mathematically very similar system is
the Hilbert space of the motion of a single particle in a harmonic trap, 
{\em e.g.} an ion stored in a Paul trap \cite{Blockley92}. Non-coherent
motional quantum states have been observed by Wineland {\em et al.} \cite
{Wineland98}. Quantum correlations have also been studied theoretically in
atomic Bose-Einstein condensates, and there are propositions on how to
create non-coherent states of BECs \cite{Ruostekoski98,Cirac98,Gordon99}.
(Note that non-coherent state BECs are not less coherent, but contain more
complicated quantum correlations than Glauber state field distributions.)
This field of investigations may be called ''quantum atom optics'' in
analogy to the field of quantum optics dealing with the non-classical
features of light.

At the interface between the macroscopic world and the microscopic quantum
world, {\it Schr\"{o}dinger cat states} are epitomized by new theories on
quantum decoherence. Schr\"{o}dinger cat states are coherent superpositions
of multi-particle quantum states. A perfect cat state can be written as $%
|N,0\rangle \pm |0,N\rangle $, {\em i.e.} {\it all particles} are in a
superposition of two states of an arbitrary degree of freedom, {\em e.g.}
coordinate, momentum or internal excitation. Because of their large scale,
mesoscopic coherent quantum objects like Bose-condensates are ideal testing
grounds for studies of fundamental questions on quantum entanglement,
quantum measurement, and decoherence. Unfortunately, big Schr\"{o}dinger
cats are extremely sensitive to decoherence. Proposals to generate such
states in BECs \cite{Ruostekoski98,Cirac98,Gordon99} have been reexamined by
Dalvit {\em et al.} \cite{Dalvit00} who also suggested several measures to
master the decoherence problem. The decoherence rate $\gamma _{dec}$
generally depends on the\ ''macroscopicity'' $N$ of the quantum state and
its contact with the environment. Thermal collisions (occurring at a rate $%
\gamma _{coll}$) are the main reason for decoherence in a BEC cat, $\gamma
_{dec}\sim N^{2}\gamma _{coll}$ \cite{Dalvit00}, but Rayleigh scattering and
three-body recombination also contribute. In the extreme case of a perfect
cat state, the coherence is destroyed by scattering of a single atom, since
its detection tells the state of all atoms. It is worth emphasizing that
Schr\"{o}dinger cat states should not be confused with the beamsplitter
states discussed in Section~\ref{SecOutputCoupling}, where {\it every single
atom} has the option of being in one of two states, $\left( |1,0\rangle \pm
|0,1\rangle \right) ^{N}$. Beamsplitter states only involve single-particle
correlations and are readily produced by Bragg scattering techniques.

The perfect cat state exhibits maximum entanglement and is, in this respect,
similar to Einstein-Podolski-Rosen (EPR) and Greenberger-Horne-Zeilinger
(GHZ) states. Such states of several entangled particles are currently
investigated in the context of quantum computation and have recently been
realized with single ions \cite{Sackett00} and with micromasers \cite
{Rauschenbeutel00}. Controlled collisions in optical lattices may offer new
opportunities for entangling neutral atoms and implementing schemes for
coherent quantum operations \cite{Jaksch98}. However, while for
computational purposes it is necessary to show up with a scheme where the
fundamental registers (called {\it qubits}) can be individually addressed,
the delocalized Bose-condensed atoms do not lend themselves to individual
manipulations. Still it is conceivable that new ideas that make use of the
mesoscopic coherence of BECs will emerge from the paradigm of coherent
entanglement and quantum control between BECs and laser modes.

The theory describing the coherent coupling of optical quantum fields and
Bose-Einstein condensates encompasses the classical domains of quantum
optics and atom optics as limiting cases. This theory is in many aspects
similar to optical cavity-QED theories, and the analogy seeds new ideas
about cavity atom optics, entanglement between atomic and laser beams, and
optical control of BECs \cite{Goldstein99}. As an example: in quantum optics
the {\it Optical Parametric Amplifier} (OPA) generates correlated
photon-photon states. Similarly, as we have seen in the superradiant
Rayleigh scattering experiments, the coherent interaction between light and
BECs creates entangled atom-photon states \cite{Moore99}. The range of
possible applications is wide and may include tests of Bell's inequality,
quantum cryptography and quantum teleportation.

\chapter{Collision Resonances}

\label{SecCollisionResonances}The mean-field interaction of ultracold dilute
atomic gases is dominated by binary $s$-wave collisions. In the shapeless
approximation, the collisions can be modelled by a single atomic constant,
the scattering length, which measures the low energy phase shift of the
relative de Broglie wave of the atoms during a collision process. The
scattering length determines the magnitude of the elastic and inelastic
collision rates.

However, the scattering length may be manipulated with optical \cite
{Fedichev96} or microwave \cite{Moerdijk96} radiation fields or, close to 
{\it Feshbach collision resonances}, with external magnetic fields \cite
{Tiesinga92}. Feshbach resonances were first predicted for nuclear systems 
\cite{Feshbach58}, and have recently regained much attention in the context
of Bose-Einstein condensation. They permitted the condensation of a new
atomic species (Section~\ref{SecBEC85Rb}) and are currently investigated in
the context of free-bound coupling and the creation of molecular BECs
(Section~\ref{SecMolecularBECs}).

\section{Feshbach Resonances in $^{85}$Rb and $^{23}$Na}

\label{SecFeshbachResonances}Feshbach resonances are collision resonances
that occur when the energy of a colliding channel coincides with the energy
of a vibrational bound state of a potential that correlates with a higher
lying asymptote (Fig.~\ref{FigFeshbachResonance}). If the bound state and
the free atoms have different magnetic momenta, the resonance condition may
be tuned via external magnetic fields exploiting the Zeeman-effect. When a
Feshbach resonance is crossed, the scattering length goes through a
singularity (Fig.~\ref{FigFeshbachScatteringLength}).

The complex spin-structure of the alkalis results from a combination of
exchange, hyperfine and Zeeman interaction and gives rise to a multitude of
energy levels, with a good chance of having one or more Feshbach resonances.
Verhaar and coworkers \cite{Moerdijk95,Vogels97} performed coupled
multi-channel calculations and found Feshbach resonances at experimentally
accessible field strengths in $^{85}$Rb and $^{23}$Na.

There are several ways to detect Feshbach resonances. The first way is using 
{\it Photoassociation Spectroscopy} (PA). PA is a frequently used tool to
explore the level structure of excited molecular states by irradiating a
laser tuned between the colliding channel and a vibrational bound level of
the excited state potential \cite{Thorsheim87}. Close to a Feshbach
resonance the colliding wavefunction amplitude is enhanced and therefore its
Franck-Condon overlap with the excited state wavefunction, and the
photoassociative transition rate gets larger. Since the excited molecular
state preferentially decays into the dissociation continuum, where the atoms
have high kinetic energy, the transition rate may be monitored via trap
losses. This method has been employed to detect a broad Feshbach resonance
in $^{85}$Rb near $B=160$~G with $6$~G width \cite{Courteille98b}.

A second method is based on the fact that the elastic cross-section and
therefore the collision rate in an atomic gas are both proportional to the
square of the scattering length in the limit of very low temperatures. This
yields a simple recipe for probing Feshbach resonances: One drives a dense
cold cloud out of thermal equilibrium and simply measures the
rethermalization speed. It takes on average three collisions per atoms to
rethermalize a sample. Close to the Feshbach resonance the rethermalization
speed should be drastically enhanced. This method has provided an improved
measurement of the location of the $^{85}$Rb Feshbach resonance \cite
{Roberts98}, which in turn has been utilized to calibrate the calculations
of the $^{85}$Rb potentials and to enhance their precision to a large
extent. On the other hand, the influence of the Feshbach resonance on the
collision rate suggests its use to control and improve evaporative cooling
of atomic clouds.

A third method is based on the influence of the scattering length on the
mean-field energy of Bose-Einstein condensates. In fact, the scattering
length is the only atomic parameter showing up in the Gross-Pitaevskii
equation. It governs the shape and size of the BEC wavefunction, the BEC
dynamics and of course all nonlinear interactions. The effect of a Feshbach
resonance on BECs has been studied in $^{23}$Na \cite{Inouye98}.
Unfortunately, this experiment also showed the occurrence of inelastic
collision processes close to the Feshbach resonance leading to a strong
depletion of the condensate. This will most likely limit the practical use
of this resonance in sodium.

\section{Bose-Einstein Condensation in $^{85}$Rb}

\label{SecBEC85Rb}The zero-field scattering length of the $^{85}$Rb isotope
in the ground-state $F=2,m_{F}=-2$ is $a_{|2,-2\rangle }\approx -400a_{B}$ 
\cite{Courteille98b}. The negative scattering length inhibits the formation
of stable Bose-Einstein condensates with this atomic species. However, in
proximity to a Feshbach resonance the scattering length is very sensitive to
ambient magnetic fields, $B$ (Fig.~\ref{FigFeshbachScatteringLength}), which
can alter its value and even its sign.

The JILA group, led by Wieman, recently reached the quantum degenerate
regime with $^{85}$Rb \cite{Cornish00} operating in a regime of positive
scattering length. Efficient evaporation is hindered by a deep notch in the
elastic scattering cross section at collision energies around $350~{\rm \mu }
$K, a peculiarity of $^{85}$Rb, and by inelastic two- and three-body
collisions being very frequent at some regimes of the scattering length.
Cornish {\em et al.} \cite{Cornish00} avoided these difficulties using a
relatively weak trap, $\bar{\omega}\approx 2\pi \times 13$~Hz, to reduce the
atomic cloud density and by following a sophisticated evaporation path. The
low density slowed down the evaporation and required a long magnetic trap
lifetime. The first evaporation step was performed in the high-field wing,
at $B=250$~G, far from the Feshbach resonance. When the sample was cooled to 
$2~{\rm \mu }$K, the sign of the scattering length was reversed and its
absolute value reduced, $a_{|2,-2\rangle }(B)=290a_{B}$, by moving the
magnetic field strength towards the Feshbach resonance to $B=162.3$~G (Fig.~%
\ref{FigFeshbachScatteringLength}). This further reduced the inelastic
collision rate. The evaporation was now pursued until the condensation
threshold was approached with $10^{6}$ atoms. While the condensate formed,
inelastic loss processes rapidly reduced the trapped atom number to $10^{4}$
at typically $15~$nK temperature and $10^{12}$~cm$^{-3}$ density. The BEC
lifetime was about $10$~s.

Size and shape of the condensate depend on the self-energy and thus on the
scattering length. Tuning the magnetic field across the Feshbach resonance
changes the condensate shape between the limits of an ideal gas Gaussian
density distribution and a Thomas-Fermi regime parabolic distribution.
Cornish {\em et al.} recorded pictures of the condensate with $1.6$~ms
absorptive time-of-flight imaging, determined the scattering length from the
shape of the BEC, and confirmed the magnetic field dependence shown in
Figure~\ref{FigFeshbachScatteringLength}. At the magnetic field strength $%
B=156.6$~G, the scattering length climbs to $a_{|2,-2\rangle }(B)>10000a_{B}$%
, and the measured condensate peak density yielded $na^{3}\approx 0.01$. In
this regime, the dilute-gas assumption $na^{3}\ll 1$ begins to break down,
and effects beyond the mean-field approximation, like characteristic shifts
in the frequencies of collective excitations, may be observed.

When the self-energy was reduced by tuning the scattering length from a
positive regime away from the Feshbach resonance into a regime where the
scattering length is negative (Fig.~\ref{FigFeshbachScatteringLength}), {\em %
i.e.}\ beyond $B=166.8$~G, the BEC exhibited an abrupt dynamical behavior.
The BEC shrank until it collapsed ejecting a burst of hot atoms. If we
compare to Hulet's experiment (Section~3.2.1) which relies on the ensemble
analysis of collapsed condensates, we find that the ability to control the
value and the onset of the $a<0$ instability greatly facilitates studies of
the collapse dynamics. Also, we expect that the successful realization of
BEC in $^{85}$Rb\ taps a whole new field of possibly very interesting
investigations with the scattering length as an additional, dynamically
tunable degree of freedom.

\section{Molecular Bose-Einstein Condensates}

\label{SecMolecularBECs}Recent theoretical investigations \cite
{Timmermans99b,Timmermans99}, \cite{Drummond98}\nocite{Hu00}-\cite{Holland00}
have shown, that the physics of Feshbach resonances is considerably richer
than that of an altered effective scattering length. Feshbach resonances
provide a free-bound coupling between the two-colliding-atoms continuum
state and a quasibound vibrational molecular state that has some analogy to 
{\it Second Harmonic Generation} (SHG). When the Feshbach resonance is
excited in a Bose-condensate, the quasi-molecules are predicted to form a 
{\it molecular BEC}. The atomic and the molecular BEC are coupled via
intercondensate tunneling of atom pairs. The system may even exhibit
Josephson oscillations as a signature of this novel type of quantum
tunneling. Free-bound coupling can alternatively be established by driving
two-photon Raman photoassociation transitions \cite{Javanainen99,Julienne98}%
. This system closely resembles the Feshbach resonance system and may
generate molecular BECs and Josephson oscillations between atomic and
molecular BECs as well.

The possibility of using {\it incoherent PA} to convert large amounts of
free atoms to low-lying vibrational levels of ultracold groundstate
molecules has been pointed out by Band {\em et al.} \cite{Band95}. Ultracold
molecules have recently been produced in such photoassociation schemes \cite
{Fioretti98}. On the other hand, the equilibrium yield of {\it coherent
Raman PA} depends on the entropies of the coupled systems \cite{Javanainen99}%
. In a thermal atomic gas, the (quasi-)continuum of dissociated atomic
states (they are still confined in a magnetic trap) has a much larger
entropy than the discrete spectrum of vibrational molecular states.
Therefore, the balance of the coherent free-bound coupling has to be on the
side of the continuum, {\em i.e.} molecules dissociate more frequently than
they associate. Quantitative estimates of the PA rates have to thermally
average over transition rates (as opposed to transition amplitudes). As a
consequence, coherent processes, even {\it STImulated Raman Adiabatic Passage%
} (STIRAP) transitions, have negligible molecular yield. (In short, STIRAP
consists of a counterintuitive pulse sequence of the two frequencies
involved in the Raman process.) However, as the atomic cloud approaches
quantum degeneracy, the dimensionality of the phase space is reduced to a
large extent (a BEC has zero entropy) and the coherent free-bound coupling
should transform a considerable amount of condensed atoms into a {\it %
molecular BEC.} In specific schemes, {\em e.g.} by quickly removing the
molecules from the interaction region \cite{Julienne98} or by using
two-photon photoassociative STIRAP pulses \cite{Mackie99}, it should be
possible to produce molecules in a controlled manner with unity yield. The
formation of molecules within a condensate can be understood as a
Bose-stimulated chemical process. The dynamics of this process is driven by
quantum statistics rather than by chemical forces between individual atoms.

A possible system to study such phenomena is $^{87}$Rb (Fig.~\ref
{FigTwoPhotonPhotoassociation}). Two-photon transitions to very weakly bound
vibrational molecular Rydberg states have been observed in a dark MOT \cite
{Courteille98c} and later even in BECs \cite{Wynar00}. A narrow linewidth of
down to $1.5$~kHz of the Raman dark resonance, only observed with quantum
degenerate gases, is a clear indication of {\it coherent} coupling. The
narrow dark resonance permitted the measurement of the binding energy of the
molecular state, $E=2\pi \hbar \times 636.0094$~MHz, with an unprecedented
resolution only limited by the inhomogeneous spatial distribution of the
atoms and their self-energy. The molecules were produced at rest, because
the Raman process does not transfer momentum. Molecular condensates may form
in the groundstate of the trap confining the condensate, provided the dark
resonance width is inferior to the trap secular frequencies. Unfortunately,
fast inelastic decay into lower vibrational states limits the lifetime of
the molecular Rydberg states to less than $1$~ms.

\chapter{Criteria of Bose-Einstein Condensation}

The phenomenon of Bose-Einstein condensation involves several rather
delicate concepts, such as coherence and gauge-symmetry breaking. In
literature, these concepts are very often misinterpreted. Therefore we feel
it is necessary to give an accurate and detailed description of the
principal notions lying in the foundation of the considered phenomenon.

One generally implies that the Bose-Einstein condensation is a {\it %
macroscopic occupation of a single quantum state}, usually of the ground
state, as it was suggested by Bose \cite{Bose24} and Einstein \cite
{Einstein24} who considered this phenomenon for ideal gases. For
noninteracting atoms, the meaning of single-particle quantum states is well
defined. This, however, is not always the case for interacting atoms. In
order to formulate more precisely what the Bose-Einstein condensation
actually is, several criteria are employed. Here we give a careful analysis
of these criteria, of their mutual interrelations, and of their relation to
the original idea \cite{Bose24,Einstein24} of a macroscopic occupation of
the ground state.

Intuitively, one expects that the condensation in a system of $N$ bosons
occupying volume $V$ can occur when the thermal wavelength $\lambda_T$
becomes much larger than the mean interatomic distance $a$, that is, 
\begin{equation}  \label{7.1}
\frac{a}{\lambda_T}\ll 1\; , \qquad \lambda_T\equiv\left ( \frac{2\pi\hbar^2 
}{m_0k_BT}\right )^{1/2} \; ,
\end{equation}
where $m_0$ is the atom mass; $T$, temperature. At the same time, the
characteristic interaction radius, $r_{int}$, has to be much smaller than
the mean interparticle distance 
\begin{equation}  \label{7.2}
\frac{r_{int}}{a}\ll 1\; , \qquad r_{int}\sim |a_s|\; ,
\end{equation}
the interaction radius being of the order of scattering length $a_s$. In the
other case, strong interaction between atoms could deplete the condensate or
even completely destroy it. For the density of particles 
\[
\rho\equiv\frac{N}{V}\sim a^{-3}\; , 
\]
the above conditions can be written as 
\begin{equation}  \label{7.3}
\rho\lambda_T^3\gg 1\; , \qquad \rho r_{int}^3\ll 1\; .
\end{equation}
Inequalities (\ref{7.1}) and (\ref{7.2}), or (\ref{7.3}), are the {\it %
expected conditions} for the occurrence of the Bose-Einstein condensation.
The discussion of sufficient conditions is given in the following
subsections.

\section{Einstein Criterion of Condensation}

The statement of a macroscopic occupation of a quantum state \cite
{Einstein24} can be formalized as follows. Let an orthonormalized basis $%
\{\varphi _{n}({\bf r})\}$ be given composed of wave functions corresponding
to single-particle quantum states labelled by a multi-index $n$. Field
operators can be expanded in this single-particle basis as 
\begin{equation}
\psi ({\bf r})=\sum_{n}\;a_{n}\;\varphi _{n}({\bf r})\;,  \label{7.4}
\end{equation}
with the coefficients 
\[
a_{n}=(\varphi _{n},\psi )\equiv \int \;\varphi _{n}^{\ast }({\bf r})\;\psi (%
{\bf r})\;d{\bf r}\;. 
\]
A physical quantity is called macroscopic if it is proportional to the
average number of particles 
\begin{equation}
N\equiv \;<\hat{N}>\;=\sum_{n}\;<a_{n}^{\dagger }a_{n}>\;,  \label{7.5}
\end{equation}
being the statistical average of the number-of-particles operator 
\[
\hat{N}\equiv \int \;\psi ^{\dagger }({\bf r})\;\psi ({\bf r})\;d{\bf r}\;. 
\]
The quantity $<a_{n}^{\dagger }a_{n}>$ is the occupation number of a quantum
state $n$. The occupation is termed macroscopic if $<a_{n}^{\dagger }a_{n}>$
is proportional to $N$. In equilibrium systems, it is the single-particle
ground state, {\em i.e.} the state of the minimal single-particle energy,
that can become macroscopically occupied, which means that the number of
particles in the ground state, 
\begin{equation}
N_{0}\equiv \;<a_{0}^{\dagger }a_{0}>\;,  \label{7.6}
\end{equation}
can become proportional to $N$. This can be stated more rigorously by means
of the limit 
\begin{equation}
\lim_{N\rightarrow \infty }\;\frac{N_{0}}{N}>0\;.  \label{7.7}
\end{equation}
Condition (\ref{7.7}) defines precisely what one actually implies when
talking about the macroscopic occupation of a quantum state.

However, the criterion of condensation (\ref{7.7}) has several weak points.
First of all, there is an ambiguity in choosing a single-particle basis $%
\{\varphi _{n}({\bf r})\}$ which all the following consideration is based
on. Such a basis naturally arises and is well defined for ideal gases \cite
{TerHaar64}, while for interacting particles it is, in general, not uniquely
defined. Hence the single-particle ground state and the related occupation
number are not well defined for a system of interacting atoms.

Some delicate problems may appear in defining the limit (\ref{7.7}), as was
shown for an exactly solvable model (Michoel and Verbeure \cite{Michoel99}).
This means the following. The number of particles in the ground state (\ref
{7.6}) can be defined as 
\[
N_{0}=\lim_{\delta \rightarrow 0}N_{\delta }\;,\qquad N_{\delta }\equiv
\sum_{n=0}^{\delta }\;<a_{n}^{\dagger }a_{n}>\;. 
\]
It happens sometimes that 
\begin{equation}
\lim_{N\rightarrow \infty }\;\lim_{\delta \rightarrow 0}N_{\delta }=0\;.
\label{7.8}
\end{equation}
although 
\begin{equation}
\lim_{\delta \rightarrow 0}\;\lim_{N\rightarrow \infty }N_{\delta }>0\;.
\label{7.9}
\end{equation}

It is also worth emphasizing that, when considering the criterion (\ref{7.7}%
), one usually tacitly assumes that the macroscopic occupation occurs solely
for one quantum level, {\em i.e.} for the ground state level. But, in
general, the situation may happen when several quantum states, or even an
infinite number of them, become macroscopically occupied so that 
\begin{equation}
\lim_{N\rightarrow \infty }\;\frac{1}{N}\;<a_{n}^{\dagger }a_{n}>\;=const>0
\label{7.10}
\end{equation}
for several quantum numbers $n$.

\section{Penrose Criterion of Condensation}

Penrose \cite{Penrose51} and Penrose and Onsager \cite{Penrose56} criticized
the criterion (\ref{7.7}) stressing that ''this criterion has meaning for
noninteracting particles only, because single-particle energy levels are not
defined for interacting particles''. They suggested a generalization of this
criterion valid for interacting particles as well. The generalization is
based on\ the definition of the eigenvalues of the first-order density
matrix 
\begin{equation}
\rho _{1}({\bf r},{\bf r}^{\prime })\equiv \;<\psi ^{\dagger }({\bf r}%
^{\prime })\psi ({\bf r})>\;.  \label{7.11}
\end{equation}
The eigenvalues of the matrix (\ref{7.11}) are given by the eigenproblem 
\[
\int \;\rho _{1}({\bf r},{\bf r}^{\prime })\;\varphi _{n}({\bf r}^{\prime
})\;d{\bf r}\;^{\prime }=\gamma _{n}\varphi _{n}({\bf r})\;. 
\]
The largest eigenvalue defines the norm 
\[
||\hat{\rho}_{1}||\equiv \sup_{n}\gamma _{n}\;. 
\]
The Penrose criterion of condensation reads 
\begin{equation}
\lim_{N\rightarrow \infty }\;\frac{||\hat{\rho}_{1}||}{N}\;>0\;.
\label{7.12}
\end{equation}

This criterion can be further generalized by introducing the notion of order
indices for reduced density matrices (Coleman and Yukalov \cite
{Coleman91,Coleman92}). For a $k$-order reduced density matrix 
\begin{equation}
\rho _{k}({\bf r}_{1}\ldots {\bf r}_{k},{\bf r}_{1}^{\prime }\ldots {\bf r}%
_{k}^{\prime })\equiv \;<\psi ^{\dagger }({\bf r}_{k}^{\prime })\ldots \psi
^{\dagger }({\bf r}_{1}^{\prime })\psi ({\bf r}_{1})\ldots \psi ({\bf r}%
_{k})>  \label{7.13}
\end{equation}
the {\it order index} is defined as 
\begin{equation}
\alpha _{k}\equiv \lim_{N\rightarrow \infty }\;\frac{\ln ||\hat{\rho}_{k}||}{%
\ln N}\;,  \label{7.14}
\end{equation}
where $||\hat{\rho}_{k}||$ is the norm of the matrix $\hat{\rho}_{k}$ with
elements (\ref{7.13}). Different types of ordering appearing in the system
of bosons can be classified \cite{Coleman93,Coleman96} as follows: 
\[
\alpha _{k}=0\;,\qquad no\;order\;or\;short\text{-}range 
\]
\[
0<\alpha _{2k}<k\;,\qquad even\;mid\text{-}range 
\]
\[
\alpha _{2k}=k\;,\qquad even\;long\text{-}range 
\]
\[
\left[ \frac{k}{2}\right] <\alpha _{k}<k\;,\qquad total\;mid\text{-}range 
\]
\begin{equation}
\alpha _{k}=k\;,\qquad total\;long\text{-}range\;,  \label{7.15}
\end{equation}
where $[x]$ is the entire part of $x$. This classification encompasses three
kinds of possible condensation: Even condensation \cite{Valatin58}-\cite
{Pashitsky99}, with $\alpha _{2k}=k$, when the groups of even numbers of
atoms are condensed but there is no single-particle condensate. Mid-range
condensation \cite{Yukalov78}-\cite{Yukalov81b}, with $[k/2]<\alpha _{k}<k$,
when there arises algebraic mid-range order but there is no long-range
order. The Bose-Einstein singe-particle condensation corresponds to the case 
$\alpha _{k}=k$.

Criteria based on the consideration of norms of reduced density matrices are
rather general. However, it is not always easy to find the eigenvalues of
the density matrices for interacting particles.

\section{Off-Diagonal Long-Range Order}

The concept of off-diagonal long-range order (Yang \cite{Yang62}) can be
formulated as follows. If the limit 
\begin{equation}
\lim_{r_{12}\rightarrow \infty }\rho _{1}({\bf r}_{1},{\bf r}_{2})\equiv
\rho _{0}>0\;,  \label{7.16}
\end{equation}
where $r_{12}\equiv |{\bf r}_{1}-{\bf r}_{2}|$ is not zero, then there
occurs Bose-Einstein condensation, and $\rho _{0}$ is the condensate
density. Really, for the first-order density matrix one may write the
spectral resolution 
\begin{equation}
\rho _{1}({\bf r}_{1},{\bf r}_{2})=\sum_{n}\gamma _{n}\varphi _{n}({\bf r}
_{1})\varphi _{n}^{\ast }({\bf r}_{2})\;,  \label{7.17}
\end{equation}
in which $\gamma _{n}$ are the eigenvalues and $\varphi _{n}({\bf r})$, the
eigenfunctions of $\hat{\rho}_{1}$. Note that $\gamma _{n}$ play the role of
the average occupation numbers of the single-particle states labelled by $n$%
. If one assumes that the considered system is uniform, then the main
contribution to the sum (\ref{7.17}), as $r_{12}\rightarrow \infty $, is
made by the term containing the largest eigenvalue $\gamma _{0}$ and the
ground-state function $\varphi _{0}({\bf r})=1/\sqrt{V}$, so that 
\begin{equation}
\rho _{1}({\bf r}_{1},{\bf r}_{2})\simeq \gamma _{0}\varphi _{0}({\bf r}%
_{1})\varphi _{0}^{\ast }({\bf r}_{2})\;,\qquad r_{12}\rightarrow \infty \;.
\label{7.18}
\end{equation}
Here it is not necessary that $\varphi _{0}$ be the average $<\psi >$ of the
field operator. To be finite, the limit (\ref{7.18}) requires that $\gamma
_{0}\sim N$. Hence the consideration is reduced to the Penrose criterion of
condensation (\ref{7.12}).

Thus, the long-range order defined by the limit (\ref{7.16}) is a sufficient
condition for the occurrence of condensation in a nonuniform system. But, in
general, this is not a necessary condition, and it is not applicable to
nonuniform systems. For example, for a system of atoms localized in a
confined region, say inside a trap, one has 
\begin{equation}  \label{7.19}
\lim_{r\rightarrow\infty}\varphi_n({\bf r}) = 0 \; .
\end{equation}
Therefore, 
\begin{equation}  \label{7.20}
\lim_{r_{12}\rightarrow\infty}\rho_1({\bf r}_1,{\bf r}_2) = 0
\end{equation}
irrespectively of the values of $\gamma_n$. Condition (\ref{7.16}) excludes
the existence of Bose-Einstein condensation in confined systems.

\section{Broken Gauge Symmetry}

The concept of broken gauge symmetry is often used as a sufficient condition
for Bose-Einstein condensation. The standard way of breaking gauge symmetry
is by means of the {\it Bogolubov prescription} \cite{Bogolubov67} for the
field operator which is presented as the sum 
\begin{equation}
\psi ({\bf r})=\psi _{0}({\bf r})+\tilde{\psi}({\bf r})  \label{7.21}
\end{equation}
of a nonoperator term $\psi _{0}$ and an operator $\tilde{\psi}$ such that 
\begin{equation}
\psi _{0}({\bf r})=\;<\psi ({\bf r})>\;,\qquad <\tilde{\psi}({\bf r})>\;=0\;.
\label{7.22}
\end{equation}
The nonoperator term $\psi _{0}$ corresponds to condensate atoms in a
single-particle ground state, while the operator $\tilde{\psi}$ describes
atoms outside the condensate. Because of Eqs.~(\ref{7.21}) and (\ref{7.22}),
the statistical average of the field operator 
\begin{equation}
<\psi ({\bf r})>\;\neq 0  \label{7.23}
\end{equation}
is not zero, which manifests the broken gauge symmetry as far as the average 
$<\psi >$ is now not invariant under the gauge transformation 
\[
\psi ({\bf r})\rightarrow e^{i\alpha }\;\psi ({\bf r})\;, 
\]
where $\alpha $ is an arbitrary real number.

In order to understand better what are the assumptions under which the
prescription (\ref{7.21}) is valid, it is useful to look attentively at the
original ideas of Bogolubov \cite{Bogolubov67}, which we shall follow below.
Let us select an orthonormalized basis $\{\varphi _{n}({\bf r})\}$ of
single-particle states. The field operator can be expanded over the chosen
basis as 
\begin{equation}
\psi ({\bf r})=\sum_{n}\;\psi _{n}({\bf r})=\sum_{n}\;a_{n}\varphi _{n}({\bf %
r})\;,  \label{7.24}
\end{equation}
with $a_{n}=(\varphi _{n},\psi )$. From the Bose commutation relations 
\[
\lbrack \psi ({\bf r}),\;\psi ^{\dagger }({\bf r}\;^{\prime })]=\delta ({\bf %
r}-{\bf r}\;^{\prime }) 
\]
one has 
\[
\lbrack a_{m},\;a_{n}^{\dagger }]=\delta _{mn}\;. 
\]
Define the condensate and noncondensate operators 
\begin{equation}
\psi _{0}({\bf r})\equiv a_{0}\varphi _{0}({\bf r})\;,\qquad \tilde{\psi}(%
{\bf r})\equiv \sum_{n\neq 0}\;a_{n}\varphi _{n}({\bf r})\;.  \label{7.25}
\end{equation}
From the commutation relations for $a_{n}$ it follows that 
\[
\lbrack \psi _{0}({\bf r}),\;\psi _{0}^{\dagger }({\bf r}\;^{\prime
})]=\varphi _{0}({\bf r})\varphi _{0}^{\ast }({\bf r}\;^{\prime })\;. 
\]
For treating $\psi _{0}$ as a nonoperator term, it is necessary that this
commutator would be asymptotically small, at least, in the thermodynamic
limit, when 
\[
N\rightarrow \infty \;,\qquad V\rightarrow \infty \;,\qquad \frac{N}{V}%
\rightarrow const\;. 
\]
This is really the case for uniform systems for which one has the basis $%
\{\varphi _{k}({\bf r})\}$ formed of plane waves 
\[
\varphi _{k}({\bf r})=\;\frac{1}{\sqrt{V}}\;e^{i{\bf k}\cdot {\bf r}}\;. 
\]
Then one gets 
\begin{equation}
\left[ \psi _{0}({\bf r}),\;\psi _{0}^{\dagger }({\bf r}\;^{\prime })\right]
=\;\frac{1}{V}\;\rightarrow \;0\qquad (V\rightarrow \infty )\;.  \label{7.26}
\end{equation}
However this is not yet sufficient for announcing $\psi _{0}$ a nonoperator
term. Consider the operator $a_{0}=(\varphi _{0},\psi )$. Taking into
account that 
\[
(\varphi _{0},\tilde{\psi})=\sum_{n\neq 0}\;a_{n}(\varphi _{0},\varphi
_{n})=0\;, 
\]
we have 
\[
a_{0}=(\varphi _{0},\psi _{0})\;. 
\]
This shows that if $\psi _{0}$ is not an operator then $a_{0}$ is also such.
Hence, one should have $[a_{0},a_{0}^{\dagger }]=0$, which contradicts the
commutation relation $[a_{0},a_{0}^{\dagger }]=1$. Then one needs to make an
assumption that the ground state is macroscopically occupied, so that 
\begin{equation}
<a_{0}^{\dagger }a_{0}>\;\sim N\;.  \label{7.27}
\end{equation}
Only after this, one can say that the finite value of the commutator $%
[a_{0},a_{0}^{\dagger }]$ is negligibly small as compared to the macroscopic
number (\ref{7.27}), 
\begin{equation}
\frac{\lbrack a_{0},a_{0}^{\dagger }]}{<a_{0}^{\dagger }a_{0}>}\;\sim \; 
\frac{1}{N}\;\rightarrow 0\qquad (N\rightarrow \infty )\;.  \label{7.28}
\end{equation}
As is evident, the assumption (\ref{7.27}) is nothing but the Einstein
criterion of condensation (\ref{7.7}). In this way, the Bogolubov
prescription (\ref{7.21}) {\it presupposes} Bose-Einstein condensation.
Moreover, this prescription assumes that the condensation occurs solely in 
{\it one} quantum state. In general, the commutator $[\psi _{n},\psi
_{n}^{\dagger }]$ can become asymptotically small for several states. For
instance, the commutator 
\[
\lbrack \psi _{k}({\bf r}),\;\psi _{k}^{\dagger }({\bf r}^{\prime })]=\;%
\frac{1}{V}\;e^{i{\bf k}\cdot ({\bf r}-{\bf r}^{\prime })}\rightarrow 0 
\]
is asymptotically, as $V\rightarrow \infty $, small for any ${\bf k}$.
Nevertheless, one does not announce that all $\psi _{k}$ are nonoperator
terms. Vice versa, all $\psi _{k}$ with ${\bf k}\neq 0$ are treated as
operators satisfying the standard Bose commutation relations. This means
that the Bogolubov prescription segregates one ground-state level that is
assumed to be macroscopically occupied, so that relation (\ref{7.27}) holds
true; and all other levels are not occupied macroscopically, so that $%
<a_{n}^{\dagger }a_{n}>\sim 1$ for $n\neq 0$.

To be practical, the Bogolubov prescription (\ref{7.21}) requires one more
assumption that is always made. One treats $\tilde\psi$ as a small
perturbation about the mean-field value $\psi_0=<\psi>$. This is equivalent
to the assumption that almost all atoms are condensed, 
\begin{equation}  \label{7.29}
\frac{N-N_0}{N} \ll 1 \; .
\end{equation}

In this way, breaking gauge symmetry by means of the Bogolubov prescription (%
\ref{7.21}) presupposes the existence of Bose-Einstein condensation. This
prescription provides a practical tool for calculations under the assumption
that almost all atoms are in the condensed state. But breaking gauge
symmetry {\it is not necessary} for the validity of the Einstein criterion (%
\ref{7.7}) or Penrose criterion (\ref{7.12}), that is, it is not necessary
for the existence of Bose-Einstein condensation.

It is also important to note that the Bogolubov prescription (\ref{7.21}) is
not applicable for strongly interacting particles whose interactions are
described by nonintegrable potentials. Breaking gauge symmetry by this
prescription requires that the interatomic interactions are given by an
integrable potential $\Phi({\bf r})$, such that 
\begin{equation}  \label{7.30}
\left | \int\; \Phi({\bf r})\; d{\bf r}\right | < \infty \; .
\end{equation}
This is necessary since using the prescription (\ref{7.21}) yields the
appearance in the Hamiltonian of the term 
\[
\frac{1}{2}\; \int |\psi_0({\bf r})|^2\; \Phi({\bf r} - {\bf r}\;
^{\prime})\; |\psi_0({\bf r}\; ^{\prime})|^2\; d{\bf r}\; d{\bf r}\;
^{\prime}\; , 
\]
with nonoperator functions $|\psi_0({\bf r})|$ and $|\psi_0({\bf r}\;
^{\prime})|$. This term diverges if the interaction potential does not
satisfy condition (\ref{7.30}), hence, gauge symmetry cannot be broken for
such systems.

\section{Condensation in Confined Systems}

When atoms are confined in a box or by means of external confining
potentials, then all single-particle functions $\varphi _{n}({\bf r})$ tend
to zero, as $r\rightarrow \infty $, because of which the limit (\ref{7.20})
of the first-order density matrix is zero, which tells that there is no
off-diagonal long-range order. However, it is possible to weaken condition (%
\ref{7.20}) considering, instead of the exact limit, an asymptotic behavior
at large $r_{12}$, when the density matrix can be approximately factorized
as 
\begin{equation}
\rho _{1}({\bf r}_{1},{\bf r}_{2})\sim \varphi _{0}({\bf r}_{1})\varphi
_{0}^{\ast }({\bf r}_{2})\;,  \label{7.31}
\end{equation}
which tells that there exists a kind of long-range order \cite
{Penrose51,Penrose56,Ginzburg50}. The factorization (\ref{7.31}) may appear
if the distance $r_{12}$ is much larger than the mean interatomic distance $%
a $, but much smaller than the effective size $l_{0}$ of the confined
system, that is, in the region 
\begin{equation}
a\ll r_{12}\ll l_{0}\;.  \label{7.32}
\end{equation}
The meaning of the inequality $r_{12}\gg a$ is evident, and the inequality $%
r_{12}\ll l_{0}$ arises because the wave function of a ground state is
always more localized than the wave functions of excited states.
Consequently, at the distance $r_{12}\sim l_{0}$ the excited-state wave
functions are much larger than $\varphi _{0}$, and the factorization (\ref
{7.31}), in general, will not occur. Thus, in confined systems, strictly
speaking, there is no long-range order but there can be quasilong-range
order, when the density matrix factorizes, as in Eq.~(\ref{7.31}), in the
region (\ref{7.32}).

Gauge symmetry in a confined system cannot be broken. Thus, for employing
the Bogolubov prescription (\ref{7.21}), one would need that the commutator $%
[\psi _{0},\psi _{0}^{+}]$ be zero. However, this is not so. For instance,
expanding the field operator, according to Eq.~(\ref{7.24}), over an
oscillator basis, we have 
\[
\varphi _{0}({\bf r})=a_{0}\left( \frac{m_{0}\omega _{0}}{\pi \hbar }\right)
^{3/4}\;\exp \left( -\;\frac{m_{0}\omega _{0}}{2\hbar }\;r^{2}\right) \;, 
\]
from where 
\begin{equation}
\lbrack \varphi _{0}({\bf r}_{1}),\;\psi _{0}^{\dagger }({\bf r}%
_{2})]=\left( \frac{m_{0}\omega _{0}}{\pi \hbar }\right) ^{3/2}\;\exp
\left\{ -\;\frac{m_{0}\omega _{0}}{2\hbar }\;(r_{1}^{2}+r_{2}^{2})\right\}
\;,  \label{7.33}
\end{equation}
which is, certainly, not zero.

The absence of long-range order and of broken gauge symmetry in confined
systems is in agreement with the known fact that there are no sharp phase
transitions in such systems, although Bose-Einstein condensation can occur
without being a sharp phase transition but a gradual crossover \cite
{Kirsten96}, \cite{Ziff77}-\cite{Kirsten98}. During this crossover all
thermodynamic characteristics behave smoothly and no discontinuities appear,
although some quantities can change very rapidly. Since all thermodynamic
characteristics change in a completely smooth way, the identification of a
specific critical temperature is problematic. It is the standard situation
for crossover phenomena that the crossover temperature is not uniquely
defined, but its definition, anyway, can be done by assigning the crossover
temperature to the maximum of one of thermodynamic functions \cite
{Yukalov97b,Yukalov97c}. In the case of Bose condensation in confined
systems, one can relate the condensation temperature to the maximum of
specific heat \cite{Kirsten96}.

If gauge symmetry in confined systems is, strictly speaking, never broken,
is it then admissible to use the Bogolubov prescription (\ref{7.21}) in some
approximate sense? It seems that when the mean interatomic distance $a$ is
much smaller than the effective system size $l_{0}$, then the confined
system could be treated as almost infinite. For atoms confined, {\em e.g.}
in a harmonic potential, the required inequality is 
\begin{equation}
\frac{a}{l_{0}}\ll 1\;,\qquad l_{0}\equiv \sqrt{\frac{\hbar }{m_{0}\omega
_{0}}}\;.  \label{7.34}
\end{equation}
The effective volume of the confined system is $V\sim l_{0}^{3}$, hence $%
l_{0}\sim N^{1/3}$. Therefore, an admissible description of the effective
thermodynamic limit in this case could be 
\begin{equation}
N\rightarrow \infty \;,\qquad l_{0}\rightarrow \infty \;,\qquad \frac{N}{%
l_{0}^{3}}\rightarrow const\;.  \label{7.35}
\end{equation}
Because of the relation (\ref{7.34}), $\omega _{0}\sim l_{0}^{-2}$, thence $%
\omega _{0}\sim N^{-2/3}$. Consequently, the thermodynamic limit (\ref{7.35}
) can be presented as 
\begin{equation}
N\rightarrow \infty \;,\qquad \omega _{0}\rightarrow 0\;,\qquad N\omega
_{0}^{3/2}\rightarrow const\;.  \label{7.36}
\end{equation}
The ground-state wave function $\varphi _{0}\sim l_{0}^{-3/2}$, that is $%
\varphi _{0}\sim N^{-1/2}$. Then the commutator (\ref{7.33}) is 
\begin{equation}
\lbrack \psi _{0}({\bf r}_{1}),\;\psi _{0}^{\dagger }({\bf r}_{2})]\sim \;%
\frac{1}{\pi ^{3/2}l_{0}^{3}}\sim \;\frac{1}{N}\;,  \label{7.37}
\end{equation}
which shows that it is asymptotically zero, as $N\rightarrow \infty $, for
any ${\bf r}_{1}$ and ${\bf r}_{2}$. This means that the Bogolubov
prescription (\ref{7.21}) can have the sense of an approximate relation for
large confined systems satisfying condition (\ref{7.34}). A slightly
different definition of the effective thermodynamic limit for trapped atoms
will be given in Section 11. It is worth recalling that breaking gauge
symmetry is a sufficient condition for Bose-Einstein condensation but not
necessary \cite{Takahashi98}. The Einstein criterion (\ref{7.7}) or Penrose
criterion (\ref{7.12}) do not require any broken symmetry. The occurrence of
Bose-Einstein condensation in a confined system, say in a trap, can be
noticed by observing the density of atoms, which can be presented as the sum 
\begin{equation}
\rho ({\bf r})=\rho _{0}({\bf r})+\tilde{\rho}({\bf r})\;,  \label{7.38}
\end{equation}
where the first and second terms correspond to the density of atoms in a
ground state and in excited states, respectively, 
\begin{equation}
\rho _{0}({\bf r})\equiv N_{0}|\varphi _{0}({\bf r})|^{2}\;,\qquad \tilde{%
\rho}({\bf r})\equiv \sum_{n\neq 0}\;N_{n}|\varphi _{n}({\bf r})|^{2}\;.
\label{7.39}
\end{equation}
According to the Einstein or Penrose criteria, condensation happens when $%
N_{0}\sim N$, which does not involve any mentioning of gauge symmetry. In
experiments, the occurrence of condensation is manifested by the appearance
of a narrow distribution $\rho _{0}(\vec{r})$ above the wider $\tilde{\rho}(%
{\bf r})$. Because of the normalization 
\begin{equation}
\int \rho ({\bf r})\;d{\bf r}=N\;,\qquad N_{0}+\sum_{n\neq 0}\;N_{n}=N\;,
\label{7.40}
\end{equation}
the ground-state density $\rho _{0}({\bf r})$ becomes noticeable when $%
N_{0}\sim N$. Although condensation in a trap is a gradual crossover, the
latter can be rather sharp reminding a phase transition occurring at a point.

\chapter{Coherent Atomic States}

One usually connects the occurrence of Bose-Einstein condensation with the
appearance of coherence in an atomic system. This sounds reasonable since
inequalities (\ref{7.1}) and (\ref{7.3}) can be interpreted as the
conditions of coherence. In order to understand better the relation between
condensation and coherence, it is necessary to give a rigorous definition of
coherent states and to study their main properties. This is done in the
following subsections that are based on Ref.~\cite{Yukalov98}.

\section{Definition and Main Properties}

We consider the field operators $\psi ({\bf r})$ and $\psi ^{\dagger }({\bf r%
})$ satisfying the Bose commutation relations and defined on the Fock space $%
{\cal F}$. A state $h\in {\cal F}$ is called a {\it coherent state} if it is
an eigenvector of the annihilation operator 
\begin{equation}
\psi ({\bf r})h=\eta ({\bf r})h\;.  \label{8.1}
\end{equation}
The function, $\eta ({\bf r})$, playing the role of an eigenvalue, can be
called the {\it coherent field}. The latter is assumed to be normalizable
with the norm $||\eta ||\equiv (\eta ,\eta )$ defined by means of the scalar
product 
\[
(\eta ,\eta ^{\prime })\equiv \int \;\eta ^{\ast }({\bf r})\;\eta ^{\prime }(%
{\bf r})\;d{\bf r}\;. 
\]
The coherent state $h$ is not an eigenvector of the creation operator. But
there is a useful property 
\begin{equation}
h^{+}\psi ^{\dagger }({\bf r})=\eta ^{\ast }({\bf r})\;h^{+}  \label{8.2}
\end{equation}
that follows from the Hermitian conjugation of Eq.~(\ref{8.1}). The state $h$%
, being a vector of the Fock space ${\cal F}$, is presentable as a column 
\begin{equation}
h=\{h_{k}({\bf r}_{1},{\bf r}_{2},\ldots ,{\bf r}_{k})|\;k=0,1,2,\ldots \}\;.
\label{8.3}
\end{equation}
From the definition (\ref{8.1}) one can derive that 
\begin{equation}
h_{k}({\bf r}_{1},{\bf r}_{2},\ldots ,{\bf r}_{k})=\;\frac{C_{0}}{\sqrt{k!}}%
\;\prod_{j=1}^{k}\;\eta ({\bf r}_{j})\;.  \label{8.4}
\end{equation}
Requiring that the state (\ref{8.3}) be normalized to unity, 
\begin{equation}
h^{+}h=\sum_{i=1}^{\infty }\;(h_{i},h_{i})=1\;,  \label{8.5}
\end{equation}
where 
\[
(h_{i},h_{i})\equiv \int \;|h_{i}({\bf r}_{1},{\bf r}_{2},\ldots ,{\bf r}%
_{i})|^{2}\;d{\bf r}_{1}d{\bf r}_{2}\ldots d{\bf r}_{i}\;, 
\]
one gets the normalization constant 
\[
|C_{0}|=\exp \left\{ -\;\frac{1}{2}(\eta ,\eta )\right\} \;. 
\]
Two different coherent states are not orthogonal since the product 
\begin{equation}
h^{+}h^{\prime }=\exp \left\{ -\;\frac{1}{2}(\eta ,\eta )+(\eta ,\eta
^{\prime })-\;\frac{1}{2}(\eta ^{\prime },\eta ^{\prime })\right\}
\label{8.6}
\end{equation}
is not zero.

It is possible to introduce time-dependent coherent states 
\begin{equation}  \label{8.7}
h(t) =\hat U(t)\; h
\end{equation}
by means of the evolution operator $\hat U(t)$, which is a unitary operator
satisfying the Schr\"odinger equation 
\begin{equation}  \label{8.8}
i\hbar\; \frac{d}{dt}\; \hat U(t) = H\hat U(t) \; ,
\end{equation}
where $H$ is a Hamiltonian depending, in general, on time. The
generalization of definition (\ref{8.1}) is the eigenproblem 
\begin{equation}  \label{8.9}
\psi({\bf r})h(t) =\eta({\bf r},t)h(t) \; .
\end{equation}
Similarly to the time-independent case, one may derive that $h(t)$ has the
structure of the column 
\begin{equation}  \label{8.10}
h(t) =\left\{ \frac{C_0(t)}{\sqrt{k!}}\; \prod_{j=1}^k \; \eta({\bf r}_j,t)
\right\} \; .
\end{equation}
The eigenproblem (\ref{8.9}) can also be presented in the form 
\begin{equation}  \label{8.11}
\psi({\bf r},t)\; h =\eta({\bf r},t)\; h
\end{equation}
involving the time-dependent field operator 
\begin{equation}  \label{8.12}
\psi({\bf r},t) =\hat U^+(t)\;\psi({\bf r})\;\hat U(t) \; .
\end{equation}

If the evolution of the system is prescribed by the Schr\"{o}dinger equation
(\ref{8.8}), then the coherent field $\eta ({\bf r},t)$ is not arbitrary.
Let us take the system Hamiltonian in the standard form 
\[
H=\int \;\psi ^{\dagger }({\bf r},t)\left[ -\;\frac{\hbar ^{2}{\bf \nabla }%
^{2}}{2m_{0}}+U({\bf r},t)\right] \;\psi ({\bf r},t)\;d{\bf r}\;+ 
\]
\begin{equation}
+\;\frac{1}{2}\;\int \;\psi ^{\dagger }({\bf r},t)\;\psi ^{\dagger }({\bf r}%
^{\prime },t)\;\Phi ({\bf r}-{\bf r}^{\prime })\;\psi ({\bf r}^{\prime
},t)\;\psi ({\bf r},t)\;d{\bf r}\;d{\bf r}^{\prime }\;,  \label{8.13}
\end{equation}
with the interaction potential $\Phi (-{\bf r})=\Phi ({\bf r})$. The
evolution prescribed by Eqs.~(\ref{8.8}) and (\ref{8.12}) yields the
Heisenberg equation 
\[
i\hbar \;\frac{\partial }{\partial t}\;\psi ({\bf r},t)=[\psi ({\bf r}%
,t),\;H]\;, 
\]
which is also equivalent to the variational equation 
\[
i\hbar \;\frac{\partial \psi ({\bf r},t)}{\partial t}=\frac{\delta H}{\delta
\psi ^{\dagger }({\bf r},t)}\;. 
\]
With the Hamiltonian (\ref{8.13}), the evolution equation for the field
operator (\ref{8.12}) is 
\begin{equation}
i\hbar \;\frac{\partial }{\partial t}\;\psi ({\bf r},t)=H(\psi )\;\psi ({\bf %
r},t)\;,  \label{8.14}
\end{equation}
where 
\begin{equation}
H(\psi )\equiv -\;\frac{\hbar ^{2}{\bf \nabla }^{2}}{2m_{0}}+U({\bf r}%
,t)+\int \;\Phi ({\bf r}-{\bf r}^{\prime })\;\psi ^{\dagger }({\bf r}%
^{\prime },t)\;\psi ({\bf r}^{\prime },t)\;d{\bf r}^{\prime }\;.
\label{8.15}
\end{equation}
Multiplying Eq. (\ref{8.14}) by $h^{+}$ from the left and by $h$ from the
right, and using definition (\ref{8.11}), yields the evolution equation for
the coherent field 
\begin{equation}
i\hbar \;\frac{\partial }{\partial t}\;\eta ({\bf r},t)=H(\eta )\;\eta ({\bf %
r},t)\;,  \label{8.16}
\end{equation}
with the effective nonlinear Hamiltonian 
\begin{equation}
H(\eta )=-\;\frac{\hbar ^{2}{\bf \nabla }^{2}}{2m_{0}}+U({\bf r},t)+\int
\;\Phi ({\bf r}-{\bf r}^{\prime })\;|\eta ({\bf r}^{\prime },t)|^{2}\;d{\bf r%
}^{\prime }\;.  \label{8.17}
\end{equation}
The {\it nonlinear Schr\"{o}dinger equation} (\ref{8.16}) is the exact
equation for the coherent field.

The norm of the coherent field has not yet been specified and, in general,
it can be arbitrary. It is convenient to introduce the coherent field $%
\varphi({\bf r},t)$ normalized to unity, so that 
\begin{equation}  \label{8.18}
\eta({\bf r},t) \equiv \sqrt{\kappa} \; \varphi({\bf r},t) \; ,
\end{equation}
where $\kappa$ is an arbitrary positive number and 
\[
(\eta,\eta)=\kappa\; , \qquad (\varphi,\varphi)=1\; . 
\]
Then Eqs. (\ref{8.16}) and (\ref{8.17}) are transformed to the {\it %
nonlinear Schr\"odinger equation} 
\begin{equation}  \label{8.19}
i\hbar\; \frac{\partial}{\partial t} \; \varphi({\bf r},t) = \hat
H(\varphi)\; \varphi({\bf r},t)
\end{equation}
with the nonlinear Hamiltonian 
\begin{equation}  \label{8.20}
\hat H(\varphi) \equiv -\; \frac{\hbar^2{\bf \nabla}^2}{2m_0} + U({\bf r},t)
+ \kappa\; \int \; \Phi({\bf r}-{\bf r}^{\prime})\; |\varphi({\bf r}%
^{\prime},t)|^2\; d{\bf r}^{\prime}\; .
\end{equation}
Notice that Eq. (\ref{8.19}) has nontrivial solutions for the coherent field
only if the interaction potential is integrable in the sense of condition (%
\ref{7.30}). A system of particles with a nonintegrable interaction
potential cannot possess coherent states.

\section{Stationary Coherent States}

If the external potential $U({\bf r},t)=U({\bf r})$ does not depend on time,
then the nonlinear Schr\"odinger equation (\ref{8.19}) has stationary
solutions of the form 
\begin{equation}  \label{8.21}
\varphi_n({\bf r},t) = \varphi_n({\bf r})\; \exp\left ( -\; \frac{i}{\hbar}
\; E_n\; t\right ) \; ,
\end{equation}
in which $\varphi_n({\bf r})$ and $E_n$ are defined by the eigenproblem 
\begin{equation}  \label{8.22}
\hat H(\varphi_n)\; \varphi_n({\bf r}) = E_n\; \varphi_n({\bf r}) \; .
\end{equation}
The stationary solutions $\varphi_n$ labelled by a multi-index $n$ can be
called {\it coherent modes}.

The Hamiltonian (\ref{8.20}) is nonlinear, hence cannot be Hermitian.
Therefore the set of solutions $\{\varphi _{n}({\bf r})\}$ to the
eigenproblem (\ref{8.22}) does not necessarily form a complete orthonormal
basis. Actually, even in the case of a Hermitian operator in an
infinite-dimensional space, the set of its eigenvectors does not always form
such a basis \cite{Jordan69,Levine91}, contrary to the case of Hermitian
operators in finite-dimensional spaces.

Nonlinear eigenproblems are usually solved by an iterative procedure, as it
is done for self-consistent mean-field problems like Hartree or Hartree-Fock
equations \cite{Thouless61,Slater63}. In the process of such solutions it is
often possible to preserve the orthogonality of eigenvectors, at least
approximately in the sense of the inequality 
\[
|(\varphi _{m},\varphi _{n})|\ll 1\;,\qquad m\neq n\;. 
\]
The latter, because of Eq. (\ref{8.22}), is equivalent to the condition 
\[
|(\hat{H}\varphi _{m},\varphi _{n})-(\varphi _{m},\hat{H}\varphi _{n})|\ll
|E_{m}-E_{n}|\;, 
\]
if $m\neq n$ and $\hat{H}\varphi \equiv \hat{H}(\varphi )\varphi $. Here,
with the Hamiltonian (\ref{8.20}), we have 
\[
(\hat{H}\varphi _{m},\varphi _{n})-(\varphi _{m},\hat{H}\varphi _{n})= 
\]
\[
=\kappa \;\int \;\varphi _{m}^{\ast }({\bf r})\;\Phi ({\bf r}-{\bf r}%
^{\prime })\;\left[ |\varphi _{m}(\vec{r}\;^{\prime })|^{2}-|\varphi _{n}(%
{\bf r}^{\prime })|^{2}\right] \;\varphi _{n}({\bf r})\;d{\bf r}\;d{\bf r}%
^{\prime }\;. 
\]
Thus, the set $\{\varphi _{n}\}$, in general, is not orthogonal although can
often be made quasiorthogonal, so that $|(\varphi _{m},\varphi _{n})\ll 1$
if $m\neq n$. The modes $\varphi _{n}$ can always be normalized to $%
||\varphi _{n}||=1$.

One may notice that the eigenproblem (\ref{8.22}), with the Hamiltonian (\ref
{8.20}), defines the coherent modes up to a phase factor $e^{i\alpha }$ with
an arbitrary real phase $\alpha $. Therefore, the general solution of the
eigenproblem (\ref{8.22}) writes 
\begin{equation}
\varphi _{n\alpha }({\bf r})=\varphi _{n}({\bf r})\;e^{i\alpha }\qquad
(0\leq \alpha \leq 2\pi )\;.  \label{8.23}
\end{equation}
The phase $\alpha $ is an {\it unobservable random variable} uniformly
distributed in the interval $[0,2\pi )$, which has to be averaged over when
evaluating the expectation values of observables \cite{Mandel95}-\cite
{Zimanyi99}. The {\it random global phase} $\alpha $ should not be confused
with a local phase of the coherent mode $\varphi _{n}$, which can arise in
the process of solution of the eigenproblem (\ref{8.22}) and which is
determined by this eigenproblem.

Keeping all numbers $\kappa ,n$, and $\alpha $ fixed, we have a {\it pure
coherent mode} 
\begin{equation}
\eta _{\kappa n\alpha }=\sqrt{\kappa }\;\varphi _{n}({\bf r})\;e^{i\alpha
}\;.  \label{8.24}
\end{equation}
Then, Eqs. (\ref{8.3}) and (\ref{8.4}) define a {\it pure coherent state} 
\begin{equation}
h_{\kappa n\alpha }=\left\{ \frac{C_{0}}{\sqrt{k!}}\;\prod_{j=1}^{k}\;\eta
_{\kappa n\alpha }({\bf r}_{j})\right\} \;,  \label{8.25}
\end{equation}
in which 
\[
|C_{0}|=\exp \left( -\;\tfrac{1}{2}\;\kappa \right) \;. 
\]
The pure coherent states are not orthogonal since 
\begin{equation}
h_{\kappa n\alpha }^{+}\;h_{\kappa ^{\prime }n^{\prime }\alpha ^{\prime
}}=\exp \left\{ -\;\tfrac{1}{2}\;(\kappa +\kappa ^{\prime })+\sqrt{\kappa
\kappa ^{\prime }}\;(\varphi _{n},\varphi _{n^{\prime }})\;e^{-i(\alpha
-\alpha ^{\prime })}\right\} \;.  \label{8.26}
\end{equation}
But they are asymptotically orthogonal if either $\kappa \rightarrow \infty $
or $\kappa ^{\prime }\rightarrow \infty $. For example, if the pure coherent
modes (\ref{8.24}) are normalized to the number of particles $N=\kappa
=\kappa ^{\prime }$, then the product (\ref{8.26}) is 
\begin{equation}
h_{Nn\alpha }^{+}\;h_{Nn^{\prime }\alpha ^{\prime }}=\exp \left\{
-N+(\varphi _{n},\varphi _{n^{\prime }})\;N\;e^{i(\alpha ^{\prime }-\alpha
)}\right\} \;.  \label{8.27}
\end{equation}
This shows that two different pure states, for which either $n\neq n^{\prime
}$ or $\alpha \neq \alpha ^{\prime }$, are asymptotically orthogonal in the
sense of the limit 
\begin{equation}
\lim_{N\rightarrow \infty }h_{Nn\alpha }^{+}\;h_{Nn^{\prime }\alpha ^{\prime
}}=\delta _{nn^{\prime }}\;\delta _{\alpha \alpha ^{\prime }}\;.
\label{8.28}
\end{equation}
Thus, the set $\{h_{Nn\alpha }\}$ of pure coherent states forms a normalized
asymptotically, with respect to $N\rightarrow \infty $, orthogonal basis.
This basis is asymptotically, as $N\rightarrow \infty $, completely
providing the resolution of unity 
\begin{equation}
\sum_{n}\;h_{Nn\alpha }\;h_{Nn\alpha }^{+}\simeq \hat{1}\qquad (N\rightarrow
\infty )\;.  \label{8.29}
\end{equation}
The latter equality is to be understood in the weak sense as an equality for
the matrix elements 
\begin{equation}
h_{Nn\alpha }^{+}\;\hat{1}\;h_{Nn^{\prime }\alpha ^{\prime }}\simeq \delta
_{nn^{\prime }}\;\delta _{\alpha \alpha ^{\prime }}\qquad (N\rightarrow
\infty )\;.  \label{8.30}
\end{equation}
Hence the set $\{h_{Nn\alpha }\}$ can be treated as an asymptotically
complete and orthonormalized basis, when $N\rightarrow \infty $.

\section{Quantum Coherent Averages}

For an operator $\hat A$, we can define the {\it pure coherent average} 
\begin{equation}  \label{8.31}
<\hat A>_{\kappa n\alpha} \equiv h^+_{\kappa n\alpha}\; \hat A \; h_{\kappa
n\alpha}
\end{equation}
with respect to the pure coherent states (\ref{8.25}). Thus for the field
operator, one has 
\begin{equation}  \label{8.32}
<\psi({\bf r})>_{\kappa n\alpha}\; = \sqrt{\kappa}\; \varphi_n({\bf r}) \;
e^{i\alpha} \; ,
\end{equation}
which tells us that the usage of pure coherent states breaks gauge symmetry.
The first-order density matrix factorizes as 
\begin{equation}  \label{8.33}
<\psi^\dagger({\bf r})\psi({\bf r}^{\prime})>_{\kappa n\alpha}\; = \;
<\psi^\dagger({\bf r})>_{\kappa n\alpha} \; <\psi({\bf r}^{\prime})>_{\kappa
n\alpha} \; .
\end{equation}
For the particle density operator 
\[
\hat n({\bf r}) \equiv \psi^\dagger({\bf r}) \; \psi({\bf r})\; , 
\]
one gets 
\begin{equation}  \label{8.34}
<\hat n({\bf r})>_{\kappa n\alpha}\; = \kappa \; |\varphi_n({\bf r})|^2 \; .
\end{equation}
And the density-density correlation function is 
\begin{equation}  \label{8.35}
<\hat n({\bf r})\;\hat n({\bf r}^{\prime})>_{\kappa n\alpha}\; =\; <\hat n(%
{\bf r})>_{\kappa n\alpha}\; <\hat n({\bf r}^{\prime})>_{\kappa n\alpha} \;
+ \; \delta({\bf r}-{\bf r}^{\prime}) \; <\hat n({\bf r})>_{\kappa n\alpha}
\; .
\end{equation}
This function, in general, is not factorized. However, there is no great
sense to consider correlations at one point. It is meaningful to consider
the correlations only for ${\bf r}\neq{\bf r}^{\prime}$. In the latter case,
the correlation function (\ref{8.35}) is factorized.

The average number of particles is 
\begin{equation}  \label{8.36}
<\hat N>_{\kappa n\alpha}\; =\kappa \; , \qquad \hat N \equiv\int\; \hat n(%
{\bf r})\; d{\bf r}\; .
\end{equation}
And the average of the Hamiltonian (\ref{8.13}) writes 
\[
<H>_{\kappa n\alpha} \; =\kappa\; \int\; \varphi_n^*({\bf r}) \left [ -\; 
\frac{\hbar^2{\bf \nabla}^2}{2m_0} + U({\bf r}) \right ] \; \varphi_n({\bf r}%
) \; d{\bf r} \; + 
\]
\begin{equation}  \label{8.37}
+ \; \frac{\kappa^2}{2}\; \int \; |\varphi_n({\bf r})|^2 \; \Phi({\bf r} -%
{\bf r}^{\prime}) \; |\varphi_n({\bf r}^{\prime})|^2\; d{\bf r} \; d{\bf r}%
\; ^{\prime}\; ,
\end{equation}
where a stationary external potential is assumed. Employing the eigenproblem
(\ref{8.22}), the average energy (\ref{8.37}) can be rewritten in two other
forms, 
\[
<H>_{\kappa n\alpha}\; =\kappa\; E_n -\; \frac{\kappa^2}{2}\; \int \;
|\varphi_n({\bf r})|^2 \; \Phi({\bf r}-{\bf r}^{\prime})| \; \varphi_n({\bf r%
}^{\prime})|^2 \; d{\bf r}\; d{\bf r}^{\prime}\; , 
\]
\begin{equation}  \label{8.38}
<H>_{\kappa n\alpha} \; = \frac{\kappa}{2}\; E_n + \frac{\kappa}{2}\; \int\;
\varphi_n^*({\bf r})\left [ -\; \frac{\hbar^2{\bf \nabla}^2}{2m_0} + U({\bf r%
})\right ] \; \varphi_n({\bf r}) \; d{\bf r}\; .
\end{equation}

As has been explained above, the phase $\alpha $ in the coherent modes (\ref
{8.23}) and (\ref{8.24}) is an unobservable random variable that has to be
averaged out when calculating the expectation values of operators. This
means that the pure coherent averages (\ref{8.31}), strictly speaking, do
not correspond to observable quantities. The latter are to be defined as the
averages over the {\it random-phase coherent state} \cite{Mandel95} 
\begin{equation}
h_{\kappa n}\equiv \left\{ h_{\kappa n\alpha }|\;\alpha \in \lbrack 0,2\pi
)\right\}  \label{8.39}
\end{equation}
being the set of all pure states $h_{\kappa n\alpha }$, with the random
variable $\alpha $. The corresponding {\it coherent average} is 
\begin{equation}
<\hat{A}>_{\kappa n}\;\equiv \frac{1}{2\pi }\;\int_{0}^{2\pi }<\hat{A}%
>_{\kappa n\alpha }d\alpha \;.  \label{8.40}
\end{equation}
Then, for instance, for the field operator one has 
\begin{equation}
<\psi ({\bf r})>_{\kappa n}\;=0\;,  \label{8.41}
\end{equation}
which shows that for the coherent state (\ref{8.39}) gauge symmetry is not
broken. This sounds rather reasonable since the field operator does not
pertain to the algebra of observables \cite{Yukalov98,Berezin66}. One more
reason for the absence of broken gauge symmetry is that Eq. (\ref{8.41}) is
in agreement with the conservation of the number of particles \cite
{Bogolubov70}. The absence of broken gauge symmetry does not preclude the
first-order density matrix from being factorized as 
\begin{equation}
<\psi ^{\dagger }({\bf r})\psi ({\bf r}^{\prime })>_{\kappa n}\;=\kappa
\;\varphi _{n}^{\ast }({\bf r})\;\varphi _{n}({\bf r}^{\prime })\;.
\label{8.42}
\end{equation}
But the form (\ref{8.33}) does not hold true, 
\begin{equation}
<\psi ^{\dagger }({\bf r})\psi ({\bf r}^{\prime })>_{\kappa n}\;\neq \;<\psi
^{\dagger }({\bf r})>_{\kappa n}\;<\psi ({\bf r}^{\prime })>_{\kappa n}\;,
\label{8.43}
\end{equation}
because of Eq. (\ref{8.41}).

The operators $\hat A$ pertaining to the algebra of observables contain the
products of even numbers of field operators, with equal numbers of creation
and annihilation operators. Therefore, for such operators, the coherent
averages (\ref{8.40}) coincide with the pure averages (\ref{8.31}). For
example, the average particle density 
\[
<\hat n({\bf r})>_{\kappa n}\; =\kappa \; |\varphi_n({\bf r})|^2 
\]
is the same as the density (\ref{8.34}). The density-density correlation
function 
\begin{equation}  \label{8.44}
<\hat n({\bf r})\; \hat n({\bf r}^{\prime})>_{\kappa n}\; =\; <\hat n({\bf r}
)>_{\kappa n}\; <\hat n({\bf r}^{\prime})>_{\kappa n}\; +\; \delta({\bf r}- 
{\bf r}^{\prime}) \; <\hat n({\bf r})>_{\kappa n}
\end{equation}
is analogous to that of (\ref{8.35}). Again considering the correlations for
different points ${\bf r}\neq{\bf r}^{\prime}$, we see that the correlator (%
\ref{8.44}) factorizes. One may also notice that the second term in the
right-hand side of Eq. (\ref{8.44}) can be omitted when $\kappa\gg 1$, since
on average it is much smaller than the first term. This follows from the
integration 
\[
\int\; <\hat n({\bf r})\; \hat n({\bf r}^{\prime})>_{\kappa n}\; d{\bf r}\;
d {\bf r}^{\prime}= \kappa^2 +\kappa \; . 
\]

For any two operators $\hat{A}({\bf r})$ and $\hat{B}({\bf r}^{\prime })$,
the correlator defined through the coherent average (\ref{8.40}) always
factorizes if ${\bf r}\neq {\bf r}^{\prime }$. When both these operators
pertain to the algebra of observables, the factorization takes the form 
\begin{equation}
<\hat{A}({\bf r})\hat{B}({\bf r}^{\prime })>_{\kappa n}\;=\;<\hat{A}({\bf r}%
)>_{\kappa n}\;<\hat{B}({\bf r}^{\prime })>_{\kappa n}\;,  \label{8.45}
\end{equation}
if ${\bf r}\neq {\bf r}^{\prime }$. However, if one of these operators is
not from the algebra of observables, then the correlator does not have the
form (\ref{8.45}), although the factorization does occur. For instance, the
density matrix (\ref{8.42}) is factorized, though $\varphi _{n}({\bf r})$ is
not proportional to $<\psi ({\bf r})>_{\kappa n}$, which is zero according
to Eq. (\ref{8.41}). The factorization properties of the coherent averages (%
\ref{8.40}) are not connected with gauge symmetry breaking. The latter
occurs only for the pure average (\ref{8.31}), with a fixed global phase.
However, it looks unphysical to fix a phase that is random and cannot be
measured.

\section{Statistical Coherent Averages}

For a system of many particles, the {\it statistical state} is presented by
a given statistical operator $\hat{\rho}$ defining the expectation values of
operators as {\it statistical averages} 
\begin{equation}
<\hat{A}>\;=\;{\rm Tr}\;\hat{\rho}\;\hat{A}\;.  \label{8.46}
\end{equation}
Since the trace operation does not depend on the chosen basis, we may take
for this purpose the basis $\{h_{\kappa n\alpha }\}$ of the coherent states (%
\ref{8.25}). Then the statistical average (\ref{8.46}) is presented as 
\begin{equation}
<\hat{A}>\;=\;\sum_{n}\;\int_{0}^{\infty }\;\int_{0}^{2\pi }<\hat{\rho}\hat{A%
}>_{\kappa n\alpha }d\kappa \;d\alpha \;,  \label{8.47}
\end{equation}
where $<\ldots >_{\kappa n\alpha }$ is the pure coherent average (\ref{8.31}%
). For an equilibrium statistical state, the Gibbs statistical operator is 
\begin{equation}
\hat{\rho}=\;\frac{\exp \{-\beta (H-\mu \hat{N})\}}{{\rm Tr}\exp \{-\beta
(H-\mu \hat{N})\}}\;,  \label{8.48}
\end{equation}
where $\beta \equiv (k_{B}T)^{-1}$ and the chemical potential $\mu $ can be
found from the condition $<\hat{N}>=N$. The statistical operator (\ref{8.48}%
) represents the grand canonical Gibbs ensemble.

In the thermodynamic limit, $N\rightarrow \infty $, the coherent states
become sharply peaked around the average number of particles \cite{Negele88}%
. If we make a reasonable assumption that, integrating over the norm $\kappa
\equiv ||\eta ||$ in the average (\ref{8.47}), the main contribution, when $%
N\rightarrow \infty $, comes from the term with $\kappa =N$, then the value (%
\ref{8.47}) is asymptotically close to the {\it statistical coherent average}
\begin{equation}
<\hat{A}>_{N}\;\equiv \;\sum_{n}\;\int_{0}^{2\pi }\;<\hat{\rho}\;\hat{A}%
>_{Nn\alpha }\;d\alpha  \label{8.49}
\end{equation}
defined as a trace over the restricted Hilbert space 
\begin{equation}
{\cal H}_{N}\equiv {\cal L}\{h_{Nn\alpha }\}  \label{8.50}
\end{equation}
being a linear envelope of the coherent basis $\{h_{Nn\alpha }\}$. Since the
observable quantities 
\begin{equation}
<\hat{N}>_{Nn\alpha }\;=N\;,\qquad <H>_{Nn\alpha }\;\equiv E_{Nn}
\label{8.51}
\end{equation}
do not depend on the unobservable random phase $\alpha $, and because of the
asymptotic orthogonality (\ref{8.28}), the average (\ref{8.49}) can be
written as 
\begin{equation}
<\hat{A}>_{N}\;\simeq \sum_{n}\;\rho _{Nn}\;<\hat{A}>_{Nn}\;,  \label{8.52}
\end{equation}
with the coherent statistical weight 
\begin{equation}
\rho _{Nn}\;\equiv \;\frac{\exp (-\beta E_{Nn})}{\sum_{n}\exp (-\beta E_{Nn})%
}  \label{8.53}
\end{equation}
and with the coherent average 
\begin{equation}
<\hat{A}>_{Nn}\;\equiv \;\frac{1}{2\pi }\;\int_{0}^{2\pi }<\hat{A}%
>_{Nn\alpha }d\alpha \;.  \label{8.54}
\end{equation}
The averaged field operator is zero, 
\begin{equation}
<\psi ({\bf r})>_{N}\;=0\;,  \label{8.55}
\end{equation}
showing that gauge symmetry is not broken. The density matrix 
\begin{equation}
<\psi ^{\dagger }({\bf r})\;\psi ({\bf r}^{\prime })>_{N}\;\simeq
N\sum_{n}\;\rho _{Nn}\varphi _{n}^{\ast }({\bf r})\varphi _{n}({\bf r}%
^{\prime })  \label{8.56}
\end{equation}
does not factorize as well as the density-density correlation function 
\begin{equation}
<\hat{n}({\bf r})\hat{n}({\bf r}^{\prime })>_{N}\simeq N^{2}\sum_{n}\;\rho
_{Nn}|\varphi _{n}({\bf r})|^{2}|\varphi _{n}({\bf r}^{\prime })|^{2}\;.
\label{8.57}
\end{equation}
But as temperature tends to zero, and $\beta \rightarrow \infty $, then the
system tends to the ground-state energy level 
\begin{equation}
E_{N0}\equiv \min_{n}E_{Nn}\;,  \label{8.58}
\end{equation}
so that the weight (\ref{8.53}) becomes 
\begin{equation}
\rho _{Nn}\rightarrow \delta _{n0}\qquad (\beta \rightarrow \infty )\;.
\label{8.59}
\end{equation}
Then the correlators (\ref{8.56}) and (\ref{8.57}) asymptotically factorize.
For example, the density-density correlator is 
\[
<\hat{n}({\bf r})\;\hat{n}({\bf r}^{\prime })>_{N}\;\simeq \;<\hat{n}({\bf r}%
)>_{N}\;<\hat{n}({\bf r}^{\prime })>_{N}\;, 
\]
when $\beta \rightarrow \infty $. The atom density 
\[
<\hat{n}({\bf r})>_{N}\;\simeq N\sum_{n}\;\rho _{Nn}\;|\varphi _{n}({\bf r}
)|^{2} 
\]
can be written as 
\begin{equation}
<\hat{n}({\bf r})>_{N}\;\simeq N\;\rho _{N0}\;|\varphi _{0}({\bf r}
)|^{2}+N\sum_{n\neq 0}|\varphi _{n}({\bf r})|^{2}\;.  \label{8.60}
\end{equation}
At high temperatures, when the first term in Eq. (\ref{8.60}) is negligible,
one may say that the system is in thermal state. The temperature $T_{c}$, at
which the coherent first term becomes noticeable, characterizes the
transition to the coherent state. Since the ground-state wave function is
better localized in space than the wave functions of excited modes, the
increase of the first term in the sum (\ref{8.60}) can be noticed as the
appearance of a narrow space distribution above the wide thermal cloud
described by the second term. With lowering temperature below $T_{c}$, the
sharp coherent peak described by the first term grows while the wide thermal
distribution corresponding to the second term diminishes. At zero
temperature, all atoms are concentrated in the ground-state coherent mode.
In this way, Bose-Einstein condensation can be understood as a transition of
atoms from excited single-state coherent modes to the ground-state coherent
mode.

Note that the coherent states (\ref{8.25}) and (\ref{8.39}) are not the
eigenvectors of the system Hamiltonian (\ref{8.13}). However, this does not
preclude the statistical average (\ref{8.46}) from satisfying the limiting
relation 
\[
\lim_{N\rightarrow\infty}\; \lim_{\beta\rightarrow\infty} \; \frac{1}{N}\;
<\hat A>\; = \; \lim_{N\rightarrow\infty}\; \frac{1}{N}\; <\hat A>_{N0} \; . 
\]
And there is no contradiction between the {\it many-particle} coherent
states (\ref{8.25}) or (\ref{8.39}) not being the eigenvectors of the system
Hamiltonian and the ability of atoms to condense into a {\it single-particle}
coherent mode (\ref{8.23}).

\section{Correlation Functions and Coherence}

As follows from the previous subsection, an equilibrium system of atoms can
become totally coherent only in the thermodynamic limit at zero temperature.
At finite temperatures or for a finite number of particles, a system of
atoms can be only partially coherent. Since the level of coherence is
directly related to the strength of correlations between atoms, this level
can be characterized by the behavior of correlation functions.

The simplest correlation function is the dimensionless first-order density
matrix 
\begin{equation}  \label{8.61}
C({\bf r},{\bf r}^{\prime})\equiv \; \frac{<\psi^\dagger({\bf r})\psi({\bf r}
^{\prime})>}{\sqrt{\rho({\bf r}) \rho({\bf r}^{\prime})}} \; ,
\end{equation}
where $\rho({\bf r})\equiv\;<\hat n({\bf r})>$. This function has the
properties 
\[
C^*({\bf r},{\bf r}^{\prime}) = C({\bf r}^{\prime},{\bf r})\; , \qquad C( 
{\bf r},{\bf r})=1 \; . 
\]
Averaging over the coherent states $h_{Nn}$ according to the definition (\ref
{8.40}), we have 
\begin{equation}  \label{8.62}
<\psi({\bf r})>_{Nn} \; = 0 \; ;
\end{equation}
and the density matrix writes 
\begin{equation}  \label{8.63}
<\psi^\dagger({\bf r})\; \psi({\bf r}^{\prime})>_{Nn} \; = N\; \varphi_n^*( 
{\bf r}) \; \varphi_n({\bf r}^{\prime}) \; .
\end{equation}
Hence, if the average in the function (\ref{8.61}) is assumed as the
coherent average (\ref{8.40}), then 
\begin{equation}  \label{8.64}
|C({\bf r},{\bf r}^{\prime})|=1 \qquad (coherence)
\end{equation}
for all ${\bf r}$ and ${\bf r}^{\prime}$. But for the statistical average (%
\ref{8.46}), in general, 
\begin{equation}  \label{8.65}
|C({\bf r},{\bf r}^{\prime})| \leq 1\; ,
\end{equation}
with the equality occurring only for ${\bf r}={\bf r}^{\prime}$.

An effective radius characterizing the length of strong correlations between
atoms defines the {\it coherence radius} 
\begin{equation}
r_{coh}\equiv \;\frac{\int r|C(0,{\bf r})|d{\bf r}}{\int |C(0,{\bf r})|d{\bf %
r}}\;.  \label{8.66}
\end{equation}
If this radius is less than or comparable with the mean interatomic
distance, 
\begin{equation}
r_{coh}\leq a\qquad (chaos)\;,  \label{8.67}
\end{equation}
then atoms are not correlated, which can be ascribed to chaotic behavior.
When the coherence radius is much larger than the distance $a$ but much
smaller than the linear size of the system, 
\begin{equation}
a\ll r_{coh}\ll L\qquad (local\;coherence)\;,  \label{8.68}
\end{equation}
then a large number of atoms are mutually correlated, although this number
is much smaller than the total number of atoms in the system. And when the
coherence radius is of the order of the system size, 
\begin{equation}
r_{coh}\sim L\qquad (global\;coherence)\;,  \label{8.69}
\end{equation}
then almost all atoms in the system are correlated and practically all
particles condense into a coherent mode. The correlation function (\ref{8.61}%
) describes the property of the system, which is called {\it first-order
coherence}.

The correlation function 
\begin{equation}  \label{8.70}
C_2({\bf r}_1,{\bf r}_2)\equiv\; \frac{<\psi^\dagger({\bf r}_1)\psi^\dagger(%
{\bf r}_2) \psi({\bf r}_2)\psi({\bf r}_1)>}{\rho({\bf r}_1)\rho({\bf r}_2)}
\end{equation}
characterizes {\it second-order coherence}. If the average here is defined
as the coherent average (\ref{8.40}), then 
\begin{equation}  \label{8.71}
C_2({\bf r}_1,{\bf r}_2) =1 \qquad (coherence)
\end{equation}
for any ${\bf r}_1$ and ${\bf r}_2$. The opposite case corresponds to the
statistical average (\ref{8.46}) under the condition that particles are not
correlated, so that the average in Eq. (\ref{8.70}) can be simplified using
the Wick decomposition. The latter yields 
\begin{equation}  \label{8.72}
C_2({\bf r}_1,{\bf r}_2) = 1 +|C({\bf r}_1,{\bf r}_2)|^2 \qquad (chaos)\; .
\end{equation}
For this chaotic state, 
\begin{equation}  \label{8.73}
C_2({\bf r}_1,{\bf r}_2) > 1 \qquad (chaos) \; .
\end{equation}
Combining Eqs. (\ref{8.71}) and (\ref{8.72}), the coinciding arguments we
have 
\begin{eqnarray}  \label{8.74}
C_2({\bf r},{\bf r}) =\left\{ 
\begin{array}{cc}
1\; , & coherence \\ 
2\; , & chaos \; .
\end{array}
\right.
\end{eqnarray}

Similarly, the {\it third-order coherence} is described by the correlation
function 
\begin{equation}
C_{3}({\bf r}_{1},{\bf r}_{2},{\bf r}_{3})\equiv \;\frac{<\psi ^{\dagger }(%
{\bf r}_{1})\psi ^{\dagger }({\bf r}_{2})\psi ^{\dagger }({\bf r}_{3})\psi (%
{\bf r}_{3})\psi ({\bf r}_{2})\psi ({\bf r}_{1})>}{\rho ({\bf r}_{1})\rho (%
{\bf r}_{2})\rho ({\bf r}_{3})}\;.  \label{8.75}
\end{equation}
In the case of the coherent average (\ref{8.40}), 
\begin{equation}
C_{3}({\bf r}_{1},{\bf r}_{2},{\bf r}_{3})=1\qquad (coherence)\;.
\label{8.76}
\end{equation}
While for the statistical average (\ref{8.46}), under the assumption that
atoms are not correlated, one may employ the Wick decomposition resulting in 
\[
C_{3}({\bf r}_{1},{\bf r}_{2},{\bf r}_{3})=1+|C({\bf r}_{1},{\bf r}%
_{2})|^{2}+|C({\bf r}_{2},{\bf r}_{3})|^{2}+|C({\bf r}_{3},{\bf r}%
_{1})|^{2}+ 
\]
\begin{equation}
+2{\rm Re}\;C({\bf r}_{1},{\bf r}_{2})C({\bf r}_{2},{\bf r}_{3})C({\bf r}%
_{3},{\bf r}_{1})\;.  \label{8.77}
\end{equation}
In the two opposite cases, we get 
\[
C_{3}({\bf r},{\bf r},{\bf r})=\left\{ 
\begin{array}{cc}
1\;, & coherence \\ 
3!\;, & chaos\;.
\end{array}
\right. 
\]

The same way can be followed for characterizing higher-order coherence by
means of the correlation function $C_{k}(\ldots )$ defined analogously to
Eqs. (\ref{8.70}) and (\ref{8.75}). For the coinciding arguments, one finds 
\[
C_{k}({\bf r},{\bf r},\ldots ,{\bf r})=\left\{ 
\begin{array}{cc}
1\;, & coherence \\ 
k!\;, & chaos\;.
\end{array}
\right. 
\]
One may notice that, if the system is coherent, this is reflected in the
correlators of all orders, so that 
\[
C_{k}({\bf r}_{1},{\bf r}_{2},\ldots ,{\bf r}_{k})=1\qquad (coherence)\;. 
\]
Thence, it is not compulsory to distinguish between different orders of
coherence, but it is sufficient to use just one word ''coherence''. In the
intermediate case, when there is neither complete coherence nor pure chaos,
the properties of correlators of different orders can be different. Then one
could distinguish between different orders of particle correlations.

Correlation functions can be defined not only for field operators but also
for any operators. The correlator for two operators of local observables, $%
\hat{A}({\bf r})$ and $\hat{B}({\bf r})$, satisfies an important limiting
property 
\begin{equation}
<\hat{A}({\bf r}_{1})\;\hat{B}({\bf r}_{2})>\;\simeq \;<\hat{A}({\bf r}%
_{1})>\;<\hat{B}({\bf r}_{2})>\;,  \label{8.80}
\end{equation}
when $r_{12}\rightarrow \infty $, which is called the {\it principle of
correlation weakening} \cite{Bogolubov70,Akhiezer81}. This property holds
only when both operators represent {\it local observables}. Since the field
operators do not correspond to observable quantities, the correlator $<\psi
^{\dagger }({\bf r}_{1})\psi ({\bf r}_{2})>$ does not need to be factorized
into the product $<\psi ^{\dagger }({\bf r}_{1})><\psi ({\bf r}_{2})>$, as $%
r_{12}\rightarrow \infty $, although it may factorize in a different form,
as in Eq. (\ref{8.42}).

\chapter{Meaning of Gross-Pitaevskii Equation}

The nonlinear Schr\"{o}dinger equation (\ref{8.19}) is an {\it exact
equation } defining the coherent field $\varphi ({\bf r},t)$ that can also
be called the coherent wave function. In the nonlinear Hamiltonian (\ref
{8.20}), one has to specify the external potential $U$ and the interaction
potential $\Phi $. For trapped atomic gases, these potentials are usually
modelled as a harmonic confining potential and a contact Fermi potential,
respectively \cite{Parkins98,Dalfovo99}. This concretization is given below,
where we also discuss the difference between the exact equation (\ref{8.19})
for the coherent wave function and some approximate equations for
broken-symmetry order parameter introduced by means of the Bogolubov
prescription (\ref{7.21}). For concreteness, we set the normalization
parameter $\kappa =N$.

\section{Coherent Wave Function}

The external potential $U({\bf r},t)$ in the nonlinear Hamiltonian (\ref
{8.20}) may, in general, consist of two terms, one describing a stationary
trapping potential that is due to the trap used and another term
corresponding to time-dependent perturbation superimposed on the stationary
part. The trapping potential is usually modelled by a harmonic oscillator, \ 
\begin{equation}  \label{9.1}
U({\bf r}) =\; \frac{m_0}{2}\; ( \omega_x^2 \; x^2 +\omega_y^2 \; y^2 +
\omega_z^2\; z^2) \; ,
\end{equation}
with the frequencies defined by the confining fields of the trap.

The density of trapped gases is always small, so that the effective range of
the two-body potential describing interatomic interactions is much smaller
compared to the interparticle distance $a\sim \rho ^{-1/3}$. Then the
interatomic potential can be assumed to act locally and be characterized
entirely by the $s$-wave scattering length $a_{s}$ \cite{Huang57}-\cite
{Baym99}. This means that, under the condition 
\begin{equation}
\frac{|a_{s}|}{a}\;\ll 1\;,\qquad \rho \;|a_{s}|^{3}\ll 1\;,  \label{9.2}
\end{equation}
the interaction potential can be presented in the Fermi form 
\begin{equation}
\Phi ({\bf r})=A\;\delta ({\bf r})\;,\qquad A\equiv 4\pi \hbar ^{2}\;\frac{%
a_{s}}{m_{0}}\;.  \label{9.3}
\end{equation}
In typical experiments with $^{87}$Rb and $^{23}$Na, one reaches the density 
$\rho \sim 10^{14}$ cm$^{-3}$, the scattering length being $a_{s}\sim
5\times 10^{-7}$ cm, hence $a_{s}/a\sim 10^{-2}$ and $\rho a_{s}^{3}\sim
10^{-5}$. In the case of $^{7}$Li, one has $\rho \sim 10^{12}$ cm $^{-3}$,
with the scattering length $a_{s}\sim -10^{-7}$ cm, so that $|a_{s}|/a\sim
10^{-3}$ and $\rho |a_{s}|^{3}\sim 10^{-9}$. In the Bose-Einstein
condensation of atomic hydrogen \cite{Fried98}, the density of condensed
atoms is $\rho \sim 2\times 10^{4}$ cm$^{-3}$, while the scattering length
of H is $a_{s}\sim 6.5\times 10^{-9}$ cm, from where $a_{s}/a\sim 10^{-4}$
and $\rho a_{s}^{3}\sim 10^{-10}$. Thus, the inequalities (\ref{9.2}) always
hold true.

With the trapping potential (\ref{9.1}) and the interaction potential (\ref
{9.3}), we get the nonlinear Hamiltonian 
\begin{equation}  \label{9.4}
\hat H(\varphi) = -\frac{\hbar^2{\bf \nabla}^2}{2m_0} \; + \; U({\bf r}) +
NA|\varphi|^2 \; .
\end{equation}
Then the equation (\ref{8.19}) for the coherent wave function takes the form 
\begin{equation}  \label{9.5}
i\hbar \; \frac{\partial\varphi}{\partial t} =\left [ \hat H(\varphi) + \hat
V\right ] \; \varphi \; ,
\end{equation}
where $\hat V=\hat V({\bf r},t)$ is a time-dependent perturbation potential.

From the mathematical point of view, Eq. (\ref{9.5}) is a nonlinear
Schr\"{o}dinger equation. This is an exact equation for the coherent wave
function. Similar equations can be derived in the mean-field approximation
for the order parameter associated with the condensate \cite{Dalfovo99}. In
the latter case, one calls such equations Gross-Ginzburg-Pitaevskii equation
or Gross-Pitaevskii equation, since such equations were considered by these
authors \cite{Gross57}-\cite{Gross63}.

\section{Condensate Order Parameter}

The order parameter associated with the Bose-Einstein condensate is commonly
defined as 
\begin{equation}  \label{9.6}
\psi_0({\bf r},t) \equiv\; <\psi({\bf r},t)> \; ,
\end{equation}
that is, as the statistical average of a field operator. This definition
implies that Bose-Einstein condensation is accompanied by broken gauge
symmetry, which is usually done by means of the Bogolubov prescription (\ref
{7.21}). Substituting the Bogolubov-shifted field operator $%
\psi=\psi_0+\tilde\psi$ into the Heisenberg equation (\ref{8.14}), and
averaging the latter, one has 
\[
i\hbar\; \frac{\partial}{\partial t}\; \psi_0 = \left ( -\; \frac{\hbar^2%
{\bf \nabla}^2}{2m_0} \; + \; U\right ) \psi_0 + 
\]
\begin{equation}  \label{9.7}
+ A \left ( |\psi_0|^2\; \psi_0 + 2\; <\tilde\psi^\dagger\; \tilde\psi>\;
\psi_0 + \; <\tilde\psi\tilde\psi>\; \psi_0^* + \;
<\tilde\psi^\dagger\;\tilde\psi\;\tilde\psi>\;\right ) \; ,
\end{equation}
where the dependence of functions on the space-time variables ${\bf r}$ and $%
t$, for brevity, is dropped.

Equation (\ref{9.7}) for the condensate order parameter (\ref{9.6}) is
exact. However, it is too complicated to be useful. To simplify it, one may
invoke the {\it mean-field approximation} 
\[
<\tilde\psi^\dagger\;\tilde\psi\;\tilde\psi>\; \cong \;
<\tilde\psi^\dagger>\;<\tilde\psi\; \tilde\psi> \; + 2 \;
<\tilde\psi^\dagger\; \tilde\psi>\; <\tilde\psi>\; , 
\]
which, because of $<\tilde\psi>\; =0$ yields 
\[
<\tilde\psi^\dagger\; \tilde\psi\; \tilde\psi>\; = 0 \; . 
\]
Then Eq. (\ref{9.7}) for the order parameter slightly simplifies becoming 
\begin{equation}  \label{9.8}
i\hbar\; \frac{\partial}{\partial t}\; \psi_0 = \left ( - \; \frac{\hbar^2 
{\bf \nabla}^2}{2m_0} \; + U\right ) \; \psi_0 + A\left ( |\psi_0|^2\;
\psi_0 + 2<\tilde\psi^\dagger\;\tilde\psi>\; \psi_0 + \;
<\tilde\psi\;\tilde\psi>\; \psi_0^*\right ) \; ,
\end{equation}
which corresponds to the {\it Hartree-Fock-Bogolubov approximation}.

Sometimes one makes an {\it ad hoc} assumption that the anomalous averages $<%
\tilde{\psi}\tilde{\psi}>$ are much smaller than the normal ones $<\tilde{
\psi}^{\dagger }\tilde{\psi}>$. Then, dropping the former in Eq. (\ref{9.8}%
), one comes to the {\it Popov approximation} 
\begin{equation}
i\hbar \;\frac{\partial }{\partial t}\;\psi _{0}=\left( -\;\frac{\hbar ^{2} 
{\bf \nabla }^{2}}{2m_{0}}\;+U\right) \;\psi _{0}+A\left( |\psi _{0}|^{2}+2< 
\tilde{\psi}^{\dagger }\;\tilde{\psi}>\;\right) \;\psi _{0}\;,  \label{9.9}
\end{equation}
considered first by Popov \cite{Popov87}.

The Hartree-Fock-Bogolubov approximation is self-consistent; however, it
leads to the appearance of a gap in the spectrum of elementary excitations 
\cite{Isoshima97}. Since in reality there is no gap, this approximation is
not satisfactory. The Popov approximation yields a gapless spectrum of
elementary excitations; but this approximation is not self-consistent
because the anomalous averages $<\tilde{\psi}\tilde{\psi}>$ are, in general,
of the order of or even much larger than the normal averages \cite
{Isoshima97,Griffin80}, thus, $<\tilde{\psi}\tilde{\psi}>$ cannot be
neglected when gauge symmetry is broken. Moreover, the Popov approximation
is unstable with respect to the formation of vortices with negative energy 
\cite{Isoshima97}. Therefore, the Popov approximation also cannot be
accepted as satisfactory.

As discussed in subsection 7.4, the Bogolubov prescription (\ref{7.21}) is
meaningful under the assumption of small condensate depletion, which is
expressed by inequality (\ref{7.29}). In the extreme case, one may assume
that all particles are condensed, so that all the averages $<\tilde{\psi}%
\tilde{\psi}>$ as well as $<\tilde{\psi}^{\dagger }\tilde{\psi}>$ can be
omitted. The procedure of neglecting all these averages, corresponding to
non-condensed atoms, is often termed the {\it Bogolubov approximation}. Then
Eq. (\ref{9.8}) becomes 
\begin{equation}
i\hbar \;\frac{\partial }{\partial t}\;\psi _{0}=\left( -\;\frac{\hbar ^{2}%
{\bf \nabla }^{2}}{2m_{0}}\;+U+A\;|\psi _{0}|^{2}\right) \;\psi _{0}\;.
\label{9.10}
\end{equation}
It is this {\it approximate} equation (\ref{9.10}) for the order parameter (%
\ref{9.6}) which is commonly called the {\it Gross-Pitaevskii equation}. If,
similarly to Eq. (\ref{8.18}), we change the normalization of the order
parameter by means of the replacement 
\[
\psi _{0}({\bf r},t)=\sqrt{N}\;\varphi _{0}({\bf r},t)\;, 
\]
then Eq. (\ref{9.10}) takes the same form as the equations (\ref{8.19}) or (%
\ref{9.5}) for the coherent wave function. The difference is that the
nonlinear Schr\"{o}dinger equation (\ref{9.5}) is an {\it exact equation for
the coherent wave function}, while the equation (\ref{9.10}) is an {\it %
approximate equation for the condensate order parameter}. Equation (\ref{9.5}%
) exists irrespectively of whether gauge symmetry is broken or not, while
Eq. (\ref{9.10}) presupposes broken gauge symmetry. The mathematical
structure of both Eqs. (\ref{9.5}) and (\ref{9.10}) is the same, being that
of the nonlinear Schr\"{o}dinger equation. What is different is their
physical interpretation. However, it is admissible to accept a generalized
point of view and to define the Gross-Pitaevskii equation as a nonlinear
Schr\"{o}dinger equation describing a system of Bose atoms, irrespectively
of the interpretation of the solution to this equation.

\section{General Anisotropic Case}

To study the properties of the Gross-Pitaevskii equation, it is convenient
to introduce some notations simplifying the following consideration. When
all frequencies in the trapping potential (\ref{9.1}) are different, we have
the general anisotropic case. With the help of the effective frequency 
\begin{equation}
\omega _{0}\equiv (\omega _{x}\;\omega _{y}\;\omega _{z})^{1/3}\;,
\label{9.11}
\end{equation}
we may define the dimensionless frequencies 
\begin{equation}
\omega _{1}\equiv \;\frac{\omega _{x}}{\omega _{0}}\;,\qquad \omega
_{2}\equiv \;\frac{\omega _{y}}{\omega _{0}}\;,\qquad \omega _{3}\equiv \; 
\frac{\omega _{z}}{\omega _{0}}\;.  \label{9.12}
\end{equation}
From these definitions, one has the property 
\begin{equation}
\omega _{1}\;\omega _{2}\;\omega _{3}=1\;.  \label{9.13}
\end{equation}
The oscillator length 
\begin{equation}
l_{0}\equiv \;\sqrt{\frac{\hbar }{m_{0}\omega _{0}}}  \label{9.14}
\end{equation}
characterizes an effective size of the trap. This length is used for
defining the dimensionless variables 
\begin{equation}
x_{1}\equiv \;\frac{r_{x}}{l_{0}}\;,\qquad x_{2}\equiv \frac{r_{y}}{l_{0}}%
\;,\qquad x_{3}\equiv \;\frac{r_{z}}{l_{0}}  \label{9.15}
\end{equation}
instead of the dimensional Cartesian vector ${\bf r}=\{r_{x},r_{y},r_{z}\}$.
The dimensionless coupling parameter 
\begin{equation}
g\equiv 4\pi \;\frac{a_{s}}{l_{0}}\;N  \label{9.16}
\end{equation}
describes the intensity of interactions between atoms in coherent state.
Introducing the dimensionless nonlinear Hamiltonian 
\begin{equation}
\hat{H}\equiv \;\frac{\hat{H}(\varphi )}{\hbar \omega _{0}}  \label{9.17}
\end{equation}
and the coherent wave function 
\begin{equation}
\psi ({\bf x})\equiv l_{0}^{3/2}\;\varphi ({\bf r})\;,  \label{9.18}
\end{equation}
we have 
\begin{equation}
\hat{H}=\;\frac{1}{2}\;\sum_{i=1}^{3}\;\left( -\;\frac{\partial ^{2}}{
\partial x_{i}^{2}}\;+\omega _{i}^{2}\;x_{i}^{2}\right) \;+g|\psi |^{2}\;.
\label{9.19}
\end{equation}
The eigenproblem (\ref{8.22}) reduces to the form 
\begin{equation}
\hat{H}\;\psi _{n}({\bf x})=E_{n}\;\psi _{n}({\bf x})\;,  \label{9.20}
\end{equation}
where the energy $E_{n}$ is measured in units of $\omega _{0}$. The
eigenfunction is assumed to satisfy the normalization 
\[
(\psi _{n},\psi _{n})\equiv \int \;|\psi _{n}({\bf x})|^{2}\;d{\bf x}=1\;, 
\]
in which the integral is evaluated over the whole domain of ${\bf x}%
=\{x_{1},x_{2},x_{3}\}$.

\section{Cylindrically Symmetric Trap}

When the trapping potential (\ref{9.1}) is cylindrically symmetric, so that
the transverse radial frequencies are equal, 
\begin{equation}
\omega _{x}=\omega _{y}=\omega _{r}\;,  \label{9.21}
\end{equation}
and the axial-to-radial asymmetry is described by the parameter 
\begin{equation}
\nu \equiv \;\frac{\omega _{z}}{\omega _{r}}\;,  \label{9.22}
\end{equation}
then it is convenient to introduce the following notation. The radial
oscillator length 
\begin{equation}
l_{r}\equiv \;\sqrt{\frac{\hbar }{m_{0}\omega _{r}}}  \label{9.23}
\end{equation}
serves to define the dimensionless cylindrical variables 
\begin{equation}
r\equiv \;\frac{\sqrt{r_{x}^{2}+r_{y}^{2}}}{l_{r}}\;,\qquad z=\;\frac{r_{z}}{%
l_{r}}\;.  \label{9.24}
\end{equation}
Then one may define the dimensionless coherent wave function 
\begin{equation}
\psi (r,\varphi ,z)\equiv l_{r}^{3/2}\;\varphi ({\bf r})\;,  \label{9.25}
\end{equation}
depending on the cylindrical variables $r\in \lbrack 0,\infty )$, $\varphi
\in \lbrack 0,2\pi ]$, $z\in (-\infty ,+\infty )$, and the Hamiltonian 
\begin{equation}
\hat{H}\equiv \;\frac{\hat{H}(\varphi ({\bf r}))}{\hbar \omega _{r}}\;.
\label{9.26}
\end{equation}
The atom-atom coupling parameter now is 
\begin{equation}
g\equiv 4\pi \;\frac{a_{s}}{l_{r}}\;N\;.  \label{9.27}
\end{equation}
The Hamiltonian (\ref{9.26}) writes 
\begin{equation}
\hat{H}=-\;\frac{1}{2}\;{\bf \nabla }^{2}+\;\frac{1}{2}\;(r^{2}+\nu
\;z^{2})+g|\psi |^{2}\;,  \label{9.28}
\end{equation}
where 
\[
{\bf \nabla }^{2}=\;\frac{\partial ^{2}}{\partial r^{2}}\;+\;\frac{1}{r}\;%
\frac{\partial }{\partial r}\;+\;\frac{1}{r^{2}}\;\frac{\partial ^{2}}{%
\partial \varphi ^{2}}\;+\;\frac{\partial ^{2}}{\partial z^{2}}\;. 
\]
In the eigenvalue problem 
\begin{equation}
\hat{H}\;\psi _{n}=E_{n}\;\psi _{n}\;,  \label{9.29}
\end{equation}
the energy $E_{n}$ is measured in units of $\omega _{r}$ and the
eigenfunction $\psi _{n}=\psi _{n}(r,\varphi ,z)$ is normalized by the
condition 
\begin{equation}
\int_{0}^{\infty }\;r\;dr\;\int_{0}^{2\pi }\;d\varphi \;\int_{-\infty
}^{+\infty }\;|\psi _{n}(r,\varphi ,z)|^{2}\;dz=1\;.  \label{9.30}
\end{equation}
It is worth recalling here that the Hamiltonian (\ref{9.28}) is nonlinear
and, thus, non-Hermitian. Thence, the set $\{\psi _{n}\}$ of the coherent
modes being the solutions of the eigenproblem (\ref{9.29}) does not
necessarily form a complete basis. And, in general, the set $\{\psi _{n}\}$
is not orthogonal. In particular, the eigenfunctions of non-Hermitian linear
operators can form complete bi-orthonormal bases \cite{Fonda66}-\cite
{Mostafazadeh99}. The situation with nonlinear operators is more
complicated: As there are no general theorems, the completeness of the
eigenfunction set is to be analyzed separately for each concrete case. For
some one-dimensional problems with nonlinear Schr\"{o}dinger Hamiltonians
the completeness of the eigenfunction set has been proved \cite
{Zhidkov97,Zhidkov00}.

\section{Thomas-Fermi Ground State}

In many cases, one is interested not in the whole set of stationary states $%
\psi_n$ but solely in the ground state, $\psi_0$, corresponding to the
minimal energy $E_0$. There exists a simple approximation that is very often
used for describing the ground state of trapped atoms. This is the
Thomas-Fermi approximation that is valid in the asymptotic limit of strong
coupling parameter $g\rightarrow\infty$. Then one neglects the kinetic term
as compared to the potential term containing $g$, which reduces the
differential Schr\"odinger equation to an algebraic equation.

Thus, considering the general anisotropic case, one neglects in the
Hamiltonian (\ref{9.19}) the kinetic differential operator 
\[
\hat K \equiv -\; \frac{1}{2}\; \sum_{i=1}^3 \; \frac{\partial^2}{\partial
x_i^2} \; . 
\]
Then, Eq. (\ref{9.20}) yields the wave function in the Thomas-Fermi
approximation 
\begin{equation}  \label{9.31}
\psi_{TF}({\bf x}) = \; \frac{1}{\sqrt{2g}} \; \left ( r_c^2 - \sum_{i=1}^3
\; \omega_i^2 \; x_i^2 \right ) \Theta\left ( r_c^2 - \sum_{i=1}^3 \;
\omega_i^2\; x_i^2 \right ) \; ,
\end{equation}
where $\Theta(\cdot)$ is the unit-step function and 
\begin{equation}  \label{9.32}
r_c^2 \equiv 2E_{TF} \; .
\end{equation}
The energy $E_{TF}$ is defined from the normalization $(\psi_{TF},
\psi_{TF})=1$. Unfortunately, there is a serious defect in the wave function
(\ref{9.31}), since the average kinetic energy 
\[
(\psi_{TF},\hat K\psi_{TF}) = \sum_{i=1}^3 \; \omega_i^2 \; \int \; \left [
1 + \; \frac{\omega_i^2 x_i^2}{\psi_{TF}^2(x)}\right ] \; d{\bf x} 
\]
logarithmically diverges.

In the cylindrically symmetric case, with the Hamiltonian (\ref{9.28}),
omitting the term $-\frac{1}{2}{\bf \nabla}^2$, we find the wave function 
\begin{equation}  \label{9.33}
\psi_{TF}(r,z) =\left ( \frac{r_c^2-r^2-\nu^2 z^2}{2g} \right )^{1/2}\;
\Theta(r_c^2 - r^2 -\nu^2z^2) \; ,
\end{equation}
where $r_c$ is given by Eq. (\ref{9.32}) and the energy $E_{TF}$ is to be
obtained from the normalization (\ref{9.30}), which yields 
\begin{equation}  \label{9.34}
E_{TF} = \; \frac{1}{2}\; \left ( \frac{15}{4\pi} \; g\nu\right )^{2/5} =
0.536689 (g\nu)^{2/5} \; .
\end{equation}
The average kinetic energy is again logarithmically divergent.

The Thomas-Fermi approximation is often used because of its simplicity.
However, this approximation has several deficiencies:

First of all, as is evident from the form of the function $\psi _{TF}$
containing a unit-step function, this approximation cannot correctly
describe the edge of the atomic cloud, since the Thomas-Fermi density $|\psi
_{TF}|^{2}$ has a sharp boundary at $r_{c}$, while in reality the density is
to be smooth \cite{Lundh97}.

Second, this approximation is not self-consistent as far as the Thomas-Fermi
energy $E_{TF}$ is defined from the normalization $||\psi_{TF}||^2=1$. But
if one retains the kinetic term in $\hat H$ then 
\begin{equation}  \label{9.35}
(\psi_{TF},\hat H\psi_{TF}) \neq E_{TF} \; ,
\end{equation}
not even approximately. And, moreover, the average energy $(\psi_{TF},\hat
H\psi_{TF})$ has no sense because of the divergence of the average kinetic
energy.

Third, this approximation is applicable for describing only the ground state
but does not permit the consideration of other coherent stationary states of
the eigenvalue problem.

Fourth, the approximation does not make a distinction between repulsive and
attractive forces, that is, between positive and negative coupling
parameters $g$. However, in the case of attractive interatomic forces, there
should exist a critical value $g_{c}<0$, such that for $g<g_{c}$ the system
becomes unstable \cite{Parkins98,Dalfovo99}, which can be manifested in the
energy becoming negative or complex.

Fifth, for the time-dependent equation (\ref{9.5}), the Thomas-Fermi
approximation gives a solution that is unstable with respect to small
perturbations \cite{Brewczyk99}.

\chapter{Spectrum of Coherent Modes}

Coherent modes are defined by the eigenproblem (\ref{8.22}). For trapped
atoms, the nonlinear Hamiltonian is given by Eq. (\ref{9.4}), with the
trapping potential (\ref{9.1}). This nonlinear eigenproblem cannot be solved
exactly. The standard perturbation theory starting with a
harmonic-oscillator approximation cannot be employed since the coupling
parameter (\ref{9.16}), or (\ref{9.27}), can be very large because of large $%
N\gg 1$. The Thomas-Fermi approximation, as discussed in section 9.5, can
give an estimate only for the ground-state energy, with $g\rightarrow\infty$%
. How would it be possible to find accurate approximate expressions for the
whole spectrum of coherent modes and for arbitrary values of the coupling
parameter? This can be achieved by means of the self-similar approximation
theory whose simplest variant, called optimized perturbation theory, is
outlined in the following section.

\section{Optimized Perturbation Theory}

Let us be interested in a function $E(g)$ of a coupling parameter $g$. We
keep in mind that $E(g)=E(g,n)$ is an energy level but, for brevity, the
dependence on the set of quantum numbers $n$ is not written down explicitly.

If one invokes the standard perturbation theory, valid for small coupling
parameters, one gets a sequence $\{ p_k(g)\}$ of perturbative approximations 
$p_k(g)$, with $k=0,1,2,\ldots$ implying approximation orders, so that 
\begin{equation}  \label{10.1}
E(g) \simeq p_k(g) \qquad (g\rightarrow 0) \; .
\end{equation}
However, the perturbative sequence $\{ p_k(g)\}$ is usually divergent for
any finite value of $g$. Moreover, the coupling parameter $g$ is often not
small, for which case the perturbative sequence $\{ p_k(g)\}$ cannot in
principle provide reasonable approximations.

In order to make perturbation theory meaningful, one has to change the
theory so that the resulting perturbative sequence be convergent. This can
be done by introducing control functions that are so called because of their
role of controlling convergence. Then, instead of a divergent sequence $%
\{p_{k}(g)\}$, one would get a convergent sequence $\{E_{k}(g,u_{k})\}$,
whose convergence is governed by control functions $u_{k}=u_{k}(g)$. The
idea of introducing control functions for making a perturbative sequence
convergent was advanced first in Ref. \cite{Yukalov76}. The introduction of
control functions can be done in different ways. A straightforward way is to
start perturbation theory with an initial approximation containing a set of
trial parameters $u$. The latter are then transformed into functions $%
u_{k}(g)$ such that the sequence $\{e_{k}(g)\}$ of the terms 
\begin{equation}
e_{k}(g)\equiv E_{k}(g,u_{k}(g))  \label{10.2}
\end{equation}
becomes convergent. Perturbation theory reorganized, in this or that way, by
introducing control functions \cite{Yukalov76} has been successfully applied
to a variety of problems in quantum mechanics, statistical physics, and
field theory \cite{Yukalov76}-\cite{Evans98}. The so reorganized
perturbation theory is called by different authors differently, for
instance, as optimized perturbation theory, controlled perturbation theory,
modified perturbation theory, renormalized perturbation theory,
oscillator-representation method, delta expansion, and so on. The method of
potential envelopes \cite{HallRL83}-\cite{HallRL93} is also close to this
approach. More references can be found in reviews \cite{Okopinska93}-\cite
{Sissakian99}.

It is only in a few simplest cases, such as zero- and one-dimensional
anharmonic oscillators \cite{Duncan93}-\cite{Guida95}, when control
functions can be chosen from the direct observation of convergence. Contrary
to this, the standard situation is when perturbative terms of arbitrarily
large orders are not available. Vice versa, for realistic problems, one
usually is able to find just a couple of perturbative terms. Because of
this, one usually defines control functions by invoking some heuristic
reasons.

The foundation for the choice of control functions can be done in the frame
of the self-similar approximation theory \cite{Yukalov89}-\cite{Yukalov99}.
These functions are to be chosen so that they provide the optimal
convergence, which means that convergence is as fast as possible. Such an
optimal choice of control functions results in the {\it optimized
perturbation theory}. To derive the concrete equations defining control
functions, it is necessary to construct a dynamical system, called the
approximation cascade \cite{Yukalov93}-\cite{Yukalov99} whose trajectory is
bijective to the approximation sequence $\{e_{k}(g)\}$. The limit of the
latter is in one-to-one correspondence with an attractive point of the
approximation cascade. Approaching the fixed point, the cascade velocity 
\begin{equation}
V_{k}(g)=E_{k+1}(g,u_{k})-E_{k}(g,u_{k})+(u_{k+1}-u_{k})\;\frac{\partial }{%
\partial u_{k}}\;E_{k}(g,u_{k})  \label{10.3}
\end{equation}
tends to zero. Hence, the closer we are to the fixed point, the smaller is
the modulus of the cascade velocity (\ref{10.3}). In other words, to provide
the fastest convergence, control functions have to minimize the cascade
velocity modulus 
\begin{equation}
|V_{k}(g)|\leq |E_{k+1}(g,u_{k})-E_{k}(g,u_{k})|+\left| (u_{k+1}-u_{k})\;%
\frac{\partial }{\partial u_{k}}\;E_{k}(g,u_{k})\right| \;.  \label{10.4}
\end{equation}
From here, two variants of the fixed-point conditions can be written down:
either the {\it minimal-difference condition} 
\begin{equation}
E_{k+1}(g,u_{k})-E_{k}(g,u_{k})=0  \label{10.5}
\end{equation}
or the {\it minimal-sensitivity condition} 
\begin{equation}
(u_{k+1}-u_{k})\;\frac{\partial }{\partial u_{k}}\;E_{k}(g,u_{k})=0\;.
\label{10.6}
\end{equation}
The latter, since in general $u_{k+1}\neq u_{k}$, reduces to the {\it %
variational condition} 
\begin{equation}
\frac{\partial }{\partial u_{k}}\;E_{k}(g,u_{k})=0\;.  \label{10.7}
\end{equation}
Both conditions, (\ref{10.5}) and (\ref{10.7}), are widely used in various
applications. When it happens that equations (\ref{10.5}) or (\ref{10.7})
have no solutions, these equations can be generalized to the condition 
\begin{equation}
\min_{u}|E_{k+1}(g,u)-E_{k}(g,u)|  \label{10.8}
\end{equation}
or, respectively, to the condition 
\begin{equation}
\min_{n}\left| \frac{\partial }{\partial u}\;E_{k}(g,u)\right| \;.
\label{10.9}
\end{equation}
The accuracy of the optimized approximants (\ref{10.2}), as compared to the
known value $E(g)$, is characterized by the percentage errors 
\begin{equation}
\varepsilon _{k}(g)\equiv \;\frac{e_{k}(g)-E(g)}{|E(g)|}\times 100\%\;.
\label{10.10}
\end{equation}

Let us emphasize the difference between the optimized perturbation theory
and the variational procedure based on the minimization of the
internal-energy functional: First, the latter has sense solely for the
ground state while the former is valid for the whole spectrum of the
eigenproblem. Second, the latter implies the case of zero temperature, while
the former is independent of temperature. Third, the minimization of the
internal energy yields an optimal value for the energy itself, but the
described method provides optimal approximants for the spectrum.

\section{Isotropic Ground State}

In general, the eigenproblem (\ref{9.20}) involves all three space
variables. The situation can be simplified when the confining potential (\ref
{9.1}) is spherically symmetric, so that $\omega_x=\omega_y=\omega_z$, and
if we are interested only in the ground state. In this case, the
ground-state wavefunction 
\begin{equation}  \label{10.11}
\psi_0({\bf x}) =\; \frac{1}{\sqrt{4\pi}\; r} \; \chi(r)
\end{equation}
depends solely on $r\equiv|{\bf x}|$. Then the eigenproblem (\ref{9.20}) can
be reduced to the effective equation 
\[
\hat H_r\;\chi = E\; \chi \; , 
\]
\begin{equation}  \label{10.12}
\hat H_r \equiv \; \frac{1}{2}\left ( -\; \frac{d^2}{dr^2} \; + \; r^2\right
) + \; \frac{g}{4\pi r^2}\; \chi^2 \; ,
\end{equation}
containing only the radial variable $r$. The radial wave function $\chi(r)$,
because of the relation (\ref{10.11}), has to satisfy the condition $%
\chi(0)=0$.

The Rayleigh-Schr\"odinger perturbation theory can be started with the
harmonic Hamiltonian 
\begin{equation}  \label{10.13}
\hat H_r^{(0)} \equiv -\; \frac{1}{2}\; \frac{d^2}{dr^2} \; + \; \frac{u^2}{2%
}\; r^2
\end{equation}
including a parameter $u$ that will later generate control functions $u_k(g)$%
. The ground-state eigenfunction of the Hamiltonian (\ref{10.13}) is 
\[
\chi^{(0)}(r) = 2\left ( \frac{u^3}{\pi}\right )^{1/4} \; r\exp\left ( -\; 
\frac{1}{2}\; ur^2\right ) \; . 
\]
The first-order approximation for the ground-state energy writes 
\begin{equation}  \label{10.14}
E^{(1)}(g,u) = \left (\chi^{(0)},\hat H_r\chi^{(0)}\right )\; .
\end{equation}
It is convenient to introduce the notation 
\begin{equation}  \label{10.15}
s\equiv \; \frac{2g}{(2\pi)^{3/2}} \; ,
\end{equation}
which characterizes an effective interaction strength. Then, Eq. (\ref{10.14}%
) yields 
\begin{equation}  \label{10.16}
E^{(1)}(g,u) =\; \frac{3}{4}\left ( u+\; \frac{1}{u}\right ) +\; \frac{s}{2}%
\; u^{3/2} \; .
\end{equation}
Using the fixed-point condition 
\begin{equation}  \label{10.17}
\frac{\partial}{\partial u} \; E^{(1)}(g,u) = 0 \; ,
\end{equation}
we get the equation 
\begin{equation}  \label{10.18}
s\; u^{5/2} + u^2 - 1 = 0
\end{equation}
defining the control function $u=u(s)$. In general, the control function
equation (\ref{10.18}) is to be solved numerically. But for weak and strong
interaction strengths, we may derive the following asymptotic expansions:
for the weak-coupling limit, $s\rightarrow 0$, 
\[
u(s) \simeq 1 -\; \frac{1}{2} \; s + \; \frac{1}{2}\; s^2 -\; \frac{39}{64}%
\; s^3 +\; \frac{105}{128}\; s^4 \; , 
\]
and for the strong-coupling limit, $s\rightarrow\infty$, 
\[
u(s) \simeq s^{-2/5} -\; \frac{2}{5}\; s^{-6/5} +\;\frac{1}{5}\; s^{-2} -\; 
\frac{12}{125} \; s^{-14/5} \; . 
\]

Substituting the control function $u(s)$ into Eq. (\ref{10.16}) gives the
first-order optimized approximant 
\begin{equation}
E(s)\equiv E_{1}(g(s),u(s))\;.  \label{10.19}
\end{equation}
Its behavior in the weak-coupling limit is 
\begin{equation}
E(s)\simeq \;\frac{3}{2}\;+\;\frac{1}{2}\;s-\;\frac{3}{16}\;s^{2}+\;\frac{9}{%
64}\;s^{3}-\;\frac{35}{256}\;s^{4}\;,  \label{10.20}
\end{equation}
where $s\rightarrow 0$; and in the strong-coupling limit, we have 
\begin{equation}
E(s)\simeq \;\frac{5}{4}\;s^{2/5}+\;\frac{3}{4}\;s^{-2/5}-\;\frac{3}{20}%
\;s^{-6/5}+\;\frac{1}{20}\;s^{-2}-\;\frac{9}{500}\;s^{-14/5}\;,
\label{10.21}
\end{equation}
as $s\rightarrow \infty $. Following the optimized perturbation theory
described in Sec.~10.1, one can obtain optimized approximants of arbitrary
orders. However, we limit ourselves here to the first-order approximants.

For atoms with negative scattering length, as in the case of $^{7}$Li or $%
^{85}$Rb, the coupling parameter (\ref{9.16}) is negative, hence the
parameter (\ref{10.15}) is negative, too. If $s<0$, the control function
equation (\ref{10.18}) has real solutions only in the interval $s_{c}<s<0$,
where 
\[
s_{c}=-\frac{4}{5^{5/4}}=-0.534992\;. 
\]
The ground-state energy (\ref{10.19}) is real in the same interval of $s\in
(s_{c},0)$ and becomes complex for $s<s_{c}$. This corresponds to the
interval $g_{c}<g<0$, with the critical coupling parameter 
\begin{equation}
g_{c}=\;\frac{(2\pi )^{3/2}}{2}\;s_{c}=-4.212960\;.  \label{10.22}
\end{equation}
The fact that there is a critical value for the coupling parameter (\ref
{9.16}) can be reformulated as the existence of a critical number of
particles 
\begin{equation}
N_{c}=\;\frac{l_{0}g_{c}}{4\pi a_{s}}  \label{10.23}
\end{equation}
that can form a coherent state. Thus, for the parameters of the experiments 
\cite{Bradley95,Bradley97} with $^{7}$Li, we get $N_{c}\sim 10^{3}$.

When the ground-state energy becomes complex, this means that the system is
unstable. The lifetime of such a system can be estimated as 
\begin{equation}
\tau (g)\equiv \;\frac{1}{\omega _{0}|{\rm Im}\;e(g)|}\;,  \label{10.24}
\end{equation}
where $e(g)=E(s(g))$. In the limit $g\rightarrow -\infty $, we have \cite
{Yukalov97} the asymptotic expansions 
\[
{\rm Re}\;e(g)\simeq
0.169198\;g^{2/5}+0.529102\;g^{-2/5}+1.443899\;g^{-6/5}+31.006277\;g^{-2}\;, 
\]
\[
{\rm Im}\;e(g)\simeq
-0.520739\;g^{2/5}+1.628409\;g^{-2/5}+1.049054\;g^{-6/5}\;. 
\]
Therefore, in this limit the lifetime (\ref{10.24}) can be estimated as 
\begin{equation}
\tau (g)\simeq \;\frac{1.920}{\omega _{0}|g|^{2/5}}\qquad (g\rightarrow
-\infty )\;.  \label{10.25}
\end{equation}

Note that if one defines the critical coupling $g_{c}$ not from the direct
solution of the eigenproblem (\ref{10.12}) but from the minimization of the
internal-energy functional \cite{Dalfovo99}, then the resulting critical
coupling is about twice as large as the value (\ref{10.22}).

\section{Anisotropic Excited States}

When the confining potential is not isotropic or when we are interested not
solely in the ground state but also in the spectrum of excited coherent
modes, we have to deal with the eigenproblem (\ref{9.20}). Then, for
developing optimized perturbation theory, we may start with the Hamiltonian 
\begin{equation}
\hat{H}_{0}=\;\frac{1}{2}\;\sum_{i=1}^{3}\;\left( -\;\frac{\partial ^{2}}{%
\partial x_{i}^{2}}\;+u_{i}^{2}\;x_{i}^{2}\right) \;,  \label{10.26}
\end{equation}
containing three trial parameters, $u_{1},\;u_{2}$, and $u_{3}$, generating
control functions. The Hamiltonian (\ref{10.26}) possesses the eigenvalues 
\begin{equation}
E_{n}^{(0)}=\sum_{i=1}^{3}\;u_{i}\;\left( n_{i}+\;\frac{1}{2}\right) \;,
\label{10.27}
\end{equation}
with the eigenfunctions 
\[
\psi _{n}^{(0)}({\bf x})=\prod_{i=1}^{3}\;\psi _{n_{i}}(x_{i})\;, 
\]
\[
\psi _{n_{i}}(x_{i})=\;\frac{(u_{i}/\pi )^{1/4}}{\sqrt{2^{n_{i}}n_{i}!}}%
\;H_{n_{i}}(\sqrt{u_{i}}\;x_{i})\;\exp \left( -\;\frac{1}{2}%
\;u_{i}x_{i}^{2}\right) \;, 
\]
where $n\equiv \{n_{1},n_{2},n_{3}\}$; $n_{i}=0,1,2,\ldots $; and $%
H_{n_{i}}(\cdot )$ is a Hermite polynomial. Perturbation theory is
accomplished \cite{Yukalov97} with respect to the perturbation 
\begin{equation}
\Delta H\equiv \hat{H}-\hat{H}_{0}=\;\frac{1}{2}\;\sum_{i=1}^{3}\;(\omega
_{i}^{2}-u_{i}^{2})\;x_{i}^{2}\;+g|\psi |^{2}\;.  \label{10.28}
\end{equation}
In the first order, one has 
\begin{equation}
E_{n}^{(1)}(g,u)=E_{n}^{(0)}+(\psi _{n},\Delta H\psi _{n})\;,  \label{10.29}
\end{equation}
where $u\equiv \{u_{1},u_{2},u_{3}\}$. Introducing the effective interaction
strength 
\begin{equation}
s\equiv 2g\;\prod_{i=1}^{3}\;I_{n_{i}}\;,  \label{10.30}
\end{equation}
in which 
\[
I_{n_{i}}\equiv \;\frac{2}{\pi (2^{n_{i}}n_{i}!)^{2}}\;\int_{0}^{\infty
}\;H_{n_{i}}^{4}(x)\;e^{-2x^{2}}\;dx\;, 
\]
where the property $|H_{n}(-x)|=|H_{n}(x)|$ of the Hermite polynomials is
used, one obtains for the energy levels (\ref{10.29}) 
\begin{equation}
E_{n}^{(1)}(g,u)=\;\frac{1}{2}\;\sum_{i=1}^{3}\;u_{i}\left( n_{i}+\;\frac{1}{%
2}\right) \left( 1+\;\frac{\omega _{i}^{2}}{u_{i}^{2}}\right) \;+\;\frac{1}{2%
}\;\sqrt{u_{1}u_{2}u_{3}}\;s\;.  \label{10.31}
\end{equation}
Control functions, playing the role of effective oscillation frequencies,
are defined by the fixed-point condition 
\begin{equation}
\frac{\partial }{\partial u_{i}}\;E_{n}^{(1)}(g,u)=0\;,  \label{10.32}
\end{equation}
which results in three equations 
\begin{equation}
\left( n_{i}+\;\frac{1}{2}\right) (u_{i}^{2}-\omega _{i}^{2})+\;\frac{u_{i}}{%
2}\;\sqrt{u_{1}u_{2}u_{3}}\;s=0  \label{10.33}
\end{equation}
defining $u_{i}=u_{i}(s)$, where $i=1,2,3$ and, for short, the quantum
multi-index $n$ is dropped. The first-order optimized approximant is given
by 
\begin{equation}
E(s)\equiv E_{n}^{(1)}(g(s),u(s))\;,  \label{10.34}
\end{equation}
where again, for brevity, the index $n$ is omitted in the left-hand side.

The control-function equations (\ref{10.33}) yield 
\[
u_i(s) \simeq \omega_i -\; \frac{s}{2(2n_i+1)} \; +\; \frac{\omega_i}{4} 
\left [ \frac{\sum_{j=1}^3(2n_j+1)\omega_j-(2n_i+1)\omega_i} {%
2\prod_{j=1}^3(2n_j+1)} \; + \; \frac{1}{(2n_i+1)^2\omega_i^2}\right ] s^2 
\]
in the weak-coupling limit $s\rightarrow 0$, and 
\[
u_i(s) \simeq \; \frac{(2n_i+1)\omega_i^2}{\prod_{j=1}^3(2n_j+1)^{1/5}} \;
s^{-2/5} 
\]
in the strong-coupling limit $s\rightarrow\infty$. For the energy levels (%
\ref{10.34}), we find 
\begin{equation}  \label{10.35}
E(s) \simeq \sum_{i=1}^3 \; \omega_i\left ( n_i +\; \frac{1}{2}\right ) + \; 
\frac{s}{2} \; - \; \frac{1}{16}\; \sum_{i=1}^3\; \frac{s^2}{(2n_i+1)\omega_i%
}
\end{equation}
in the weak-coupling limit $s\rightarrow 0$, and 
\begin{equation}  \label{10.36}
E(s) \simeq\; \frac{5}{4}\; \prod_{j=1}^3 \; (2n_j+1)^{1/5} \; s^{2/5}
\end{equation}
in the strong-coupling limit $s\rightarrow\infty$.

For the ground state, when $n_i=0$, the coupling strength (\ref{10.30})
reduces to Eq. (\ref{10.15}). Then for the control functions, we have 
\[
u_i(s) \simeq \omega_i -\; \frac{s}{2} \; + \; \frac{\omega_i}{4}\; \left [ 
\frac{1}{2}\left ( \sum_{j=1}^3 \omega_j -\omega_i\right ) +\; \frac{1}{
\omega_i^2}\right ] \; s^2 \; , 
\]
if $s\rightarrow 0$, and 
\[
u_i(s) \simeq \omega_i^2\; s^{-2/5} \; , 
\]
when $s\rightarrow\infty$. In these two limits, the ground-state energy is 
\begin{equation}  \label{10.37}
E_0(s) \simeq \; \frac{1}{2}\; (\omega_1+\omega_2+\omega_3) +\; \frac{s}{2}%
\; - \; \frac{1}{16}\; \left ( \frac{1}{\omega_1}\; +\; \frac{1}{\omega_2}\;
+ \; \frac{1}{\omega_3}\right ) \; s^2 \; ,
\end{equation}
for $s\rightarrow 0$, and 
\begin{equation}  \label{10.38}
E_0(s) \simeq \; \frac{5}{4}\; s^{2/5} = 0.547538\; g^{2/5} \; ,
\end{equation}
as $s\rightarrow\infty$. In the isotropic case, we return to Eqs. (\ref
{10.20}) and (\ref{10.21}).

The arrangement of the energy levels in the weak-coupling and
strong-coupling limits is, in general, different. This can be illustrated by
considering several first energy levels $e_{n}(g)\equiv E_{n}(s(g))$. For
example, 
\[
e_{000}(g)\simeq \;\frac{3}{2}\;+0.063\;g\;\qquad e_{100}(g)\simeq \;\frac{5%
}{2}\;+0.048\;g\;, 
\]
\[
e_{110}(g)\simeq \;\frac{7}{2}\;+0.036\;g\;,\qquad e_{200}(g)\simeq \;\frac{7%
}{2}\;+0.041\;g\;, 
\]
in the weak-coupling limit $g\rightarrow 0$, and 
\[
e_{000}(g)\simeq 0.547\;g^{2/5}\;,\qquad e_{100}(g)\simeq 0.608\;g^{2/5}\;, 
\]
\[
e_{200}(g)\simeq 0.632\;g^{2/5}\;,\qquad e_{110}(g)\simeq 0.675\;g^{2/5} 
\]
for $g\rightarrow \infty $. As is seen from here, 
\begin{equation}
e_{110}(g)<e_{200}(g)\qquad (g\rightarrow 0) \; ,  \label{10.39}
\end{equation}
but 
\begin{equation}
e_{110}(g)>e_{200}(g)\qquad (g\rightarrow \infty )\;.  \label{10.40}
\end{equation}
This effect is called {\it level crossing} \cite{Yukalov97}.

In the case of negative scattering lengths, when $g<0$, the situation is
analogous to that studied in Sec.~10.2. For each given energy level,
labelled by $n$, there exists a critical value of the coupling parameter $%
g=g_{c}$ when the corresponding energy becomes complex. Then the lifetime of
an energy level, with the complex energy $e_{n}(g)$, can be estimated as 
\begin{equation}
\tau _{n}(g)\equiv \;\frac{1}{\omega _{0}|{\rm Im}\;e_{n}(g)|}\;.
\label{10.41}
\end{equation}

The spatial shape of the cloud of trapped atoms is characterized by the {\it %
aspect ratio} 
\begin{equation}  \label{10.42}
R_i\equiv \left ( \frac{<x_i^2>_n}{<x_3^2>_n}\right )^{1/2} \qquad (i=1,2)
\; ,
\end{equation}
in which $<x_i^2>_n$ implies $(\psi_n,x_i^2\psi_n)$. For the function $%
\psi_n^{(0)}({\bf x})$, this gives 
\begin{equation}  \label{10.43}
R_i =\left [ \frac{(2n_i+1)u_3}{(2n_3+1)u_i}\right ]^{1/2} \; .
\end{equation}
In the weak-coupling limit, 
\begin{equation}  \label{10.44}
R_i \simeq \left [\frac{(2n_i+1)\omega_3}{(2n_3+1)\omega_i} \right %
]^{1/2}\left\{ 1 + \left [ \frac{1}{(2n_i+1)\omega_i}\; -\; \frac{1}{%
(2n_3+1)\omega_3}\right ]\; \frac{g}{2(2\pi)^{3/2}} \right \} \; ,
\end{equation}
as $g\rightarrow 0$, while in the strong-coupling limit, 
\begin{equation}  \label{10.45}
R_i \simeq \; \frac{\omega_3}{\omega_i} \qquad (g\rightarrow\infty) \; .
\end{equation}
This tells us that the shape of different coherent modes essentially depends
on the quantum numbers $n_i$, if $g\rightarrow 0$, but for large coupling
parameters, the shape of different modes is practically the same, tending to
that of Eq. (\ref{10.45}).

\section{Cylindric Trapping Potential}

For a cylindrically symmetric trap, it is convenient to use the notations
introduced in Sec.~9.4. Cylindrical traps are often employed in experiments,
therefore we shall pay more attention to this case.

To solve the eigenproblem (\ref{9.29}), we may again invoke the optimized
perturbation theory of Sec.~10.1, starting with the initial Hamiltonian 
\begin{equation}
\hat{H}_{0}=-\;\frac{1}{2}{\bf \nabla }^{2}+\;\frac{1}{2}%
\;(u^{2}r^{2}+v^{2}z^{2})  \label{10.46}
\end{equation}
containing two control parameters, $u$ and $v$. The eigenvalues of the
operator (\ref{10.46}) are 
\begin{equation}
E_{nmk}^{(0)}=(2n+|m|+1)\;u+\left( k+\;\frac{1}{2}\right) \;v\;,
\label{10.47}
\end{equation}
with the quantum numbers 
\[
n=0,1,2,\ldots ;\qquad m=0,\pm 1,\pm 2,\ldots ;\qquad k=0,1,2,\ldots 
\]
The related eigenfunctions are 
\[
\psi _{nmk}^{(0)}(r,\varphi ,z)=\left[ \frac{2n!\;u^{|m|+1}}{(n+|m|)!}\right]
^{1/2}\;r^{|m|}\;\exp \left( -\;\frac{1}{2}\;ur^{2}\right)
\;L_{n}^{|m|}(ur^{2})\times 
\]
\[
\times \frac{e^{im\varphi }}{\sqrt{2\pi }}\;\frac{(v/\pi )^{1/4}}{\sqrt{%
2^{k}k!}}\;\exp \left( -\;\frac{1}{2}\;vz^{2}\right) \;H_{k}(\sqrt{v}\;z)\;, 
\]
where $L_{n}^{m}(\cdot )$ is a Laguerre polynomial and $H_{k}(\cdot )$ is a
Hermite polynomial.

In the first order, we have 
\begin{equation}  \label{10.48}
E_{nmk}^{(1)}(g,u,v) = (\psi_{nmk}^{(0)},\hat H\psi_{nmk}^{(0)})\; .
\end{equation}
To write down this integral explicitly, it is convenient to use the notation 
\[
I_{nmk} \equiv \; \frac{1}{u\sqrt{v}} \; \int \; |\psi_{nmk}^{(0)}({\bf r}%
)|^4\;d{\bf r} \; , 
\]
in which ${\bf r}=\{ r,\varphi,z\}$ is the dimensionless space variable in
cylindrical coordinates. Then we get 
\[
I_{nmk} =\; \frac{2}{\pi^2} \left [ \frac{n!}{(n+|m|)!\; 2^k\; k!}\right ]^2
\; \int_0^\infty x^{2|m|} \; e^{-2x} \; \left [\; L_n^{|m|}(x) \; \right ]^4
dx\; \int_0^\infty e^{-2t^2} H_k^4(t)\; dt \; . 
\]
It is also convenient to introduce the notation 
\begin{equation}  \label{10.49}
p\equiv 2n+|m|+1 \; , \qquad q\equiv 2k+1 \; .
\end{equation}
In this way, the energy levels (\ref{10.48}) can be written as 
\begin{equation}  \label{10.50}
E^{(1)}(g,u,v) =\; \frac{p}{2}\; \left ( u + \; \frac{1}{u}\right ) + \; 
\frac{q}{4}\;\left ( v+\; \frac{\nu^2}{v}\right ) + I_{nmk}\; u\; \sqrt{v}\;
g \; ,
\end{equation}
where, for simplicity, the quantum indices $n,\;m$, and $k$ in the left-hand
side are dropped.

The fixed-point conditions are 
\begin{equation}  \label{10.51}
\frac{\partial}{\partial u}\; E^{(1)}(g,u,v) = 0 \; , \qquad \frac{\partial}{%
\partial v}\; E^{(1)}(g,u,v) = 0 \; .
\end{equation}
These yield the control-function equations 
\begin{equation}  \label{10.52}
p\left ( 1 -\;\frac{1}{u^2}\right ) + \; \frac{s}{p\nu}\; \sqrt{\frac{v}{q}}%
\; = 0 \; , \qquad q\left ( 1 - \; \frac{\nu^2}{v^2}\right ) + \; \frac{s}{%
p\nu\sqrt{\nu q}}\; = 0 \; ,
\end{equation}
in which the effective interaction strength 
\begin{equation}  \label{10.53}
s\equiv 2p\sqrt{q}\; I_{nmk}\;\nu g
\end{equation}
is introduced. Substituting the control functions $u=u(s)$ and $v=v(s)$,
defined by Eqs. (\ref{10.52}), into Eq. (\ref{10.50}), we obtain the
optimized approximant 
\begin{equation}  \label{10.54}
E(s) \equiv E^{(1)}(g(s),u(s),v(s)) \; ,
\end{equation}
where $g(s)$ is given by the relation (\ref{10.53}).

Similarly to the previous sections, it is instructive to analyze the
weak-coupling and strong-coupling limits in detail. In the weak-coupling
limit $s\rightarrow 0$, Eqs. (\ref{10.52}) give the radial control function 
\begin{equation}  \label{10.55}
u(s) \simeq 1 -\; \frac{1}{2p^2(q\nu)^{1/2}} \; s + \; \frac{p+3q\nu}{%
8p^4(q\nu)^2}\; s^2 - \; \frac{3p^2+16pq\nu+20(q\nu)^2}{64p^6(q\nu)^{7/2}}\;
s^3
\end{equation}
and, respectively, the axial control function 
\begin{equation}  \label{10.56}
v(s) \simeq \nu -\; \frac{\nu}{2p(q\nu)^{3/2}}\; s + \; \frac{\nu(p+q\nu)}{%
4p^3(q\nu)^3}\; s^2 - \; \nu\; \frac{7p^2+20pq\nu+12(q\nu)^2}{%
64p^5(q\nu)^{9/2}} \; s^3 \; .
\end{equation}
In the strong-coupling limit $s\rightarrow\infty$, the radial control
function is 
\begin{equation}  \label{10.57}
u(s) \simeq u_1s^{-2/5} + u_2s^{-6/5}+u_3s^{-2}+u_4s^{-14/5} + u_5s^{-18/5}
\; ,
\end{equation}
where 
\[
u_1=p\; , \quad \frac{5}{p}\; u_2 = -3p^2 +(q\nu)^2\; , \quad \frac{5}{p}\;
u_3=3p^4-p^2(q\nu)^2-(q\nu)^4 \; , 
\]
\[
\frac{125}{p}\; u_4 = -88p^6+33p^4(q\nu)^2 + 4p^2(q\nu)^4 + 39(q\nu)^6 \; , 
\]
\[
\frac{625}{p}\; u_5=561p^8 - 238p^6(q\nu)^2 - 21p^4(q\nu)^4 + 88p^2(q\nu)^6
- 364 (q\nu)^8 \; . 
\]
And for the axial control function, we get 
\begin{equation}  \label{10.58}
v(s) \simeq v_1s^{-2/5} + v_2s^{-6/5} + v_3 s^{-2} + v_4 s^{-14/5} + v_5
s^{-18/5} \; ,
\end{equation}
where $s\rightarrow\infty$ and 
\[
v_1 = q\nu^2\; , \quad \frac{5}{2q\nu^2}\; v_2 = p^2-2(q\nu)^2\; , \quad 
\frac{5}{q\nu^2}\; v_3 = -p^4-4p^2(q\nu)^2+6(q\nu)^4 \; , 
\]
\[
\frac{125}{2q\nu^2}\; v_4 = 11 p^6+4p^4(q\nu)^2+117p^2(q\nu)^4-
138(q\nu)^6\; , 
\]
\[
\frac{625}{q\nu^2}\; v_5 = -119p^8 - 28p^6(q\nu)^2 + 264p^4(q\nu)^4 -
2912p^2(q\nu)^6 + 2821(q\nu)^8\; . 
\]

The weak-coupling limit for the energy (\ref{10.54}) becomes 
\begin{equation}  \label{10.59}
E(s) \simeq a_0 + a_1s + a_2 s^2 + a_3s^3 \; ,
\end{equation}
as $s\rightarrow 0$, where 
\[
a_0 = p+\; \frac{q\nu}{2}\; , \quad a_1 =\; \frac{1}{2p(q\nu)^{1/2}} \; ,
\quad a_2 = -\; \frac{p+2q\nu}{16p^3(q\nu)^2} \; , \quad a_3=\; \frac{%
(p+2q\nu)^2}{64p^5(q\nu)^{7/2}} \; . 
\]
And in the strong-coupling limit, we find 
\begin{equation}  \label{10.60}
E(s)\simeq b_0 s^{2/5} + b_1s^{-2/5} + b_2s^{-6/5} + b_3s^{-2} +
b_4s^{-14/5} + b_5 s^{-18/5} \; ,
\end{equation}
as $s\rightarrow\infty$, with 
\[
b_0 =\frac{5}{4} \; , \quad 4b_1=2p^2+(q\nu)^2\; , \quad
20b_2=-3p^4+2p^2(q\nu)^2-2(q\nu)^4 \; , 
\]
\[
20b_3 =2p^6-p^4(q\nu)^2 - 2p^2(q\nu)^4 + 2(q\nu)^6 \; , 
\]
\[
500b_4 = -44p^8 + 22p^6(q\nu)^2 + 2p^4(q\nu)^4 + 78p^2(q\nu)^6 - 69(q\nu)^8
\; , 
\]
\[
12500b_5 = 1122p^{10} - 595p^8(q\nu)^2 - 70p^6(q\nu)^4 + 440p^4(q\nu)^6 -
3640p^2(q\nu)^8 + 2821 (q\nu)^{10} \; . 
\]
The derived expressions (\ref{10.59}) and (\ref{10.60}) are valid for any
combination of quantum numbers.

\section{Cloud Shape and Lifetime}

The shape of an atomic cloud in a cylindrical trap can be characterized by
the mean-square radial and axial lengths, respectively, 
\begin{equation}  \label{10.61}
r_0 \equiv (<r^2>_{nmk})^{1/2} \; , \qquad z_0\equiv (<z^2>_{nmk})^{1/2} \; ,
\end{equation}
where $<\cdot>_{nmk}$ is a quantum-mechanical average over the wave function 
$\psi_{nmk}$. Taking these averages with respect to the function $%
\psi_{nmk}^{(0)}$, we have 
\begin{equation}  \label{10.62}
r_0 =\sqrt{\frac{p}{u}}\; , \qquad z_0 =\sqrt{\frac{q}{2v}} \; .
\end{equation}
In the weak-coupling limit $s\rightarrow 0$, the radial mean-square
deviation is 
\begin{equation}  \label{10.63}
r_0 \simeq\sqrt{p}\left ( 1 +\; \frac{1}{4p^2(q\nu)^{1/2}} \; s -\; \frac{%
2p+3q\nu}{32p^4(q\nu)^2}\; s^2 + \; \frac{3p^2+10pq\nu+7(q\nu)^2}{%
128p^6(q\nu)^{7/2}}\; s^3\right )\; ,
\end{equation}
and the axial mean-square length is 
\begin{equation}  \label{10.64}
z_0 \simeq \sqrt{\frac{q}{2\nu}}\; \left ( 1 +\; \frac{1}{4p(q\nu)^{3/2}}\;
s -\; \frac{p+4q\nu}{32p^3(q\nu)^3}\; s^2 + \; \frac{2p+3q\nu}{%
32p^5(q\nu)^{7/2}}\; s^3\right ) \; .
\end{equation}
In the strong-coupling limit $s\rightarrow\infty$, for the radial and axial
averages (\ref{10.62}) we find 
\begin{equation}  \label{10.65}
r_0 \simeq s^{1/5} +\; \frac{3p^2-(q\nu)^2}{10} \; s^{-3/5} - \; \frac{%
33p^4-2p^2(q\nu)^2-23(q\nu)^4}{200}\; s^{-7/5}
\end{equation}
and, respectively, 
\begin{equation}  \label{10.66}
z_0 \simeq\; \frac{1}{\sqrt{2}\;\nu} \; \left ( s^{1/5} -\; \frac{%
p^2-2(q\nu)^2}{5}\; s^{-3/5} + \; \frac{4p^4+4p^2(q\nu)^2-9(q\nu)^4}{25}\;
s^{-7/5}\right ) \; .
\end{equation}

For the aspect ratio 
\begin{equation}  \label{10.67}
R_r\equiv\left ( \frac{<r^2>_{nmk}}{<z^2>_{nmk}}\right )^{1/2} = \; \frac{r_0%
}{z_0}
\end{equation}
we have 
\begin{equation}  \label{10.68}
R_r =\sqrt{\frac{2p\;v}{q\; u}}\; .
\end{equation}
This gives in the weak-coupling limit 
\begin{equation}  \label{10.69}
R_r(s) \simeq \sqrt{\frac{2p\nu}{q}}\;\left ( 1 +\alpha_1 s +\alpha_2 s^2
+\alpha_3 s^3 +\alpha_4 s^4\right ) \; ,
\end{equation}
where $s\rightarrow 0$, and 
\[
4p^2(q\nu)^{3/2}\alpha_1=-p+q\nu \; , \qquad 32p^4(q\nu)^3\alpha_2 =
3p^2-3(q\nu)^2 \; , 
\]
\[
128p^6(q\nu)^{9/2}\alpha_3 = - 4p^3 -8p^2q\nu + 5p(q\nu)^2 + 7(q\nu)^3 \; , 
\]
\[
2048p^8(q\nu)^6\alpha_4 =15p^4+112p^3q\nu+ 70p^2(q\nu)^2 - 120p(q\nu)^3
-77(q\nu)^4 \; . 
\]
In the strong-coupling limit, the aspect ratio (\ref{10.68}) is 
\begin{equation}  \label{10.70}
R_r(s) \simeq \sqrt{2}\;\nu\; \left ( 1 +\beta_1s^{-4/5} + \beta_2s^{-8/5}
+\beta_3 s^{-12/5} +\beta_4 s^{-16/5} \right )\;,
\end{equation}
where $s\rightarrow\infty$, and 
\[
2\beta_1=p^2-(q\nu)^2 \; , \qquad \frac{40}{9}\;\beta_2 = -p^4
-2p^2(q\nu)^2+ 3(q\nu)^4 \; , 
\]
\[
400\beta_3 = 77p^6 +5p^4(q\nu)^2 + 391p^2(q\nu)^4 - 473(q\nu)^6\; , 
\]
\[
\frac{16000}{13}\;\beta_4 = -253 p^8 - 4p^6(q\nu)^2 + 266p^4(q\nu)^4 -
2900p^2(q\nu)^6 + 2891(q\nu)^8 \; . 
\]
These expansions confirm that the dependence of the aspect ratio on quantum
numbers diminishes in the strong-coupling limit, so that 
\[
\lim_{s\rightarrow\infty} R_r(s) = \sqrt{2}\;\nu \; , 
\]
in agreement with Eq. (\ref{10.45}).

Considering the stationary properties of coherent modes, we should not
forget that in real traps atoms cannot be confined infinitely long. This is
because the trapping of neutral atoms requires their special spin
polarization which can be lost during atomic collisions \cite{Weiner99}. One
usually considers binary and triple depolarizing collisions \cite
{Moerdijk96b,Edwards96b}. The corresponding loss rate of atoms in a coherent
mode $\psi _{nmk}$ can be written as 
\begin{equation}
\Gamma _{2}\equiv \lambda _{2}\;N^{2}\;\gamma _{nmk}\;,\qquad \Gamma
_{3}\equiv \lambda _{3}\;N^{3}\;\delta _{nmk}\;,  \label{10.71}
\end{equation}
where $\lambda _{2}$ is a two-body dipolar loss rate coefficient, $\lambda
_{3}$ is a three-body recombination loss rate coefficient, and 
\begin{equation}
\gamma _{nmk}\equiv \int \;|\psi _{nmk}({\bf r})|^{4}\;d{\bf r}\;,\qquad
\delta _{nmk}\equiv \int \;|\psi _{nmk}({\bf r})|^{6}\;d{\bf r}\;.
\label{10.72}
\end{equation}
For alkali atoms \cite{Moerdijk96b,Edwards96b}, the two-body loss rate
coefficient is $\lambda _{2}\sim 10^{-16}-10^{-15}$ cm$^{3}$/s and the
three-body one is $\lambda _{3}\sim 10^{-30}-10^{-28}$ cm$^{6}$/s, the
lifetime of atoms in a trap is on the order of $1-100$ s.

The integrals (\ref{10.72}), with $\psi _{nmk}^{(0)}$, take the form 
\begin{equation}
\gamma _{nmk}=u\sqrt{v}\;I_{nmk}\;,\qquad \delta _{nmk}=u^{2}v\;J_{nmk}\;,
\label{10.73}
\end{equation}
where $I_{nmk}$ is the same integral as in Sec.~10.4 and 
\[
J_{nmk}\equiv \;\frac{2}{\pi ^{7/2}}\left[ \frac{n!}{(n+|m|)!\;2^{k}k!}%
\right] ^{3}\;\int_{0}^{\infty }x^{3|m|}\;e^{-3x}\;\left[ L_{n}^{|m|}(x)%
\right] ^{6}dx\;\int_{0}^{\infty }\;e^{-3t^{2}}H_{k}^{6}(t)\;dt\;. 
\]
In the weak-coupling limit, when $s\rightarrow 0$, the quantities (\ref
{10.73}) are 
\begin{equation}
\gamma _{nmk}(s)\simeq \sqrt{\nu }\;I_{nmk}\;,\qquad \delta _{nmk}(s)\simeq
\nu \;J_{nmk}\;,  \label{10.74}
\end{equation}
and in the strong-coupling limit, we have 
\begin{equation}
\gamma _{nmk}(s)\simeq p\sqrt{q}\;\nu \;I_{nmk}\;s^{-3/5}\;,\qquad \delta
_{nmk}\simeq p^{2}q\nu ^{2}\;J_{nmk}\;s^{-6/5}\;,  \label{10.75}
\end{equation}
as $s\rightarrow \infty $. Since $s\sim N$, the quantities (\ref{10.75})
decrease with increasing number of particles. However, the loss rates (\ref
{10.71}) increase with $N$ according to the laws $\Gamma _{2}\sim N^{7/5}$
and $\Gamma _{3}\sim N^{9/5}$.

To compare the loss rates of excited coherent modes with those of the ground
state, we may analyze the reduced loss rates 
\[
\overline{\gamma }_{nmk}(s)\equiv \;\frac{\gamma _{nmk}(s)}{\gamma _{000}(s)}%
\;,\qquad \overline{\delta }_{nmk}(s)\equiv \;\frac{\delta _{nmk}(s)}{\delta
_{000}(s)}\;. 
\]
The latter possess the properties 
\[
\overline{\gamma }_{nmk}(0)\leq \overline{\gamma }_{nmk}(s)\leq \overline{%
\gamma }_{nmk}(\infty )\;,\qquad \overline{\delta }_{nmk}(0)\leq \overline{%
\delta }_{nmk}(s)\leq \overline{\delta }_{nmk}(\infty )\;. 
\]
In order to get a feeling in what range the reduced loss rates vary, we may
consider several first states. For this purpose, we need to calculate the
corresponding integrals $I_{nmk}$ and $J_{nmk}$. For instance, for the
ground state, 
\[
I_{000}=\;\frac{1}{(2\pi )^{3/2}}\;=0.063494\;,\qquad J_{000}=\;\frac{1}{%
(3\pi ^{2})^{3/2}}\;=0.006207\;. 
\]
And for several first excited states, 
\[
I_{100}=0.031747\;,\qquad J_{100}=0.002222\;, 
\]
\[
I_{010}=0.031747\;,\qquad J_{010}=0.001379\;, 
\]
\[
I_{001}=0.047620\;,\qquad J_{001}=0.003448\;. 
\]
In the case of $n=k=0$, but arbitrary $m$, 
\[
I_{0m0}=\;\frac{(2|m|)!}{(2\pi )^{3/2}(2^{|m|}|m|!)^{2}}\;\qquad J_{0m0}=\;%
\frac{(3|m|)!}{(3\pi ^{2})^{3/2}(3^{|m|}|m|)^{3}}\;. 
\]
In this way, we find 
\[
0.5\leq \overline{\gamma }_{100}(s)\leq 1.18\;,\qquad 0.36\leq \overline{%
\delta }_{100}(s)\leq 1.98\;, 
\]
\[
0.5\leq \overline{\gamma }_{010}(s)\leq 1\;,\qquad 0.22\leq \overline{\delta 
}_{010}(s)\leq 0.88\;, 
\]
\[
0.75\leq \overline{\gamma }_{001}(s)\leq 1.11\;\qquad 0.56\leq \overline{%
\delta }_{001}(s)\leq 1.22\;. 
\]
Thus, the loss rates of excited states are close to those of the ground
state.

\chapter{Weak-to-Strong Coupling Crossover}

When considering the properties of trapped atoms at arbitrary coupling
parameters, one usually needs to invoke computer calculations. Analytical
expressions can be available only in the weak-coupling and strong-coupling
limits. Nevertheless, there is a method permitting one to reconstruct an
analytical formula, valid for the whole region of coupling parameters, for a
function whose asymptotic expansions in the weak-coupling and
strong-coupling limits are known. Here we briefly delineate this method and
then apply it for describing several properties of trapped atoms.

\section{Self-Similar Crossover Approximants}

Assume that we are interested in the behavior of a function $f(s)$ of the
coupling parameter $s$. Let this function be defined by a complicated
equation that can be solved only numerically. But we can find the asymptotic
expansion 
\begin{equation}
f(s)\simeq a_{0}+a_{1}s+a_{2}s^{2}+\ldots \qquad (s\rightarrow 0)
\label{11.1}
\end{equation}
in the weak-coupling limit. And we can often analytically derive the
asymptotic expansion 
\begin{equation}
f(s)\simeq b_{0}s^{\beta _{0}}+b_{1}s^{\beta _{1}}+b_{2}s^{\beta
_{2}}+\ldots \qquad (s\rightarrow \infty )  \label{11.2}
\end{equation}
in the strong-coupling limit, where the powers $\beta _{j}$ are arranged in
the decreasing order, $\beta _{j}>\beta _{j+1}$.

Introducing into series (\ref{11.1}) control functions by means of an
algebraic transformation \cite{Yukalov97d}-\cite{Yukalov97e} and using the
self-similar approximation theory \cite{Yukalov89}-\cite{Yukalov99}, we
obtain \cite{Yukalov98b,Gluzman98} the self-similar root approximant 
\begin{equation}
f_{k}^{\ast }(s)=a_{0}\left( \ldots \left\{ \left[
(1+A_{1}s)^{n_{1}}+A_{2}s^{2}\right] ^{n_{2}}+A_{3}s^{3}\right\}
^{n_{3}}+\ldots +A_{k}s^{k}\right) ^{n_{k}}\;,  \label{11.3}
\end{equation}
where $k$ is the order of the approximation taken. The coefficients $A_{j}$
and powers $n_{j}$ are to be defined by considering the strong-coupling
limit of the approximant (\ref{11.3}) and equating it to the strong-coupling
expansion (\ref{11.2}). This way can be called the {\it left-to-right
crossover}.

In general, it could be possible to go the opposite way, from the right to
the left. That is, we could construct a nested-root approximant starting
from the strong-coupling asymptotic form (\ref{11.2}) and then define the
corresponding coefficients and powers by equating the approximant expansion
in the weak-coupling limit to the asymptotic expansion (\ref{11.1}).
However, the right-to-left crossover results in approximants that usually
are less accurate than the left-to-right crossover formulas. This is
connected to the fact that the weak-coupling expansions have, as a rule,
zero radius of convergence, while the strong-coupling ones have a finite
radius of convergence. The accuracy of the left-to-right crossover
approximants is usually better than that of the right-to-left ones because
of the larger region of applicability of the strong-coupling expansion (\ref
{11.2}) as compared to the region of validity of the weak-coupling expansion
(\ref{11.1}). In fact, the latter can be valid for $s\ll 1$, hence its
region of validity is inside the interval $[0,1)$. In contrast, the
strong-coupling form, derived for $s\gg 1$, has the region of applicability
inside the interval $(1,\infty)$. Therefore the self-similar crossover
approximant has to be fitted to the asymptotic expansion that possesses the
larger region of validity.

When considering the strong-coupling limit $s\rightarrow\infty$ for the
approximant (\ref{11.3}), we need to know the relation between the powers $%
n_j$ and the numbers $j=1,2,\ldots$. Among all possible relations, we have
to choose that one which is the most general, imposing no restrictions on
the powers $\beta_j$. It is possible to show that the condition 
\begin{equation}  \label{11.4}
jn_j < j+1 \qquad (j=1,2,\ldots, k-1)
\end{equation}
provides a general way of expanding the form (\ref{11.3}), valid for any $%
k=1,2,\ldots$ and arbitrary $\beta_j$.

Under the criterion (\ref{11.4}), and rewriting the approximant (\ref{11.3})
in the form 
\[
f_k^*(s) = a_0(A_ks^k)^{n_k} \left ( 1 +\; \frac{A_{k-1}^{n_{k-1}}}{A_k} \;
x^{k-(k-1)n_{k-1}}\left\{ 1+\; \frac{A_{k-2}^{n_{k-2}}}{A_{k-1}}\;
x^{k-1-(k-2)n_{k-2}}\left ( 1 + \ldots \right.\right.\right. 
\]
\[
\ldots + \frac{A_2^{n_2}}{A_3}\; x^{3-2n_2} \left.\left.\left. \left [ 1 +\; 
\frac{A_1^{n_1}}{A_2}\; x^{2-n_1}\left ( 1+\; \frac{x}{A_1}%
\right
)^{n_1}\right ]^{n_2}\right )^{n_3} \ldots\right\}^{n_{k-1}}
\right
)^{n_k} \; , 
\]
where $x\equiv s^{-1}$, it is easy to expand the latter in powers of $x$.
Comparing the resulting expansion with the strong-coupling limit (\ref{11.2}%
), we obtain 
\[
k\; n_k =\beta_0 \; , 
\]
\begin{equation}  \label{11.5}
(k-j)n_{k-j} = \beta_j - \beta_{j-1} + k-j+1 \; ,
\end{equation}
with $1\leq j\leq k-1$. The values of $n_j$, defined by Eqs. (\ref{11.5}),
are in compliance with the criterion (\ref{11.4}) because of the inequality $%
\beta_j-\beta_{j-1}<0$.

The first-order self-similar approximant (\ref{11.3}) is 
\[
f_1^*(s) = a_0(1+As)^{n_1} \; , 
\]
where 
\[
A^{n_1} =\frac{b_0}{a_0}\; , \qquad n_1 =\beta_0 \; . 
\]
The second-order approximant (\ref{11.3}) takes the form 
\[
f_2^*(s) = a_0 \left [ \left ( 1 +A_1s\right )^{n_1} + A_2 s^2\right ]^{n_2}
\; , 
\]
in which 
\[
A_1^{n_1n_2} =\frac{b_0}{a_0} \left (\frac{b_1}{n_2b_0}\right )^{n_2} \; ,
\qquad A_2^{n_2} = \frac{b_0}{a_0} \; , 
\]
\[
n_1=\beta_1-\beta_0 + 2\; , \qquad 2n_2=\beta_0 \; . 
\]
In the third order, we find 
\[
f_3^*(s) = a_0\left\{\left [\left ( 1 + B_1s\right )^{n_1} + B_2s^2\right %
]^{n_2} + B_3s^3\right\}^{n_3} \; , 
\]
where 
\[
B_1^{n_1n_2n_3} =\frac{b_0}{a_0}\left ( \frac{b_1}{n_3b_0}%
\right
)^{n_3}\left ( \frac{b_2}{n_2b_1}\; - \; \frac{n_3-1}{2n_2n_3}\; 
\frac{b_1}{b_0}\right )^{n_2n_3} \; , 
\]
\[
B_2^{n_2n_3} = \frac{b_0}{a_0}\left ( \frac{b_1}{n_3b_0}\right )^{n_3} \; ,
\qquad B_3^{n_3} =\frac{b_0}{a_0} \; , 
\]
\[
n_1=\beta_2-\beta_1+2\; , \qquad 2n_2=\beta_1-\beta_0 + 3\; , \qquad
3n_3=\beta_0 \; . 
\]

The method of constructing self-similar crossover formulas is also
applicable to asymptotic expansions more general than Eq. (\ref{11.1}), for
instance, to series 
\begin{equation}  \label{11.6}
f(s) = a_0 + a_1 s^{\alpha_1} + a_2 s^{\alpha_2} + \ldots \qquad
(s\rightarrow 0) \; ,
\end{equation}
in which $\alpha_j$ are arbitrary positive powers arranged in the increasing
order as 
\begin{equation}  \label{11.7}
0 <\alpha_j <\alpha_{j+1} \; .
\end{equation}
Then, instead of Eq. (\ref{11.3}), we obtain the self-similar approximant 
\begin{equation}  \label{11.8}
f_k^*(s) = a_0\left ( \ldots\left\{ \left [\left ( 1+ A_1 s^{\alpha_1}\right
)^{n_1} + A_2s^{\alpha_2}\right ]^{n_2} + A_3s^{\alpha_3}\right\}^{n_3} +
\ldots + A_ks^{\alpha_k} \right )^{n_k} \; .
\end{equation}
The criterion (\ref{11.4}) transforms to the inequality 
\begin{equation}  \label{11.9}
\alpha_j\; n_j < \alpha_{j+1} \; .
\end{equation}
And, in the place of Eqs. (\ref{11.5}), we find 
\[
\alpha_k\; n_k =\beta_0 \; , 
\]
\begin{equation}  \label{11.10}
\alpha_j\; n_j = \alpha_{j+1} +\beta_{k-j}-\beta_{k-j-1} \; ,
\end{equation}
with $j=1,2,\ldots,k-1$.

The described method makes it possible to construct analytical interpolative
formulas for the whole range of the coupling parameter. The method can also
be used for interpolating any functions of other variables, provided the
corresponding asymptotic expansions are available.

\section{One-Dimensional Confined System}

To illustrate the method presented in the previous section, let us consider
a model case of a one-dimensional system of trapped atoms \cite{Yukalov98b}.
This means that in the eigenproblem $\hat{H}\psi =E\psi$, we consider the
nonlinear Hamiltonian 
\begin{equation}
\hat{H}=-\;\frac{1}{2}\;\frac{d^{2}}{dx^{2}}\;+\;\frac{1}{2}\;x^{2}+g|\psi
|^{2}\;,  \label{11.11}
\end{equation}
in which $x\in (-\infty ,+\infty )$.

In order to derive the weak-coupling and strong-coupling asymptotic
expansions, we may resort to the optimized perturbation theory of Sec.~10.1.
To this end, we start with the trial Hamiltonian 
\begin{equation}
\hat{H}_{0}=-\;\frac{1}{2}\;\frac{d^{2}}{dx^{2}}\;+\;\frac{u^{2}}{2}%
\;x^{2}\;,  \label{11.12}
\end{equation}
containing a control parameter $u$, possessing the eigenvalue 
\[
E_{n}^{(0)}=\left( n+\frac{1}{2}\right) \;u\;, 
\]
and having the eigenfunction 
\[
\psi _{n}^{(0)}(x)=\frac{(u/\pi )^{1/4}}{\sqrt{2^{n}\;n!}}\;H_{n}(\sqrt{u}%
\;x)\;\exp \left( -\;\frac{1}{2}\;ux^{2}\right) \;, 
\]
where $n=0,1,2,\ldots $.

The first-order approximation gives 
\begin{equation}  \label{11.13}
E_n^{(1)}(g,u) =\frac{u}{2}\; \left ( n +\frac{1}{2}\right ) \left ( 1 +%
\frac{1}{u^2}\right ) + \; \sqrt{u}\; I_n g \; ,
\end{equation}
with the notation 
\[
I_n\equiv \frac{2}{\pi(2^n\; n!)^2} \; \int_0^\infty H_n^4(x) \; e^{-2x^2}\;
dx \; . 
\]
The variational condition for Eq. (\ref{11.13}) yields the equation 
\begin{equation}  \label{11.14}
u^2 +\alpha\; u^{3/2} - 1 = 0
\end{equation}
for the control function $u=u(\alpha)$, where 
\begin{equation}  \label{11.15}
\alpha\equiv \frac{2I_ng}{2n+1} \; .
\end{equation}
For the optimized approximant 
\begin{equation}  \label{11.16}
E(\alpha) \equiv E_n^{(1)}(g(\alpha),u(\alpha)) \; ,
\end{equation}
we have 
\begin{equation}  \label{11.17}
E(\alpha) =\frac{1}{2}\left ( n +\frac{1}{2}\right ) \left ( \frac{3}{u}\; -
u\right ) \; .
\end{equation}
Expression (\ref{11.17}), together with the control-function equation (\ref
{11.14}), results in the weak-coupling expansion 
\begin{equation}  \label{11.18}
E(\alpha) \simeq \left ( n +\frac{1}{2}\right ) \left ( 1 + \alpha -\; \frac{%
1}{8}\; \alpha^2 +\frac{1}{32}\; \alpha^3 -\; \frac{1}{128}\; \alpha^4 + 
\frac{3}{2048}\; \alpha^5\right ) \; ,
\end{equation}
for $\alpha\rightarrow 0$, and in the strong-coupling expansion 
\begin{equation}  \label{11.19}
E(\alpha) \simeq \left ( n +\frac{1}{2}\right ) \left ( \frac{3}{2}%
\;\alpha^{2/3} +\frac{1}{2}\;\alpha^{-2/3} -\; \frac{1}{6}\; \alpha^{-2}+%
\frac{7}{54}\; \alpha^{-10/3} \right ) \; ,
\end{equation}
as $\alpha\rightarrow\infty$.

Following Sec.~11.1, we find the self-similar crossover approximants. In the
first order, this gives 
\begin{equation}
E_{1}^{\ast }(\alpha )=\left( n+\frac{1}{2}\right) \left( 1+A\alpha \right)
^{2/3}\;,  \label{11.20}
\end{equation}
with $A=\frac{3}{4}\sqrt{6}$. In the second order, we find 
\begin{equation}
E_{2}^{\ast }(\alpha )=\left( n+\frac{1}{2}\right) \left[ \left(
1+A_{1}\alpha \right) ^{2/3}+A_{2}\alpha ^{2}\right] ^{1/3}\;,  \label{11.21}
\end{equation}
where 
\[
A_{1}=\frac{81}{32}\;\sqrt{6}\;,\qquad A_{2}=\frac{27}{8}\;. 
\]
Formulas (\ref{11.20}) and (\ref{11.21}) interpolate between the
weak-coupling expansion (\ref{11.18}) and the strong-coupling limit (\ref
{11.19}).

A similar interpolation procedure can be applied for constructing an
analytical expression for the ground-state wave function \cite{Yukalov98b}.
The latter, at small space variable, has an expansion 
\begin{equation}
\psi (x)\simeq c_{0}+c_{2}x^{2}+c_{4}x^{4}\qquad (x\rightarrow 0)\;,
\label{11.22}
\end{equation}
and in the opposite limit 
\begin{equation}
\psi (x)\simeq C\;\exp \left( -\;\frac{1}{2}\;x^{2}\right) \qquad
(x\rightarrow \infty )\;.  \label{11.23}
\end{equation}
The interpolating formula, sewing the limits (\ref{11.22}) and (\ref{11.23}%
), can be constructed invoking the self-similar exponential approximants 
\cite{Yukalov98c}, which results in 
\begin{equation}
\psi _{\ast }(x)=C\;\exp \left\{ -\;\frac{1}{2}\;x^{2}+ax^{2}\;\exp
(-bx^{2})\right\} \;.  \label{11.24}
\end{equation}
Here, the coefficients $a$ and $b$ are to be defined by expanding the
function (\ref{11.24}) in powers of $x\rightarrow 0$ and substituting this
expansion into the eigenproblem $\hat{H}\psi _{\ast }=E\psi _{\ast }$, which
yields 
\[
a=\frac{1}{2}+gC^{2}-E\;,\qquad b=\frac{2E-4aE-1}{12a}\;. 
\]
The normalization constant $C$ and energy $E$ are defined by the equations 
\begin{equation}
(\psi _{\ast },\psi _{\ast })=1\;,\qquad E_{\ast }=(\psi _{\ast },\hat{H}
\psi _{\ast })\;,  \label{11.25}
\end{equation}
where $E\equiv E_{\ast }$.

The self-similar approximant (\ref{11.24}) for the wave function is
different from the ground-state wave function 
\begin{equation}  \label{11.26}
\psi_0^{(0)}(x) =\left (\frac{u}{\pi}\right )^{1/4} \; \exp\left (-\; \frac{u%
}{2}\; x^2\right ) \; ,
\end{equation}
with the control function $u(\alpha)$ defined by the variational equation (%
\ref{11.14}), with $\alpha=\sqrt{(2/\pi)}\; g$. And in the Thomas-Fermi
approximation, we have the wave function 
\begin{equation}  \label{11.27}
\psi_{TF}(x) =\left (\frac{x_c^2-x^2}{2g}\right )^{1/2}\; \Theta(x_c-|x|) \;
,
\end{equation}
with the energy 
\begin{equation}  \label{11.28}
E_{TF}(g) =\frac{x_c^2}{2} =\frac{1}{2}\left ( \frac{3}{2}\; g \right
)^{2/3} \; .
\end{equation}

To compare these different approximations, we consider the properties of the
ground-state energy $E(g)$, as a function of the coupling parameter $g$,
presented by the optimized approximant $E_{opt}(g)\equiv E(\alpha(g))$ given
by Eq. (\ref{11.17}), by the crossover approximant $E_2^*(g)$ from Eq. (\ref
{11.21}), by the energy $E_*(g)$ defined in Eq. (\ref{11.25}), and by the
Thomas-Fermi energy $E_{TF}(g)$ in Eq. (\ref{11.28}). We also compare the
shape of the density 
\begin{equation}  \label{11.29}
n(x) \equiv |\psi(x)|^2 \; ,
\end{equation}
defined for the corresponding functions (\ref{11.24}), (\ref{11.26}), and (%
\ref{11.27}). The accuracy of the approximations can be characterized by
substituting the wave functions into the Schr\"odinger equation and
calculating the residual $R(x)$ and dispersion $\sigma^2$, given by the
equations 
\begin{equation}  \label{11.30}
R(x) \equiv (\hat H -E)\psi(x) \; , \qquad \sigma^2\equiv
\int_{-\infty}^{+\infty} |R(x)|^2\; dx \; .
\end{equation}

Figure~\ref{Fig23} shows the energies $E_{\ast }(g)$, $E_{2}^{\ast }(g)$,
and $E_{TF}(g)$. The first two energies are almost indistinguishable from
each other. The optimized approximant $E_{opt}(g)$ is not shown since it
practically coincides with $E_{\ast }(g)$. The Thomas-Fermi energy $%
E_{TF}(g) $ has an incorrect weak-coupling limit and becomes a reasonable
approximation for $g\geq 7$.

The density (\ref{11.29}) for the self-similar wave function (\ref{11.24}),
Gaussian function (\ref{11.26}) and Thomas-Fermi wave function (\ref{11.27})
is presented in Fig. \ref{Fig24} for different coupling parameters. As is
seen, the self-similar function (\ref{11.24}) has the correct behavior in
both weak-coupling as well as strong-coupling limits, while the Gaussian
function (\ref{11.26}) does not present a good approximation in the
strong-coupling limit and the Thomas-Fermi function (\ref{11.27}) is not
correct in the weak-coupling limit. In addition, the latter function is
always incorrect at the boundary of an atomic cloud.

\begin{sloppypar}
The accuracy of the corresponding approximate solutions to the
nonlinear Schr\"odinger equation is well illustrated by the
residual $R(x)$, which is shown in Fig. 28. We also  calculated
the dispersion $\sigma^2$ for different coupling parameters within 
the region $0\leq g\leq 100$. The maximal, with respect to $g$,
dispersion for the self-similar function (\ref{11.24}) is of order
one, for the Gaussian function (\ref{11.26}) it is about 400, and
for the Thomas-Fermi function (\ref{11.27}) it is divergent. This
clearly proves that the self-similar function (\ref{11.24}) is the
most accurate solution to the nonlinear Schr\"odinger equation for
small as well as for large coupling parameters.
\end{sloppypar}

\section{Spherically Symmetric Trap}

Similarly to the model one-dimensional case considered above, we can
construct self-similar crossover approximants for the realistic
three-dimensional situation. We shall illustrate this for a spherically
symmetric trap. To this end, let us consider the isotropic ground state
studied in Sec.~10.2.

Using the weak-coupling and strong-coupling expansions, (\ref{10.20}) and
respectively (\ref{10.21}), for the ground-state energy (\ref{10.19}), we
construct the crossover formulas of first order, 
\begin{equation}
E_{1}^{\ast }(s)=\frac{3}{2}(1+As)^{2/5}\;,  \label{11.31}
\end{equation}
where $A=0.633938$; of second order, 
\begin{equation}
E_{2}^{\ast }(s)=\frac{3}{2}\left[ \left( 1+A_{1}s\right) ^{6/5}+A_{2}s^{2}%
\right] ^{1/5}\;,  \label{11.32}
\end{equation}
with $A_{1}=1.168636$ and $A_{2}=0.401878$; of third order, 
\begin{equation}
E_{3}^{\ast }(s)=\frac{3}{2}\left\{ \left[ \left( 1+B_{1}s\right)
^{6/5}+B_{2}s^{2}\right] ^{11/10}+B_{3}s^{3}\right\} ^{2/15}\;,
\label{11.33}
\end{equation}
where $B_{1}=1.633061$, $B_{2}=1.132289$, and $B_{3}=0.254766$; of fourth
order, 
\begin{equation}
E_{4}^{\ast }(s)=\left( \left\{ \left[ \left( 1+C_{1}s\right)
^{6/5}+C_{2}s^{2}\right] ^{11/10}+C_{3}s^{3}\right\}
^{16/15}+C_{4}s^{4}\right) ^{1/10}\;,  \label{11.34}
\end{equation}
with $C_{1}=2.066398$, $C_{2}=2.111737$, $C_{3}=0.970940$, $C_{4}=0.161506$;
and of fifth order, 
\begin{equation}
E_{5}^{\ast }(s)=\frac{3}{2}\left[ \left( \left\{ \left[ \left(
1+D_{1}s\right) ^{6/5}+D_{2}s^{2}\right] ^{11/10}+D_{3}s^{3}\right\}
^{16/15}+D_{4}s^{4}\right) ^{21/20}+D_{5}s^{5}\right] ^{2/25}\;,
\label{11.35}
\end{equation}
where the coefficients are $D_{1}=2.479006$, $D_{2}=3.311734$, $%
D_{3}=2.278301$, $D_{4}=0.777603$, and $D_{5}=0.102385$. The variable $s$ is
defined in Eq. (\ref{10.15}). Note that $A>A_{2}>B_{3}>C_{4}>D_{5}$, which
suggests that the accuracy of $E_{k}^{\ast }(s)$ should increase with
increasing $k$. The accuracy of the crossover approximants $E_{k}^{\ast }(s)$
can be characterized by the percentage errors 
\[
\varepsilon _{k}^{\ast }(s)\equiv \frac{E_{k}^{\ast }(s)-E(s)}{|E(s)|}\times
100\%\;, 
\]
calculated with respect to the optimized approximant (\ref{10.19}). Even
more instructive are the maximal errors 
\[
\varepsilon _{k}^{\ast }\equiv \max_{s}\varepsilon _{k}^{\ast }(s)\qquad
(0\leq s\leq \infty )\;. 
\]
For the latter, we find 
\[
\varepsilon _{1}^{\ast }=3.7\%\;,\quad \varepsilon _{2}^{\ast
}=1.4\%\;,\quad \varepsilon _{3}^{\ast }=0.8\%\;,\quad \varepsilon
_{4}^{\ast }=0.6\%\;,\quad \varepsilon _{5}^{\ast }=0.4\%\;, 
\]
which demonstrates good numerical convergence of the crossover approximants
for the ground-state energy.

A crossover approximant for the radial wave function satisfying Eq. (\ref
{11.12}) can also be constructed by sewing the small-radius limit 
\begin{equation}
\chi (r)\simeq c_{1}r+c_{3}r^{3}+c_{5}r^{5}\qquad (r\rightarrow 0)
\label{11.36}
\end{equation}
and the large-distance asymptotic form 
\begin{equation}
\chi (r)\simeq Cr\exp \left( -\;\frac{1}{2}\;r^{2}\right) \qquad
(r\rightarrow \infty )\;.  \label{11.37}
\end{equation}
The self-similar crossover formula is 
\begin{equation}
\chi _{\ast }(r)=Cr\exp \left\{ -\;\frac{1}{2}\;r^{2}+ar^{2}\exp
(-br^{2})\right\} \;.  \label{11.38}
\end{equation}
Here the coefficients $a$ and $b$ are to be found by expanding Eq. (\ref
{11.38}) in powers of $r$ and substituting this expansion into the equation (%
\ref{10.12}). Equating the coefficients at the like powers of $r$, we get 
\[
a=\frac{1}{2}+\frac{gC^{2}-4\pi E}{12\pi }\;,\qquad b=\frac{%
2(1-2a)E-2(1-2a)^{2}-1}{20a}\;. 
\]
The normalization coefficient $C$ and energy $E$ are defined by the
equations 
\begin{equation}
E_{\ast }=\int_{0}^{\infty }\chi _{\ast }(r)\hat{H}_{r}\chi (r)\;dr\;,\qquad
\int_{0}^{\infty }\chi _{\ast }^{2}(r)\;dr=1\;,  \label{11.39}
\end{equation}
where $E_{\ast }\equiv E$.

In this way, we obtain several representations for the ground-state energy:
The self-similar approximant $E_{\ast }$, given in Eq. (\ref{11.39}), the
crossover approximants $E_{k}^{\ast }$ in Eqs. (\ref{11.31}) to (\ref{11.35}%
), the optimized approximant $E(s)$ defined by Eq. (\ref{10.19}), and the
Thomas-Fermi approximation $E_{TF}$ from Eq. (\ref{9.34}), with $\nu =1$.
Similarly to the previous section, we may analyze the behavior of these
approximations as functions of the coupling parameter $g$. The analysis
shows that the self-similar approximant $E_{\ast }(g)$, defined in Eq. (\ref
{11.39}), gives the best approximation, valid for the whole range of the
parameter $g\in \lbrack 0,\infty )$, correctly interpolating the
weak-coupling expansion and the Thomas-Fermi limit. The latter gives a good
approximation only for $g>300$, essentially deviating from the weak-coupling
form, as is seen in Fig. \ref{Fig26}. The percentage errors $\varepsilon
_{k}^{\ast }(s)$ of different crossover approximants $E_{k}^{\ast }(s)$ are
presented in Figs. \ref{Fig27} and \ref{Fig28}. The maximal error occurs
around $g\sim 1$.

The spatial density 
\begin{equation}
n(r)\equiv \frac{\chi ^{2}(r)}{r^{2}}  \label{11.40}
\end{equation}
is expressed through the corresponding radial function, for which one can
take either the self-similar form $\chi _{\ast }(r)$ in Eq.~(\ref{11.38}),
or the Gaussian approximation $\chi ^{(0)}(r)$ from Sec.~10, or the
Thomas-Fermi wave function 
\[
\chi _{TF}(r)=r\;\sqrt{\frac{2\pi }{g}\;\left( r_{c}^{2}-r^{2}\right) }%
\;\Theta \left( r_{c}^{2}-r^{2}\right) \;,\qquad r_{c}^{2}=2E_{TF}\;,\qquad
E_{TF}=\frac{1}{2}\left( \frac{15}{4\pi }\;g\right) ^{2/5}\;. 
\]
The accuracy of the considered approximate solutions can again be
characterized by the residual 
\[
R(r)\equiv (\hat{H}_{r}-E)\chi (r) 
\]
and the dispersion 
\[
\sigma ^{2}\equiv \int_{0}^{\infty }|R(r)|^{2}\;dr\;. 
\]
The analysis here is similar to the previous section, and again the
self-similar form (\ref{11.38}) turns out to be the best approximation,
valid for all coupling parameters. Thus, the residual as well as the
dispersion diverge for the Thomas-Fermi approximation at any $g$. The
accuracy of the Gaussian variational approximation is good for small $g$ but
decreases with increasing $g$. For instance, the dispersion $\sigma ^{2}$
for the Gaussian approximation monotonically rises with $g$, being {\em e.g.}
at $g=2513$, equal to $\sigma ^{2}=13.2$. At the same time, the dispersion
for the self-similar approximant (\ref{11.38}) reaches the maximum of $%
\sigma ^{2}=4.1$ at $g=2411$ and then again diminishes to $\sigma ^{2}=1.1$
at $g=2513$.

\section{Traps of Cylindrical Shape}

Self-similar crossover approximants can also be constructed for cylindrical
traps, using the expansions of Sec.~10.4. Recall that these expansions,
being done in terms of the variable 
\[
s\equiv 2p\sqrt{q}\;I_{nmk}\;\nu g\qquad (p\equiv 2n+|m|+1,\;q\equiv
2k+1)\;, 
\]
are valid for arbitrary excited coherent modes labelled by the quantum
numbers $n,m$, and $k$.

Being based on the weak-coupling, (\ref{10.59}), and strong-coupling, (\ref
{10.60}), expansions for the energy levels, and employing the technique of
Sec.~11.1, we obtain the crossover approximants $E_{k}^{\ast }(s)$. Thus, in
the first order, we have 
\begin{equation}
E_{1}^{\ast }(s)=a_{0}\left( 1+As\right) ^{2/5}\;,  \label{11.41}
\end{equation}
where 
\[
a_{0}=p+\frac{q\nu }{2}\;,\qquad A=\frac{1.746928}{a_{0}^{5/2}}\;. 
\]
The second order yields 
\begin{equation}
E_{2}^{\ast }(s)=a_{0}\left[ \left( 1+A_{1}s\right) ^{6/5}+A_{2}s^{2}\right]
^{1/5}\;,  \label{11.42}
\end{equation}
with the same $a_{0}$ and with 
\[
A_{1}=\frac{2.533913}{a_{0}^{25/6}}\left[ 2p^{2}+(q\nu )^{2}\right]
^{5/6}\;,\qquad A_{2}=\frac{3.051758}{a_{0}^{5}}\;. 
\]
In the third order, we get 
\begin{equation}
E_{3}^{\ast }(s)=a_{0}\left\{ \left[ \left( 1+B_{1}s\right) ^{6/5}+B_{2}s^{2}%
\right] ^{11/10}+B_{3}s^{3}\right\} ^{2/15}\;,  \label{11.43}
\end{equation}
where 
\[
B_{1}=\frac{1.405455}{a_{0}^{125/22}[2p^{2}+(q\nu )^{2}]^{5/66}}\;\left[
8p^{4}+12p^{2}(q\nu )^{2}+(q\nu )^{4}\right] ^{5/6}\;, 
\]
\[
B_{2}=\frac{6.619620}{a_{0}^{75/11}}\;\left[ 2p^{2}+(q\nu )^{2}\right]
^{10/11}\;\qquad B_{3}=\frac{5.331202}{a_{0}^{15/2}}\;. 
\]
Similarly, we find $E_{4}^{\ast }(s)$ and $E_{5}^{\ast }(s)$, although we do
not write them down explicitly.

To check the accuracy of the crossover approximants $E_k^*(s)$, we calculate
the percentage errors $\varepsilon_k^*(s)$ comparing $E_k^*(s)$ with the
optimized approximant (\ref{10.54}). We have calculated the maximal errors $%
\varepsilon_k^*\equiv\max_s\varepsilon_k^*(s)$ for the anisotropy parameter $%
\nu$, defined in Eq. (\ref{9.22}), in the range $0.1\leq\nu\leq 100$ for the
ground state and for ten first excited states. For example, for the ground
state, with $n=m=k=0$, and for $\nu=1$, we find 
\[
\varepsilon_1^*=3.7\%\; , \qquad \varepsilon_2^*=1.4\% \; , \qquad
\varepsilon_3^*=0.8\% \; , \qquad \varepsilon_4^*=0.6\% \; , \qquad
\varepsilon_5^*=0.4\% \; , 
\]
which demonstrates good convergence. In the case of a cigar-shape trap, with 
$\nu=0.1$, we obtain 
\[
\varepsilon_1^*=8\%\; , \qquad \varepsilon_2^*=3.5\% \; , \qquad
\varepsilon_3^*=2\%\; , \qquad \varepsilon_4^*=1.2\% \; , \qquad
\varepsilon_5^*=0.8\% \; . 
\]
For a disk-shape trap, with $\nu=10$, we have 
\[
\varepsilon_1^*=12.5\%\; , \qquad \varepsilon_2^*=3\%\; \qquad
\varepsilon_3^*=2.8\% \; , \qquad \varepsilon_4^*=-1.8\% \; . 
\]
The same good convergence occurs for excited states with different quantum
numbers and for various anisotropy parameters. The standard situation is
such that $\varepsilon_1^*\approx 4-12\%$, $\varepsilon_2^*\approx 2-5\%$,
and already the third-order approximant has $\varepsilon_3^*\sim 1\%$.

To illustrate in more detail the accuracy of the crossover approximants $%
E_{k}^{\ast }$ as functions of the coupling parameter $g$, we show in Fig.~%
\ref{Fig29} the percentage errors of $E_{1}^{\ast }$, $E_{2}^{\ast }$, and $%
E_{3}^{\ast }$ for several levels and different anisotropy parameters. The
errors are calculated with respect to the optimized approximant (\ref{10.54}%
) whose ground-state behavior is presented in Fig.~\ref{Fig30}, where the
Thomas-Fermi energy is also given for comparison.

In the same way, we may construct the crossover approximants for the aspect
ratio (\ref{10.68}), being based on the asymptotic expansions (\ref{10.69})
and (\ref{10.70}). Here, it is more convenient to deal with the quantity 
\begin{equation}
\overline{R}_{r}(s)\equiv \frac{R_{r}(s)}{\sqrt{2}\;\nu }\;-1\;,
\label{11.44}
\end{equation}
for which Eqs. (\ref{10.69}) and (\ref{10.70}) transform to the expansions 
\[
\overline{R}_{r}(s)\simeq \left( \sqrt{\frac{p}{q\nu }}\;-1\right) +\sqrt{%
\frac{p}{q\nu }}\;\left( \alpha _{1}s+\alpha _{2}s^{2}+\ldots \right) 
\]
in the weak-coupling limit $s\rightarrow 0$, and to 
\[
\overline{R}_{r}(s)\simeq \beta _{1}s^{-4/5}+\beta _{2}s^{-8/5}+\ldots 
\]
in the strong-coupling limit $s\rightarrow \infty $. Constructing the
self-similar approximant $\overline{R}_{r}^{\ast }(s)$, we then return to
the aspect ratio 
\begin{equation}
R_{r}^{\ast }(s)=\sqrt{2}\;\nu \left[ 1+\overline{R}_{r}^{\ast }(s)\right]\;.
\label{11.45}
\end{equation}
We have compared the accuracy of the crossover approximants, corresponding
to Eq.~(\ref{11.45}), with the value (\ref{10.68}) for the anisotropy
parameter $\nu $ in the range $0<\nu \leq 100$, and for the first ten energy
levels. The results are similar to those obtained for the energy levels
themselves.

\section{Strong-Coupling and Thermodynamic Limits}

The atom-atom coupling (\ref{9.27}) is proportional to the number of atoms $%
N $, which suggests that the strong-coupling limit $g\rightarrow\infty$ has
to be related to the thermodynamic limit $N\rightarrow\infty$. The averages
of observable quantities $\hat A$ should behave in the thermodynamic limit
so that 
\begin{equation}  \label{11.46}
\lim_{N\rightarrow\infty} \; \left |\frac{1}{N}\; <\hat A>\right | < \infty
\; .
\end{equation}
Let us check this property for the coherent averages of the Hamiltonian (\ref
{8.13}). For the latter, the coherent average (\ref{8.54}) coincides with
the pure coherent average (\ref{8.31}), that is with (\ref{8.38}). For the
normalization (\ref{8.51}), we have 
\begin{equation}  \label{11.47}
<H>_{Nn} \; = \hbar\omega_rN \left ( S_n +\frac{1}{2}\; g\gamma_n \right )
\; ,
\end{equation}
where $n$ implies the whole set of quantum numbers $n,m$, and $k$; the mean
single-particle energy is 
\begin{equation}  \label{11.48}
S_n \equiv \int\; \psi_n^*({\bf r}) \left [ -\; \frac{1}{2}\; {\bf \nabla}^2
+\frac{1}{2}\; (r^2 +\nu^2 z^2) \right ] \psi_n({\bf r})\; d{\bf r} \; ,
\end{equation}
with dimensionless ${\bf r}$ measured in units of $l_r$, defined in Eq. (\ref
{9.23}); and $\gamma_n$ being the same as $\gamma_{nmk}$ in Eq. (\ref{10.72}%
). From the eigenproblem (\ref{9.29}) it follows that 
\[
E_n\equiv (\psi_n,\hat H\psi_n) = S_n + g\gamma_n \; . 
\]
Hence, the average energy (\ref{11.47}) of a coherent state can also be
presented in two other forms as 
\begin{equation}  \label{11.49}
<H>_{Nn}\; = \frac{1}{2}\;\hbar\omega_r N(E_n+ S_n) \; , \qquad <H>_{Nn}\; =
\hbar\omega_r N(E_n-\; \frac{1}{2}\; g\gamma_n) \; .
\end{equation}
Note that one should not confuse here $H$, which is the system Hamiltonian (%
\ref{8.13}), with $\hat H$, which is the Schr\"odinger Hamiltonian (\ref
{9.28}).

Consider the strong-coupling limit $g\rightarrow\infty$ for the coherent
average energy (\ref{11.47}) or (\ref{11.49}). According to the notation (%
\ref{10.53}), this corresponds to $s\rightarrow\infty$. Then, Eqs. (\ref
{10.60}) and (\ref{10.75}) yield 
\[
E_n(s) \simeq \frac{5}{4}\; s^{2/5} \; , \qquad \gamma_n(s) \simeq \frac{1}{%
2g}\; s^{2/5} \qquad (s\rightarrow\infty) \; . 
\]
In this limit, the coherent single-particle energy (\ref{11.48}) is 
\begin{equation}  \label{11.50}
S_n(s) = E_n(s) -g\gamma_n \simeq \frac{3}{4}\; s^{2/5} \; .
\end{equation}
Therefore, 
\begin{equation}  \label{11.51}
<H>_{Nn} \; \simeq \hbar\omega_r Ns^{2/5} \qquad (s\rightarrow\infty)\; .
\end{equation}
From the definition of the coupling (\ref{9.27}), we have 
\begin{equation}  \label{11.52}
g=4\pi\sqrt{\frac{\omega_r}{\varepsilon}}\; N \; , \qquad \varepsilon\equiv 
\frac{\hbar}{m_0a_s^2} \; .
\end{equation}
And the relation (\ref{10.53}) between $g$ and $s$ gives 
\begin{equation}  \label{11.53}
s=C_n^{5/2}\left ( \frac{\omega_r}{\varepsilon}\right )^{1/2}\; N \; ,
\qquad C_n\equiv \left ( 8\pi \sqrt{q}\; I_n\; \nu \right )^{2/5} \; .
\end{equation}
Then, the average energy (\ref{11.51}) becomes 
\begin{equation}  \label{11.54}
<H>_{Nn}\simeq C_n\hbar\varepsilon\left ( \frac{\omega_r}{\varepsilon}\right
)^{6/5}\; N^{7/5} \; ,
\end{equation}
as $N\rightarrow\infty$.

Note that in the strong-coupling limit, when $s\rightarrow\infty$, the
average kinetic energy 
\[
K_n\equiv \left ( \psi_n, -\;\frac{1}{2}{\bf \nabla}^2\psi_n \right ) 
\]
becomes negligible as compared to the average potential energy and the mean
interaction energy. To show this, we may write the mean single-particle
energy (\ref{11.48}) as 
\[
S_n = K_n +\frac{1}{2}\; r_0^2 +\frac{1}{2}\; \nu^2z_0^2 \; . 
\]
From the asymptotic expansions (\ref{10.65}) for $r_0$ and (\ref{10.66}) for 
$z_0$, we have 
\[
\frac{1}{2}\; r_0^2 +\frac{1}{2}\; \nu^2z_0^2 \simeq \frac{3}{4}\; s^{2/5}
\qquad (s\rightarrow\infty) \; . 
\]
Hence the average kinetic energy 
\[
K_n \simeq S_n -\; \frac{3}{4}\; s^{2/5} \rightarrow 0 \qquad
(s\rightarrow\infty) 
\]
tends to zero according to the limit (\ref{11.50}).

If we consider the thermodynamic limit, keeping the frequency $\omega _{r}$
fixed, then 
\[
\frac{1}{N}\;<H>_{Nn}\;\sim N^{2/5}\rightarrow \infty \qquad (N\rightarrow
\infty )\;. 
\]
Then the property (\ref{11.46}) is not valid for the Hamiltonian $H$, which
implies that such a system is thermodynamically unstable. The $N^{7/5}$-law
of divergence of the average energy (\ref{11.54}) is the same as that found 
\cite{Lieb90} for the ground-state energy of bosons interacting through
Coulomb forces.

Another possibility could be to resort to the thermodynamic limit as defined
in Eq. (\ref{7.35}), when $l_r\sim N^{1/3}$ and $\omega_r\sim N^{-2/3}$.
Then 
\[
\frac{1}{N}\; <H>_{Nn}\; \sim N^{-2/5} \rightarrow 0 \qquad
(N\rightarrow\infty) \; , 
\]
which means that the average energy becomes negligible.

Finally, we may ask the question, how we should change the frequency $\omega
_{r}$ in order to satisfy the condition of thermodynamic stability (\ref
{11.46}) so that the average energy (\ref{11.54}) would give a finite value, 
\[
\frac{1}{N}\;<H>_{Nn}\;\rightarrow \;const\qquad (N\rightarrow \infty )\;? 
\]
The latter is satisfied for $\omega _{r}\sim N^{-1/3}$ and, respectively, $%
l_{r}\sim N^{1/6}$. This suggests the definition of the thermodynamic limit
as 
\begin{equation}
N\rightarrow \infty \;,\qquad \omega _{r}\rightarrow 0\;,\qquad N\omega
_{r}^{3}\rightarrow \;const\;.  \label{11.55}
\end{equation}
It is interesting that the same definition of the thermodynamic limit
follows from a quite different condition \cite{Dalfovo99} requiring the
finiteness of the Bose-condensation temperature for an ideal gas.

\chapter{Vortices in Trapped Condensates}

Vortex states in trapped atomic clouds have been considered theoretically by
several authors \cite{Rokhsar97,Butts99,Lundh97,Edwards96b}, \cite{Lundh98}- 
\cite{Castin99}. Vortex production appears to be a common consequence of
mechanically disturbing a condensate. A variety of methods have been
suggested by which vortices could be formed. A straightforward way would be
by rotating the trap. However, since such a rotation is difficult to
realize, other techniques have been proposed: population transfer via a
Raman transition into an angular momentum state \cite{Dum98,Marzlin97};
creation of circulating states in traps with a multiply connected geometry,
such as a toroidal trap or a magnetic trap pinched by a blue-detuned laser 
\cite{Petrosyan99}; stirring the condensate by means of laser beams \cite
{JacksonB98,Caradoc-Davies99}. The possibility of creating different
topological modes, including the vortex ones, by imposing resonance fields
has been advanced first in Ref. \cite{Yukalov97} and studied later in Ref. 
\cite{Marzlin98}. Recently, vortices were created in a two-component
condensate \cite{Matthews99b} by combining a microwave field inducing
interconversion between the two components at a laser beam rotating with a
resonant frequency \cite{Williams99}.

\section{Vortex Transition Frequencies}

To transfer a coherent cloud of atoms from the ground state with the energy
per particle $E_{000}$ to another coherent state having the energy $E_{nmk}$%
, one needs to pump into the system the energy per particle 
\begin{equation}
\Omega _{nmk}=E_{nmk}-E_{000}\;.  \label{12.1}
\end{equation}
To estimate the difference (\ref{12.1}), we may use the optimized
approximants of Section 10. These show that for the strong coupling $g\gg 1$
two principally different situations can occur. Since the energy of a
coherent state, labelled by the indices $n$, $m$, and $k$, grows with $g$ as 
$E_{nmk}\sim g^{2/5}$, the difference (\ref{12.1}) also grows in the same
way, $\Omega _{nmk}\sim g^{2/5}$, except for the case with the selection
rule 
\begin{equation}
(2\pi )^{3/2}\sqrt{2k+1}\;(2n+|m|+1)\;I_{nmk}=1\;,  \label{12.2}
\end{equation}
when the difference (\ref{12.1}) diminishes with $g$. The selection rule (%
\ref{12.2}) is satisfied for the sole state with $n=0$, $m=1$, and $k=0$,
which corresponds to the vortex state with the winding number $m=1$.
Vortices with higher winding numbers have essentially higher energies that
increase with $g$. The behavior of the vortex energies $\Omega _{nmk}$,
where $m\neq 0$, as functions of $\nu g$, is pictured in Fig. \ref{Fig31}.
As is seen, the energy $\Omega _{010}$ of the {\it basic vortex state} with
the minimal winding number $m=1$ decreases with $\nu g$ while the energy $%
\Omega _{020}$ of the vortex state with the winding number $m=2$ first
decreases with $\nu g$ and then increases. The qualitatively different
behavior of the energy $\Omega _{010}$ of the basic vortex as compared to
the energies of other vortex states suggests the following criterion of {\it %
Energetic Stability}: For a given orbital momentum $\hbar |m|$, at large $%
\nu g$, the creation of $m$ basic vortices is energetically more profitable
than the formation of one or several vortices with higher winding numbers
giving in total the same orbital momentum. This is in agreement with the
thermodynamic stability of vortices studied in Ref. \cite{Castin99}.

To form a vortex in a rotating trap, one has to reach the critical rotation
frequency that in dimensionless units reads $\hbar \Omega _{nmk}/|L_{z}|$,
where $L_{z}=\hbar m$ is an eigenvalue of the orbital momentum operator $%
\hat{L}_{z}=-i\hbar \partial /\partial \varphi $. For the basic vortex with
the winding number $m=1$, the critical frequency is 
\begin{equation}
\Omega _{c}\equiv \Omega _{010}=E_{010}-E_{000}\;.  \label{12.3}
\end{equation}
To consider the dependence of this frequency on the coupling $g$ it is
convenient to employ the notation 
\[
s\equiv \frac{2\nu g}{(2\pi )^{3/2}} 
\]
and to use the results of Sec.~10. Then, in the weak-coupling limit, we have 
\begin{equation}
\Omega _{c}\simeq 1-\;\frac{1}{4\nu ^{1/2}}\;s+\frac{3+7\nu }{64\nu ^{2}}%
\;s^{2}-\;\frac{7+30\nu +31\nu ^{2}}{512\nu ^{5/2}}\;s^{3}  \label{12.4}
\end{equation}
as $s\rightarrow 0$, and in the strong coupling limit, we get 
\begin{equation}
\Omega _{c}\simeq \frac{3}{2}\;s^{-2/5}-\;\frac{3}{20}\;(15-2\nu
^{2})\;s^{-6/5}+\frac{3}{20}\;(42-5\nu ^{2}-2\nu ^{4})\;s^{-2}  \label{12.5}
\end{equation}
as $s\rightarrow \infty $. Invoking the expansion (\ref{10.65}) for the
mean-square radius, according to which $r_{0}\simeq s^{1/5}$ as $%
s\rightarrow \infty $, we may write 
\begin{equation}
\Omega _{c}\simeq \frac{3}{2r_{0}^{2}}\qquad (r_{0}\rightarrow \infty )\;.
\label{12.6}
\end{equation}
As a function of $g$, this reduces to 
\begin{equation}
\Omega _{c}\simeq \frac{3(2\pi ^{3})^{1/5}}{2(\nu g)^{2/5}}=\frac{3.424}{%
(\nu g)^{2/5}}  \label{12.7}
\end{equation}
for $\nu g\rightarrow \infty $.

Invoking for the critical rotation frequency the Thomas-Fermi approximation,
combined with a hydrodynamic approximation, one finds \cite
{Lundh97,Edwards96b,Castin99,Feder99} the value 
\begin{equation}
\Omega _{c}\approx \frac{5}{2r_{c}^{2}}\;\ln \left( 0.7\;\frac{r_{c}}{\xi
_{0}}\right) \;,  \label{12.8}
\end{equation}
where 
\begin{equation}
r_{c}^{2}=2E_{TF}=\left( \frac{15}{4\pi }\;\nu g\right) ^{2/5}  \label{12.9}
\end{equation}
is the Thomas-Fermi radius and $\xi _{0}\approx 1/r_{c}$ is the coherence
length. From here 
\begin{equation}
\Omega _{c}\approx \frac{0.932}{(\nu g)^{2/5}}\;\ln (0.8\nu g)\;.
\label{12.10}
\end{equation}
The expressions (\ref{12.7}) and (\ref{12.10}) are close to each other in
the region $1\ll \nu g<10^{3}$. For instance, when $\nu g=100$, their
difference is about $10\%$.

\section{Effective Radial Equation}

To analyze the radial cross-section of a vortex, it is convenient to derive
an effective radial equation not containing the axial variable $z$. To this
end, let us substitute the function 
\begin{equation}  \label{12.11}
\psi(r,\varphi,z) =\chi(r,z) \; \frac{e^{im\varphi}}{\sqrt{2\pi}} \; ,
\end{equation}
where $m=0,\pm 1,\pm 2,\ldots$ and $\chi$ is real, into the eigenproblem (%
\ref{9.29}) with the Hamiltonian (\ref{9.28}). Then we have 
\begin{equation}  \label{12.12}
-\; \frac{1}{2}\left ( \frac{\partial^2}{\partial r^2} + \frac{1}{r}\; \frac{%
\partial}{\partial r} + \frac{\partial^2}{\partial z^2}\right ) \; \chi + 
\frac{1}{2}\left ( r^2 +\nu^2z^2 +\frac{m^2}{r^2}\right ) \chi + \frac{g}{%
2\pi} \; \chi^3 = E\; \chi \; .
\end{equation}
Let us present $\chi$ as a product 
\begin{equation}  \label{12.13}
\chi(r,z) = F(r,z)\; h(z) \; ,
\end{equation}
in which $F(r,z)$ is a slow function of $z$, such that 
\begin{equation}  \label{12.14}
\left | \frac{\partial F}{\partial z}\; h\right | \ll \left | F\; \frac{dh}{
dz}\right | \; ,
\end{equation}
and where $h$ is normalized according to the condition 
\[
\int_{-\infty}^{+\infty} \; h^2(z) \; dz = 1 \; . 
\]
Define the radial, $E_r$, and axial, $E_z$, energies by the relation 
\begin{equation}  \label{12.15}
E\equiv E_r + E_z \; , \qquad E_z \equiv \frac{1}{2}\;
\int_{-\infty}^{+\infty} \; h(z)\left ( -\; \frac{\partial^2}{\partial z^2}
+ \nu^2z^2 \right )\; h(z)\; dz \; .
\end{equation}
Then from Eqs. (\ref{12.12}) and (\ref{12.13}), taking account of the
inequality (\ref{12.14}), we find 
\begin{equation}  \label{12.16}
-\; \frac{1}{2}\left ( \frac{\partial^2}{\partial r^2} + \frac{1}{r}\; \frac{%
\partial}{\partial r}\right ) \; F h + \frac{1}{2}\left ( r^2 +\frac{m^2}{r^2%
}\right ) F h + \frac{g}{2\pi}\; F^3 h^3 = E_r\; F h \; .
\end{equation}

Introduce the function 
\begin{equation}  \label{12.17}
f(r) \equiv \int_{-\infty}^{+\infty} \; F(r,z) \; h^2(z)\; dz\; .
\end{equation}
Keeping in mind that $F(r,z)$ is a slow function of $z$, we may use the
approximation 
\[
\int_{-\infty}^{+\infty} \; F^3(r,z)\; h^4(z) \; dz \cong f^3(r) \;
\int_{-\infty}^{+\infty} \; h^4(z) \; dz \; . 
\]
Multiplying Eq. (\ref{12.16}) by $h$, integrating over $z$, and defining the
radial coupling 
\begin{equation}  \label{12.18}
\alpha \equiv \frac{g}{2\pi}\; \int_{-\infty}^{+\infty} \; h^4(z) \; dz
\end{equation}
and the nonlinear radial Hamiltonian 
\begin{equation}  \label{12.19}
\hat H_r(f) \equiv -\;\frac{1}{2}\left ( \frac{d^2}{dr^2} + \frac{1}{r}\; 
\frac{d}{dr}\right ) + \frac{1}{2}\left ( r^2 +\frac{m^2}{r^2}\right ) +
\alpha f^2 \; ,
\end{equation}
we come to the effective radial equation 
\begin{equation}  \label{12.20}
\hat H_r(f) \; f = E_r \; f \; .
\end{equation}

As an example of the function $h$, let us take the harmonic-oscillator wave
function 
\[
h^{(0)}(z)=\frac{(\nu /\pi )^{1/4}}{\sqrt{2^{k}\;k!}}\;\exp \left( -\;\frac{%
\nu }{2}\;z^{2}\right) \;H_{k}(\sqrt{\nu }\;z)\;. 
\]
Then the axial energy is 
\[
E_{z}^{(0)}=\left( k+\frac{1}{2}\right) \;\nu 
\]
and the radial coupling (\ref{12.18}) becomes 
\[
\alpha =\frac{\sqrt{\nu }}{2\pi }\;I_{k}\;g\;, 
\]
where 
\[
I_{k}\equiv \frac{2}{\pi (2^{k}\;k!)^{2}}\;\int_{0}^{\infty
}\;e^{-2z^{2}}\;H_{k}^{4}(z)\;dz\;. 
\]
The latter integral decreases with $k$, {\em e.g.} 
\[
I_{0}=\frac{1}{\sqrt{2\pi }}=0.398942\;,\qquad I_{1}=\frac{3}{4\sqrt{2\pi }}%
=0.299207\;,\qquad I_{2}=\frac{41}{64\sqrt{2\pi }}=0.255572\;. 
\]
Therefore, the radial coupling $\alpha $ diminishes for higher excited
states.

The radial equation (\ref{12.20}) describes the radial profile of a vortex.
The angle dependence of the latter, given by Eq. (\ref{12.11}), defines the
circulation velocity 
\[
{\bf v} =\frac{\hbar m}{m_0 r}\; {\bf e}_\varphi \; , 
\]
where ${\bf e}_\varphi$ is the unit vector corresponding to the polar angle $%
\varphi$. To be finite, this velocity requires that the winding number be
nonzero, $m\neq 0$.

Note that the Thomas-Fermi approximation is not directly applicable for
solving Eq. (\ref{12.20}) in the case of vortex states. This is because the
corresponding solution 
\[
f^2_{TF}(r) \simeq -\; \frac{m^2}{2\alpha r^2} \qquad (r\rightarrow 0) 
\]
diverges at $r=0$ for $m\neq 0$.

\section{Vortex Wave Function}

The structure of the radial Hamiltonian (\ref{12.19}) shows that there are
two qualitatively different regions where either the nonlinear term or the
harmonic one is more important as compared to each other. These regions are 
\[
r\ll \sqrt{\alpha} \qquad (nonlinear) \; , 
\]
\begin{equation}  \label{12.21}
r \gg \sqrt{\alpha} \qquad (harmonic) \; .
\end{equation}
When $r\rightarrow\infty$, the harmonic term always prevails. To find an
approximate analytic solution to the radial equation (\ref{12.20}), let us
consider two cases, when the coupling is not large and when $%
\alpha\rightarrow\infty$.

In the first case, when $\alpha $ is not large, say of order one or less,
the nonlinear region is small. The radial energy $E_{r}$ can be obtained by
the optimized perturbation theory of section 10.1. As the initial
approximation, we may take the harmonic Hamiltonian 
\begin{equation}
\hat{H}_{0}=-\;\frac{1}{2}\left( \frac{d^{2}}{dr^{2}}+\frac{1}{r}\;\frac{d}{%
dr}\right) +\frac{1}{2}\left( u^{2}r^{2}+\frac{m^{2}}{r^{2}}\right) \;,
\label{12.22}
\end{equation}
with the eigenvalue 
\begin{equation}
E_{nm}^{(0)}=pu\;,\qquad p\equiv 2n+|m|+1  \label{12.23}
\end{equation}
and the eigenfunction 
\begin{equation}
f_{nm}^{(0)}(r)=\left[ \frac{2n!\;u^{|m|+1}}{(n+|m|)!}\right]
^{1/2}\;r^{|m|}\;\exp \left( -\;\frac{1}{2}\;ur^{2}\right)
\;L_{n}^{|m|}(ur^{2})\;.  \label{12.24}
\end{equation}
For the first-order approximation 
\begin{equation}
E_{nm}^{(1)}(\alpha ,u)=(f_{nm}^{(0)},\hat{H}_{r}f_{nm}^{(0)})\;,
\label{12.25}
\end{equation}
we find 
\begin{equation}
E_{nm}^{(1)}(\alpha ,u)=\frac{p}{2}\left( u+\frac{1}{u}\right)
+u\;I_{nm}\;\alpha \;,  \label{12.26}
\end{equation}
where 
\[
I_{nm}\equiv \frac{1}{u}\;\int_{0}^{\infty }\;\left[ f_{nm}^{(0)}(r)\right]
^{4}\;rdr=2\left[ \frac{n!}{(n+|m|)!}\right] ^{2}\;\int_{0}^{\infty
}\;x^{2|m|}\;e^{-2x}\;\left[ L_{n}^{|m|}(x)\right] ^{4}\;dx\;. 
\]
From the fixed-point condition 
\begin{equation}
\frac{\partial }{\partial u}\;E_{nm}^{(1)}(\alpha ,u)=0  \label{12.27}
\end{equation}
we define the control function 
\begin{equation}
u(s)=\sqrt{\frac{p}{p+s}}\;,\qquad s\equiv 2I_{nm}\alpha \;.  \label{12.28}
\end{equation}
For the optimized approximant 
\begin{equation}
E_{r}(s)\equiv E_{nm}^{(1)}(\alpha (s),u(s))  \label{12.29}
\end{equation}
we obtain 
\begin{equation}
E_{r}(s)=\sqrt{p(p+s)}\;.  \label{12.30}
\end{equation}
In the weak-coupling limit, Eq. (\ref{12.30}) gives 
\begin{equation}
E_{r}(s)\simeq p+\frac{1}{2}\;s-\;\frac{1}{p}\;s^{2}\qquad (s\rightarrow 0)
\label{12.31}
\end{equation}
and in the strong-coupling limit, one has 
\begin{equation}
E_{r}(s)\simeq \sqrt{p}\;\left( s^{1/2}+\frac{p}{2}\;s^{-1/2}-\;\frac{p^{2}}{%
8}\;s^{-3/2}\right)  \label{12.32}
\end{equation}
as $s\rightarrow \infty $. It is interesting that if, being based on these
asymptotic expansions, we construct the self-similar crossover approximant 
\begin{equation}
E_{1}^{\ast }(s)=a_{0}\left( 1+As\right) ^{n_{1}}\;,  \label{12.33}
\end{equation}
as is explained in section 11.1, then 
\[
a_{0}=p\;,\qquad A=\frac{1}{p}\;,\qquad n_{1}=\frac{1}{2}\;, 
\]
and the crossover formula (\ref{12.33}) coincides with the energy (\ref
{12.30}).

To find the energy of the basic vortex with the quantum numbers $n=0$ and $%
m=1$, we note that $I_{00}=1$ and $I_{01}=0.5$. The related control
functions, given by Eq. (\ref{12.28}), are 
\[
u_{00}=\frac{1}{\sqrt{1+2\alpha }}\;,\qquad u_{01}=\sqrt{\frac{2}{2+\alpha }}%
\;. 
\]
The corresponding radial energies are 
\[
E_{00}=\sqrt{1+2\alpha }\;,\qquad E_{01}=\sqrt{4+2\alpha }\;. 
\]
Then the vortex energy in the strong-coupling limit is 
\[
\Omega _{01}\equiv E_{01}-E_{00}\simeq \frac{3}{2\sqrt{2}}\;\alpha
^{-1/2}\qquad (\alpha \rightarrow \infty )\;. 
\]
This is to be compared with the critical rotation frequency 
\[
\Omega _{c}\approx \frac{2}{r_{c}^{2}}\;\ln \left( 0.9\;\frac{r_{c}}{\xi _{0}%
}\right) 
\]
obtained in the Thomas-Fermi plus hydrodynamic approximations \cite{Lundh97}
for a two-dimensional vortex.

From the radial equation (\ref{12.20}), it follows that its solution behaves
at small distance as 
\begin{equation}
f(r)\simeq Cr^{|m|}\left( 1+c_{2}r^{2}+c_{4}r^{4}\right) \qquad
(r\rightarrow 0)  \label{12.34}
\end{equation}
and at large distance as 
\begin{equation}
f(r)\sim r^{|m|}\;\exp \left( -\;\frac{1}{2}\;r^{2}\right)
\;L_{n}^{|m|}(r^{2})\qquad (r\rightarrow \infty )\;.  \label{12.35}
\end{equation}
For the case when $n=0$, we have 
\begin{equation}
f(r)\sim r^{|m|}\;\exp \left( -\;\frac{1}{2}\;r^{2}\right) \qquad
(r\rightarrow \infty )\;.  \label{12.36}
\end{equation}
The crossover approximant, sewing the asymptotic expansion (\ref{12.34}) and
(\ref{12.36}), is 
\begin{equation}
f_{\ast }(r)=Cr^{|m|}\;\exp \left\{ -\;\frac{1}{2}\;r^{2}+ar^{2}\;\exp
(-br^{2})\right\} \;,  \label{12.37}
\end{equation}
where $a$ and $b$ are calculated from Eq. (\ref{12.20}), after the form (\ref
{12.37}) is expanded in powers of $r$ and substituted into this equation.
This makes it possible to express the coefficients $a$ and $b$ through $%
E_{r} $ and $C$. The latter are defined by the equations 
\begin{equation}
E_{r}^{\ast }=(f_{\ast },\hat{H}_{r}f_{\ast })\;,\qquad (f_{\ast },f_{\ast
})=1\;.  \label{12.38}
\end{equation}
The accuracy of approximate solutions to Eq. (\ref{12.20}) can be
characterized by the residual 
\begin{equation}
R(r)\equiv (\hat{H}_{r}-E_{r})f(r)  \label{12.39}
\end{equation}
and the dispersion 
\begin{equation}
\sigma ^{2}\equiv \int_{0}^{\infty }\;|R(r)|^{2}\;rdr\;.  \label{12.40}
\end{equation}
Considering the nonrotating case, with $m=0$, we get 
\[
a=\frac{1}{2}\left( 1+\alpha C^{2}-E_{r}\right) \;,\qquad b=\frac{%
E_{r}^{2}-\alpha ^{2}C^{4}-1}{16a}\;. 
\]
The dispersion (\ref{12.40}) for the crossover formula (\ref{12.37}) is
smaller than that for the variational function (\ref{12.24}) when $\alpha
<70 $.

In the case of the basic vortex with the winding number $m=1$, we have 
\[
a=1 -\; \frac{E_r}{2}\; , \qquad b=\frac{E_r^2-8\alpha C^2 -4}{48a} \; . 
\]
The crossover function (\ref{12.37}) is a better approximation than the
variational function (\ref{12.24}) for $\alpha<15$.

Thus, the crossover approximant (\ref{12.37}) is a reasonable approximation
for a vortex wave function if the coupling $\alpha <10$. For large $\alpha
\gg 10$, the error of the approximant (\ref{12.37}), characterized by the
residual (\ref{12.39}) and dispersion (\ref{12.40}), quickly grows. The
reason for this is clear: In constructing the crossover formula (\ref{12.37}%
), we have used the information on the behavior of the solution to Eq. (\ref
{12.20}) at small distance, when $r\rightarrow 0$, which is described by the
form (\ref{12.34}), and at large distance, when $r\rightarrow \infty $,
where the harmonic term prevails, so that the asymptotic solution is given
by Eq. (\ref{12.36}). At the same time, there is an additional
characteristic scale $r\sim \sqrt{\alpha }$ defining the distance at which
the dominance of the nonlinear term in the Hamiltonian (\ref{12.19}) changes
to that of the harmonic term. The peculiarity in the behavior of a solution,
due to this additional crossover, can be neglected only if $\alpha $ is not
large, so that the region $1\ll r\ll \sqrt{\alpha }$ squeezes to a small
interval or practically disappears. The value $\alpha \approx 10$ is exactly
that critical value.

In order to analyze the behavior of the solution to Eq. (\ref{12.20}) for
large coupling $\alpha \gg 10$, let us consider the case, opposite to the
previous one, when there exists a wide region $1\ll r\ll \sqrt{\alpha }$,
where the nonlinear term is dominant as compared to the harmonic term. In
this region, Eq. (\ref{12.20}) may be written as 
\begin{equation}
-\;\frac{1}{2}\;\left( \frac{d^{2}f}{dr^{2}}+\frac{1}{r}\;\frac{df}{dr}%
\right) +\frac{m^{2}}{2r^{2}}\;f+\alpha f^{3}=E_{r}\;f\;,  \label{12.41}
\end{equation}
the harmonic term being omitted. To simplify the analysis of Eq. (\ref{12.41}%
), we scale it so that it reduces to the equation 
\begin{equation}
\frac{d^{2}f}{dr^{2}}+\frac{1}{r}\;\frac{df}{dr}-\;\frac{m^{2}}{r^{2}}%
\;f+f-f^{3}=0\;.  \label{12.42}
\end{equation}
The return from Eq. (\ref{12.42}) back to Eq. (\ref{12.41}) can be done by
the scaling 
\[
r\rightarrow \sqrt{2E_{r}}\;r\;,\qquad f\rightarrow \sqrt{\frac{\alpha }{%
E_{r}}}\;f\;. 
\]

To construct a crossover solution in the region $0\leq r<\sqrt{\alpha}$,
when $\alpha\gg 10$, we need an asymptotic expansion for $f(r)$ at $%
r\rightarrow 0$ and another expansion for $r\gg 1$, but $r<\sqrt{\alpha}$.
For example, the ground-state solution, with $m=0$, behaves as 
\[
f(r) \simeq C\left ( 1 + c_2r^2 +c_4 r^4 \right ) \qquad (r\rightarrow 0) 
\]
at small distance, the coefficients being 
\[
c_2 =\frac{1}{4}\left ( C^2 - 1\right )\; , \qquad c_4 =\frac{1}{64}\left (
3C^4 - 4C^2 + 1\right ) \; . 
\]
And for $r\gg 1$, the solution tends to $f(r)\simeq 1$.

Let us concentrate on the vortex solution with $m=1$. Then at small
distance, we have 
\begin{equation}
f(r)\simeq Cr\left( 1+c_{2}r^{2}+c_{4}r^{4}+c_{6}r^{6}\right) \qquad
(r\rightarrow 0)\;,  \label{12.43}
\end{equation}
where 
\[
c_{2}=-\;\frac{1}{8}\;,\qquad c_{4}=\frac{8C^{2}+1}{192}\;,\qquad c_{6}=-\;%
\frac{80C^{2}+1}{9216}\;. 
\]
At large distance, we find 
\begin{equation}
f(r)\simeq 1-\;\frac{1}{2}\;r^{-2}-\;\frac{9}{8}\;r^{-4}-\;\frac{161}{16}%
\;r^{-6}\qquad (r\gg 1)\;.  \label{12.44}
\end{equation}
Employing the method of section 11.1, we construct \cite{Yukalov98b} the
self-similar crossover approximants 
\[
f_{1}^{\ast }(r)=\frac{1}{2}\;r\left( 1+\frac{1}{4}\;r^{2}\right) ^{-1/2}\;, 
\]
\[
f_{2}^{\ast }(r)=\frac{1}{\sqrt{2}}\;r\left( 1+\frac{1}{2}\;r^{2}+\frac{1}{4}%
\;r^{4}\right) ^{-1/4}\;, 
\]
\[
f_{3}^{\ast }(r)=\frac{1}{4^{1/3}}\;r\left( 1+\frac{3}{4}\;r^{2}+\frac{3}{16}%
\;r^{4}+\frac{1}{16}\;r^{6}\right) ^{-1/6}\;, 
\]
\begin{equation}
f_{4}^{\ast }(r)=\frac{1}{136^{1/8}}\;r\left( 1+r^{2}+\frac{9}{68}\;r^{4}+%
\frac{1}{34}\;r^{6}+\frac{1}{136}\;r^{8}\right) ^{-1/8}\;.  \label{12.45}
\end{equation}
The accuracy of the approximants (\ref{12.45}) can be checked by comparing
them with the exact numerical solution \cite{Ginzburg58,Quist99} of the
vortex equation (\ref{12.42}). This comparison is presented in Fig. \ref
{Fig32}, where it is seen that the approximant $f_{4}^{\ast }(r)$
practically coincides with the numerical solution.

\chapter{Elementary Collective Excitations}

Following the experimental realization of Bose-Einstein condensate in
trapped atomic gases, there has been an intensive study, both experimental
and theoretical, of elementary excitations in these systems \cite
{Parkins98,Dalfovo99}. For the theoretical description of elementary
excitations one usually employs two equivalent schemes. One of them is based
on the diagonalization of the Hamiltonian in the Bogolubov approximation 
\cite{Bogolubov67,Bogolubov70}. Another approach relies on the linearization
of evolution equations. The latter approach can be accomplished in several
ways which we illustrate below.

It is worth noting that collective excitations of trapped atoms have many
common features with collective excitations in nuclei, that are also finite
systems where nucleons are trapped by means of self-consistent potentials 
\cite{Birbrair76,Negele82}, and with collective excitations in metallic
clusters \cite{Nesterenko92}-\cite{Nesterenko99}.

\section{Linearization of Gross-Pitaevskii Equation}

The Gross-Pitaevskii equation (\ref{9.5}), when there are no external
time-dependent forces, reads 
\begin{equation}
i\hbar \;\frac{\partial \varphi }{\partial t}=\hat{H}(\varphi )\;\varphi \;,
\label{13.1}
\end{equation}
with the nonlinear Hamiltonian 
\begin{equation}
\hat{H}(\varphi )=-\;\frac{\hbar ^{2}{\bf \nabla }^{2}}{2m_{0}}+U({\bf r}%
)+NA|\varphi |^{2}\;.  \label{13.2}
\end{equation}
Recall that Eq. (\ref{13.1}) is an {\it exact} equation for the {\it %
coherent wave function} \cite{Yukalov98}. The similar equation (\ref{9.10})
is an {\it approximate} equation for the {\it mean-field order parameter} 
\cite{Dalfovo99}.

Collective excitations are described by small oscillations around a
stationary solution given by the stationary equation 
\begin{equation}  \label{13.3}
\hat H(\varphi_n)\; \varphi_n({\bf r}) = E_n\; \varphi_n({\bf r}) \; .
\end{equation}
One usually considers small fluctuations around the ground-state function $%
\varphi_0({\bf r})$, though, in general, one may consider oscillations
around a chosen stationary state $\varphi_n({\bf r})$.

Let us look for the solution of Eq. (\ref{13.1}) describing small deviations
from a given stationary solution $\varphi _{n}({\bf r})$. To this end, we
substitute the function 
\begin{equation}
\varphi ({\bf r},t)=\left[ \varphi _{n}({\bf r})+u({\bf r})\;e^{-i\omega
t}+v^{\ast }({\bf r})\;e^{i\omega t}\right] \;\exp \left( -\;\tfrac{i}{\hbar 
}\;E_{n}t\right)  \label{13.4}
\end{equation}
into Eq. (\ref{13.1}) and linearize the latter with respect to $u({\bf r})$
and $v({\bf r})$. Equating the like terms at the exponentials $\exp (\mp
i\omega t)$, we get 
\[
\left[ \hat{H}(\varphi _{n})-E_{n}+NA|\varphi _{n}|^{2}-\hbar \omega \right]
\;u+NA\varphi _{n}^{2}\;v=0\;, 
\]
\begin{equation}
\left[ \hat{H}(\varphi _{n})-E_{n}+NA|\varphi _{n}|^{2}+\hbar \omega \right]
\;v+NA(\varphi _{n}^{\ast })^{2}\;u=0\;.  \label{13.5}
\end{equation}
This system of coupled equations, sometimes called the Bogolubov-De Gennes
equations, defines the eigenvalues $\hbar \omega $ that are the energies of
the elementary excitations. For trapped atoms, these equations are usually
solved numerically \cite{Parkins98,Dalfovo99}.

As an illustration, let us consider the case of a uniform potential $U({\bf r%
})=U=const$. For the ground state wave function 
\begin{equation}  \label{13.6}
\varphi_0=\sqrt{n_0}\; , \qquad n_0 \equiv |\varphi_0(0)|^2\; ,
\end{equation}
the stationary equation (\ref{13.3}) gives the energy 
\begin{equation}  \label{13.7}
E_0 = U+\rho_0\; A \; , \qquad \rho_0 \equiv n_0\; N \; .
\end{equation}
The solutions $u$ and $v$ for the Bogolubov-De Gennes equations (\ref{13.5})
are plane waves of the form $\exp(i{\bf k}\cdot{\bf r})$. Then Eq. (\ref
{13.5}) yields 
\[
\hbar^2\omega^2 =\frac{\hbar^2 k^2}{2m_0} \; \left ( \frac{\hbar^2k^2}{2m_0}
+ 2A\rho_0\right ) \; . 
\]
This results in the Bogolubov spectrum 
\begin{equation}  \label{13.8}
\omega_B(k) =\sqrt{c^2k^2 +\hbar^2 \left (\frac{k^2}{2m_0}\right )^2} \; ,
\end{equation}
in which 
\begin{equation}  \label{13.9}
c\equiv \sqrt{\frac{\rho_0}{m_0}\; A}
\end{equation}
is the sound velocity. In the long-wave limit, the spectrum (\ref{13.8})
reduces to the acoustic form 
\begin{equation}  \label{13.10}
\omega_B(k) \simeq ck \qquad (k\rightarrow 0) \; .
\end{equation}

When the potential $U({\bf r})$ is not a constant, the procedure of
calculating the spectrum of elementary excitations is essentially more
complicated and is usually done numerically. But by their physical meaning,
the corresponding excitations are the analog of phonons.

\section{Linearization of Hydrodynamic Equations}

The Gross-Pitaevskii equation (\ref{13.1}) can be rewritten in the form of
hydrodynamic equations. To this purpose, one can present the coherent wave
function in terms of a modulus and a phase as 
\begin{equation}  \label{13.11}
\varphi({\bf r},t) =\sqrt{n({\bf r},t)}\; \exp\left\{ i\; S({\bf r}%
,t)\right\} \; ,
\end{equation}
where the phase $S$ is real and 
\begin{equation}  \label{13.12}
n({\bf r},t) =|\varphi({\bf r},t)|^2 \; .
\end{equation}
The phase defines the velocity 
\begin{equation}  \label{13.13}
{\bf v}({\bf r},t) \equiv \frac{\hbar}{m_0}\; {\bf \nabla} S({\bf r},t) \; ,
\end{equation}
so that the density current is 
\begin{equation}  \label{13.14}
n{\bf v} = -\;\frac{i\hbar}{2m_0} \; \left ( \varphi^* {\bf \nabla}\varphi -
\varphi{\bf \nabla}\varphi^* \right ) \; .
\end{equation}
Substituting the presentation (\ref{13.11}) into Eq. (\ref{13.1}) and
separating the real and imaginary parts, one obtains the continuity equation 
\begin{equation}  \label{13.15}
\frac{\partial n}{\partial t} + {\bf \nabla}\cdot(n{\bf v}) = 0
\end{equation}
and the velocity-field equation 
\begin{equation}  \label{13.16}
m_0\; \frac{\partial{\bf v}}{\partial t} +{\bf \nabla} \left ( U + NA n -\; 
\frac{\hbar^2}{2m_0\sqrt{n}}\; {\bf \nabla}^2\sqrt{n} + \frac{m{\bf v}^2}{2}%
\right ) = 0 \; .
\end{equation}
Equations (\ref{13.15}) and (\ref{13.16}) are completely equivalent to the
Gross-Pitaevskii equation (\ref{13.1}) and are termed the hydrodynamic
representation of the latter. If we are interested in the stationary
ground-state solutions, then Eqs. (\ref{13.15}) and (\ref{13.16}) reduce to 
\begin{equation}  \label{13.17}
\frac{\partial n_0}{\partial t} = 0 \; , \qquad {\bf v}_0 = 0 \; ,
\end{equation}
where 
\[
n_0 =n_0({\bf r})\equiv |\varphi_0({\bf r})|^2 
\]
and $\varphi_0$ satisfies the equation 
\[
-\;\frac{\hbar^2}{2m_0} \; {\bf \nabla}^2\varphi_0 + \left ( U +
NA|\varphi_0|^2\right ) \; \varphi_0 = E\;\varphi_0\; . 
\]

To analyze small deviations from the ground-state solutions $n_{0}$ and $%
{\bf v}_{0}$, one writes 
\begin{equation}
n=n_{0}+\delta n\;,\qquad {\bf v}={\bf v}_{0}+\delta {\bf v}\;.
\label{13.18}
\end{equation}
Linearizing Eq. (\ref{13.15}), one gets 
\begin{equation}
\frac{\partial }{\partial t}\;\delta n+{\bf \nabla }(n_{0}\delta {\bf v}%
)=0\;.  \label{13.19}
\end{equation}
Linearizing Eq. (\ref{13.16}), one assumes that $n_{0}$ changes in space
much slower than $\delta n$, 
\begin{equation}
|{\bf \nabla }n_{0}|\ll |{\bf \nabla }\delta n|\;.  \label{13.20}
\end{equation}
Then one finds 
\begin{equation}
m_{0}\;\frac{\partial }{\partial t}\;\delta {\bf v}+{\bf \nabla }\;\left(
NA-\;\frac{\hbar ^{2}}{4m_{0}n_{0}}\;{\bf \nabla }^{2}\right) \;\delta n=0\;.
\label{13.21}
\end{equation}
Combining Eqs. (\ref{13.19}) and (\ref{13.21}), one comes to the equation 
\begin{equation}
\frac{\partial ^{2}}{\partial t^{2}}\;\delta n={\bf \nabla }\left( c^{2}{\bf %
\nabla }\delta n\right) -\left( \frac{\hbar }{2m_{0}}\;{\bf \nabla }%
^{2}\right) ^{2}\;\delta n\;,  \label{13.22}
\end{equation}
in which 
\begin{equation}
c({\bf r})\equiv \sqrt{\frac{n_{0}({\bf r})}{m_{0}}\;NA}  \label{13.23}
\end{equation}
is a local sound velocity. For the harmonically oscillating $\delta n$, say,
as $\cos \omega t$, one has 
\begin{equation}
\omega ^{2}\delta n+{\bf \nabla }\left( c^{2}{\bf \nabla }\delta n\right)
=\left( \frac{\hbar }{2m_{0}}\;{\bf \nabla }^{2}\right) ^{2}\;\delta n\;.
\label{13.24}
\end{equation}
Note that for the uniform case, when $c=const$ and $\delta n\sim \cos {\bf k}%
\cdot {\bf r}$, we return back to the Bogolubov spectrum (\ref{13.8}).
However, for the nonuniform case, the local sound velocity (\ref{13.23})
depends on the space variable ${\bf r}$. For low-lying excitations, one may
neglect the right-hand side of Eq. (\ref{13.24}), which gives 
\[
\omega ^{2}\delta n+{\bf \nabla }\left( c^{2}{\bf \nabla }\delta n\right)
=0\;. 
\]
For a spherical trap, and using the Thomas-Fermi approximation for $n_{0}(%
{\bf r})$, the solutions to this equation can be presented as 
\[
\delta n({\bf r})=P_{2n}^{l}(r)\;r^{l}\;Y_{lm}(\vartheta ,\varphi )\;, 
\]
where $P_{2n}^{l}$ are even polynomials of degree $2n$; $Y_{lm}$ are
spherical functions; and $n$, $l$, $m$ are quantum numbers. The dispersion
law is given by the Stringari spectrum \cite{Stringari96} 
\begin{equation}
\omega _{nl}=\omega _{0}\left( 2n^{2}+2nl+3n+l\right) ^{1/2}\;.
\label{13.25}
\end{equation}
For cylindrically symmetric traps, analytical solutions for the spectrum of
elementary excitations are available only for some particular modes \cite
{Dalfovo99}. For instance, the scissors mode, generated by a sudden rotation
of the confining trap \cite{Guery-Odelin99}, has the frequency 
\[
\omega =\sqrt{2}\;\omega _{0}\;, 
\]
with the Thomas-Fermi approximation being again involved.

\section{Lagrangian Variational Technique}

For solving complicated nonlinear differential equations in partial
derivatives, a variational technique has been widely used \cite{Konotop94},
which provides approximate solutions to such equations. The basic idea of
this variational method is to take a trial function with a fixed shape but
some free parameters in order to reduce the infinite-dimensional dynamical
system of partial differential equations to a set of ordinary differential
equations for the variational parameters that characterize the solution.
This technique has also been applied \cite{Perez-Garcia97} to solve the
time-dependent Gross-Pitaevskii equation and to calculate
collective-excitation frequencies.

The first step of the method is to formulate a variational problem that
yields the considered differential equation. This can be formulated as the
problem of extremizing an action 
\begin{equation}  \label{13.26}
\delta \int L(t)\; dt = 0 \; ,
\end{equation}
in which the Lagrangian 
\begin{equation}  \label{13.27}
L(t) =\int {\cal L}({\bf r},t)\; d{\bf r}
\end{equation}
is expressed through the Lagrangian density. In our case, the latter is 
\begin{equation}  \label{13.28}
{\cal L}({\bf r},t) = -\; \frac{i}{2}\; \hbar \left ( \varphi^*\;\frac{%
\partial\varphi}{\partial t} \; - \; \frac{\partial\varphi^*}{\partial t}\;
\varphi\right ) + \frac{\hbar^2}{2m_0}\; |{\bf \nabla}\varphi|^2 +
U|\varphi|^2 + \frac{1}{2}\; NA|\varphi|^4 \; .
\end{equation}
As is easy to check, the extremum condition (\ref{13.26}), resulting in the
Lagrangian equation 
\begin{equation}  \label{13.29}
\frac{d}{dt}\; \frac{\delta L}{\delta{\dot\varphi}}\; - \; \frac{\delta L}{%
\delta\varphi} = 0 \; ,
\end{equation}
where $\dot{\varphi}\equiv d\varphi/dt$, for the Lagrangian density (\ref
{13.28}), yields the Gross-Pitaevskii equation (\ref{13.1}).

For the general anisotropic confining potential, it is convenient to pass to
dimensionless quantities as defined in section 9.3 and also to measure time
in units of $\omega_0^{-1}$. The return to the dimensional notation is done
by the substitution 
\[
{\cal L}({\bf x},t) \rightarrow \frac{{\cal L}({\bf r},t)}{\hbar\omega_0} \;
, \qquad t\rightarrow \omega_0 t \; . 
\]
The dimensionless Lagrangian density is 
\begin{equation}  \label{13.30}
{\cal L}({\bf x},t) = -\; \frac{i}{2} \left ( \psi^*\; \frac{\partial\psi}{%
\partial t} \; - \; \frac{\partial\psi^*}{\partial t}\; \psi\right ) + \frac{%
1}{2}\; \sum_{i=1}^3\; \left ( \left | \frac{\partial\psi}{\partial x_i}%
\right |^2 + \omega^2_i x_i^2 \right ) + \frac{1}{2}\; g |\psi|^4 \; .
\end{equation}
For a while, there were no approximations, so that all transformations are
exact.

Now, instead of varying the action with respect to $\psi$ and $\psi^*$, let
us present the sought solution in the Gaussian form 
\[
\psi({\bf x},t) = \prod_{i=1}^3 \; \psi_i(x_i,t) \; , 
\]
\begin{equation}  \label{13.31}
\psi_i(x_i,t) = C_i(t) \; \exp\left\{ -\; \frac{u_i(t)}{2}\; \left [ x_i -
a_i(t)\right ]^2 + i\alpha_i(t)x_i + i\beta_i(t) x_i^2 \right\} \; .
\end{equation}
From the normalization $(\psi_i,\psi_i)=1$, we have 
\[
|C_i(t)| =\left [ \frac{u_i(t)}{\pi} \right ]^{1/4} \; . 
\]
Then, with the ansatz (\ref{13.31}), we calculate the Lagrangian (\ref{13.27}%
), which can be done explicitly because of the Gaussian dependence of the
trial function (\ref{13.31}) on the space variable $x_i$. After this, we 
{\it assume} that the set of yet unknown trial functions $u_i(t)$, $a_i(t)$, 
$\alpha_i(t)$, and $\beta_i(t)$, where $i=1,2,3$, satisfies the Lagrange
equation 
\begin{equation}  \label{13.32}
\frac{d}{dt}\; \frac{\partial L}{\partial{\dot q}}\; - \; \frac{\partial L}{
\partial q} = 0 \; ,
\end{equation}
in which $q(t)$ is any function from the given trial set. This assumption
reduces the infinite-dimensional problem of solving Eq. (\ref{13.1}) to a
finite-dimensional problem of ordinary differential equations. As is clear,
the ansatz (\ref{13.31}), together with the assumption (\ref{13.32}),
defines approximate solutions to Eq. (\ref{13.1}), whose accuracy cannot be
controlled.

Note that the described reduction of the partial differential equations
could be done as well for a time-dependent trapping potential $U({\bf r},t)$
in the Hamiltonian (\ref{13.2}). Since, till now, we have nowhere used any
linearization procedure, the reduced set of equations can, in general,
describe nonlinear motion.

From the set of equations (\ref{13.32}), one can derive the equations 
\begin{equation}  \label{13.33}
\ddot{a}_i + \omega_i^2 a_i = 0
\end{equation}
for the center-of-condensate variables that harmonically oscillate with the
bare frequencies $\omega_i$. The oscillations of the atomic-cloud shape are
characterized by the frequencies $u_i$, for which we get the equations 
\begin{equation}  \label{13.34}
\ddot{u}_i -\; \frac{3\dot{u}_i^2}{2u_i} + 2u_i (u_i^2 -\omega_i^2) + su_i^2%
\sqrt{u_1u_2u_3} = 0 \; ,
\end{equation}
where the standard notation 
\[
s\equiv \frac{2g}{(2\pi)^{3/2}} 
\]
is employed. Introducing the effective cloud widths 
\begin{equation}  \label{13.35}
b_i(t) \equiv \frac{1}{\sqrt{u_i(t)}} \; ,
\end{equation}
one may transform Eq. (\ref{13.34}) to the form 
\begin{equation}  \label{13.36}
\ddot{b}_i +\omega_i^2\; b_i = \frac{s}{2b_i\; b_1b_2b_3} + \frac{1}{b_i^3}
\; .
\end{equation}
The stationary solutions to Eq. (\ref{13.36}) are given by the equation 
\begin{equation}  \label{13.37}
\omega_i^4(b_i^*)^4 = \frac{s(b_i^*)^2}{2b_1^*b_2^*b_3^*} + 1\; .
\end{equation}

In order to find the frequencies of collective excitations, one has to
consider small deviations of the variables $b_i$ near their stationary
points $b_i^*$. To this end, one substitutes 
\begin{equation}  \label{13.38}
b_i = b_i^* +\delta b_i
\end{equation}
into Eq. (\ref{13.36}) and linearizes the latter with respect to $\delta b_i$%
. This results in a system of three differential linear equations whose
harmonic solutions, say of the form $\cos\omega t$, give an algebraic system
of equations. Equating the determinant of the latter system to zero, one
comes to an equation for the spectrum of collective excitations. For
example, following this procedure in the case of an isotropic trapping
potential, when $\omega_i=1$, $b_i^*=b^*$, and neglecting the term $1$ in
the right-hand side of Eq. (\ref{13.37}), which assumes that $s\gg 1$, so
that 
\begin{equation}  \label{13.39}
b^* =\left (\frac{s}{2}\right )^{1/5} \; ,
\end{equation}
we obtain the system 
\[
(\omega^2 - 3)\delta b_1 -\delta b_2 -\delta b_3 = 0 \; , 
\]
\[
\delta b_1 -(\omega^2-3)\delta b_2 +\delta b_3 = 0 \; , 
\]
\[
\delta b_1 +\delta b_2 -(\omega^2 -3)\delta b_3 = 0 \; . 
\]
Equating the determinant to zero yields 
\[
(\omega^2-3)^2 - 3(\omega^2-3) - 2 = 0 \; . 
\]
This leads to the spectrum 
\begin{equation}  \label{13.40}
\varepsilon_{1,2}=\sqrt{2} \; , \quad \varepsilon_3=\sqrt{5}\; .
\end{equation}

The equations for the effective cloud widths $b_{i}$, similar to Eq. (\ref
{13.36}), can also be derived \cite{Dalfovo99} from the hydrodynamic
equations (\ref{13.15}) and (\ref{13.16}), by assuming the harmonic
dependence of the density $n({\bf r},t)$ on the space coordinates and a
special form of the velocity ${\bf v}({\bf r},t)$.

\chapter{Multicomponent Bose Mixtures}

Multicomponent systems of trapped Bose-Einstein condensates have been
realized for rubidium in a magnetic trap \cite{Myatt97,Hall98} and for
sodium in an optical trap \cite{Stamper-Kurn98,Miesner99,Stamper-Kurn99}.
There exists a number of works with theoretical treatment of such systems
(see \cite{Myatt97}, \cite{Ohberg99}-\cite{Bashkin97} and references
therein).

\section{Coherent States of Mixtures}

The Hamiltonian of a multicomponent Bose mixture has the form 
\[
H=\sum_i\int\; \psi_i^\dagger({\bf r},t)\left [ -\; \frac{\hbar^2{\bf \nabla}%
^2}{2m_i} + U_i({\bf r},t)\right ]\; \psi_i({\bf r},t)\; d{\bf r} \; + 
\]
\begin{equation}  \label{14.1}
+\; \frac{1}{2}\; \sum_{ij}\int\; \psi_i^\dagger({\bf r},t) \psi_j^\dagger(%
{\bf r}^{\prime},t)\Phi_{ij}({\bf r} -{\bf r}^{\prime}) \psi_j({\bf r}%
^{\prime},t)\psi_i({\bf r},t)\; d{\bf r}\; d{\bf r}^{\prime}\; ,
\end{equation}
in which the index $i=1,2\ldots$ enumerate the components; $m_i$ is a mass; $%
U_i({\bf r},t)$ is an external field including the trapping potential; the
interaction potential $\Phi_{ij}$ has the symmetry properties 
\[
\Phi_{ij}({\bf r}) =\Phi_{ij}(-{\bf r}) = \Phi_{ji}({\bf r})\; ; 
\]
and $\psi_i({\bf r},t)$ are field operators satisfying the Bose commutation
relations, 
\[
\left [ \psi_i({\bf r},t),\; \psi_j^\dagger({\bf r}^{\prime},t)\right ] =
\delta_{ij}\; \delta({\bf r}-{\bf r}^{\prime}) \; . 
\]
The evolution equations for the field operators are given by the Heisenberg
equations that can be written in one of two equivalent forms: in the
commutator form 
\[
i\hbar\; \frac{\partial}{\partial t} \; \psi_j({\bf r},t) = [\psi_j({\bf r}%
,t),\; H] 
\]
or in the variational representation 
\[
i\hbar\; \frac{\partial}{\partial t} \; \psi_j({\bf r},t) = \frac{\delta H}{%
\delta\psi_j^\dagger({\bf r},t)} \; . 
\]
Any of these representations yield the same equation 
\begin{equation}  \label{14.2}
i\hbar\; \frac{\partial}{\partial t} \; \psi_j({\bf r},t) = H_j(\psi)\;
\psi_j({\bf r},t) \; ,
\end{equation}
in which 
\begin{equation}  \label{14.3}
H_j(\psi) = -\; \frac{\hbar^2{\bf \nabla}^2}{2m_j} + U_j({\bf r},t)
+\sum_i\int\; \Phi_{ij}({\bf r}-{\bf r}^{\prime}) \psi_i^\dagger({\bf r}
^{\prime},t)\psi_i({\bf r}^{\prime},t)\; d{\bf r}^{\prime}\; .
\end{equation}

Coherent states can be defined as is described in Section 8, with a
straightforward generalization for a mixture. The coherent state $h_{i}$ for
the $i$-component is an eigenvalue of the destruction operator $\psi _{i}$,
so that 
\begin{equation}
\psi _{i}({\bf r},t)h_{i}=\eta _{i}({\bf r},t)h_{i}\;.  \label{14.4}
\end{equation}
The coherent state for a multicomponent system is given by the tensor
product 
\[
h=\tbigotimes\nolimits_{i}h_{i}\;. 
\]
The action of an operator $\psi _{i}$ on the state $h$ is defined as 
\[
\psi _{i}({\bf r},t)\tbigotimes\nolimits_{j}\hat{1}_{j}h=\eta _{i}({\bf r}%
,t)h\;. 
\]
Multiplying Eq. (\ref{14.2}) from the left by $h^{+}$ and from the right by $%
h$, we obtain the evolution equation 
\begin{equation}
i\hbar \;\frac{\partial }{\partial t}\;\eta _{j}({\bf r},t)=H_{j}(\eta )\eta
_{j}({\bf r},t)  \label{14.5}
\end{equation}
for the coherent field $\eta _{j}({\bf r},t)$, with the effective nonlinear
Hamiltonian 
\begin{equation}
H_{j}(\eta )=-\;\frac{\hbar ^{2}{\nabla }^{2}}{2m_{j}}+U_{j}({\bf r}
,t)+\sum_{i}\int \;\Phi _{ij}({\bf r}-{\bf r}^{\prime })\;|\eta _{i}({\bf r}
^{\prime },t)|^{2}\;d{\bf r}^{\prime }\;.  \label{14.6}
\end{equation}
By means of the notation 
\begin{equation}
\eta _{i}({\bf r},t)=\sqrt{N_{i}}\;\varphi _{i}({\bf r},t)\;,  \label{14.7}
\end{equation}
we may introduce the coherent field $\varphi _{i}$ normalized to unity, $%
(\varphi _{i},\varphi _{i})=1$, so that $N_{i}$ plays the role of the number
of particles in the $i$-component. Then, for the normalized coherent field,
the evolution equation is 
\begin{equation}
i\hbar \;\frac{\partial }{\partial t}\;\varphi _{j}({\bf r},t)=\hat{H}%
_{j}(\varphi )\;\varphi _{j}({\bf r},t)\;,  \label{14.8}
\end{equation}
with the nonlinear Hamiltonian 
\begin{equation}
\hat{H}_{j}(\varphi )=-\;\frac{\hbar ^{2}{\bf \nabla }^{2}}{2m_{j}}+U_{j}( 
{\bf r},t)+\sum_{i}\;N_{i}\int \Phi _{ij}({\bf r}-{\bf r}^{\prime
})\;|\varphi _{i}({\bf r}^{\prime },t)|^{2}\;d{\bf r}^{\prime }\;.
\label{14.9}
\end{equation}

In the case of a dilute system of atoms, whose scattering lengths $a_{ij}$
satisfy the inequality 
\begin{equation}  \label{14.10}
\frac{|a_{ij}|}{a} \ll 1 \; ,
\end{equation}
in which $a$ is the mean interatomic distance, one may model the interaction
potential by the Fermi pseudopotential 
\begin{equation}  \label{14.11}
\Phi_{ij}({\bf r}) = A_{ij}\delta({\bf r}) \; ,
\end{equation}
with the interaction amplitude 
\begin{equation}  \label{14.12}
A_{ij}=2\pi\hbar^2\; \frac{a_{ij}}{m_{ij}} \; , \qquad m_{ij} \equiv \frac{%
m_im_j}{m_i+m_j} \; .
\end{equation}
By assumption, $a_{ij}=a_{ji}$, hence $A_{ij}=A_{ji}$. Then the nonlinear
Hamiltonian (\ref{14.9}) becomes 
\begin{equation}  \label{14.13}
\hat H_j(\varphi) = -\; \frac{\hbar^2{\bf \nabla}^2}{2m_j} + U_j({\bf r},t)
+ \sum_i\; N_i A_{ij} |\varphi_i|^2 \; .
\end{equation}

The stationary solutions of Eq. (\ref{14.8}), when the external potential $%
U_{j}({\bf r})$ does not depend on time, can be presented as 
\begin{equation}
\varphi _{j}({\bf r},t)=\varphi _{j}({\bf r})\;\exp \left( -\;\tfrac{i}{%
\hbar }\;E_{j}\;t\right) \;,  \label{14.14}
\end{equation}
with the coherent modes $\varphi _{j}({\bf r})$ defined by the eigenproblem 
\begin{equation}
\hat{H}_{j}(\varphi )\varphi _{j}({\bf r})=E_{j}\;\varphi _{j}({\bf r})\;.
\label{14.15}
\end{equation}

Note that, in the same way as for single-component systems, the evolution
equation (\ref{14.8}) is an exact equation for the coherent field $\varphi_j(%
{\bf r},t)$. So is the eigenproblem (\ref{14.15}) for the coherent mode $%
\varphi_j({\bf r})$ in the stationary case.

\section{Branching of Excitation Spectrum}

The spectrum of collective excitations for a multicomponent Bose mixture can
be defined by means of the same methods as for a one-component system, as is
described in Chapter 13. For instance, one may linearize the nonlinear
equation (\ref{14.8}) after substituting there 
\begin{equation}
\varphi _{j}({\bf r},t)=\left[ \varphi _{j}({\bf r})+u_{j}({\bf r}%
)\;e^{-i\omega t}+v_{j}^{\ast }({\bf r})\;e^{i\omega t}\right] \;\exp \left(
-\;\tfrac{i}{\hbar }\;E_{j}\;t\right) \;.  \label{14.16}
\end{equation}
The linearization with respect to the functions $u_{j}$ and $v_{j}$ yields
the system of equations 
\[
\left( \hbar \omega -\hat{H}_{j}(\varphi )+E_{j}\right)
u_{j}-\sum_{i}\;N_{i}A_{ij}\left( \varphi _{i}^{\ast }u_{i}+\varphi
_{i}v_{i}\right) \varphi _{j}=0\;, 
\]
\begin{equation}
\left( \hbar \omega +\hat{H}_{j}(\varphi )-E_{j}\right)
v_{j}+\sum_{i}\;N_{i}A_{ij}\left( \varphi _{i}^{\ast }u_{i}+\varphi
_{i}v_{i}\right) \varphi _{j}^{\ast }=0\;,  \label{14.17}
\end{equation}
where $\hat{H}_{j}(\varphi )$ is defined on stationary solutions $\varphi
_{i}({\bf r})$.

For a nonuniform system, Eqs. (\ref{14.17}) are to be solved numerically. In
order to demonstrate the main specific features distinguishing the case of a
mixture from that of a single-component system, let us consider the uniform
case, when the external potential $U_j=const$. This case can be treated as a
uniform approximation for a large trap. In this approximation, the
ground-state stationary solutions of Eq. (\ref{14.15}) can be written as 
\begin{equation}  \label{14.18}
\varphi_j({\bf r}) = \sqrt{n_j} \; , \qquad n_j \equiv |\varphi_j(0)|^2 \; ,
\end{equation}
and the corresponding energies as 
\begin{equation}  \label{14.19}
E_j = U_j +\sum_i \; n_i N_i A_{ij} \; .
\end{equation}
Then $u_j$ and $v_j$ are presented by plane waves of the form $e^{i{\bf k}%
\cdot{\bf r}}$. Therefore, we have 
\[
\left ( \hat H_j(\varphi) - E_j\right ) \; u_j = K_j u_j \; , \qquad K_j
\equiv \frac{\hbar^2 k^2}{2m_j} \; , 
\]
and a similar relation for $v_j$. Using this, we may present Eqs. (\ref
{14.17}) in the form 
\[
(\hbar\omega - K_j)\; u_j -\sum_i \; \sqrt{n_in_j} \; N_i A_{ij} (u_i+v_i) =
0 \; , 
\]
\begin{equation}  \label{14.20}
(\hbar\omega + K_j)\; v_j +\sum_i \; \sqrt{n_in_j} \; N_i A_{ij} (u_i+v_i) =
0 \; .
\end{equation}

To simplify the consideration, let us analyze the case of a two-component
mixture. Then Eqs. (\ref{14.20}) form a system of four linear algebraic
equations. The condition of having nontrivial solutions is that the
determinant be zero. This condition can be presented as 
\begin{equation}  \label{14.21}
| \hat D -\hbar\omega\hat 1| = 0 \; ,
\end{equation}
where $\hat D$ is the dynamical matrix, whose elements are 
\[
D_{11} = K_1 + n_1N_1A_{11}\; , \quad D_{12} =\sqrt{n_1n_2}\; N_2 A_{12} \;
, \quad D_{13} = n_1N_1A_{11} \; , \quad D_{14}= D_{12} \; , 
\]
\[
D_{21}=\sqrt{n_1n_2}\; N_1A_{12} \; , \quad D_{22}= K_2+ n_2N_2A_{22} \; ,
\quad D_{23}= D_{21} \; , \quad D_{24} = n_2N_2 A_{22} \; , 
\]
\[
D_{31}=-D_{13} \; , \qquad D_{32}=-D_{12} \; , \qquad D_{33}= - D_{11} \; ,
\qquad D_{34} = - D_{12} \; , 
\]
\[
D_{41} = - D_{21} \; , \qquad D_{42} = - D_{24} \; , \qquad D_{43} = -
D_{21}\; , \qquad D_{44}= - D_{22} \; . 
\]
To write down Eq. (\ref{14.21}) explicitly, it is convenient to introduce
the following notation: 
\begin{equation}  \label{14.22}
\omega_i^2(k) \equiv c_i^2 k^2 +\hbar^2 \left ( \frac{k^2}{2m_i}\right )^2
\; ,
\end{equation}
which would define a single-component spectrum with the sound velocity 
\begin{equation}  \label{14.23}
c_i \equiv \sqrt{\frac{\rho_i}{m_i}\; A_{ii}}\; , \qquad \rho_i\equiv n_i
N_i \; ;
\end{equation}
and let us also denote 
\begin{equation}  \label{14.24}
\omega_{12}^2(k) \equiv c_{12}^2 k^2
\end{equation}
with 
\begin{equation}  \label{14.25}
c_{12}^2 \equiv \sqrt{\frac{\rho_1\rho_2}{m_1m_2}} \; A_{12} \; .
\end{equation}
Then from Eq. (\ref{14.21}), we find 
\begin{equation}  \label{14.26}
(\omega^2-\omega_1^2)(\omega^2 -\omega_2^2) =\omega_{12}^4 \; .
\end{equation}
The latter equation yields for the spectrum 
\begin{equation}  \label{14.27}
\omega_\pm^2(k) =\frac{1}{2} \left [ \omega_1^2 +\omega_2^2 \pm \sqrt{%
(\omega_1^2-\omega_2^2)^2 + 4\omega_{12}^4} \right ] \; .
\end{equation}
This means that, instead of one branch (\ref{13.8}) for the spectrum of
collective excitations of a single-component system, we now have a
two-branch spectrum, given by Eq. (\ref{14.27}) for a two-component mixture.
In general, for an $n$ -component mixture, we should have $n$ branches of
the excitation spectrum.

In the long-wave limit, Eq. (\ref{14.27}) gives two acoustic branches 
\begin{equation}  \label{14.28}
\omega_\pm(k) \simeq c_\pm k \qquad (k\rightarrow 0) \; ,
\end{equation}
with the corresponding sound velocities $c_\pm$ defined by the expression 
\begin{equation}  \label{14.29}
c_\pm^2 \equiv \frac{1}{2}\; \left [ c_1^2 + c_2^2 \pm \sqrt{(c_1^2
-c_2^2)^2 + 4c_{12}^4}\right ] \; .
\end{equation}
In the short-wave limit, one has two single-particle branches 
\begin{equation}  \label{14.30}
\omega_\pm(k) \simeq \frac{k^2}{2m_j} \qquad (k\rightarrow\infty) \; .
\end{equation}

The two branches of the collective spectrum of a two-component mixture can
be interpreted in the following way. One branch, $\omega_+(k)$, describes
the oscillation of the total density of the mixture, when both components
move together. And another branch, $\omega_-(k)$, characterizes the relative
fluctuations of the components with respect to each other. It is worth
noting that neither of the sound velocities $c_\pm$ coincides with the
hydrodynamic sound velocity $c$ defined by the derivative $c^2\equiv\partial
P/\partial\rho_m$, in which $P$ is pressure and $\rho_m$ is mass density.

\section{Dynamic and Thermodynamic Stability}

The mixture of trapped atoms is not always stable and it may stratify into
subsystems of pure one-component phases \cite{Myatt97}, \cite{Ohberg99} 
\nocite{Sinatra00}-\cite{Bashkin97} in the same way as it happens for
uniform Bose mixtures \cite{Nepomnyashchy74}\nocite{Colson78}-\cite
{Yukalov80b}. The criteria of stability can be separated into dynamic and
thermodynamic ones.

The system is dynamically stable if its spectrum of collective excitations
is positive everywhere, except for a countable number of points where it is
zero. For a uniform two-component Bose mixture, the {\it condition of
dynamic stability} is 
\begin{equation}
\omega _{\pm }(k)>0\;.  \label{14.31}
\end{equation}
With the spectrum (\ref{14.27}), this gives 
\begin{equation}
\omega _{1}^{2}(k)\omega _{2}^{2}(k)>\omega _{12}^{4}(k)\;.  \label{14.32}
\end{equation}
Since $\omega _{-}(k)\leq \omega _{+}(k)$, we actually need to analyze
stability only for the branch $\omega _{-}(k)$. This branch describes
relative oscillations of components with respect to each other. When $\omega
_{-}(k)$ becomes negative, the mixture is unstable with respect to the
stratification of the components.

In the long-wave limit, $k\rightarrow 0$, the inequality (\ref{14.32})
reduces to 
\begin{equation}
c_{1}^{2}c_{2}^{2}>c_{12}^{4}\;.  \label{14.33}
\end{equation}
Using Eqs. (\ref{14.23}) and (\ref{14.25}), we have 
\begin{equation}
A_{11}A_{22}>A_{12}^{2}\;.  \label{14.34}
\end{equation}
Taking account of the relation (\ref{14.12}), we find the condition of
dynamic stability for the scattering lengths, 
\begin{equation}
a_{11}a_{22}>\;\frac{(m_{1}+m_{2})^{2}}{4m_{1}m_{2}}\;a_{12}^{2}\;.
\label{14.35}
\end{equation}
The equality 
\begin{equation}
\frac{a_{11}a_{22}}{a_{12}^{2}}=\frac{(m_{1}+m_{2})^{2}}{4m_{1}m_{2}}
\label{14.36}
\end{equation}
defines the {\it stratification boundary}. It is interesting that the
condition of dynamic stability is the same for uniform as well as for
trapped mixtures \cite{Sinatra00,Bashkin97}.

The system is thermodynamically stable if its free energy is minimal. At
zero temperature, free energy coincides with internal energy. Here, we
consider the case when the whole system is in a coherent state, so that we
need to compare the average energies of the given coherent state for the
mixture and for the stratified system. The Hamiltonian of the latter writes 
\[
H_{str} = \sum_i \; H_i \; \qquad V=\sum_i \; V_i \; , 
\]
\[
H_i=\int_{V_i}\; \psi_i^\dagger({\bf r},t) \left [ -\; \frac{\hbar^2{\bf %
\nabla}^2}{2m_i} + U_i({\bf r},t)\right ] \; \psi_i({\bf r},t) \; d{\bf r}
\; + 
\]
\begin{equation}  \label{14.37}
+ \; \frac{1}{2} \; \int_{V_i} \; \psi_i^\dagger({\bf r},t) \;
\psi_i^\dagger({\bf r}^{\prime},t)\; \Phi_{ii}({\bf r}-{\bf r}^{\prime})\;
\psi_i({\bf r}^{\prime},t)\; \psi_i({\bf r},t)\; d{\bf r}\; d{\bf r}%
^{\prime}\; .
\end{equation}
This describes independent pure components separated in their own regions of
space.

For simplicity, we again employ the uniform approximation when the coherent
field for each component can be written as 
\begin{eqnarray}  \label{14.38}
\varphi_i({\bf r}) =\left\{ 
\begin{array}{cc}
V^{-1/2} & (mixed) \\ 
V_i^{-1/2} & (stratified)
\end{array}
\right.
\end{eqnarray}
according to the case of a mixed or stratified system, respectively. For a
two-component system, one has 
\begin{equation}  \label{14.39}
N=N_1 + N_2 \; , \qquad V = V_1 + V_2 \; .
\end{equation}
We assume that all particles $N_1$ and $N_2$ are in their coherent states
characterized by the ground-state coherent fields (\ref{14.38}). Then the
quantum coherent average of the Hamiltonian (\ref{14.1}) gives the energy of
the mixture 
\begin{equation}  \label{14.40}
<H>_N\; = N_1U_1 + N_2 U_2 + \frac{N_1^2A_{11}+N_2^2A_{22}+2N_1N_2A_{12}}{%
2(V_1+V_2)}\; .
\end{equation}
And the quantum coherent average of the Hamiltonian (\ref{14.37}) yields the
energy 
\begin{equation}  \label{14.41}
<H_{str}>_N\; = N_1U_1 + N_2U_2 + \frac{N_1^2A_{11}}{2V_1} + \frac{%
N_2^2A_{22}}{2V_2}
\end{equation}
of the stratified system.

The {\it condition of thermodynamic stability} is 
\begin{equation}  \label{14.42}
\left ( <H>_N\; - \; <H_{str}>_N\right ) < 0 \; .
\end{equation}
From here, we find 
\begin{equation}  \label{14.43}
N_1^2V_2^2 A_{11} + N_2^2 V_1^2 A_{22} - 2N_1N_2 V_1 V_2 A_{12}>0 \; .
\end{equation}
In an important case, when $A_{11}$ and $A_{22}$ are positive, condition (%
\ref{14.43}) reduces to 
\begin{equation}  \label{14.44}
\left ( N_1 V_2 \sqrt{A_{11}}\; - N_2V_1\sqrt{A_{22}}\right )^2 + 2N_1N_2
V_1V_2\left ( \sqrt{A_{11}A_{22}}\; - A_{12}\right )>0\; .
\end{equation}

To be stable in general, any system has to be both dynamically as well as
thermodynamically stable, so that both conditions (\ref{14.34}) and (\ref
{14.43}) be valid. Let us analyze the relation between these conditions for
different particular cases:

\vskip 2mm

(i) $A_{11},\; A_{22},\; A_{12}\; >0$. The condition of dynamic stability (%
\ref{14.34}) is stronger than that of thermodynamic stability (\ref{14.43})
in the sense that from the former the latter follows. Dynamic stability is
sufficient for thermodynamic stability, although not necessary. A system can
be thermodynamically stable, but not dynamically stable. Conversely,
thermodynamic instability yields dynamic instability.

\vskip 2mm

(ii) $A_{11},\; A_{22} >0, \; A_{12}<0$. The mixture is always
thermodynamically stable, but not necessarily dynamically stable.

\vskip 2mm

(iii) $A_{11}$ and $A_{22}$ are of different signs, while $A_{12}$ is of
arbitrary sign. The system is never dynamically stable, although it can be
thermodynamically stable.

\vskip 2mm

(iv) $A_{11},\; A_{22}<0,\; A_{12}>0$. The system is never thermodynamically
stable, but can be dynamically stable.

\vskip 2mm

(v) $A_{11},\; A_{22},\; A_{12}<0$. Then inequality (\ref{14.43}) can be
transformed into 
\[
\left ( N_1V_2\sqrt{|A_{11}|} \; - N_2V_1 \sqrt{|A_{22}|}\right )^2 +
2N_1N_2V_1V_2\left ( \sqrt{|A_{11}A_{22}|}\; - |A_{12}|\right ) < 0 \; . 
\]
The latter inequality cannot be compatible with condition (\ref{14.34}), so
that dynamic stability leads to thermodynamic instability and thermodynamic
stability provokes dynamic instability.

Summarizing this analysis, we conclude that the two-component Bose mixture
is both dynamically and thermodynamically stable provided that 
\begin{equation}  \label{14.45}
A_{11}>0 \; , \qquad A_{22}>0 \; ,
\end{equation}
and condition (\ref{14.34}) is valid.

Here, we have analyzed the conditions of stability for a coherent mixture.
The same conditions can be obtained for a liquid mixture with broken gauge
symmetry, when only a fraction of atoms are in the Bose condensate \cite
{Nepomnyashchy74,Colson78}, and also for a normal mixture of Bose liquids
without broken gauge symmetry \cite{Yukalov80b}.

The first experiments \cite{Myatt97,Hall98} involving multiple-species
condensates were performed with $^{87}$Rb atoms evaporatively cooled in the $%
|F=2,\;m_{F}=2>$ and $|1,-1>$ spin states. The scattering lengths, known at
the $1\%$ level, are in the proportion $a_{11}:a_{12}:a_{22}=1.03:1:0.97$,
with the average of three being $55\;\AA $ \cite{Matthews98}. For equal
masses $m_{1}=m_{2}$, the stability condition (\ref{14.35}) reads $%
a_{11}a_{22}>a_{12}^{2}$. In the case of $^{87}$Rb, one has $%
a_{11}a_{22}/a_{12}^{2}=0.9991<1$. Hence, these two condensates cannot
compose a uniform mixture. The Bose condensates of sodium \cite
{Stamper-Kurn98,Miesner99,Stamper-Kurn99} in two different internal states $%
|F=1,\;m_{F}=0>$ and $|1,1>$ have the scattering lengths $%
a_{11}=a_{12}=2.75\times 10^{-9}$ cm and $a_{22}=2.65\times 10^{-9}$ cm.
From here, one has $a_{11}a_{22}/a_{12}^{2}=0.964<1$, thence these
condensates cannot be mixed.

\section{Stratification of Moving Components}

It is possible to experimentally create a binary mixture of trapped
Bose-Einstein condensates with a relative motion of components \cite{Hall98}%
. The presence of such a motion should impose additional constraints on the
stability of a mixture.

Consider a coherent mixture of components, each of which moves with a
constant linear velocity ${\bf V}_{j}$. For the sake of simplicity, let us
again invoke the uniform approximation, when the ground-state coherent field
of an immovable component is $\sqrt{n_{j}}$. Then for a moving component,
because of the Galilean transformation, one has 
\begin{equation}
\varphi _{j}({\bf r})=\sqrt{n_{j}}\;e^{i{\bf k}_{j}\cdot {\bf r}}\;,\qquad 
{\bf k}_{j}\equiv \frac{m_{j}}{\hbar }\;{\bf V}_{j}\;.  \label{14.46}
\end{equation}
The eigenvalue of Eq. (\ref{14.15}) becomes 
\begin{equation}
E_{j}=U_{j}+\sum_{i}\;\rho _{i}A_{ij}\;+\frac{1}{2}\;m_{j}{\bf V}_{j}^{2}\;,
\label{14.47}
\end{equation}
where $\rho _{i}\equiv n_{i}N_{i}$. As compared to the ground-state energy (%
\ref{14.19}), we have here an additional term corresponding to the kinetic
energy of motion. Similarly, for the coherent averages of the Hamiltonians (%
\ref{14.1}) and (\ref{14.37}), we would obtain the expressions (\ref{14.40})
and (\ref{14.41}) with additive terms $\frac{1}{2}\sum_{j}m_{j}V_{j}^{2}$.
These terms cancel each other when analyzing the condition (\ref{14.42}).
Hence, thermodynamic stability is not affected by such a motion of
components.

To check dynamic stability, we need to find the spectrum of collective
excitations given by the frequencies $\omega$ satisfying Eqs. (\ref{14.17}).
The solutions $u_j$ and $v_j$ to the latter equations, in the case of the
coherent fields (\ref{14.46}), depend on the space variable as 
\[
u_j \sim e^{i({\bf k}+{\bf k}_j)\cdot{\bf r}} \; , \qquad v_j\sim e^{i({\bf k%
} -{\bf k}_j)\cdot{\bf r}} \; , 
\]
which is required for $\varphi_j({\bf r},t)$ to have the same Galilean
transformation as $\varphi_j({\bf r})$. Then it follows that 
\[
\left (\hat H_j(\varphi) - E_j\right ) u_j = \left ( K_j + \hbar{\bf k}\cdot 
{\bf V}_j \right ) u_j \; , 
\]
\[
\left (\hat H_j(\varphi) - E_j\right ) v_j = \left ( K_j - \hbar{\bf k}\cdot 
{\bf V}_j \right ) v_j \; . 
\]
Therefore, Eqs. (\ref{14.17}) retain practically the same form, but with the
frequency $\omega$ shifted as 
\[
\omega\rightarrow \omega -\varepsilon_j \; , \qquad \varepsilon_j \equiv 
{\bf k}\cdot{\bf V}_j \; , 
\]
which corresponds to the Doppler shift.

Consider now a binary mixture. Without loss of generality, we may connect
the system of coordinates with one of the components, say with the first
one, so that ${\bf V}_{1}=0$. Then ${\bf V}_{2}\equiv {\bf v}$ is the
velocity of the second component with respect to the first. The dynamical
matrix $\hat{D}=[D_{ij}]$ in the spectral equation (\ref{14.21}) has the
same elements except 
\[
D_{22}=-D_{11}+\hbar {\bf k}\cdot {\bf v}\;,\qquad D_{44}=-D_{22}+\hbar {\bf %
k}\cdot {\bf v}\;. 
\]
In general, instead of Eq. (\ref{14.26}), we now have 
\begin{equation}
\left[ (\omega -\varepsilon _{1})^{2}-\omega _{1}^{2}\right] \left[ (\omega
-\varepsilon _{2})^{2}-\omega _{2}^{2}\right] =\omega _{12}^{4}\;.
\label{14.48}
\end{equation}
In the chosen system of coordinates, connected with the first component, Eq.
(\ref{14.48}) simplifies to 
\begin{equation}
(\omega -\omega _{1}^{2})\left[ (\omega -\varepsilon )^{2}-\omega _{2}^{2}%
\right] =\omega _{12}^{4}\;,  \label{14.49}
\end{equation}
where 
\begin{equation}
\varepsilon \equiv {\bf k}\cdot {\bf v}=kv\cos \vartheta \;.  \label{14.50}
\end{equation}
The spectral equation (\ref{14.49}) is the fourth-order algebraic equation 
\begin{equation}
\omega ^{4}-2\varepsilon \omega ^{3}-(\omega _{1}^{2}+\omega
_{2}^{2}-\varepsilon ^{2})\omega ^{2}+2\varepsilon \omega _{1}^{2}\omega
+\omega _{1}^{2}\omega _{2}^{2}-\omega _{12}^{4}-\varepsilon ^{2}\omega
_{1}^{2}=0\;.  \label{14.51}
\end{equation}
As we know, this equation can define not more than two stable, that is
positive, solutions for the spectrum $\omega _{\pm }(k)$. By Descartes'
theorem, the necessary condition for Eq. (\ref{14.51}) to possess two
positive solutions for arbitrary $\varepsilon \in \lbrack -kv,+kv]$ is 
\begin{equation}
\omega _{1}^{2}\omega _{2}^{2}>\omega _{12}^{4}+\varepsilon ^{2}\omega
_{1}^{2}\;.  \label{14.52}
\end{equation}
Since this inequality is to be true for all $\vartheta \in \lbrack 0,\pi ]$,
we can put here the maximal $\varepsilon =kv$. In this way, we obtain 
\begin{equation}
\omega _{1}^{2}\omega _{2}^{2}>\omega _{12}^{4}+(\omega _{1}kv)^{2}\;.
\label{14.53}
\end{equation}
In the long-wave limit, $k\rightarrow 0$, this gives 
\begin{equation}
c_{1}^{2}c_{2}^{2}>c_{12}^{4}+c_{1}^{2}v^{2}\;.  \label{14.54}
\end{equation}
Even if the immovable components do mix, they stratify as soon as the
relative velocity reaches the critical value 
\begin{equation}
v_{c}=\frac{\sqrt{c_{1}^{2}c_{2}^{2}-c_{12}^{4}}}{c_{1}}\;.  \label{14.55}
\end{equation}
The stratification appears first inside the cone of the angle 
\begin{equation}
\vartheta _{c}={\rm arccos}\;\frac{v_{c}}{v}\;.  \label{14.56}
\end{equation}
With the notations (\ref{14.23}) and (\ref{14.25}), the condition (\ref
{14.54}) becomes 
\begin{equation}
A_{11}A_{22}>A_{12}^{2}+\frac{m_{2}v^{2}}{\rho _{2}}\;A_{11}\;,
\label{14.57}
\end{equation}
while for the critical velocity (\ref{14.55}) one gets 
\begin{equation}
v_{c}^{2}=\frac{\rho _{2}}{m_{2}A_{11}}\;\left(
A_{11}A_{22}-A_{12}^{2}\right) \;.  \label{14.58}
\end{equation}
Invoking the relation (\ref{14.12}), we find the stability condition 
\begin{equation}
\frac{a_{11}a_{22}}{a_{12}^{2}}\;>\;\frac{(m_{1}+m_{2})^{2}}{4m_{1}m_{2}}+%
\frac{m_{2}^{2}v^{2}a_{11}}{4\pi \hbar ^{2}\rho _{2}a_{12}^{2}}
\label{14.59}
\end{equation}
expressed through the atomic scattering lengths. And the squared critical
velocity (\ref{14.58}) takes the form 
\begin{equation}
v_{c}^{2}=\frac{4\pi \hbar ^{2}\rho _{2}a_{12}^{2}}{m_{2}^{2}a_{11}}\;\left[ 
\frac{a_{11}a_{22}}{a_{12}^{2}}\;-\;\frac{(m_{1}+m_{2})^{2}}{4m_{1}m_{2}}%
\right] \;.  \label{14.60}
\end{equation}
The value of the critical velocity (\ref{14.60}) depends on the parameters
of the species involved. For instance, in the case of alkali atoms, $%
m_{2}/\hbar \sim 10^{5}$ s$/$cm$^{2}$, $a_{11}\sim a_{12}\approx 5.5\times
10^{-7}$ cm, from where 
\[
\frac{4\pi \hbar ^{2}a_{12}^{2}}{m_{2}^{2}}\sim 10^{-15}\;\frac{{\rm cm}^{5}%
}{{\rm s}^{2}}\;. 
\]
Typical atomic-trap condensate densities are $\rho _{2}\sim 10^{12}-10^{15}$
cm$^{-3}$. But what is needed first of all in order to have a finite
critical velocity is that immovable components could be mixed, which
requires the positiveness of the expression in the square bracket of Eq. (%
\ref{14.60}).

\section{Mixing by Feshbach Resonance}

By definition, Feshbach resonances involve intermediate states that are
quasi-bound \cite{Feshbach92}. These intermediate states are not bound in
the true sense, as they acquire a finite lifetime due to the interaction
with continuum states of other scattering channels. The metastable objects,
formed in the process of the Feshbach resonance atom-atom scattering, are
molecules with electronic and nuclear spins that have been rearranged from
the spins of the colliding atoms by virtue of the hyperfine interaction. It
is important that the difference of the initial and intermediate state
energies can be varied by means of an external magnetic field. The effective
scattering length that describes the low-energy binary collisions similarly
varies with the near-resonant magnetic field. Thus, employing the Feshbach
resonances, it is possible to create two-component mixtures consisting of
atoms and of molecules formed by these atoms. Since the overlapping
components can be either stable or unstable with respect to stratification,
depending on the relation between their scattering lengths, one could render
the components miscible or immiscible by varying their scattering lengths.
The Feshbach resonances were recently observed in ultra-cold atomic gases of 
$^{87}$Rb \cite{Courteille98b} and $^{23}$Na \cite{Inouye98}. An important
feature of the experiment \cite{Inouye98} is that the Feshbach resonances
were observed in an atomic Bose-condensed system. In this way, it looks
feasible to create a two-component system of Bose-condensed atoms and
molecules, with rather rich properties and with a variety of applications 
\cite{Timmermans99b}.

When two atoms of mass $m_1$ each form a Feshbach quasi-molecule, the mass
of the latter is 
\begin{equation}  \label{14.61}
m_2 \cong 2m_1 \; .
\end{equation}
Starting with a total number of atoms in a trap, $N$, one can form, via the
magnetically controlled Feshbach resonance, $N_2$ molecules coexisting with $%
N_1$ unbound atoms. Then, between the number of unbound atoms and that of
molecules, there is the relation 
\begin{equation}  \label{14.62}
N_1 + 2N_2 = N \; .
\end{equation}
This conservation law for the total number of atoms imposes the relation 
\begin{equation}  \label{14.63}
\mu_2 = 2\mu_1
\end{equation}
between the chemical potentials of molecules, $\mu_2$, and of unbound atoms, 
$\mu_1$.

The Hamiltonian of an atomic-molecular mixture can be presented as 
\[
H = \sum_{i=1}^2 \; \int\; \psi_i^\dagger({\bf r}) \left ( -\; \frac{\hbar^2 
{\bf \nabla}^2}{2m_i} \; - \mu_i + U_i\right )\; \psi_i({\bf r})\; d{\bf r}
\; + 
\]
\[
+\; \frac{1}{2}\; \sum_{ij}^2\; \int\; \psi_i^\dagger({\bf r})\;
\psi_j^\dagger({\bf r}^{\prime})\; \Phi_{ij}({\bf r}-{\bf r}^{\prime})\;
\psi_j({\bf r}^{\prime}) \; \psi_i({\bf r})\; d{\bf r}\; d{\bf r}^{\prime}\;
+ 
\]
\[
+ \; \frac{1}{2}\; \int\; \psi_1^\dagger({\bf r})\; \psi_1^\dagger({\bf r}%
)\; \Theta_{12}({\bf r}-{\bf r}^{\prime})\; \psi_2({\bf r}^{\prime})\; d{\bf %
r}\; d{\bf r}^{\prime}\; + 
\]
\begin{equation}  \label{14.64}
+\; \frac{1}{2}\; \int\; \psi_2^\dagger({\bf r})\; \Theta_{21}({\bf r}-{\bf r%
}^{\prime})\; \psi_1({\bf r}^{\prime})\; \psi_1({\bf r}^{\prime}) \; d{\bf r}%
\; d{\bf r}^{\prime}\; ,
\end{equation}
where the last two terms describe the atom-molecule reaction with a
transition amplitude having the symmetry properties 
\begin{equation}  \label{14.65}
\Theta_{12}({\bf r}) = \Theta_{12}(-{\bf r}) = \Theta_{21}({\bf r}) \; .
\end{equation}

Note that the global gauge symmetry, connected with the transformation $%
\psi_j\rightarrow e^{i\alpha}\psi_j$ is broken for the Hamiltonian (\ref
{14.64}). However, it possesses the gauge symmetry with respect to the
transformation 
\begin{equation}  \label{14.66}
\psi_j\rightarrow e^{i\alpha_j}\; \psi_j \qquad (2\alpha_1= \alpha_2) \; ,
\end{equation}
which is related to the atom-number conservation law (\ref{14.62}). The
anomalous averages 
\begin{equation}  \label{14.67}
<\psi_j>=<\psi_i\psi_j>\; =0 \; , \qquad <\psi_1\psi_1\psi_2>\; = \;
<\psi_2\psi_2\psi_1>\; = 0 \; ,
\end{equation}
and alike, which are not invariant with respect to the gauge transformation (%
\ref{14.66}), are zero. But the averages like 
\begin{equation}  \label{14.68}
<\psi_1^\dagger\psi_1^\dagger\psi_2>\; \neq 0 \; ,
\end{equation}
that are invariant with respect to the transformation (\ref{14.66}), are not
zero.

The Heisenberg equations for the atomic, $\psi_1$, and molecular, $\psi_2$,
field operators can be written as 
\[
i\hbar\; \frac{\partial}{\partial t}\; \psi_1({\bf r},t) = H_1(\psi)\;
\psi_1({\bf r},t) + \psi_1^\dagger({\bf r},t) \int \Theta_{12}({\bf r}-{\bf r%
}^{\prime})\; \psi_2({\bf r}^{\prime},t)\; d{\bf r}^{\prime}\; , 
\]
\begin{equation}  \label{14.69}
i\hbar\; \frac{\partial}{\partial t}\; \psi_2({\bf r},t) = H_2(\psi)\;
\psi_2({\bf r},t) + \frac{1}{2} \; \int \Theta_{12}({\bf r}-{\bf r}%
^{\prime})\; \psi_1({\bf r}^{\prime},t)\; \psi_1({\bf r}^{\prime},t) \; d%
{\bf r}^{\prime}\; ,
\end{equation}
where the notation 
\begin{equation}  \label{14.70}
H_i(\psi) \equiv -\; \frac{\hbar^2{\bf \nabla}^2}{2m_i}\; -\mu_i + U_i({\bf %
r },t) +\sum_j \int \Phi_{ij}({\bf r}-{\bf r}^{\prime})\; \psi_j^\dagger(%
{\bf r }^{\prime},t)\; \psi_j({\bf r}^{\prime},t)\; d{\bf r}^{\prime}
\end{equation}
is used.

Assuming dilute gases, one models the interaction potentials and transition
amplitudes by local functions 
\begin{equation}  \label{14.71}
\Phi_{ij}({\bf r}) = A_{ij}\delta({\bf r}) \; , \qquad \Theta_{12}({\bf r})=
B_{12}\delta({\bf r}) \; .
\end{equation}
Supposing that the whole mixture is in a coherent state, one has for the
corresponding coherent fields 
\[
i\hbar\; \frac{\partial}{\partial t} \; \varphi_1 = \left [ \hat
H_1(\varphi) -\mu_1\right ] \; \varphi_1 + \sqrt{N_2}\;
B_{12}\varphi_1^*\varphi_2 \; , 
\]
\begin{equation}  \label{14.72}
i\hbar\; \frac{\partial}{\partial t} \; \varphi_2 = \left [ \hat
H_2(\varphi) -\mu_2\right ] \; \varphi_2 + \frac{N_1}{2 \sqrt{N_2}}\;
B_{12}\varphi_1^2 ,
\end{equation}
where $\rho_j\equiv N_j/V$ and 
\begin{equation}  \label{14.73}
\hat H_i(\varphi) = -\; \frac{\hbar^2{\bf \nabla}^2}{2m_i} + U_i({\bf r},t)
+ \sum_{j=1}^2 N_j A_{ij}|\varphi_j|^2 \; .
\end{equation}

Looking for stationary solutions of Eqs. (\ref{14.72}) in the standard form (%
\ref{14.14}), we see that the related energies $E_j$ are connected with each
other by the relation 
\begin{equation}  \label{14.74}
E_2 = 2E_1 \; .
\end{equation}
Introducing the notation 
\begin{equation}  \label{14.75}
E\equiv E_1 +\mu_1 \; ,
\end{equation}
we obtain the following equations for the stationary coherent fields: 
\[
\hat H_1(\varphi)\varphi_1({\bf r}) +\sqrt{N_2}\; B_{12} \varphi_1^*({\bf r}%
)\varphi_2({\bf r}) = E\; \varphi_1({\bf r})\; , 
\]
\begin{equation}  \label{14.76}
\hat H_2(\varphi)\varphi_2({\bf r}) +\frac{N_1}{2\sqrt{N_2}}\; B_{12}
\varphi_1^2({\bf r}) = 2E\; \varphi_2({\bf r})\; .
\end{equation}

Wanting to study collective excitations in this reacting mixture, we may
linearize Eqs. (\ref{14.72}) after substituting there the form (\ref{14.16}%
). This results in the system of four equations 
\[
\left ( \hbar\omega -\hat H_1(\varphi) + E \right ) u_1 - \sqrt{N_2}\;
B_{12}\left ( \varphi_1^*u_2 +\varphi_2 v_1 \right ) - \sum_{i=1}^2 \; N_i
A_{i1} \left ( \varphi_i^* u_i + \varphi_i v_i\right ) \varphi_1 = 0 \; , 
\]
\[
\left ( \hbar\omega + \hat H_1(\varphi) - E \right ) v_1 + \sqrt{N_2}\;
B_{12}\left ( \varphi_1 v_2 +\varphi_2^* u_1 \right ) + \sum_{i=1}^2 \; N_i
A_{i1} \left ( \varphi_i^* u_i + \varphi_i v_i\right ) \varphi_1^* = 0 \; , 
\]
\[
\left ( \hbar\omega -\hat H_2(\varphi) + 2E \right ) u_2 - \frac{N_1}{\sqrt{%
N_2}}\; B_{12}\; \varphi_1 u_1 - \sum_{i=1}^2 \; N_i A_{i2} \left (
\varphi_i^* u_i + \varphi_i v_i\right ) \varphi_2 = 0 \; , 
\]
\begin{equation}  \label{14.77}
\left ( \hbar\omega + \hat H_2(\varphi) - 2E \right ) v_2 + \frac{N_1}{\sqrt{%
N_2}}\; B_{12}\; \varphi_1^* v_1 + \sum_{i=1}^2 \; N_i A_{i2} \left (
\varphi_i^* u_i + \varphi_i v_i\right ) \varphi_2^* = 0 \; ,
\end{equation}
where $\hat H_i(\varphi)$ are defined for the stationary states $\varphi_i(%
{\bf r})$.

In the uniform approximation (\ref{14.18}), the stationary equations (\ref
{14.76}) yield 
\[
E= U_1 +\rho_1 A_{11} +\rho_2 A_{12} +\sqrt{\rho_2}\; B_{12} \; , 
\]
\begin{equation}  \label{14.78}
2E= U_2 +\rho_1 A_{12} +\rho_2 A_{22} + \frac{\rho_1}{2\sqrt{\rho_2}}\;
B_{12} \; .
\end{equation}
Under given values for $A_{11},\; A_{12},\; A_{22}$, and $B_{12}$, Eq. (\ref
{14.78}) defines the relation between the atomic density $\rho_1$ and the
molecular density $\rho_2$. Since $\rho_i\equiv n_iN_i$, this means that the
number of unbound atoms $N_1$ and that of molecules $N_2$ are not arbitrary
but are related with each other through Eqs. (\ref{14.78}).

Keeping in mind the uniform approximation, when $n_{i}=1/V$, and looking for
the solutions of Eqs. (\ref{14.77}) in the form of plane waves $e^{i{\bf k}%
\cdot {\bf r}}$, we come to the equations 
\[
(\hbar \omega -K_{1})u_{1}-\sqrt{\rho _{2}}\;B_{12}(u_{2}+v_{1})-%
\sum_{i=1}^{2}\;\rho _{i}A_{i1}(u_{i}+v_{i})=0\;, 
\]
\[
(\hbar \omega -K_{2})u_{2}-\frac{\rho _{1}}{\sqrt{\rho _{2}}}%
\;B_{12}u_{1}-\sum_{i=1}^{2}\;\rho _{i}A_{i2}(u_{i}+v_{i})=0\;, 
\]
\[
(\hbar \omega +K_{1})v_{1}+\sqrt{\rho _{2}}\;B_{12}(v_{2}+u_{1})+%
\sum_{i=1}^{2}\;\rho _{i}A_{i1}(u_{i}+v_{i})=0\;, 
\]
\begin{equation}
(\hbar \omega +K_{2})v_{2}+\frac{\rho _{1}}{\sqrt{\rho _{2}}}
\;B_{12}v_{1}+\sum_{i=1}^{2}\;\rho _{i}A_{i2}(u_{i}+v_{i})=0\; .
\label{14.79}
\end{equation}
The spectral equation can be presented as in Eq. (\ref{14.21}), but with the
dynamic matrix having the elements 
\[
D_{11}=K_{1}+\rho _{1}A_{11},\;\;D_{12}=\rho _{2}A_{12}+\sqrt{\rho _{2}}%
\;B_{12},\;\;D_{13}=\rho _{1}A_{11}+\sqrt{\rho _{2}}\;B_{12},\;\;D_{14}=\rho
_{2}A_{12}, 
\]
\[
D_{21}=\rho _{1}A_{12}+\frac{\rho _{1}}{\sqrt{\rho _{2}}}\;B_{12}\;,\quad
D_{22}=K_{2}+\rho _{2}A_{22}\;,\quad D_{23}=\rho _{1}A_{12}\;,\quad
D_{24}=\rho _{2}A_{22}\;, 
\]
\[
D_{31}=-D_{13}\;,\qquad D_{32}=-D_{14}\;,\qquad D_{33}=-D_{11}\;,\qquad
D_{34}=-D_{12}\;, 
\]
\[
D_{41}=-D_{23}\;,\qquad D_{42}=-D_{24}\;,\qquad D_{43}=-D_{21}\;,\qquad
D_{44}=-D_{22}\; . 
\]
The general form of the spectral equation is rather cumbersome, so we shall
not write it down. Some particular cases have been studied in Ref. \cite
{Timmermans99b}.

Similarly to Sec.~14.3, we can study the stability conditions for the
atom-molecule mixture. For example, to find the condition of dynamic
stability, we need to find the inequality guaranteeing the positiveness of
the collective excitation spectrum corresponding to the oscillation of
components with respect to each other. To derive the condition of
thermodynamic stability, we have to compare the coherent average of the
mixture Hamiltonian (\ref{14.64}), which in the uniform approximation is 
\[
<H>_{N}\;=N_{1}(U_{1}-\mu _{1})+N_{2}(U_{2}-\mu _{2})+\frac{%
N_{1}^{2}A_{11}+2N_{1}N_{2}A_{12}+N_{2}^{2}A_{22}}{2(V_{1}+V_{2})}\;+N_{1}%
\sqrt{\rho _{2}}\;B_{12}\;, 
\]
with the energy (\ref{14.41}) of a stratified system, where $U_{i}$ is to be
replaced by $U_{i}-\mu _{i}$. Then the condition of thermodynamic stability (%
\ref{14.42}) becomes 
\[
N_{1}^{2}V_{2}^{2}A_{11}+N_{2}^{2}V_{1}^{2}A_{22}-2N_{1}N_{2}V_{1}V_{2}A_{12}-2N_{1}%
\sqrt{N_{2}V}\;V_{1}V_{2}B_{12}>0\;. 
\]
The sufficient stability conditions for each component separately are $%
A_{11}>0$ and $A_{22}>0$. If so, the condition of thermodynamic stability of
the mixture reads 
\[
(N_{1}V_{2}\sqrt{A_{11}}-N_{2}V_{1}\sqrt{A_{22}})^{2}+2N_{1}N_{2}V_{1}V_{2}(%
\sqrt{A_{11}A_{22}}\;-A_{12})-2N_{1}\sqrt{N_{2}V}\;V_{1}V_{2}B_{12}>0\;. 
\]
From here, the sufficient condition of thermodynamic stability is 
\begin{equation}
A_{11}A_{22}>\left( A_{12}+\frac{B_{12}}{\sqrt{\rho _{2}}}\right) ^{2}\;,
\label{14.80}
\end{equation}
where $\rho _{2}\equiv N_{2}/V$. Since the low-energy Feshbach resonances
make it feasible to vary the effective scattering lengths by a near-resonant
external magnetic field, one could realize different experiments with stable
mixtures as well as with stratifying components.

\chapter{Topological Coherent Modes}

Coherent modes are defined in Chapter 8 as stationary solutions of the
Gross-Pitaevskii equation. The ground-state coherent mode, with a
single-particle energy $E_0$, corresponds to the Bose-Einstein condensate.
In an equilibrium statistical system, the Bose-condensed state is always the
ground single-particle state.

An intriguing question is whether one could create non-groundstate
condensates of Bose atoms, that is, a macroscopic occupation of a non-ground
single-particle state. Clearly, if this is possible, this could be done only
in a non-equilibrium system. Second, in order to transfer atoms from a
single-particle ground state, with an energy $E_{0}$, to another state of
higher energy $E_{j}$, one should subject the system to the action of a
resonance field with a frequency close to the transition frequency $%
(E_{j}-E_{0})/\hbar $. Hence, this is to be a resonance process.

The possibility of the resonance formation of non-groundstate condensates of
Bose atoms was first advanced in Ref. \cite{Yukalov97}, these condensates
being associated with excited coherent modes. Such nonlinear coherent modes
have also been considered recently in Refs. \cite
{Williams99b,Kivshar99,Ostrovskaya99}. One often terms these excited
coherent modes as topological in order to stress their distinction from
elementary collective excitations. The latter correspond to small linear
oscillations around a state, thence these small oscillations do not change
the macroscopic density distribution in space. But different coherent modes
have principally different space dependence, because of which they are
termed topological.

\section{Resonance Field Modulation}

The Gross-Pitaevskii equation, describing a coherent field $\varphi=\varphi(%
{\bf r},t)$, is 
\begin{equation}  \label{15.1}
i\hbar\; \frac{\partial\varphi}{\partial t} = \left [ \hat H(\varphi) +\hat
V_{res}\right ] \varphi \; ,
\end{equation}
where, in addition to the nonlinear Hamiltonian 
\begin{equation}  \label{15.2}
\hat H(\varphi) = -\; \frac{\hbar^2}{2m_0} \; {\bf \nabla}^2 +U({\bf r}) +
NA|\varphi|^2 \; ,
\end{equation}
we include a resonant field 
\begin{equation}  \label{15.3}
\hat V_{res} = V_1({\bf r})\cos\omega t + V_2({\bf r}) \sin\omega t \; .
\end{equation}
We assume that at the initial time the system is Bose-condensed to the
ground state 
\begin{equation}  \label{15.4}
\varphi({\bf r},0) =\varphi_0({\bf r}) \; ,
\end{equation}
characterized by the energy $E_0$.

The transition frequencies between coherent modes are given by the equation 
\begin{equation}
\hbar \omega _{mn}\equiv E_{m}-E_{n}\;,  \label{15.5}
\end{equation}
in which the spectrum of coherent modes is defined by the eigenproblem 
\[
\hat{H}(\varphi _{n})\varphi _{n}=E_{n}\;\varphi _{n}\;. 
\]
Suppose that our aim is to transfer atoms from the ground state $\varphi
_{0} $ to a chosen state $\varphi _{j}$. Therefore, we require that the
frequency of the modulating field (\ref{15.3}) be close to the transition
frequency 
\begin{equation}
\omega _{j}\equiv \frac{E_{j}-E_{0}}{\hbar }\;.  \label{15.6}
\end{equation}
The closeness implies the quasiresonance condition 
\begin{equation}
\left| \frac{\Delta \omega }{\omega }\right| \ll 1\;,\qquad \Delta \omega
\equiv \omega -\omega _{j}\;.  \label{15.7}
\end{equation}
Another important requirement is that the spectrum of coherent modes be not
equidistant \cite{Yukalov97}. In fact, if that were the case, then the
pumping of atoms from the ground state to the chosen particular state would,
at the same time, induce transitions from the latter to another equidistant
state and from the latter to even higher equidistant states. Thus, all atoms
would be dispersed over all states making it impossible to achieve a
macroscopic population of one of them. Fortunately, as is shown in Chapters
10 and 11, the spectrum of coherent modes is not equidistant because of the
nonlinearity induced by atomic interactions.

Let us look for the solution of Eq. (\ref{15.1}) in the form of the sum 
\begin{equation}  \label{15.8}
\varphi({\bf r},t) =\sum_n \; c_n(t)\varphi_n({\bf r},t)
\end{equation}
over the coherent modes 
\[
\varphi_n({\bf r},t) =\varphi_n({\bf r})\;\exp\left ( -\; \frac{i}{\hbar}\;
E_n\; t\right ) \; . 
\]
It is worth noting that the presentation (\ref{15.8}) does not require the
set $\{\varphi_n({\bf r})\}$ to form a complete basis. As can be checked in
any textbook on Quantum Mechanics or Functional Analysis, the property of
completeness of a basis presupposes that an {\it arbitrary} function from
the considered Hilbert space could be presented as an expansion over this
basis. We do not require here such a restrictive property for {\it all}
functions, but we invoke just the {\it sole} expansion, looking for a
solution in the form of (\ref{15.8}).

What we need in the following is the assumption that the coefficients $%
c_n(t) $ in the sum (\ref{15.8}) vary in time slower than the exponentials
in $\varphi_n({\bf r},t)$, that is, 
\begin{equation}  \label{15.9}
\left | \frac{dc_n}{dt}\right | \ll E_n \; .
\end{equation}

Looking for a solution in the form (\ref{15.8}), one has to substitute it
into Eq. (\ref{15.1}). To obtain equations for the coefficients $c_{n}(t)$,
one may invoke the averaging techniques \cite{Bogolubov61}. This is possible
because, according to the inequality (\ref{15.9}), the functions $c_{n}(t)$
can be classified as slow, compared to the fast functions $\exp
(-iE_{n}t/\hbar )$. Thus, $c_{n}(t)$ can be treated as quasi-invariants.
Multiplying Eq. (\ref{15.1}) by $\varphi _{n}^{\ast }({\bf r},t)$ and
integrating the result over ${\bf r}$ and averaging over time as 
\begin{equation}
\lim_{\tau \rightarrow \infty }\;\frac{1}{\tau }\;\int_{0}^{\tau }F(t)\;dt\;,
\label{15.10}
\end{equation}
keeping $c_{n}$ as quasi-invariants, one obtains an equation describing the 
{\it guiding centers} for $c_{n}(t)$. Averaging over time, one uses the
equalities 
\[
\lim_{\tau \rightarrow \infty }\;\frac{1}{\tau }\;\int_{0}^{\tau }\exp
\left\{ \frac{i}{\hbar }\;(E_{m}-E_{n})t\right\} \;dt=\delta _{mn}\;, 
\]
\[
\lim_{\tau \rightarrow \infty }\;\frac{1}{\tau }\;\int_{0}^{\tau }\exp
\left\{ \frac{i}{\hbar }\;(E_{m}+E_{k}-E_{n}-E_{l})\;t\right\} \;dt=\delta
_{mn}\;\delta _{kl}+\delta _{ml}\;\delta _{nk}-\delta _{mk}\;\delta
_{kn}\;\delta _{nl}\;. 
\]
The latter assume that the spectrum $E_{n}$ is nondegenerate. Generally, in
the presence of the nonlinear term in the Hamiltonian (\ref{15.2}), this is
true. But even if the spectrum $E_{n}$ were degenerate, one could avoid
complications in the following way. One may add to the Hamiltonian (\ref
{15.2}) a term lifting the degeneracy and to set this term to zero at the
end of the calculations.

Note that normalizing the function (\ref{15.8}) as $(\varphi,\varphi)=1$,
one gets 
\[
\sum_{mn} \; c_m^*(t)\; c_n(t)\; (\varphi_m,\varphi_n)\; e^{-i\omega_{mn}t}
= 1 \; . 
\]
Averaging this over time, according to the rule (\ref{15.10}) and invoking
condition (\ref{15.9}), gives 
\begin{equation}  \label{15.11}
\sum_n |c_n(t)|^2 = 1 \; .
\end{equation}
From here a useful relation follows: 
\[
|c_n(t)|^2 = 1 -\sum_{m(\neq n)}\; |c_m(t)|^2 \; . 
\]

Substituting the form (\ref{15.8}) into Eq. (\ref{15.1}) and accomplishing
the described time-averaging procedure results in the equation 
\begin{equation}  \label{15.12}
\frac{dc_n}{dt} = - i\sum_{m(\neq n)} \alpha_{nm}|c_{m}|^2 c_n -\; \frac{i}{2%
}\; \delta_{n0}\;\beta\; c_j\; e^{i\Delta\omega t} -\; \frac{i}{2}%
\;\delta_{nj}\;\beta^*\; c_0\; e^{-i \Delta\omega t} \;,
\end{equation}
in which the amplitude 
\begin{equation}  \label{15.13}
\alpha_{nm} \equiv A\; \frac{N}{\hbar}\; \int \; |\varphi_n({\bf r})|^2
\left ( 2|\varphi_m({\bf r})|^2 - |\varphi_n({\bf r})|^2\right )\; d{\bf r}
\end{equation}
is due to the nonlinear term in the Hamiltonian (\ref{15.2}), while the
transition amplitude 
\begin{equation}  \label{15.14}
\beta \equiv \frac{1}{\hbar} \; \int \varphi_0^*({\bf r})\; V({\bf r})\;
\varphi_j({\bf r})\; d{\bf r} \; , \qquad V({\bf r}) \equiv V_1({\bf r}) - i
V_2({\bf r}) \; ,
\end{equation}
is related to the resonant modulating field (\ref{15.3}). In the process of
the time averaging (\ref{15.10}), the function $\exp(i\Delta\omega t)$ is
also treated as slow because of the quasiresonance condition (\ref{15.7}).

Equation (\ref{15.12}) shows that the resonant field induces transitions
only between the ground-state and a chosen $j$-level. At first glance, the
nonlinear term, being nonresonant, could induce transitions between all
levels, changing the corresponding fractional populations 
\begin{equation}  \label{15.15}
n_m(t) \equiv |c_m(t)|^2 \; .
\end{equation}
However, from Eq. (\ref{15.12}) it follows that for all levels $m\neq 0,j$,
except the two selected resonant levels, one has 
\[
\frac{d}{dt}\; n_m(t) = 0 \qquad (m\neq 0,j) \; . 
\]
This, with the initial condition $n_m(0)=0$, yields 
\[
n_m(t) = 0 \qquad (m\neq 0,j)\; . 
\]
Similarly, $c_m(t)=0$ for all $m\neq 0,j$. Therefore, Eq. (\ref{15.12}) is
equivalent to the system of two equations 
\[
\frac{dc_0}{dt} = -i\alpha_{0j}\; n_j\; c_0 -\; \frac{i}{2}\; \beta\; c_j\;
e^{i\Delta\omega t} \; , 
\]
\begin{equation}  \label{15.16}
\frac{dc_j}{dt} = -i\alpha_{j0}\; n_0\; c_j -\; \frac{i}{2}\; \beta^*\;
c_0\; e^{-i\Delta\omega t} \; .
\end{equation}
The initial conditions to these equations, according to Eq. (\ref{15.4}),
are 
\begin{equation}  \label{15.17}
c_0(0)= 1 \; , \qquad c_j(0) = 0 \; .
\end{equation}
The equations for the fractional populations (\ref{15.15}) immediately
follow from Eqs. (\ref{15.16}) giving 
\begin{equation}  \label{15.18}
\frac{dn_0}{dt} = {\rm Im}\; \left ( \beta\; e^{i\Delta\omega t} \; c_0^*\;
c_j\right ) \; , \qquad \frac{dn_j}{dt} = {\rm Im}\; \left ( \beta^*\;
e^{-i\Delta\omega t} \; c_j^*\; c_0\right ) \; ,
\end{equation}
with the corresponding initial conditions 
\[
n_0(0)=1 \; , \qquad n_j(0) = 0 \; , 
\]
resulting from the conditions (\ref{15.17}). The normalization (\ref{15.11})
for the fractional populations reduces to the equation 
\begin{equation}  \label{15.19}
n_0(t) + n_j(t) = 1\; .
\end{equation}
In what follows, for the simplicity of notation, we write 
\begin{equation}  \label{15.20}
\alpha\equiv \alpha_{0j}
\end{equation}
and set $\alpha_{0j}=\alpha_{j0}$.

Note that in deriving Eqs. (\ref{15.16}), the orthogonality of the coherent
modes $\varphi _{m}({\bf r})$ and $\varphi _{n}({\bf r})$, for $m\neq n$,
has not been assumed. What is used is the condition (\ref{15.9}) permitting
one to invoke the averaging technique \cite{Bogolubov61}. In addition,
employing for these modes the solutions of Chapter 10, one can check that $%
|(\varphi _{m},\varphi _{n})|$ are less or of order $0.1$ if $m\neq n$.
Hence the coherent modes can be treated as approximately orthogonal since $%
|(\varphi _{m},\varphi _{n})|\ll 1$ for $m\neq n$.

The solutions to Eqs. (\ref{15.16}) and (\ref{15.18}) can be obtained
analytically, provided the inequality 
\begin{equation}
\left| \frac{\beta }{\alpha }\right| \ll 1  \label{15.21}
\end{equation}
holds true. In that case, one can again resort to the averaging technique 
\cite{Bogolubov61}, being based on the fact that the functions $c_{0}(t)$
and $c_{j}(t)$ can be classified as fast, compared to the slow functions $%
n_{0}(t)$ and $n_{j}(t)$. With the slow functions treated as
quasi-invariants, Eqs. (\ref{15.15}) are linear equations with respect to $%
c_{0}(t)$ and $c_{j}(t)$, which gives 
\[
c_{0}=\left[ \cos \;\frac{\Omega t}{2}+i\;\frac{\alpha (n_{0}-n_{j})-\Delta
\omega }{\Omega }\;\sin \;\frac{\Omega t}{2}\right] \;\exp \left\{ -\;\frac{i%
}{2}\;(\alpha -\Delta \omega )\;t\right\} \;, 
\]
\begin{equation}
c_{j}=-i\;\frac{\beta ^{\ast }}{\Omega }\;\sin \;\frac{\Omega t}{2}\;\exp
\left\{ -\;\frac{i}{2}\;(\alpha +\Delta \omega )\;t\right\} \;,
\label{15.22}
\end{equation}
where the {\it collective frequency} $\Omega $, defined by the equality 
\begin{equation}
\Omega ^{2}\equiv \left[ \alpha (n_{0}-n_{j})-\Delta \omega \right]
^{2}+|\beta |^{2}\;,  \label{15.23}
\end{equation}
is introduced. Comparing our case with the resonant excitation of atoms in
optics \cite{Mandel95}, we see that $|\beta |$ plays the role of the Rabi
frequency, while $\sqrt{|\beta |^{2}+(\Delta \omega )^{2}}$ is what is
called the effective Rabi frequency. The quantity $\Omega $, defined in Eq. (%
\ref{15.23}), differs from the effective Rabi frequency by the presence of
the term containing the interaction amplitude $\alpha $. Because of this, in
our case, $\Omega $ could be called the {\it collective Rabi frequency},
although it is not a parameter but a function depending on time through the
fractional populations $n_{0}(t)$ and $n_{j}(t)$. For the latter, we get 
\begin{equation}
n_{0}=1-\;\frac{|\beta |^{2}}{\Omega ^{2}}\;\sin ^{2}\frac{\Omega t}{2}
\;,\qquad n_{j}=\frac{|\beta |^{2}}{\Omega ^{2}}\;\sin ^{2} \frac{\Omega t}{2%
}\;.  \label{15.24}
\end{equation}

If at some finite time $t_0$, the modulation field (\ref{15.3}) is switched
off, then, as follows from Eqs. (\ref{15.18}), the fractional populations
stand constant, with the values $n_0(t_0)$ and $n_j(t_0)$. Then, we have a
mixture of two topological modes. This mixture will, of course, not exist
for ever, but during the lifetime of the corresponding modes limited by
atomic collisions. For instance, the loss rates caused by binary
depolarizing collisions can be estimated as 
\[
\gamma_0 =\lambda_{00} N^2 n_0^2 \int|\varphi_0({\bf r})|^4\; d{\bf r} +
\lambda_{0j} N^2 n_0n_j \int |\varphi_0({\bf r})|^2 |\varphi_j({\bf r})|^2\;
d{\bf r}\; , 
\]
\begin{equation}  \label{15.25}
\gamma_j =\lambda_{jj} N^2 n_j^2 \int|\varphi_j({\bf r})|^4\; d{\bf r} +
\lambda_{j0} N^2 n_jn_0 \int |\varphi_j({\bf r})|^2 |\varphi_0({\bf r})|^2\;
d{\bf r}\; ,
\end{equation}
where $\lambda_{ij}$ are the related loss-rate coefficients and $%
n_0=n_0(t_0) $ and $n_j=n_j(t_0)$.

A modulating field (\ref{15.3}) that is not monochromatic but characterized
by a frequency distribution $\rho (\omega )$ that is centered at $\omega
_{j} $ will cause heating of the system due to nonresonant transitions \cite
{Savard97}. The corresponding heating rate can be expressed as the sum 
\[
\Gamma _{heat}=2\pi \sum_{n(\neq j)}|\beta _{n0}|^{2}\rho (\omega _{n0})\;, 
\]
in which 
\[
\beta _{n0}\equiv \frac{1}{\hbar }\int \varphi _{n}^{\ast }({\bf r})V({\bf r}%
)\varphi _{0}({\bf r})\;d{\bf r}\;. 
\]
If the density of frequencies $\rho (\omega )$ is sharply centered, say as $%
\rho (\omega )\approx \delta (\omega -\omega _{j0})$, then the heating rate
is close to zero.

\section{Critical Dynamic Effects}

The solution (\ref{15.22}) to the evolution equations (\ref{15.16}) has been
obtained by using the averaging technique \cite{Bogolubov61}, which requires
the inequality (\ref{15.21}). Wanting to analyze the behavior of solutions
to Eqs. (\ref{15.16}) under an arbitrary relation between $\alpha $ and $%
\beta $, we have to solve these equations numerically. This behavior turned
out to be surprisingly rich exhibiting unexpected critical effects \cite
{Yukalov00}.

For the numerical analysis of Eqs. (\ref{15.1}), it is convenient to
introduce the dimensionless parameters 
\begin{equation}
b\equiv \frac{|\beta |}{\alpha }\;,\qquad \delta \equiv \frac{\Delta \omega 
}{\alpha }  \label{15.26}
\end{equation}
and to perform a scaling, measuring time in units of $\alpha ^{-1}$. To
return back to dimensional time, one has to make the substitution 
\begin{equation}
t\rightarrow \;\frac{t}{\alpha }\;.  \label{15.27}
\end{equation}
We solve \cite{Yukalov00} the system of equations (\ref{15.16}) for
different values of the parameters (\ref{15.26}), keeping in mind that the
dimensionless detuning is small, 
\begin{equation}
\delta \ll 1\;.  \label{15.28}
\end{equation}
Varying the parameters (\ref{15.26}), we find \cite{Yukalov00} that there
exists a bifurcation line, described by the relation 
\begin{equation}
b+\delta \approx 0.5\;,  \label{15.29}
\end{equation}
at which the qualitative behavior of solutions changes abruptly.

To illustrate the drastic change in the behavior of solutions, when crossing
the bifurcation line (\ref{15.29}), let us first fix $b=0.4999$ and vary the
detuning $\delta $. Figure \ref{Fig33} presents the fractional populations $%
n_{0}(t)$ and $n_{j}(t)$, defined as in Eq. (\ref{15.15}), time being
measured in units of $\alpha ^{-1}$. In Fig. \ref{Fig33}(a), the detuning is
zero, $\delta =0$, and the behavior of the fractional populations
approximately follows the law (\ref{15.24}). Slightly changing the detuning
to $\delta =0.0001$ essentially transforms the behavior to that in Fig. \ref
{Fig33}(b), where the top of $n_{j}(t)$ and the bottom of $n_{0}(t)$ become
flat, touching each other, while the oscillation period is approximately
doubled. Shifting the detuning by a tiny portion to $\delta =0.0001001$
results in Fig. \ref{Fig33}(c), where the period is again doubled, and there
appear the upward cusps of $n_{j}(t)$ and the downward cusps of $n_{0}(t)$.
Increasing a little the detuning to $\delta =0.00011$ squeezes the
oscillation period twice, as is shown in Fig. \ref{Fig33}(d). Similar
changes happen when crossing the bifurcation line under a fixed detuning and
a varied transition amplitude, as illustrated in Fig. \ref{Fig34}.

The unusual behavior of the fractional populations is due to the
nonlinearity of the evolution equations (\ref{15.16}). It is known that
systems of nonlinear differential equations can possess qualitatively
different solutions for parameters differing by infinitesimally small
values. The transfer from one type of solutions to another qualitatively
different type is termed in the theory of dynamical systems bifurcation \cite
{Guckenheimer86}. At a bifurcation point, a dynamical system is structurally
unstable. Bifurcations in dynamical systems are somewhat analogous to phase
transitions and critical phenomena in equilibrium statistical systems \cite
{Yukalov90c}. To elucidate this analogy for the present case, we have to
consider the time-averaged properties of the system, which can be done as
follows. First, it is necessary to define an effective Hamiltonian
generating the evolution equations (\ref{15.16}). This can be done by
noticing that these equations can be presented in the Hamiltonian form 
\begin{equation}
i\;\frac{dc_{0}}{dt}=\frac{\partial H_{eff}}{\partial c_{0}^{\ast }}%
\;,\qquad i\;\frac{dc_{j}}{dt}=\frac{\partial H_{eff}}{\partial c_{j}^{\ast }%
}\;,  \label{15.30}
\end{equation}
with the effective Hamiltonian 
\begin{equation}
H_{eff}=\alpha n_{0}n_{j}+{\rm Re}\;\left( \beta \;e^{i\Delta \omega
t}\;c_{0}^{\ast }\;c_{j}\right) \;.  \label{15.31}
\end{equation}
Substituting here the approximate solutions (\ref{15.22}) yields 
\[
H_{eff}=\alpha n_{0}n_{j}-n_{j}\left[ \alpha (n_{0}-n_{j})-\Delta \omega %
\right] \;. 
\]
The latter, with the normalization (\ref{15.20}), gives 
\begin{equation}
H_{eff}=\alpha \;n_{j}^{2}+n_{j}\;\Delta \omega \;.  \label{15.32}
\end{equation}
Averaging the fractional populations (\ref{15.24}) over the explicit time
and using this averaged quantity in the collective frequency (\ref{15.23}),
one has the averaged population 
\begin{equation}
\overline{n}_{j}=\frac{|\beta |^{2}}{2\;{\overline{\Omega }}^{2}}\;,
\label{15.33}
\end{equation}
in which the averaged collective frequency is given by the equality 
\begin{equation}
\overline{\Omega }^{2}=\left[ \alpha (1-2\;\overline{n}_{j})-\Delta \omega %
\right] ^{2}+|\beta |^{2}\;.  \label{15.34}
\end{equation}
The effective average energy can be defined by taking the effective
Hamiltonian (\ref{15.32}) with $n_{j}$ replaced by the averaged population (%
\ref{15.33}), which gives 
\begin{equation}
E_{eff}=\alpha \;\overline{n}_{j}^{2}+\overline{n}_{j}\Delta \omega \;.
\label{15.35}
\end{equation}

To study a kind of thermodynamics of the so defined effective system, it is
possible to introduce the following characteristics. The {\it pumping
capacity} 
\begin{equation}  \label{15.36}
C_\beta \equiv \frac{\partial E_{eff}}{\partial|\beta|}
\end{equation}
describes the capacity of the system to store the energy pumped into it by
the modulating field. The {\it order parameter} 
\begin{equation}  \label{15.37}
\eta \equiv \overline n_0 - \overline n_j = 1 - 2\; \overline n_j
\end{equation}
characterizes the level of excitation, being $\eta=1$ for a system in the
ground state and $\eta = -1$ for a completely excited system. The {\it %
detuning susceptibility} 
\begin{equation}  \label{15.38}
\chi_\delta \equiv \left | \frac{\partial\eta}{\partial\delta} \right |
\end{equation}
defines how a variation of detuning influences the order parameter.

It is convenient to pass again to the dimensionless quantities (\ref{15.26})
and to introduce the dimensionless average collective frequency 
\begin{equation}  \label{15.39}
\varepsilon \equiv \frac{\overline\Omega}{\alpha} \; .
\end{equation}
Then Eq. (\ref{15.34}) takes the form 
\begin{equation}  \label{15.40}
\varepsilon^4(\varepsilon^2 - b^2) = (\varepsilon^2 - b^2
-\varepsilon^2\delta)^2 \; .
\end{equation}
The average energy (\ref{15.35}) becomes 
\begin{equation}  \label{15.41}
E_{eff} =\frac{\alpha b^2}{2\varepsilon} \left ( \frac{b^2}{2\varepsilon^2}
+\delta\right ) \; .
\end{equation}
And the order parameter (\ref{15.37}) is 
\begin{equation}  \label{15.42}
\eta = 1 -\; \frac{b^2}{\varepsilon^2} \; .
\end{equation}
Taking into account the smallness of the detuning (\ref{15.28}), one can
simplify the above expressions.

Looking for a positive solution of Eq. (\ref{15.40}), one can notice that
there is the critical value 
\begin{equation}
b_{c}=\frac{1}{2}\;-\delta \;,  \label{15.43}
\end{equation}
at which the average collective frequency (\ref{15.39}) has a jump, so that 
\begin{equation}
\varepsilon =\frac{1}{\sqrt{2}}\;\left[ 1-2\delta +\sqrt{(1-2\delta
)^{2}-4b^{2}}\right] ^{1/2}  \label{15.44}
\end{equation}
for $0\leq b\leq b_{c}$, but 
\begin{equation}
\varepsilon =b\qquad (b>b_{c})\;.  \label{15.45}
\end{equation}
The frequency (\ref{15.44}) changes from 
\[
\varepsilon =1-2\delta \qquad (b=0) 
\]
to the critical frequency 
\[
\varepsilon _{c}=\sqrt{\frac{1}{2}\;-\delta }\qquad (b=b_{c})\;, 
\]
with the jump at $b=b_{c}$ being 
\[
\varepsilon (b_{c}+0)-\varepsilon (b_{c}-0)=\frac{1-\sqrt{2}}{2}\;-\left(
1-\;\frac{1}{\sqrt{2}}\right) \;\delta \;. 
\]
A sudden decrease in the frequency implies an abrupt increase in the
oscillation period. The order parameter (\ref{15.42}) varies from $\eta =1$
at $b=0$ to the critical value 
\[
\eta _{c}=\frac{1}{2}+\delta \qquad (b=b_{c})\;, 
\]
becoming zero for $b>b_{c}$. The average energy (\ref{15.41}), above $b_{c}$%
, does not depend on the value of $b$, 
\[
E_{eff}=\frac{\alpha }{2}\left( \frac{1}{2}+\delta \right) \qquad
(b>b_{c})\;. 
\]
Thus, the pumping capacity (\ref{15.36}), order parameter (\ref{15.37}), and
detuning susceptibility (\ref{15.38}) all are zero above $b_{c}$, 
\begin{equation}
C_{\beta }=0\;,\qquad \eta =0\;,\qquad \chi _{\delta }=0\qquad (b>b_{c})\;.
\label{15.46}
\end{equation}
The behavior of these characteristics in the asymptotic vicinity of the
critical line (\ref{15.43}), below the critical pumping $b_{c}$, is as
follows. With the notation 
\[
\tau \equiv \frac{b_{c}-b}{b_{c}}\qquad (b\leq b_{c})\;, 
\]
the pumping capacity is 
\begin{equation}
C_{\beta }\simeq \frac{\sqrt{2}}{8}\;\tau ^{-1/2}-\;\frac{1}{4}\;-\;\frac{3%
\sqrt{2}}{32}\;\tau ^{1/2}\;,  \label{15.47}
\end{equation}
the order parameter becomes 
\begin{equation}
\eta \simeq \frac{1}{2}+\delta +\frac{\sqrt{2}}{2}\;\left( 1-2\delta \right)
\;\tau ^{1/2}\;,  \label{15.48}
\end{equation}
and the detuning susceptibility takes the form 
\begin{equation}
\chi _{\delta }\simeq \frac{1}{\sqrt{2}}\;\tau ^{-1/2}-1+\frac{\sqrt{2}}{8}%
\;\tau ^{1/2}\;.  \label{15.49}
\end{equation}
This shows that the pumping capacity and detuning susceptibility diverge as $%
\tau \rightarrow 0$. Hence, Eq. (\ref{15.43}) really defines a critical line
where critical phenomena occur. The critical indices for all characteristics
are $1/2$, satisfying the scaling relation 
\begin{equation}
{\rm ind}\;C_{\beta }+2\;{\rm ind}\;\chi +{\rm ind}\;\chi _{\delta }=2\;,
\label{15.50}
\end{equation}
known for critical phenomena \cite{Yukalov90c}; here $ind$ is the
abbreviation for index. The critical line (\ref{15.43}) coincides with the
bifurcation line (\ref{15.29}).

\section{Spatio-Temporal Evolution of Density}

In the two-level picture of section 15.1 the coherent field (\ref{15.8}) is 
\begin{equation}
\varphi ({\bf r},t)=c_{0}(t)\varphi _{0}({\bf r})\;\exp \left( -\;\tfrac{i}{
\hbar }\;E_{0}t\right) +c_{j}(t)\varphi _{j}({\bf r})\;\exp \left( -\;\tfrac{
i}{\hbar }\;E_{j}t\right) \;.  \label{15.51}
\end{equation}
To study the spatio-temporal properties of an atomic cloud, it is convenient
to average the density of atoms $N|\varphi ({\bf r},t)|^{2}$ over the period 
$2\pi /\omega _{j}$ of fast oscillations, treating $c_{0}$ and $c_{j}$ as
slow functions of time. The result is the {\it envelope density} 
\begin{equation}
\overline{\rho }({\bf r},t)=\overline{\rho }_{0}({\bf r},t)+\overline{\rho }
_{j}({\bf r},t)\;,  \label{15.52}
\end{equation}
in which 
\[
\overline{\rho }_{p}({\bf r})\equiv N\;n_{p}(t)\;|\varphi _{p}({\bf r}
)|^{2}\;, \qquad p=0,j\; . 
\]
The density (\ref{15.52}) is normalized to the number of atoms, 
\[
\int \rho ({\bf r},t)\;d{\bf r}=N\;. 
\]

For a cylindrical trap, one may pass to the dimensionless notation of
section 9.4 and define the dimensionless densities 
\begin{equation}  \label{15.53}
\rho(r,\varphi,z,t) \equiv \frac{l_r^3}{N}\; \overline\rho({\bf r},t) \; ,
\qquad \rho_n(r,\varphi,z,t) \equiv \frac{l_r^3}{N}\; \overline\rho_n({\bf r}%
,t)
\end{equation}
depending on the dimensionless space variables (\ref{9.24}). The introduced
dimensionless density is normalized as 
\[
\int_0^\infty \int_0^{2\pi} \int_{-\infty}^{+\infty} \;
\rho(r,\varphi,z,t)\; rdr\; d\varphi\; dz = 1 
\]
and is given by the sum 
\begin{equation}  \label{15.54}
\rho(r,\varphi,z,t) = \rho_0(r,\varphi,z,t) + \rho_j(r,\varphi,z,t) \; ,
\end{equation}
in which 
\begin{equation}  \label{15.55}
\rho_0(r,\varphi,z,t) \equiv n_0(t)\; |\psi_0(r,\varphi,z)|^2 \; , \qquad
\rho_j(r,\varphi,z,t) \equiv n_j(t)\; |\psi_j(r,\varphi,z)|^2 \; ,
\end{equation}
with $\psi_0$ and $\psi_j$ being the dimensionless coherent modes.

In the optimized approximation of Chapter 10, the ground state mode can be
written as 
\begin{equation}  \label{15.56}
\psi_{000}(r,\varphi,z) =\left ( \frac{u^2v}{\pi^3} \right )^{1/4}
\exp\left\{ -\frac{1}{2}\; (ur^2 +vz^2) \right\}\; ,
\end{equation}
where the control functions $u=u_{000}$ and $v=v_{000}$, in the
strong-coupling limit, are 
\[
u_{000}=\frac{(2\pi^3)^{1/5}}{(\nu g)^{2/5}}\; , \qquad v_{000} =\frac{%
(2\pi^3)^{1/5}\nu^2}{(\nu g)^{2/5}} \; . 
\]
The ground-state energy is 
\begin{equation}  \label{15.57}
E_{000} =\frac{5(\nu g)^{2/5}}{4(2\pi^3)^{1/5}} = 0.547538\; (\nu g)^{2/5}
\; ,
\end{equation}
where the indices mean $n=0,\; m=0,\; k=0$, and again the strong-coupling
condition $\nu g\gg 1$ is assumed.

The radial dipole mode, with the quantum numbers $n=1,\; m=k=0$, is
presented by 
\begin{equation}  \label{15.58}
\psi_{100}(r,\varphi,z) =\left ( \frac{u^2v}{\pi^3} \right )^{1/4}\; (ur^2
-1)\;\exp\left\{ -\; \frac{1}{2}\; (ur^2+vz^2)\right\} \; ,
\end{equation}
where $u=u_{100}$ and $v=v_{100}$ are 
\[
u_{100} =\frac{(6\pi)^{3/5}}{(\nu g)^{2/5}} \; , \qquad v_{100} = \frac{%
(6\pi)^{3/5}\nu^2}{3(\nu g)^{2/5}}\; . 
\]
The corresponding energy writes 
\begin{equation}  \label{15.59}
E_{100} =\frac{5}{8}\left ( \frac{36}{\pi^3}\right )^{1/5} (\nu g)^{2/5} =
0.643948\; (\nu g)^{2/5} \; .
\end{equation}
Therefore, the transition frequency is 
\begin{equation}  \label{15.60}
\omega_{100} \equiv E_{100} - E_{000} = 0.096410\; (\nu g)^{2/5} \; .
\end{equation}
Here and in what follows the strong-coupling limit $\nu g\gg 1$ is again
supposed.

The vortex mode, with the quantum numbers $n=0,\; m=1,\; k=0$, is of the
form 
\begin{equation}  \label{15.61}
\psi_{010}(r,\varphi,z) = u\left ( \frac{v}{\pi^3}\right )^{1/4} r\;
e^{i\varphi}\; \exp\left\{ -\; \frac{1}{2}\; (ur^2 + vz^2) \right\} \; ,
\end{equation}
where $u=u_{010}$, $v=v_{010}$ are 
\[
u_{010} =\frac{2(2\pi^3)^{1/5}}{(\nu g)^{2/5}}\; , \qquad v_{010} = \frac{%
(2\pi^3)^{1/5}\nu^2}{(\nu g)^{2/5}}\; . 
\]
The transition frequency from the ground to the vortex state is 
\begin{equation}  \label{15.62}
\omega_{010}\equiv E_{010} -E_{000} = \frac{3.424}{(\nu g)^{2/5}} \; .
\end{equation}

The axial dipole mode, with the quantum numbers $n=0,\; m=0,\; k=1$ reads 
\begin{equation}  \label{15.63}
\psi_{001}(r,\varphi,z) =\left ( \frac{4u^2v^3}{\pi^3} \right )^{1/4} z\;
\exp\left\{ -\; \frac{1}{2}\; (ur^2 + vz^2)\right\} \; ,
\end{equation}
where $u=u_{001}$, $v=v_{001}$ are 
\[
u_{001} = \frac{2(\pi/3)^{3/5}}{(\nu g)^{2/5}} \; , \qquad v_{001} = \frac{
6(\pi/3)^{3/5}\nu^2}{(\nu g)^{2/5}} \; . 
\]
The related energy is 
\begin{equation}  \label{15.64}
E_{001} =\frac{5}{8}\left ( \frac{3}{\pi}\right )^{3/5} \; (\nu g)^{2/5} =
0.607943\; (\nu g)^{2/5} \; .
\end{equation}
Hence, for the transition frequency, one has 
\begin{equation}  \label{15.65}
\omega_{001} \equiv E_{001} - E_{000} = 0.060405\; (\nu g)^{2/5} \; .
\end{equation}

The spatio-temporal behavior of the densities (\ref{15.55}) for low-lying
modes is illustrated in Figs. \ref{Fig35} to \ref{Fig37}. The corresponding
wave functions are taken from Eqs. (\ref{15.56}), (\ref{15.58}), (\ref{15.61}%
), and (\ref{15.63}). The fractional populations (\ref{15.15}) are
calculated by solving Eqs. (\ref{15.16}).

\section{Resonance Formation of Vortices}

To form a vortex, the resonance field (\ref{15.3}) must depend on the radial
angle, so that the corresponding transition amplitude (\ref{15.12}) be
nonzero. For the latter, employing the dimensionless cylindrical variables,
one has 
\begin{equation}
\beta _{nmk}=\frac{1}{\hbar }\;\int_{0}^{\infty }rdr\;\int_{0}^{2\pi
}d\varphi \;\int_{-\infty }^{+\infty }dz\;\psi _{0}(r,\varphi
,z)\;V(r,\varphi ,z)\;\psi _{nmk}(r,\varphi ,z)\;.  \label{15.66}
\end{equation}
In the case of a pure vortex, with $n=k=0$ and $m\neq 0$, using for the
ground state the function (\ref{15.56}) and for the vortex mode 
\begin{equation}
\psi _{0m0}(r,\varphi ,z)=\left( \frac{u_{m}^{|m|+1}}{\pi |m|!}\right)
^{1/2}\left( \frac{v_{m}}{\pi }\right) ^{1/4}r^{|m|}\;\exp \left\{ -\;\frac{1%
}{2}\;(u_{m}r^{2}+v_{m}z^{2})+im\varphi \right\} \;,  \label{15.67}
\end{equation}
with the control functions $u_{m}\equiv u_{0m0}$ and $v_{m}\equiv v_{0m0}$,
one finds 
\[
\beta _{0m0}=\frac{1}{\pi \hbar }\left( \frac{2u_{0}u_{m}^{|m|+1}}{|m|!}%
\right) ^{1/2}\;\frac{(v_{0}v_{m})^{1/4}}{(v_{0}+v_{m})^{1/2}}\times 
\]
\begin{equation}
\times \int_{0}^{\infty }\;rdr\;\int_{0}^{2\pi }\;d\varphi \;V(r,\varphi
)\;r^{|m|}\;\exp \left\{ -\;\frac{1}{2}\;(u_{0}+u_{m})r^{2}+im\varphi
\right\} \;;  \label{15.68}
\end{equation}
the resonant field being assumed to depend only on $r$ and $\varphi $.
Taking this field in the form 
\begin{equation}
V(r,\varphi )=\frac{\kappa }{p!}\;\hbar \omega _{r}r^{p}\exp (-im^{\prime
}\varphi )\;,  \label{15.69}
\end{equation}
which corresponds to the rotating potential (\ref{15.3}) with $V_{1}({\bf r}%
)\sim \cos m^{\prime }\varphi $ and $V_{2}({\bf r})\sim \sin m^{\prime
}\varphi $, one obtains 
\begin{equation}
\beta _{0m0}=\delta _{mm^{\prime }}\;\frac{\kappa \omega _{r}}{p!}\;\Gamma
\left( 1+\frac{p+|m|}{2}\right) \left( \frac{u_{0}\;u_{m}^{|m|+1}}{|m|!\;{%
\overline{u}}^{p+|m|+2}}\right) ^{1/2}\left( \frac{v_{0}v_{m}}{{\overline{v}}%
^{2}}\right) ^{1/4}\;,  \label{15.70}
\end{equation}
where 
\[
\overline{u}\equiv \frac{1}{2}\;(u_{0}+u_{m})\;,\qquad \overline{v}\equiv 
\frac{1}{2}\;(v_{0}+v_{m})\;. 
\]
For some particular cases, when $m^{\prime }=1$ and $p=0,1,2$, we get 
\[
\beta _{0m0}=\delta _{m1}\;0.964\;\kappa \omega _{r}\qquad (p=0)\;, 
\]
\[
\beta _{0m0}=\delta _{m1}\;0.588\;\kappa \omega _{r}(\nu g)^{1/5}\qquad
(p=1)\;, 
\]
\[
\beta _{0m0}=\delta _{m1}\;0.598\;\kappa \omega _{r}(\nu g)^{2/5}\qquad
(p=2)\;. 
\]

Expression (\ref{15.70}) shows that the modulating field (\ref{15.69}) will
excite vortices with the winding numbers $m$.

\section{Problems in Resonance Excitation}

In deriving the evolution equations (\ref{15.16}) for an effective two-level
system, an essential assumption was made that the coefficients $c_{n}$ are
slow functions of time, so that the inequality (\ref{15.9}) holds true. The
transition amplitude (\ref{15.20}) of the resonant field can always be taken
so that $\beta <\alpha $. Then Eqs. (\ref{15.15}) show that the time
variation $dc_{n}/dt$ is of order $\alpha $. Hence, it should be that $%
|\alpha |\ll E_{j}$. From the definition (\ref{15.13}) it follows that 
\begin{equation}
\alpha =2g\omega _{r}\int_{0}^{\infty }rdr\;\int_{0}^{2\pi }d\varphi
\;\int_{-\infty }^{+\infty }dz\;|\psi _{0}(r,\varphi ,z)|^{2}\left( 2|\psi
_{j}(r,\varphi ,z)|^{2}-|\psi _{0}(r,\varphi ,z)|^{2}\right) \;,
\label{15.71}
\end{equation}
where the index $j$ implies the triplet of quantum numbers $n,\;m,\;k$.
Calculations show that for $g\nu \gg 1$ the value of $\alpha $ can become of
order $E_{j}$. This means that the two-level picture can be a rather rough
approximation for $g\nu \gg 1$, and one would expect the ground state to be
coupled to more than one excited mode. Such a situation is analogous to the
effect of power broadening in optics \cite{Mandel95}. In order for the
two-level picture to be a good approximation, one should choose $g\nu $ not
too large. The atom-atom coupling parameter $g$, defined in Eq. (\ref{9.27}%
), is proportional to the number of atoms $N$. If the number of atoms in the
coherent state is large, $N\gg 1$, then it may be that $g\gg 1$. Hence, to
reduce the value of the product $g\nu $, one has to take small $\nu $,
making $g\nu \sim 1$. Small $\nu $ implies that the trap should have the
shape of a long cylinder.

In order to check directly that the two-level picture is a reasonable first
approximation, it is necessary to solve numerically the time-dependent
equation (\ref{15.1}). The latter, in the dimensionless units of section
9.4, acquires the form 
\begin{equation}  \label{15.72}
i\; \frac{\partial\psi}{\partial t} =\left ( \hat H + \hat V_{res}\right )
\psi \; ,
\end{equation}
where $\psi=\psi({\bf r},t)$; time is measured in units of $\omega_r^{-1}$; $%
{\bf r}=\{ r,\varphi,z\}$, with the dimensionless cylindrical variables $%
r\in[0,\infty)$, $\varphi\in [0,2\pi]$, and $z\in(-\infty,+\infty)$. The
Hamiltonian reads 
\begin{equation}  \label{15.73}
\hat H = -\; \frac{1}{2}\nabla^2 +\frac{1}{2}\; (r^2 +\nu^2 z^2) + g|\psi|^2
\; ,
\end{equation}
where 
\[
\nabla^2 = \frac{\partial^2}{\partial r^2} + \frac{1}{r}\; \frac{\partial}{%
\partial r} + \frac{1}{r^2}\; \frac{\partial^2}{\partial \varphi^2} + \frac{%
\partial^2}{\partial z^2} \; . 
\]

The resonant field $\hat V_{res}$, measured in units of $\hbar\omega_r$, can
be taken in one of the following forms, depending on the type of a
topological mode one would wish to excite: Thus, the modulating field 
\begin{equation}  \label{15.74}
\hat V_{res} = \kappa r\cos \omega t
\end{equation}
is needed for exciting the radial dipole mode. The field 
\begin{equation}  \label{15.75}
\hat V_{res} = \frac{\kappa\; r^p}{p!} [\cos(m\varphi)\cos \omega t +
\sin(m\varphi)\sin\omega t]\; ,
\end{equation}
with $p=0,1,2,\ldots$, is sufficient for exciting vortices with the winding
number $m$. And the resonance field 
\begin{equation}  \label{15.76}
\hat V_{res} =\kappa z\cos \omega t
\end{equation}
will excite the axial dipole mode. Choosing the appropriate modulating
field, one can create the related topological mode.

As initial condition to Eq. (\ref{15.72}), one has to take the ground-state
mode that can be approximately presented as 
\begin{equation}  \label{15.77}
\psi({\bf r},t) =\left ( \frac{u^2 v}{\pi^3}\right )^{1/4} \; \exp\left\{
-\; \frac{1}{2}\; (ur^2 + vz^2 ) \right\} \; .
\end{equation}
Here, the control functions $u$ and $v$ are defined by Eqs. (\ref{10.52}),
which, for the ground-state case, reduce to 
\[
\left ( 1 -\; \frac{1}{u^2}\right ) + \frac{s}{\nu}\; \sqrt{v} = 0 \; ,
\qquad \left ( 1 -\; \frac{\nu^2}{v^2}\right ) + \frac{s}{\nu\sqrt{v}} = 0\;
, 
\]
with the variable $s=2\nu g/(2\pi)^{3/2}$.

The resonance effect in the two-level picture can be noticed as follows. One
may observe the spatio-temporal behavior of the dimensionless density 
\begin{equation}
n(r,\varphi ,z,t)\equiv |\psi (r,\varphi ,z,t)|^{2}\;,  \label{15.78}
\end{equation}
studying the radial and axial cross-sections, $n(r,0,0,t)$ and $n(0,0,z,t)$.
The appearance of excited topological modes, with the spatial shape
qualitatively different from the ground-state mode (\ref{15.77}), should be
noticed in the corresponding cross-sections of the density (\ref{15.78}).

The formation of a vortex can also be noticed by studying the angular
orbital momentum 
\begin{equation}
L_{z}=-i\;\int \;\psi ^{\ast }\frac{\partial }{\partial \varphi }\;\psi \;d%
{\bf r}\;.  \label{15.79}
\end{equation}
If there are no vortices, $L_{z}=0$, while when there appears a vortex with
the winding number $m$, then $L_{z}=m$. Because of the oscillatory character
of the problem, the orbital momentum will also oscillate since $%
L_{z}=L_{z}(t)$ is a function of time. One may consider the temporal
behavior of $L_{z}(t)$. If in some moments of time the latter reaches an
integer value $m$, this would mean that there occurs the formation of a
vortex with the winding number $m$.

The problem of numerical solution can be simplified in three particular
cases. The first case is when one is interested in exciting the radial
dipole mode in a long cylindrical trap, for which $\nu \ll 1$. Then one can
limit oneself to the consideration of the wave function behavior at the
center $z=0$, assuming that at this center, the wave function practically
does not depend on $z$ and $\varphi $. This permits one to neglect the
derivatives over $z$ and $\varphi $ in Eq. (\ref{15.72}), which yields 
\[
i\;\frac{\partial \psi }{\partial t}=\left( \hat{H}_{r}+\hat{V}_{res}\right)
\psi \;, 
\]
where $\psi =\psi (r,t)\equiv \psi (r,0,0,t)$ and the radial nonlinear
Hamiltonian is 
\[
\hat{H}_{r}=-\;\frac{1}{2}\left( \frac{\partial ^{2}}{\partial r^{2}}+\frac{1%
}{r}\;\frac{\partial }{\partial r}\right) +\frac{1}{2}\;r^{2}+g|\psi
|^{2}\;. 
\]
To excite the radial dipole mode, one has to use the modulating field (\ref
{15.74}).

Another case of simplifying the computational problem is when one
investigates the excitation of a vortex mode in a long cylindrical trap, so
that again $\nu\ll 1$. Then considering, as previously, the problem at the
center $z=0$, one may assume that the wave function slowly changes. The
latter allows one to omit the derivative over $z$, which simplifies Eq. (\ref
{15.72}) to 
\[
i\; \frac{\partial \psi}{\partial t} =\left ( \hat H_\perp + \hat
V_{res}\right )\; \psi \; , 
\]
with the function $\psi=\psi(r,\varphi,t) \equiv \psi(r,\varphi,0,t)$ and
the transverse Hamiltonian 
\[
\hat H_\perp = -\; \frac{1}{2}\; \nabla_\perp^2 + \frac{1}{2}\; r^2 +
g|\psi|^2 \; , 
\]
in which 
\[
\nabla_\perp^2 =\frac{\partial^2}{\partial r^2} + \frac{1}{r}\; \frac{%
\partial}{\partial r} + \frac{1}{r^2} \; \frac{\partial^2}{\partial\varphi^2}
\; . 
\]
For the excitation of the vortex mode, one should take the resonant field (%
\ref{15.75}).

Finally, one may consider the excitation of the axial dipole mode in a
disk-shaped trap, with $\nu\gg 1$. It is then admissible to analyze the
situation at the axis $r=0$, assuming the slow dependence of the wave
function on $r$ and $\varphi$. This changes Eq. (\ref{15.72}) to 
\[
i\; \frac{\partial\psi}{\partial t} =\left ( \hat H_z + \hat V_{res}\right )
\; \psi \; , 
\]
with $\psi=\psi(z,t)\equiv\psi(0,0,z,t)$ and the axial Hamiltonian 
\[
\hat H_z = -\; \frac{1}{2} \; \frac{\partial^2}{\partial z^2} + \frac{1}{2}
\; \nu^2 z^2 + g|\psi|^2 \; . 
\]
The axial dipole mode is to be excited by the modulating resonant field (\ref
{15.76}).

In all these simplified cases, as well as for the general equation (\ref
{15.72}), one has to set boundary conditions in addition to the considered
initial conditions. These boundary conditions are rather obvious, and for
the general wave function $\psi(r,\varphi,z,t)$ they write 
\[
\lim_{r\rightarrow\infty}\; \psi(r,\varphi,z,t) = 0 \; , \qquad
\lim_{z\rightarrow\pm\infty}\; \psi(r,\varphi,z,t) = 0\; , 
\]
\begin{equation}  \label{15.80}
\psi(r,\varphi+2\pi,z,t) =\psi(r,\varphi,z,t) \; .
\end{equation}

Numerical investigation of the equations discussed in this section is yet in
process. However, the validity of the two-level picture has been proved by
direct numerical calculations for several similar problems \cite
{Williams99b,Caradoc-Davies99,Marzlin98,Ostrovskaya99,Caradoc-Davies00},
where the nonlinear Rabi-type oscillations between the ground-state and an
excited mode have been clearly observed.

Another interesting problem would be to study the possibility and
peculiarity of the resonance formation of coherent topological modes in
Bose-Einstein condensates with attractive interactions. Such condensates
exhibit oscillatory collective collapse \cite{Sackett99}. Being subject to a
resonant modulating field, the condensate should also show the nonlinear
Rabi oscillations. These two kinds of oscillations should interfere
resulting in a rather intricate behavior. It could, probably, be possible to
regulate the oscillating collapse by means of a resonant field.

\chapter{Coherence and Atom Lasers}

The possibility of realizing Bose-Einstein condensation in trapped dilute
gases demonstrates, that a macroscopic number of bosons can be produced in a
single quantum state of trapped atoms. The occupation of a single quantum
state by a large number of bosons is the matter-wave analog of the storage
of photons in a single mode of a laser cavity. A device that could emit
coherent beams of Bose atoms, similarly to the emission of photon rays by
light lasers, can be called {\it atom laser} \cite{Wiseman95}-\cite{Kneer98}
. Briefly speaking, an atom laser can be defined as a {\it device emitting
highly-directional beams of coherent atoms}. Therefore, there are two
principal questions related to the realization of atom lasers: whether the
stored bosons are prepared in a coherent state and how to form a
well-collimated beam of atoms in any desired direction.

\section{Interference and Josephson Effect}

As shown in Chapter 8, Bose-Einstein condensation in dilute gases of trapped
atoms can be understood as the macroscopic occupation of the ground-state
coherent mode. An important consequence of coherence is the occurrence of
interference phenomena. These have been observed in a nice experiment \cite
{Andrews97}, which confirms that Bose-Einstein condensed trapped atoms are
in a coherent state. In this experiment, a laser beam was used to cut a
cigar-shaped atomic cloud into two spatially separated parts. After
switching off the confining potential and the laser, the two independent
atomic clouds fall down because of gravity, expand because of atomic
repulsion, and eventually overlap. Clean interference patterns have been
observed in the overlapping region.

The appearance of interference patterns can be easily explained as follows.
Imagine that a cloud of atoms is separated into two parts whose locations
are centered at ${\bf l}_1$ and ${\bf l}_2$. Being released from the trap,
these parts move with the corresponding velocities ${\bf V}_1$ and ${\bf V}%
_2 $. Then the field operator can be presented as 
\begin{equation}  \label{16.1}
\psi({\bf r},t) = \psi_1({\bf r},t)\; e^{i{\bf k}_1\cdot{\bf r}} + \psi_2(%
{\bf r},t)\; e^{i{\bf k}_2\cdot{\bf r}} \; ,
\end{equation}
where $\psi_1$ and $\psi_2$ are the field operators of the separated
immovable parts, and $\hbar{\bf k}_j\equiv m_0{\bf V}_j$.

The interference pattern can be described by the function 
\begin{equation}  \label{16.2}
I({\bf r},t) \equiv \rho({\bf r},t) -\rho_1({\bf r},t) -\rho_2({\bf r},t) \;
,
\end{equation}
in which 
\begin{equation}  \label{16.3}
\rho({\bf r},t)\equiv\; <\psi^\dagger({\bf r},t)\; \psi({\bf r},t)>\; ,
\qquad \rho_j({\bf r},t) \equiv\; <\psi_j^\dagger({\bf r},t)\; \psi_j({\bf r}%
,t)>\; .
\end{equation}
From Eqs. (\ref{16.1}) and (\ref{16.2}) it follows that 
\begin{equation}  \label{16.4}
I({\bf r},t) = 2{\rm Re}\;\rho_{12}({\bf r},t)\; e^{-{\bf k}_{12}\cdot{\bf r}%
} \; ,
\end{equation}
where ${\bf k}_{12}\equiv{\bf k}_1 -{\bf k}_2$ and 
\begin{equation}  \label{16.5}
\rho_{12}({\bf r},t) \equiv\; <\psi_1^\dagger({\bf r},t)\;\psi_2({\bf r}%
,t)>\; .
\end{equation}
The initial separation of the cloud parts is assumed to be much larger than
the mean interatomic distance, 
\begin{equation}  \label{16.6}
l_{12}\gg a\; , \qquad l_{12} \equiv |{\bf l}_1 -{\bf l}_2|\; .
\end{equation}
If atoms are not in a coherent state, so that the coherence length is small, 
$r_{coh}\leq a$, then the correlation function (\ref{16.5}) is practically
zero, together with the interference function (\ref{16.4}). Hence, no
interference can be observed. When there exists local coherence, so that $%
r_{coh}\gg a$, then the correlation function (\ref{16.5}) does not decay so
fast, and the observation of interference becomes possible. If almost the
whole system was initially in a coherent state, so that $r_{coh}\sim
L>l_{12} $, then the correlation function (\ref{16.5}) takes the form 
\begin{equation}  \label{16.7}
\rho_{12}({\bf r},t) = \rho_{12}({\bf r})\; e^{i\omega_{12}t} \; ,
\end{equation}
in which $\hbar\omega_{12}\equiv E_1-E_2$, with $E_j$ being the energy of
the coherent mode related to a $j$-part, and 
\begin{equation}  \label{16.8}
\rho_{12}({\bf r}) \equiv N\varphi_1^*({\bf r})\varphi_2({\bf r})\; ,
\end{equation}
where $\varphi_j$ is a coherent mode located at ${\bf l}_j$. If atoms are in
the ground state, then the modes $\varphi_j$ are real. Consequently, $%
\rho_{12}({\bf r})$ is also real. Therefore, the interference function (\ref
{16.4}) becomes 
\begin{equation}  \label{16.9}
I({\bf r},t) = 2\rho_{12}({\bf r})\cos\left ( {\bf k}_{12}\cdot {\bf r}
-\omega_{12}t\right ) \; ,
\end{equation}
and one can observe clean interference patterns. In general, these patterns
will display collapses and revivals with the period 
\begin{equation}  \label{16.10}
\Delta t = \frac{2\pi}{\omega_{12}} \; .
\end{equation}
But if the energies of both separated parts are the same, $E_1=E_2$, then $%
\omega_{12}=0$, and the interference pattern is stationary, with the
interference function 
\begin{equation}  \label{16.11}
I({\bf r},t) = 2\rho_{12}({\bf r}) \cos\left ({\bf k}_{12}\cdot {\bf r}%
\right )\; .
\end{equation}

The spatial interference can be characterized by the interference fringe
spacing. Considering, say the $x$-direction, one may define the fringe
period $\Delta x=2\pi /k_{12}$. With the evident renotation 
\[
\hbar k_{12}=m_{0}V_{12}\;,\qquad V_{12}=\frac{l_{12}}{t}\;, 
\]
one obtains the fringe period 
\begin{equation}
\Delta x=2\pi \;\frac{\hbar t}{m_{0}l_{12}}\;,  \label{16.12}
\end{equation}
in agreement with the experiment \cite{Andrews97} and with the discussion in
Ref. \cite{Rohrl97}.

It is worth emphasizing that coherence is the necessary and sufficient
condition for interference. And this requires no breaking of gauge symmetry,
so that one can set $<\psi>=0$, as is elucidated in Chapter 8. The
assumption of broken gauge symmetry is only a sufficient condition for
interpreting interference, but it is not a necessary condition. Therefore it
is not correct to state, as many do, that the observation of interference
proves the existence of broken gauge symmetry. Such a statement is wrong,
since one presupposes what is alleged to be proved.

Another manifestation of coherence in trapped condensates could be the
possible occurrence of Josephson-type effects, in analogy with the known
properties of Josephson junctions in superconductors and superfluids. To
work out the physical idea, we consider again two separated condensates
confined in a double-well trap which erects a barrier between them. Then the
field operator can be written as 
\begin{equation}
\psi ({\bf r},t)=\psi _{1}({\bf r},t)+\psi _{2}({\bf r},t)\;,  \label{16.13}
\end{equation}
similarly to Eq. (\ref{16.1}), but with ${\bf k}_{j}=0$, if the condensate
on average does not moves. The physical meaning of the Josephson effect is
the manifestation of interference in the atomic current, equivalently to the
manifestation of interference in the atomic density, which is described by
the interference density function (\ref{16.2}). Thus, the interference
current is defined as 
\begin{equation}
{\bf J}({\bf r},t)=\;<{\bf j}({\bf r},t)>\;-\;<{\bf j}_{1}({\bf r},t)>\;-\;<%
{\bf j}_{2}({\bf r},t)>\;,  \label{16.14}
\end{equation}
where 
\begin{equation}
{\bf j}({\bf r},t)\equiv -\;\frac{i\hbar }{2m_{0}}\left[ \psi ^{\dagger }\;%
{\bf \nabla }\psi -({\bf \nabla }\psi ^{\dagger })\;\psi \right] \;,\qquad 
{\bf j}_{i}({\bf r},t)\equiv -\;\frac{i\hbar }{2m_{0}}\left[ \psi
_{i}^{\dagger }\;{\bf \nabla }\psi _{i}-({\bf \nabla }\psi _{i}^{\dagger
})\;\psi _{i}\right] \;.  \label{16.15}
\end{equation}
With the split operator (\ref{16.13}), the current (\ref{16.14}) is 
\begin{equation}
{\bf J}({\bf r},t)=2{\rm Re}\;<{\bf j}_{12}({\bf r},t)>\;,  \label{16.16}
\end{equation}
where 
\begin{equation}
{\bf j}_{12}({\bf r},t)\equiv -\;\frac{i\hbar }{2m_{0}}\left[ \psi
_{1}^{\dagger }\;{\bf \nabla }\psi _{2}-({\bf \nabla }\psi _{1}^{\dagger
})\;\psi _{2}\right] \;.  \label{16.17}
\end{equation}

The following argumentation is the same as when considering the density
interference. One assumes that the potential wells, separating the
condensate in two parts, are located sufficiently far from each other, in
the sense of the inequality (\ref{16.6}). Then, if the system is not
coherent, the correlation function $<{\bf j}_{12}>$ is negligible, and there
is no interference current. However, if the atomic system is in a coherent
state, then the coherent average gives 
\begin{equation}  \label{16.18}
<{\bf j}_{12}({\bf r},t)>_N \; = - i{\bf J}_{12}({\bf r})\;
e^{i\omega_{12}t} \; ,
\end{equation}
with $\hbar\omega_{12}\equiv E_1-E_2$ and 
\begin{equation}  \label{16.19}
{\bf J}_{12}({\bf r}) \equiv \frac{\hbar N}{2m_0}\left [ \varphi_1^*({\bf r}%
)\;{\bf \nabla}\varphi_2({\bf r}) - \varphi_2({\bf r})\;{\bf \nabla}%
\varphi_1^*({\bf r})\right ] \; .
\end{equation}
For the ground state, $\varphi_j$ are real, so is ${\bf J}_{12}({\bf r})$.
Therefore, the interference current (\ref{16.16}) becomes 
\begin{equation}  \label{16.20}
{\bf J}({\bf r},t) = 2 {\bf J}_{12}({\bf r})\sin\omega_{12} t\; ,
\end{equation}
which is the typical Josephson form. This current depends on time only if $%
\omega_{12}\neq 0$, so that the energies $E_1$ and $E_2$ should be different.

Note that for a coherent state, the average $<{\bf j}_i>_N$ does not depend
on time, and for the ground state, when $\varphi_i$ are real, $<{\bf j}%
_i>_N=0$. But the interference current has been defined as in Eq. (\ref
{16.14}) for generality and for closer analogy with the interference density
(\ref{16.2}).

\section{Conditions on Atom Lasers}

Defining an atom laser as a device emitting highly-directional beams of
coherent atoms, one always assumes \cite{Wiseman97} that the very first
condition on a laser is that its output is a well-collimated beam that can
be pointed in an arbitrary direction:

\vskip 2mm

(1) {\it Highly-directional beam}.

\vskip 1mm

This condition allows one to distinguish a longitudinal direction of
propagation and two transverse directions of diffraction. Good collimation
implies the smallness of the aspect ratio 
\begin{equation}  \label{16.21}
\left | \frac{r(t)}{z(t)}\right | \ll 1 \; ,
\end{equation}
in which $r(t)$ is the average transverse radius of a beam and $z(t)$, its
length. The directionality also supposes that it can be chosen arbitrarily
in space.

Characterizing the coherence of a laser, it is useful to slightly generalize
the notions introduced in section 8.5. Coherence is intimately related to
strong interatomic correlations. The information about the latter is hidden
in the correlation function 
\begin{equation}  \label{16.22}
C({\bf r},{\bf r}^{\prime},t) \equiv \frac{<\psi^\dagger({\bf r},t)\psi({\bf %
r}^{\prime},0)>} {\sqrt{\rho({\bf r},0)\rho({\bf r}^{\prime},0)}}\; ,
\end{equation}
where the density $\rho({\bf r},t)$ is the same as in Eq. (\ref{16.3}).

The {\it coherence length} can be defined as 
\begin{equation}  \label{16.23}
r_{coh}(t) \equiv \frac{\int r |C({\bf r},0,t)|d{\bf r}} {\int |C({\bf r}%
,0,t)|d{\bf r}} \; ,
\end{equation}
and the {\it coherence time} as 
\begin{equation}  \label{16.24}
\tau_{coh}({\bf r})\equiv \int_0^\infty \; |C({\bf r},0,t)|\; dt \; .
\end{equation}
As is seen, the coherence length is, generally, a function of time, while
the coherence time depends on the spatial variable. One may distinguish
spatial and temporal coherence.

\vskip 2mm

(2) {\it Spatial coherence}.

\vskip 1mm

This requires that for some period of time, 
\begin{equation}  \label{16.25}
r_{coh}(t)\gg a \; ,
\end{equation}
where $a$ is the mean interatomic distance. It is not compulsory that the
inequality (\ref{16.25}) be valid for all times, but it is sufficient that
it holds true during the time of beam emission. Thus, for a pulsing laser,
this should be the time of emitting one beam.

\vskip 2mm

(3) {\it Temporal coherence}.

\vskip 1mm

This is the condition on the coherence time, 
\begin{equation}  \label{16.26}
\tau_{coh}({\bf r}) \gg \gamma^{-1} \; ,
\end{equation}
where $\gamma$ is a spectrum linewidth. Temporal coherence is related to the
condition of {\it monochromaticity}, 
\begin{equation}  \label{16.27}
\gamma\ll \omega \; ,
\end{equation}
with $\hbar\omega$ being a characteristic atomic energy.

A simple model for an atom laser can be formulated as follows \cite{Kneer98}%
. Assume that not all atoms of the system are in a coherence state, but only 
$N_{coh}$ of them, so that the coherent field $\eta $ is normalized as 
\begin{equation}
N_{coh}=(\eta ,\eta )\;.  \label{16.28}
\end{equation}
A part of atoms, $N_{inc}$, is not coherent, for example because of
fluctuations \cite{Idziaszek99} or because of depolarizing collisions \cite
{Weiner99}. In order to take into account that the number of atoms in a trap
is not conserved, one should add to the equation for the coherent field the
terms describing atom loss and gain. This can be done by adding to the
nonlinear Hamiltonian (\ref{8.17}) the corresponding terms 
\begin{equation}
H_{gain}\equiv \frac{i}{2}\;\hbar \gamma _{+}\;N_{inc}\;,\qquad
H_{loss}\equiv -\;\frac{i}{2}\;\hbar \gamma _{-}\;.  \label{16.29}
\end{equation}
Then the evolution equation (\ref{8.16}) for the coherent field transforms
to 
\begin{equation}
i\hbar \;\frac{\partial \eta }{\partial t}=\left( -\;\frac{\hbar ^{2}}{2m_{0}%
}\;{\bf \nabla }^{2}+U\right) \eta +A|\eta |^{2}\eta +\frac{i}{2}\;\hbar
\left( \gamma _{+}\;N_{inc}-\gamma _{-}\right) \eta \;.  \label{16.30}
\end{equation}
From here, with the normalization (\ref{16.28}), it is straightforward to
get the rate equation 
\begin{equation}
\frac{d}{dt}\;N_{coh}=\left( \gamma _{+}\;N_{inc}-\gamma _{-}\right)
N_{coh}\;.  \label{16.31}
\end{equation}
The latter is to be complemented by the rate equation for incoherent atoms,
which can be taken in the form 
\begin{equation}
\frac{d}{dt}\;N_{inc}=P-(\gamma _{+}\;N_{inc}+\Gamma )N_{inc}\;,
\label{16.32}
\end{equation}
where $P$ is a pumping or generation rate and $\Gamma $ is a loss rate.

Analyzing the stationary solutions to the rate equations (\ref{16.31}) and (%
\ref{16.32}), one finds that two regimes exist, depending on the value of
the generation rate $P$ as compared to the critical threshold quantity 
\begin{equation}  \label{16.33}
P_c \equiv \frac{\gamma_-}{\gamma_+}\; \Gamma \; .
\end{equation}
For low generation rates, the stable stationary solutions are 
\begin{equation}  \label{16.34}
N_{coh}^*=0 \; , \qquad N_{inc}^*=\frac{P}{\Gamma} \qquad (P < P_c) \; ,
\end{equation}
hence, there is no stationary generation of coherent atoms. But as soon as
the generation rate $P$ exceeds the threshold (\ref{16.33}), the stable
stationary solutions become 
\begin{equation}  \label{16.35}
N_{coh}^*=\frac{P-P_c}{\gamma_-} \; , \qquad N_{inc}^*= \frac{\gamma_-}{%
\gamma_+} \qquad (P>P_c) \; .
\end{equation}
Then the steady-state number of atoms in the condensate grows linearly with
the pump rate $P$. This situation reminds the lasing threshold for
generation in photon lasers.

The model outlined above has not addressed details of the output coupling,
simply assuming the existence of a loss mechanism from the lasing mode. But
output coupling obviously constitutes a vital element of an atom laser. The
general idea of realizing output coupling is to transfer atoms, via a
radiofrequency or microwave field, from a trapped state to an untrapped
state. Being transferred to a state that is not confined by magnetic fields,
the atoms would fly out in all directions, if gravity would not force them
to fall down.

Mewes {\em et al.} \cite{Mewes97} have experimentally demonstrated precisely
such an output coupler for Bose-condensed sodium atoms. Using short resonant
pulses of radiofrequency radiation, an arbitrary percentage of the atomic
population could be transferred in a controllable manner to the output
state. Atoms in the output state simply fall down from the trap under the
action of gravity. Bloch {\em et al.} \cite{Bloch99} have demonstrated a
continuous output coupler for magnetically trapped rubidium atoms. Over a
period of up to 100 ms, atoms could be continuously extracted from
condensate by a weak radiofrequency field inducing spin flips between
trapped and untrapped states. In the untrapped state, the atoms leak out of
the trap, experiencing the action of gravity. Hagley {\em et al.} \cite
{Hagley99} extracted sodium atoms from a trapped condensate using stimulated
Raman transitions between magnetic sublevels. In the latter experiment \cite
{Hagley99}, contrary to the previous ones \cite{Mewes97,Bloch99}, optical
Raman pulses drove transitions between trapped and untrapped magnetic
sublevels, giving the output-coupled fraction of atoms a well-defined
momentum kick from the photon recoil. Because of this, atoms exited the trap
in a well-defined beam whose direction could be varied via the details of
the Raman lasers. This technique produced a device that could really be
called an atom laser, since the orientation of the laser beam did not rely
on gravity but could be selected \cite{Helmerson99}.

\section{Nonadiabatic Dynamics of Atoms}

The motion of trapped atoms is usually described as being governed by an
effective confining potential. Such a picture is equivalent to the adiabatic
approximation that is applicable for describing the stationary motion of
atoms. But when atoms escape from a trap, their motion is, certainly, not
stationary and hence, in general, it is not necessarily adiabatic. The study
of nonadiabatic dynamics of atoms is not only useful because this gives a
more general picture of atomic motion, but also because in this more general
picture some novel dynamical regimes could be found, suggesting new
mechanisms for creating highly-directional beams from atom lasers.

To derive general equations of atomic motion in a trap, one should start not
with an ad hoc introduced effective confining potential but with the
consideration of the real forces in the trap. For this purpose, one can
invoke a quantum-mechanical description based on the Hamiltonian 
\begin{equation}
\hat{H}_{N}=\sum_{i=1}^{N}\left( \frac{{\bf p}_{i}^{2}}{2m_{0}}\;-\mu _{0}%
{\bf S}_{i}\cdot {\bf B}_{i}-m_{0}{\bf g}\cdot {\bf r}_{i}\right) +\frac{1}{2%
}\;\sum_{i\neq j}\Phi _{ij}  \label{16.36}
\end{equation}
for a system of $N$ atoms, in which $\mu _{0}$ is an atomic magnetic moment, 
${\bf p}_{i}=-i\hbar {\bf \nabla }$ is a momentum operator, ${\bf S}_{i}$ is
a spin operator, ${\bf B}_{i}$ is the magnetic field formed by the trap, $%
{\bf g}$ is the gravitational acceleration, and $\Phi _{ij}$ is an
interaction potential. The evolution of this system is given by the wave
function $\Psi _{N}=\Psi _{N}({\bf r}_{1},{\bf r}_{2},\ldots ,{\bf r}_{N},t)$
satisfying the Schr\"{o}dinger equation 
\[
i\hbar \;\frac{\partial }{\partial t}\;\Psi _{N}=\hat{H}_{N}\;\Psi _{N}\;. 
\]
Note that $\Psi _{N}=[\Psi _{N}^{\sigma }]$ is a column in the space of spin
variables. For an operator $\hat{A}$ from the algebra of observables, the
quantum-mechanical average is given by the scalar product 
\[
<\hat{A}>\;\equiv \left( \Psi _{N},\hat{A}\Psi _{N}\right) \;. 
\]
The temporal behavior of this average follows from the Schr\"{o}dinger
equation giving 
\begin{equation}
\frac{d}{dt}\;<\hat{A}>\;=\;<\frac{\partial \hat{A}}{\partial t}>\;+\frac{i}{%
\hbar }\;<[\hat{H}_{N},\hat{A}]>\;.  \label{16.37}
\end{equation}
In particular cases, this yields the so-called Ehrenfest equations, that is
the equations for the mean space variable, 
\begin{equation}
\frac{d}{dt}\;<{\bf r}_{i}>\;=\frac{1}{m_{0}}\;<{\bf p}_{i}>\;,
\label{16.38}
\end{equation}
and for the mean momentum of one atom, 
\begin{equation}
\frac{d}{dt}\;<p_{i}^{\alpha }>\;=\mu _{0}<{\bf S}_{i}\cdot \frac{\partial 
{\bf B}_{i}}{\partial r_{i}^{\alpha }}>\;+\mu _{0}g^{\alpha }+f^{\alpha }\;,
\label{16.39}
\end{equation}
where $\alpha =x,y,z$ and 
\begin{equation}
{\bf f}\equiv -\sum_{j(\neq i)}^{N}\;<{\bf \nabla }_{i}\Phi _{ij}>\;.
\label{16.40}
\end{equation}
For the mean spin, one gets 
\begin{equation}
\frac{d}{dt}\;<{\bf S}_{i}>\;=\frac{\mu _{0}}{\hbar }\;<{\bf S}_{i}\times 
{\bf B}_{i}>\;.  \label{16.41}
\end{equation}

It is convenient to introduce the notation 
\begin{equation}
{\bf r}\equiv \;<{\bf r}_{i}>\;,\qquad {\bf S}\equiv \;<{\bf S}%
_{i}>\;,\qquad {\bf B}\equiv \;<{\bf B}_{i}>\;.  \label{16.42}
\end{equation}
To render the system of equations closed, one employs the semiclassical
approximation 
\begin{equation}
<S_{i}^{\alpha }B_{i}^{\beta }>\;=S^{\alpha }B^{\beta }\;,\qquad <\frac{%
\partial {\bf B}_{i}}{\partial r_{i}^{\alpha }}>\;=\frac{\partial {\bf B}}{%
\partial r^{\alpha }}\;.  \label{16.43}
\end{equation}
Then Eqs. (\ref{16.38}) and (\ref{16.39}) yield 
\begin{equation}
\frac{d^{2}r^{\alpha }}{dt^{2}}=\frac{\mu _{0}}{m_{0}}\;{\bf S}\cdot \frac{%
\partial {\bf B}}{\partial r^{\alpha }}+g^{\alpha }+\frac{f^{\alpha }}{m_{0}}%
\;,  \label{16.44}
\end{equation}
and Eq. (\ref{16.41}) results in 
\begin{equation}
\frac{d{\bf S}}{dt}=\frac{\mu _{0}}{\hbar }\;{\bf S}\times {\bf B}\;.
\label{16.45}
\end{equation}
The system of equations (\ref{16.44}) and (\ref{16.45}) is basic for
considering the dynamics of atoms in nonuniform magnetic fields \cite
{Yukalov97f}.

The total magnetic field of the trap can be taken as the sum 
\begin{equation}  \label{16.46}
{\bf B} = {\bf B}_1 + {\bf B}_2 \; ,
\end{equation}
in which the first term is the quadrupole field 
\begin{equation}  \label{16.47}
{\bf B}_1 = B_1^{\prime}(x{\bf e}_x + y{\bf e}_y +\lambda z{\bf e}_z) \; ,
\end{equation}
typical of magnetic traps, where $\lambda$ is an anisotropy parameter. If
the quadrupole field is formed by one pair of magnetic coils, then one has $%
{\bf \nabla}\cdot{\bf B}_1=0$, which gives $\lambda=-2$. However, in
general, the anisotropy parameter $\lambda $ can be varied. The second term 
\begin{equation}  \label{16.48}
{\bf B}_2 = B_2 (h_x{\bf e}_x + h_y{\bf e}_y) \; ,
\end{equation}
where $h_\alpha=h_\alpha(t)$ and 
\[
h_x^2(t) + h_y^2(t) = 1 \; , 
\]
is a transverse field often employed in magnetic traps to stabilize the
motion of atoms.

In what follows, it is convenient to switch to the dimensionless space
variable ${\bf r}=\{ x,y,z\}$ measured in units of the characteristic length 
\begin{equation}  \label{16.49}
R_0 \equiv \frac{B_2}{B_1^{\prime}} \; .
\end{equation}
To return to the dimensional Cartesian vector, one has to make the
substitution 
\[
{\bf r}\rightarrow \frac{{\bf r}}{R_0} \; . 
\]
Let us define the characteristic frequencies $\omega_1$ and $\omega_2$ by
the equalities 
\begin{equation}  \label{16.50}
\omega_1^2 \equiv \frac{\mu_0B_1^{\prime}}{m_0R_0} \; , \qquad \omega_2
\equiv \frac{\mu_0B_2}{\hbar}
\end{equation}
and introduce an effective frequency 
\begin{equation}  \label{16.51}
\omega \equiv \max_t \left | \frac{d}{dt}\; {\bf h}(t)\right | \; ,
\end{equation}
where ${\bf h}=\{ h_x,h_y,0\}$. Also, the notation 
\begin{equation}  \label{16.52}
\gamma{\vec\xi}\equiv \frac{{\bf f}}{m_0R_0}\; , \qquad {\bf G}\equiv \frac{%
{\bf g}}{R_0\omega_1^2}
\end{equation}
will be used, with $\gamma$ being a collision rate.

The force (\ref{16.40}) caused by pair interactions can be modelled by a
random force due to pair collisions. Then ${\vec\xi}$ in Eq. (\ref{16.52})
is treated as a random variable defined by the stochastic averages 
\begin{equation}  \label{16.53}
\ll\xi_\alpha(t)\gg\; = 0 \; , \qquad \ll
\xi_\alpha(t)\xi_\beta(t^{\prime})\gg \; = 2 D_\alpha\;
\delta_{\alpha\beta}\; \delta(t-t^{\prime}) \; ,
\end{equation}
where $D_\alpha$ is a diffusion rate.

In this way, Eq. (\ref{16.44}) transforms to the stochastic differential
equation 
\begin{equation}  \label{16.54}
\frac{d^2{\bf r}}{dt^2} =\omega_1^2 \left ( S^x{\bf e}_x + S^y{\bf e}_y +
\lambda S^z{\bf e}_z +{\bf G}\right ) + \gamma{\vec\xi} \; ,
\end{equation}
and Eq. (\ref{16.45}) is written in the form 
\begin{equation}  \label{16.55}
\frac{d{\bf S}}{dt} = \omega_2\hat A\; {\bf S} \; ,
\end{equation}
where $\hat A=[A_{\alpha\beta}]$ is an antisymmetric matrix with the
elements 
\[
A_{\alpha\alpha} =0 \; , \qquad A_{\alpha\beta} = -A_{\beta\alpha}\; , 
\]
\[
A_{12}=\alpha z\; , \qquad A_{23} = x+ h_x \; , \qquad A_{31}=y + h_y \; . 
\]

If one invokes for Eqs. (\ref{16.54}) and (\ref{16.55}) the adiabatic
approximation, one finds an effective confining potential being harmonic
near the trap center. In fact, the adiabatic approximation here assumes that
the spin adiabatically follows the magnetic field, which implies that $d{\bf %
S}/dt=0$. The latter leads to the equality $\hat{A}{\bf S}=0$ or ${\bf B}%
\times {\bf S}=0$. Consequently, ${\bf S}$ is aligned along ${\bf B}$, so
that one can put ${\bf S}=({\bf S}(0)\cdot {\bf B}){\bf B}/{\bf B}^{2}$.
Substituting this in Eq. (\ref{16.54}), one finds that the motion is
approximately harmonic around the trap center. But recall that the adiabatic
approximation has sense only for describing a stationary regime, when atoms
are permanently trapped. And such an approximation is, in general, invalid
for treating nonstationary regimes, e.g. when atoms fly out of the trap.

\section{Scale Separation Approach}

The evolution equations (\ref{16.54}) and (\ref{16.55}) can be treated by
employing the Scale Separation Approach \cite{Yukalov93c}-\cite{Yukalov98d},
which is a generalization of the averaging method \cite{Bogolubov61} to the
system of stochastic equations. To this end, it is necessary, first, to
classify the functional variables onto fast and slow. The latter can be done
by assuming the existence of the following small parameters 
\begin{equation}
\left| \frac{\omega _{1}}{\omega _{2}}\right| \ll 1\;,\qquad \left| \frac{%
\omega }{\omega _{2}}\right| \ll 1\;,\qquad \left| \frac{\gamma }{\omega _{2}%
}\right| \ll 1\;.  \label{16.56}
\end{equation}
Then from Eqs. (\ref{16.54}) and (\ref{16.55}) it follows that ${\bf r}$ and 
${\bf h}$ can be treated as slow, while ${\bf S}$ as fast. This permits one
to solve Eq. (\ref{16.55}) for the fast function, keeping the slow functions 
${\bf r}$ and ${\bf h}$ as quasi-invariants, that is, the matrix $\hat{A}$
can also be kept as a quasi-invariant.

For the matrix $\hat A$, one can solve the eigenproblem 
\[
\hat A{\bf b}_j = a_j{\bf b}_j \qquad (j=1,2,3) \; , 
\]
obtaining the eigenvalues 
\[
a_{1,2} = \pm i\sqrt{A_{12}^2 + A_{23}^2 +A_{31}^2} \; , \qquad a_3 = 0 
\]
and the eigenvectors 
\[
{\bf b}_j = \frac{(A_{12}A_{23}-a_jA_{31}){\bf e}_x +(A_{12}A_{31} +
a_jA_{23}){\bf e}_y +(A_{12}^2+a_j^2){\bf e}_z} {%
[(A_{12}^2-|a_j|^2)^2+(A_{12}^2+|a_j|^2)(A_{23}^2+A_{31}^2)]^{1/2}} \; . 
\]
The latter possess the properties 
\[
{\bf b}_1^*={\bf b}_2 \; , \qquad {\bf b}_3^* ={\bf b}_3 \; , \qquad {\bf b}%
_1^2={\bf b}_2^2 = 0 \; , \qquad {\bf b}_3^2 = 1 
\]
and they form an orthonormalized basis, 
\[
{\bf b}_i^*{\bf b}_j =\delta_{ij} \; , \qquad |{\bf b}_i|^2 = 1 \; . 
\]
With the matrix $\hat A$ treated as a quasi-invariant, the solution to Eq. (%
\ref{16.55}) reads 
\begin{equation}  \label{16.57}
{\bf S}(t) =\sum_{j=1}^3 \; C_j{\bf b}_j(t)\exp\{\omega_2 a_j(t)\; t\} \; ,
\end{equation}
where the coefficients 
\[
C_j ={\bf S}(0)\cdot{\bf b}_j(0) 
\]
are defined by initial conditions.

The solution (\ref{16.57}) is to be substituted in the right-hand side of
Eq. (\ref{16.54}) for the slow variable, averaging this right-hand side over
time and over the stochastic variable. In the process of the averaging, the
functions ${\bf r}$ and ${\bf h}$ should be distinguished between each other
due to the inequality 
\begin{equation}  \label{16.58}
\left | \frac{\omega_1}{\omega}\right | \ll 1
\end{equation}
that usually holds true. Then ${\bf r}$ is to be considered as slow,
compared to the fast function ${\bf h}$. Thus, the double averaging
procedure for a function $f({\bf r},{\bf h},\vec\xi,t)$ is defined as 
\begin{equation}  \label{16.59}
\lim_{\tau\rightarrow\infty} \; \frac{1}{\tau} \; \int_0^\tau\; \ll f({\bf r}%
,{\bf h}(t),\vec\xi,t)\gg \; dt \; ,
\end{equation}
where the slow variable ${\bf r}$ is kept fixed. Accomplishing this
procedure leads to the equation 
\begin{equation}  \label{16.60}
\frac{d^2{\bf r}}{dt^2} =\omega_1^2 \left ( {\bf F} + {\bf G}\right ) \; ,
\end{equation}
with the effective force 
\begin{equation}  \label{16.61}
{\bf F} = C_3\; \lim_{\tau\rightarrow\infty} \; \frac{1}{\tau} \int_0^\tau
\; \left ( b_3^x{\bf e}_x + b_3^y{\bf e}_y + \lambda b_3^z{\bf e}_z\right
)\; dt \; ,
\end{equation}
in which 
\[
C_3 =\frac{(x+h_x^0)S_0^x +(y+h_y^0)S_0^y +\lambda zS_0^z} {[(x+h_x^0)^2
+(y+h_y^0)^2 +\lambda^2 z^2]^{1/2}} \; , 
\]
\[
{\bf b}_3 =\frac{(x+h_x^0){\bf e}_x +(y+h_y^0){\bf e}_y +\lambda z{\bf e}_z%
} {[(x+h_x^0)^2 +(y+h_y^0)^2 +\lambda^2 z^2]^{1/2}} \; , 
\]
where $h_\alpha^0\equiv h_\alpha(0)$ and $S_0^\alpha\equiv S^\alpha(0)$.

As an example of the transverse field (\ref{16.48}) let us take the rotating
field, used in some traps \cite{Petrich95}, when 
\begin{equation}
h_{x}=\cos \omega t\;,\qquad h_{y}=\sin \omega t\;.  \label{16.62}
\end{equation}
Then the effective force (\ref{16.61}) becomes 
\begin{equation}
{\bf F}=\frac{[(1+x)S_{0}^{x}+yS_{0}^{y}+\lambda zS_{0}^{z}](x{\bf e}_{x}+y%
{\bf e}_{y}+2\lambda ^{2}z{\bf e}_{z})}{2[(1+2x+x^{2}+y^{2}+\lambda
^{2}z^{2})(1+x^{2}+y^{2}+\lambda ^{2}z^{2})]^{1/2}}\;.  \label{16.63}
\end{equation}

If initial conditions for the spin polarization are chosen so that $%
S_0^x\neq 0$ and $S_0^y=S_0^z=0$, then the force (\ref{16.63}) at $|{\bf r}%
|\ll 1$ reduces to the harmonic form. For $S_0^x<0$, atoms are confined in
the trap, oscillating in an effective harmonic potential. The presence of
gravity does not change much this motion, simply shifting the equilibrium
position from the trap center. This picture describes the standard motion of
trapped atoms.

Suppose that, after atoms have been trapped, their spin polarization is
prepared in the initial state 
\begin{equation}
S_{0}^{x}=S_{0}^{y}=0\;,\qquad S_{0}^{z}=S\;.  \label{16.64}
\end{equation}
This can be done, for instance, by means of a short pulse of magnetic field.
In quantum mechanics such a process is termed sudden perturbation \cite
{Fong62,AndersonEE71}. If the spin of trapped atoms was aligned along ${\bf B%
}_{2}$, then the duration of a magnetic pulse, turning spins to the
polarization (\ref{16.64}), has to be shorter than $\omega _{2}^{-1}$, and
its amplitude larger than $B_{2}$. The initial spin polarization (\ref{16.64}%
) is such that the spins are not aligned along the magnetic field ${\bf B}$.
Therefore, Eq. (\ref{16.64}) corresponds to nonadiabatic initial conditions.
Consequently, the following dynamics will also be nonadiabatic, and atoms
will not be necessarily confined, but will escape from the trap. The finite
size of the latter can be taken into account by introducing the trap shape
factor 
\begin{equation}
\Xi ({\bf r})\equiv 1-\Theta \left( x^{2}+y^{2}-R^{2}\right) \;\Theta \left(
|z|-\;\frac{L}{2}\right) \;,  \label{16.65}
\end{equation}
where the trap is assumed to have the shape of a cylinder of radius $R$ and
length $L$, with $\Theta (\cdot )$ being a unit step function. Since the
magnetic fields of the trap are supposed to act on atoms only inside the
trap, the force (\ref{16.63}), caused by these magnetic fields, should be
nonzero only inside the trap. This is easy to take into account by
multiplying (\ref{16.63}) by the shape factor (\ref{16.65}). Thus, the
effective force of the trap magnetic fields, under the initial spin
polarization (\ref{16.64}), acquires the form 
\begin{equation}
{\bf F}=\frac{1}{2}\;\lambda Suz\left( x{\bf e}_{x}+y{\bf e}_{y}+2\lambda
^{2}z{\bf e}_{z}\right) \;,  \label{16.66}
\end{equation}
in which $u=u({\bf r})$, 
\begin{equation}
u({\bf r})\equiv \frac{\Xi ({\bf r})}{[(1+2x+x^{2}+y^{2}+\lambda
^{2}z^{2})(1+x^{2}+y^{2}+\lambda ^{2}z^{2})]^{1/2}}\;.  \label{16.67}
\end{equation}
The evolution equation (\ref{16.60}), with the effective force (\ref{16.66}%
), possesses the property of invariance under the transformation 
\begin{equation}
\lambda S\rightarrow -\lambda S\;,\qquad {\bf r}\rightarrow -{\bf r}%
\;,\qquad {\bf G}\rightarrow -{\bf G}\;.  \label{16.68}
\end{equation}
Therefore, it is sufficient to consider the case of a fixed sign of $\lambda
S$, say one can fix $\lambda S>0$.

For convenience, let us introduce the dimensionless gravitational force 
\begin{equation}  \label{16.69}
{\bf \delta} \equiv \frac{{\bf G}}{\lambda S} = \frac{{\bf g}}{\lambda
SR_0\omega_1^2}\; ,
\end{equation}
and let us pass to the dimensionless time measured in units of $(\sqrt{%
\lambda S}\; \omega_1)^{-1}$. To return to the dimensional time, one has to
make the replacement 
\[
t\rightarrow \sqrt{\lambda S}\; \omega_1 t \; . 
\]
Then Eq. (\ref{16.60}) yields 
\begin{equation}  \label{16.70}
\frac{d^2x}{dt^2} =\frac{1}{2}\; uzx +\delta_x \; , \qquad \frac{d^2z}{dt^2}
=\lambda^2 uz^2 +\delta_z \; ,
\end{equation}
where the equation for the $y$-component is not written down, being of the
same form as the equation for the $x$-component.

Before analyzing Eqs. (\ref{16.70}), it is useful to give estimates for the
parameters typical of magnetic traps \cite{Han98,Petrich95}. The
characteristic frequency of the atomic motion $\omega _{1}\sim 10^{2}-10^{3}$
s$^{-1}$, the frequency of the spin motion $\omega _{2}\sim 10^{7}-10^{8}$ s$%
^{-1}$. The frequency of the transverse rotating field is $\omega \sim
10^{4}-10^{5}$ s$^{-1}$. The collision rate is $\gamma \sim 10$ s$^{-1}$.
From here 
\[
\frac{\omega _{1}}{\omega _{2}}\sim 10^{-5}\;,\qquad \frac{\omega }{\omega
_{2}}\sim 10^{-3}\;,\qquad \frac{\gamma }{\omega _{2}}\sim 10^{-6}\;,\qquad 
\frac{\omega _{1}}{\omega }\sim 10^{-2}\;, 
\]
which shows that the inequalities (\ref{16.56}) are valid, as well as the
inequality (\ref{16.58}). The characteristic length (\ref{16.49}) is $%
R_{0}\sim 0.1-0.5$ cm. For $S\sim 1$ and the gravitational acceleration $%
g\sim 10^{3}$ cm$/$s$^{2}$, by choosing appropriate $\lambda $ and $\omega
_{1}$, one can always make the dimensionless gravitational force (\ref{16.69}%
) small, so that $|\delta _{\alpha }|\ll 1$. Thus, for the typical values $%
\lambda \sim 2$, $S\sim 1$, $R_{0}\sim 0.5$ cm and $\omega _{1}\sim
10^{2}-10^{3}$ s$^{-1}$, one has $\delta _{\alpha }\sim 10^{-3}-10^{-1}$.

\section{Magnetic Semiconfinement of Atoms}

The evolution equations (\ref{16.70}) possess solutions displaying an
interesting regime of semiconfined motion, when atoms are confined from one
side of the axis $z$ but are not confined from another side \cite{Yukalov97f}%
, \cite{Yukalov97g}-\cite{Yukalov99d}. This semiconfinement is realized by
means of only magnetic fields, without involving additional laser beams
kicking atoms out and without mechanical collimators. The existence of such
a magnetic semiconfinement can be demonstrated both analytically and
numerically.

First, let us demonstrate the occurrence of semiconfinement analytically.
Since $|\delta _{\alpha }|\ll 1$, the presence of gravity does not
drastically shift the center of the atomic cloud from the trap center. So
that for atoms in the middle of the trap one can put $|{\bf r}|\ll 1$. Then
the function (\ref{16.67}) reduces to 
\begin{equation}
u({\bf r})\simeq 1\qquad (|{\bf r}|\ll 1)\;.  \label{16.71}
\end{equation}
Using this, the second of Eqs. (\ref{16.70}) can be transformed to the form 
\begin{equation}
\left( \frac{dz}{dt}\right) ^{2}=\frac{2}{3}\;\lambda ^{2}\left(
z^{3}-z_{0}^{3}\right) +2\delta _{z}(z-z_{0})+\dot{z}_{0}^{2}\;,
\label{16.72}
\end{equation}
where $z_{0}=z(0)$ and $\dot{z}_{0}=\dot{z}(0)$ are initial conditions for
the location and velocity in the $z$-direction, respectively. With the
notation 
\begin{equation}
z(t)=\frac{6}{\lambda ^{2}}\;{\cal P}(t-t_{0})\;,  \label{16.73}
\end{equation}
in which $t_{0}$ is an integration constant defined by the initial condition 
\[
z(0)=\frac{6}{\lambda ^{2}}\;{\cal P}(-t_{0})=z_{0}\;, 
\]
the form (\ref{16.72}) reduces to the Weierstrass equation 
\begin{equation}
\left( \frac{d{\cal P}}{dt}\right) ^{2}=4{\cal P}^{3}-g_{2}{\cal P}-g_{3}
\label{16.74}
\end{equation}
with the invariants 
\[
g_{2}\equiv -\;\frac{1}{3}\;\lambda ^{2}\delta _{z}\;,\qquad g_{3}\equiv 
\frac{\lambda ^{4}}{54}\;\left( \lambda ^{2}z_{0}^{3}+3\delta _{z}z_{0}-\;%
\frac{3}{2}\;\dot{z}_{0}^{2}\right) \;. 
\]
The solution to Eq. (\ref{16.74}) is called the elliptic Weierstrass
function \cite{Abramowitz72}. To analyze possible regimes of motion, it is
useful to introduce the characteristic roots $e_{i}$ ($i=1,2,3$) defined by
the equation 
\begin{equation}
4e_{i}^{3}-g_{2}e_{i}-g_{3}=0\;,  \label{16.75}
\end{equation}
whose solutions are 
\[
e_{1,2}=-\;\frac{1}{4}\;\left( C+\frac{g_{2}}{3C}\right) \pm \frac{\sqrt{3}}{%
4}\;i\left( C-\;\frac{g_{2}}{3C}\right) \;, 
\]
\begin{equation}
e_{3}=\frac{1}{2}\left( C+\frac{g_{2}}{3C}\right) \;,\qquad C^{3}\equiv
g_{3}+\sqrt{g_{3}^{2}-\;\frac{g_{2}^{3}}{27}}\;.  \label{16.76}
\end{equation}
Then Eq. (\ref{16.74}) can be written as 
\begin{equation}
\left( \frac{d{\cal P}}{dt}\right) ^{2}=4({\cal P}-e_{1})({\cal P}-e_{2})(%
{\cal P}-e_{3})\;.  \label{16.77}
\end{equation}

The properties of the characteristic roots (\ref{16.76}) depend on the sign
of the determinant $g_2^3/27-g_3^2$. There are three different cases:

\vskip 1mm

When $g_2^3<27g_3^2$, the roots $e_1$ and $e_2$ are complex conjugate, while 
$e_3$ is real. The right-hand side of Eq. (\ref{16.77}) can be presented as $%
4|{\cal P}-e_1|^2({\cal P}-e_3)$. As far as the left-hand side of Eq. (\ref
{16.77}) is nonnegative, it follows that ${\cal P}\geq e_3$ or, according to
the relation (\ref{16.73}), one gets $z\geq 6e_3/\lambda^2$.

If $g_2^3=27g_3^2$, then all three characteristic roots are real and are 
\[
e_1=e_2 =-\; \frac{1}{2}\; g_3^{1/3} \; , \qquad e_3=g_3^{1/3} \; . 
\]
Again, admissible solutions are to be such that the right-hand side of Eq. (%
\ref{16.77}) be nonnegative. This gives, as in the previous case, ${\cal P}
\geq e_3$, hence $z\geq 6e_3/\lambda^2$. Thus, in both cases considered, one
has $z\geq z_{min}$, with 
\begin{equation}  \label{16.78}
z_{min} \equiv \frac{6}{\lambda^2}\; e_3 \; .
\end{equation}
That is, the motion along the $z$-axis is confined from below by the minimal
value (\ref{16.78}), but it is not confined from above. This means nothing
but semiconfined motion. Such a type of motion is realized for $g_2^3\leq
27g_3^2$, which yields the inequality 
\[
-\delta_z^3 \leq \frac{\lambda^2}{4}\; \left ( \lambda^2 z_0^3 + 3\delta_z
z_0 -\; \frac{3}{2}\; \dot{z}_0^2\right )^2 \; . 
\]
The latter would always hold true if there were no gravity or when the trap
axis is directed along the gravitational force, that is when $\delta_z\geq 0$%
. However, if $\delta_z<0$, this inequality is valid not for all initial
conditions, though for the majority of them, since $|\delta_z|\ll 1$.

For the case $g_2^3>27g_3^2$, which is possible only for $g_2\geq 0$, hence $%
\delta_z\leq 0$, the characteristic roots (\ref{16.76}) are real and can be
written as 
\[
e_{1,2}=-\; \frac{1}{2}\sqrt{\frac{g_2}{3}}\left ( \cos\;\frac{\varphi}{3}%
\pm \sqrt{3}\sin\; \frac{\varphi}{3}\right ) \; , 
\]
\[
e_3 = \sqrt{\frac{g_2}{3}}\; \cos\; \frac{\varphi}{3}\; , \qquad
\varphi\equiv {\rm arctg}\left ( \frac{g_2^3}{27g_3^2}\; - 1\right )^{1/2}
\; . 
\]
The roots are arranged so that $e_1<e_2\leq 0<e_3$. There are two admissible
kinds of motion. One kind corresponds to $z\geq z_{min}$, with the same
minimal $z$ as in Eq. (\ref{16.78}), which is again the semiconfined motion.
And the other type corresponds to a motion confined between $e_1$ and $e_2$,
so that $e_1\leq z\leq e_2$. This means that in the whole phase space of
initial conditions, the fraction of atoms that remain confined is of order $%
e_2-e_1$, while all other atoms are semiconfined.

To estimate the fraction of atoms that remain confined, one can take into
account that $|\delta_\alpha|\ll 1$, thence $g_2\ll 1$, which shows that $%
g_3\simeq 0$. From here $\varphi\simeq \pi/2$, and the related
characteristic roots are 
\[
e_1 \simeq -\; \frac{\sqrt{g_2}}{2}\; , \qquad e_2\simeq 0 \; , \qquad
e_3\simeq \frac{\sqrt{g_2}}{2}\; . 
\]
This results in 
\[
e_2 -e_1 =\frac{\sqrt{g_2}}{2}=\frac{\lambda}{2\sqrt{3}}\; |\delta_z|^{1/2}
\; , 
\]
which for the typical values of the parameters considered above gives $%
e_2-e_1\sim 10^{-2}-10^{-1}$. Therefore the fraction of atoms that remain
confined is less than $10\%$ and can be made as small as $1\%$.

The elliptic Weierstrass function, being the solution of Eq. (\ref{16.74}),
diverges at $t\rightarrow t_{0}$, which results in the divergence of the $z$%
-variable (\ref{16.73}) according to the law 
\[
z(t)\rightarrow \frac{6}{\lambda ^{2}|t-t_{0}|^{2}}\qquad (t\rightarrow
t_{0})\;. 
\]
The characteristic time 
\begin{equation}
t_{0}=\int_{p_{0}}^{\infty }\;\frac{dp}{\sqrt{4p^{3}-g_{2}p-g_{3}}}\qquad
\left( p_{0}\equiv \frac{\lambda ^{2}}{6}\;z_{0}\right)  \label{16.79}
\end{equation}
can serve as an estimate for the {\it escape time}, that is the time after
which an atom, starting at the location $z_{0}$ at $t=0$, leaves the trap.
The estimates for the typical trap parameters give \cite{Yukalov97f} an
escape time of order $0.1$ s.

The existence of semiconfinement has been confirmed \cite
{Yukalov98e,Yukalov99b,Yukalov99d} by direct numerical solution of Eqs. (\ref
{16.70}). Several typical trajectories for the cross-section $x(t)-z(t)$ and
the related phase portraits for the velocities $\dot{x}(t)$ and $\dot{z}(t)$
are presented in Figs. \ref{Fig38} to \ref{Fig39}, for the trap axis
inclined by the 45 degrees to the horizon. The influence of gravity, as is
seen, results in curving the trajectories, similarly to what happens to
cannon shells. Stronger gravitational force bends the trajectories stronger.
But the semiconfining regime remains.

To consider the role of random collisions, described by the term $\gamma{%
\vec\xi}$ in Eq. (\ref{16.54}), one has, after substituting the fast
solution (\ref{16.57}) into Eq. (\ref{16.54}), to average the right-hand
side of the latter over time, as in the definition (\ref{16.59}), but
without averaging over the variable ${\vec\xi}$. The rotating transverse
field (\ref{16.62}), initial spin polarization (\ref{16.64}), trap shape
factor (\ref{16.65}), and all notations are assumed to be as before. Then,
instead of Eqs. (\ref{16.70}), one obtains 
\begin{equation}  \label{16.80}
\frac{d^2x}{dt^2} =\frac{1}{2}\; uzx +\delta_x + \frac{\gamma}{\lambda
S\omega_1^2}\; \xi_x \; , \qquad \frac{d^2z}{dt^2} =\lambda^2 uz^2 +\delta_z
+ \frac{\gamma}{\lambda S\omega_1^2}\; \xi_z\; .
\end{equation}
The random variables $\xi_\alpha$ are characterized by the stochastic
averages (\ref{16.53}), the second of which, for the dimensionless time used
in Eq. (\ref{16.80}), writes 
\[
\ll \xi_\alpha(t)\xi_\beta(t^{\prime})\gg \; = 2D_\alpha\;
\delta_{\alpha\beta}\; \sqrt{\lambda S}\; \omega_1\; \delta(t-t^{\prime}) \;
, 
\]
just because the time here is measured in units of $(\sqrt{\lambda S}%
\omega_1)^{-1}$.

If the main behavior of the system were governed by intensive random
collisions, then, as is evident, no ordered semiconfining regime of motion
could exist. The disorganized chaotic motion of atoms is of no interest for
the present investigation. What is important is to find conditions under
which random collisions would not much disturb the semiconfinement of atoms.
Therefore the terms in Eqs. (\ref{16.80}), which are related to random
collisions, can be treated as weak perturbations. To this end, the solutions
to Eqs. (\ref{16.80}) may be presented as the sums 
\begin{equation}
x=x_{1}+x_{2}\;,\qquad z=z_{1}+z_{2}\;,  \label{16.81}
\end{equation}
in which $x_{1}$ and $z_{1}$ are the solutions to Eq. (\ref{16.70}), while $%
x_{2}$ and $z_{2}$ are given by the linearized equations. The latter, for $|%
{\bf r}|\ll 1$, when $u({\bf r})\approx 1$, are 
\begin{equation}
\frac{d^{2}x_{2}}{dt^{2}}=\frac{1}{2}\;(z_{1}x_{2}+x_{1}z_{2})+\frac{\gamma 
}{\lambda S\omega _{1}^{2}}\;\xi _{x}\;,\qquad \frac{d^{2}z_{2}}{dt^{2}}%
=2\lambda ^{2}z_{1}z_{2}+\frac{\gamma }{\lambda S\omega _{1}^{2}}\;\xi
_{z}\;.  \label{16.82}
\end{equation}
As earlier, the equation for the $y$-component is not written down, since it
has the same form as that for the $x$-component. The solutions to Eqs. (\ref
{16.82}) can be presented as 
\[
x_{2}(t)=\int_{0}^{t}\;G_{x}(t-t^{\prime })\left[ \frac{\gamma }{\lambda
S\omega _{1}^{2}}\;\xi _{x}(t^{\prime })+\frac{1}{2}\;x_{1}z_{2}(t^{\prime })%
\right] \;dt^{\prime }\;, 
\]
\begin{equation}
z_{2}(t)=\int_{0}^{t}\;G_{z}(t-t^{\prime })\;\frac{\gamma }{\lambda S\omega
_{1}^{2}}\;\xi _{z}(t^{\prime })\;dt^{\prime }\;,  \label{16.83}
\end{equation}
where 
\[
G_{\alpha }(t)\equiv \frac{{\rm sinh}(\varepsilon _{\alpha }t)}{\varepsilon
_{\alpha }}\;,\qquad \varepsilon _{x}^{2}\equiv \frac{1}{2}\;z_{1}\;,\qquad
\varepsilon _{z}^{2}\equiv 2\lambda ^{2}z_{1}\;. 
\]
According to the properties of the random variables $\xi _{\alpha }$, one
has 
\[
\ll x_{2}\gg \;=\;\ll z_{2}\gg \;=0\;. 
\]
Calculating the mean-square deviations, one can treat $x_{1}$ and $z_{1}$ as
slow variables, keeping them as quasi-invariants. Then one obtains 
\[
\ll x^{2}\gg =\frac{\gamma ^{2}D_{x}t}{(\lambda S)^{3/2}\omega
_{1}^{3}\varepsilon _{x}^{2}}\left[ \frac{{\rm sinh}(2\varepsilon _{x}t)}{%
2\varepsilon _{x}t}\;-1\right] +\frac{\gamma ^{2}D_{x}x_{1}^{2}\varepsilon
_{x}^{4}t}{(\lambda S)^{3/2}\omega _{1}^{3}z_{1}^{2}\varepsilon
_{z}^{2}(\varepsilon _{z}^{2}-\varepsilon _{x}^{2})}\times 
\]
\[
\times \left\{ \frac{{\rm sinh}(\varepsilon _{z}t)}{\varepsilon _{z}t}\left[ 
{\rm cosh}(\varepsilon _{z}t)-{\rm cosh}(\varepsilon _{x}t)\right] -\;\frac{%
\varepsilon _{z}}{\varepsilon _{x}}\;{\rm sinh}(\varepsilon _{z}t){\rm sinh}%
(\varepsilon _{x}t)+{\rm cosh}(\varepsilon _{z}t){\rm cosh}(\varepsilon
_{x}t)-1\right\} \;, 
\]
\[
\ll z^{2}\gg \;=\frac{\gamma ^{2}D_{z}t}{(\lambda S)^{3/2}\omega
_{1}^{2}\varepsilon _{z}^{2}}\left[ \frac{{\rm sinh}(2\varepsilon _{z}t)}{%
2\varepsilon _{z}t}\;-1\right] \;. 
\]
These solutions show that the small parameter here is 
\begin{equation}
\frac{\gamma ^{2}D}{(\lambda S)^{3/2}\omega _{1}^{3}}\ll 1\;,\qquad D\equiv
\sup_{\alpha }\{D_{\alpha }\}\;.  \label{16.84}
\end{equation}
Under the inequality (\ref{16.84}), random collisions can be considered as a
weak perturbation not essentially disturbing the semiconfined motion of
atoms. Taking, for estimates, the collision rate as $\gamma \sim \hbar \rho
a_{s}/m_{0}$, where $\rho $ is the density of atoms, and the diffusion rate
as $D\sim k_{B}T/\hbar $, where $T$ is temperature, one gets from Eq. (\ref
{16.84}) the condition 
\begin{equation}
\frac{k_{B}T\hbar \rho ^{2}a_{s}^{2}}{m_{0}^{2}(\lambda S)^{3/2}\omega
_{1}^{3}}\ll 1\;.  \label{16.85}
\end{equation}
If one takes the parameters typical of experiments with $^{87}$Rb and $^{23}$
Na, that is $m_{0}\sim 10^{-22}$~g, $a_{s}\sim 5\times 10^{-7}$~cm, $\omega
_{1}\sim 10^{3}$~s$^{-1}$, $\lambda S\sim 1$, and $\rho \sim 10^{12}-10^{14}$%
~cm$^{-3}$, then the condition (\ref{16.85}) requires $T\ll T_{0}\sim
10^{-5}-10^{-1}$ K. Such temperatures are essentially higher than the
Bose-condensation temperatures for the corresponding atoms. Hence, the
Bose-condensed trapped atoms can be coupled out of the trap in the regime of
semiconfined motion. This mechanism can be employed for creating
well-collimated beams from atom lasers in arbitrary direction. Such
highly-directional beams can be formed by means of only magnetic fields.
That is why the described effect has been named the magnetic semiconfinement
of atoms \cite{Yukalov97f}, \cite{Yukalov97g}-\cite{Yukalov99d}.

\chapter{Bose-Einstein Condensate in Liquids}

In the previous chapters, Bose-Einstein condensate in trapped atomic gases
has been considered. Similar types of condensates can appear in other gases
with sufficiently weak interparticle interaction. For example, Bose-Einstein
condensation of excitons in CuCl and Cu$_{2}$O has been studied both
theoretically and experimentally \cite{Snoke90}-\cite{Hasuo93}. It has been
predicted \cite{Yukalov97b,Yukalov97c} that in dense nuclear matter the Bose
condensation of dibaryons can happen \cite{Chizhov86}-\cite{Faessler98},
which suggests the possibility of creating dibaryon lasers \cite{Yukalov98f}.

As has been demonstrated by Bogolubov \cite{Bogolubov67}, Bose-Einstein
condensation does generally occur in weakly nonideal Bose {\it gases}. But
an important question is whether the condensation remains in Bose {\it %
liquids}, that is in the systems of strongly interacting atoms. The most
known and intensively studied such liquid is superfluid $^{4}$He. Since
London \cite{London38b} and Tisza \cite{Tisza38}, it is commonly believed
that superfluidity in helium is somehow connected to Bose condensation,
although an explicit relation between the superfluid and condensate fraction
is unknown till nowadays. In this Chapter, we shall briefly touch some
problems in the theoretical description of strongly interacting systems and
will discuss the most accurate experiments aiming at measuring the
condensate fraction in superfluid helium. It is not our goal to give here a
detailed review of these topics which voluminous literature is devoted to,
but we shall sketch only some, to our mind, most interesting points, paying
attention to differences and similarities in the features of liquids and
gases.

\section{Differences between Liquids and Gases}

There are several important differences that are immediately noticeable when
comparing liquids with gases. For concreteness, liquid $^{4}$He at saturated
vapor pressure will be considered in what follows. With the density $\rho
=0.0218\AA ^{-3}$, the mean interatomic distance is $a=3.58\AA $. The
superfluid transition temperature is $T_{c}=2.17$ K. For this temperature
and mass $m_{0}=6.64\times 10^{-24}$~g, the thermal wavelength is $\lambda
_{T}=5.93$~\AA .

Here, one may notice the first difference, making it clear that at $T_c$ the
ratio $a/\lambda_T=0.6$ is not much less than one, so that inequality (\ref
{7.1}) is not valid. Respectively, $\rho\lambda_T^3=4.6$. This, however,
does not look yet too dangerous, since by lowering temperature, one always
can reach the point when $\lambda_T\gg a$.

The more warning sign is that inequality (\ref{7.2}) never holds true. To
make this transparent, one needs to define the interaction radius. For an
interaction potential $\Phi(r)$, with a hard core of radius $\sigma$, the
interaction radius is defined as 
\begin{equation}  \label{17.1}
r_{int}\equiv \sigma + \frac{\int_\sigma^\infty r\Phi(r)\;r^2dr}{%
\int_\sigma^\infty\Phi(r)\; r^2dr}\; .
\end{equation}
In the case of soft-core potentials, one can put $\sigma\rightarrow 0$. But
the interaction for helium atoms is usually described by hard-core
potentials.

The most popular is the Lennard-Jones potential 
\begin{equation}
\Phi (r)=4\varepsilon \left[ \left( \frac{\sigma }{r}\right) ^{12}-\left( 
\frac{\sigma }{r}\right) ^{6}\right] \;,  \label{17.2}
\end{equation}
in which 
\[
\varepsilon =10.22~\text{K}\qquad \sigma =2.556\;\text{\AA }\;. 
\]
Here, $\varepsilon $ is given in the Kelvin scale. The minimum of this
potential, given by the expression 
\[
r_{m}=2^{1/6}\sigma =2.87~\text{\AA }\;,\qquad \Phi (r_{m})=-\varepsilon \;, 
\]
is located at the point that is smaller than the interatomic distance $a$.
There exist also several other potentials \cite{Croxton74}. One of the best
representations of the helium interaction is produced by the Aziz potential 
\cite{Aziz79}-\cite{Aziz92} having the form 
\begin{equation}
\Phi (r)=\varepsilon \left[ A\;e^{-\alpha x-\beta x^{2}}-F(x)\left( \frac{%
c_{6}}{x^{6}}+\frac{c_{8}}{x^{8}}+\frac{c_{10}}{x^{10}}\right) \right] \;,
\label{17.3}
\end{equation}
in which 
\[
F(x)=\left\{ 
\begin{array}{cc}
\exp \left[ -\left( \frac{\delta }{x}\;-1\right) ^{2}\right] \;, & x\leq
\delta \\ 
\nonumber1\;, & x\geq \delta
\end{array}
\right. 
\]
and the dimensionless variable 
\[
x\equiv \frac{r}{r_{m}}\;,\qquad \Phi (r_{m})=-\varepsilon \;, 
\]
defines the radius normalized to the point of minimum $r_{m}$, so that 
\[
\varepsilon =10.94\;{\rm K}\;,\qquad r_{m}=2.97\;\AA \;. 
\]
The other parameters, according to the last version \cite{Aziz92}, are 
\[
A=1.922\times 10^{5}\;,\qquad \alpha =10.735\;,\qquad \beta =1.893\;,\qquad
\delta =1.414\;, 
\]
\[
c_{6}=1.349\;,\qquad c_{8}=0.414\;,\qquad c_{10}=0.171\;. 
\]

Calculating the interaction radius (\ref{17.1}) is more convenient for the
simpler Lennard-Jones potential (\ref{17.2}). This gives $%
r_{int}=2.69\sigma=6.88\AA$. Comparing it with the interatomic distance, one
has $r_{int}/a=1.9$, from where $\rho r_{int}^3=7.1$. Hence, Eq. (\ref{7.2})
is not valid, as well as the second of Eqs. (\ref{7.3}), since 
\begin{equation}  \label{17.4}
\rho r_{int}^3 \gg 1 \; .
\end{equation}
Therefore, the condensate in a liquid, where Eq. (\ref{17.4}) holds true,
should be rather depleted, if condensation can occur at all.

The third peculiarity results from the fact that inequality (\ref{17.4})
contradicts Eqs. (\ref{9.2}), because of which one cannot simplify the
consideration resorting to the cartoon potential (\ref{9.3}), but one is
doomed to operate with the full potentials like those above.

One more problem immediately arises from the previous, due to the sad
circumstance that the hard-core potentials, as the Lennard-Jones one, are
not integrable, {\em i.e.} they do not satisfy condition (\ref{7.30}).
Because of this, it is impossible to break gauge symmetry by means of the
Bogolubov prescription, as is discussed in section 7.4. Nontrivial coherent
states also do not exist for nonintegrable potentials, as is explained in
Chapter 8. Thus, atoms cannot be in pure coherent states, but can be only
partially coherent. To cope with the nonintegrability of the interaction
potentials, one has to accurately take into account interatomic
correlations, especially short-range ones. For this purpose, without
breaking gauge symmetry, one employs \cite{Manousakis85}-\cite{Clements93}
the Jastrow-type variational functions 
\begin{equation}
\Psi ({\bf r}_{1},{\bf r}_{2},\ldots ,{\bf r}_{N})=\prod_{i<j}f(r_{ij})\;%
\prod_{i<j<k}f_{3}(r_{ij},r_{jk},r_{ki})\;,  \label{17.5}
\end{equation}
in which $r_{ij}\equiv |{\bf r}_{i}-{\bf r}_{j}|$; $f(\cdot )$ is a pair
correlation function, and $f_{3}(\cdot )$ is a triplet correlation function.
The pair correlation function behaves, at short distance, as 
\begin{equation}
f(r)\sim \exp \left\{ -\;\frac{1}{2}\left( \frac{b}{r}\right) ^{5}\right\}
\qquad (r\rightarrow 0)\;,  \label{17.6}
\end{equation}
and at large distance, it has the asymptotic behavior 
\begin{equation}
f(r)\simeq 1-\left( \frac{mc}{2\pi ^{2}\hbar \rho }\right) \;\frac{1}{r^{2}}%
\qquad (r\rightarrow \infty )\;,  \label{17.7}
\end{equation}
where $c$ is the velocity of sound. Exponentially tending to zero as $%
r\rightarrow 0$, the correlation function (\ref{17.6}) smooths the
divergence of the interaction potential making the {\it smoothed potential} 
\begin{equation}
\overline{\Phi }(r)\equiv f(r)\;\Phi (r)  \label{17.8}
\end{equation}
integrable. Note that, although the Aziz potential (\ref{17.3}) is formally
finite at $r=0$, its value $\Phi (0)\sim 10^{6}$~K is so large that this
potential is also to be considered as a hard-core potential, necessarily
needing to take into account interatomic correlations smoothing its sharp
rise at $r=0$. The {\it smoothing radius} $b$ in Eq. (\ref{17.6}) can be
treated as a variational parameter or can be determined from the
Schr\"{o}dinger equation for a pair wave function \cite{Pandharipande77}- 
\cite{Yukalov89b}, from where 
\begin{equation}
b=\left( \frac{4}{5\Lambda }\right) ^{1/5}\sigma \;;  \label{17.9}
\end{equation}
here the {\it DeBoer parameter} 
\begin{equation}
\Lambda \equiv \frac{\hbar }{\sigma \sqrt{m_{0}k_{B}\varepsilon }}\;.
\label{17.10}
\end{equation}
For helium, $\Lambda =0.426$ and $b=1.13\sigma =2.89~$\AA . Correlation
functions can also be found by invoking a cumulent-type expansion in the
frame of the method of collective variables \cite{Vakarchuk79b}-\cite
{Vakarchuk90}. These functions can be optimized by solving the
Euler-Lagrange equations \cite{Manousakis85}-\cite{Manousakis91}. The
large-distance behavior of the correlation function (\ref{17.7}) is a
consequence of the existence of long-wavelength phonons \cite{Reatto66}.
After the smoothed potential (\ref{17.8}) is defined, it is possible to
develop a systematic iterative procedure for Green function equations \cite
{Yukalov98,Yukalov90b}.

The necessity of taking account of strong interatomic correlations at the
very first step of any iterative procedure is dictated by two reasons. One,
as is explained above, is the nonintegrability of the hard-core interaction
potentials, because of which the Fourier transforms of such potentials do
not exist. Another reason is that the application of simple perturbation
theory, without an appropriate account of correlations, can lead to
senseless results. As an example, we may try to calculate, by using
perturbation theory \cite{Lifshits78}, the density of Bose-condensed atoms
at zero temperature, which yields 
\[
\rho _{0}=\rho -\;\frac{1}{3\pi ^{2}\hbar ^{3}}\left[ \rho m_{0}\tilde{\Phi}%
(0)\right] ^{3/2}\;, 
\]
where $\tilde{\Phi}(k)$ is the Fourier transform of the interaction
potential. As is said above, such a transform does not exist for
nonintegrable potentials. But even assuming a soft-core potential, one has $%
\tilde{\Phi}(0)\approx k_{B}\varepsilon /\rho $, which for helium results in 
$\rho _{0}/\rho \approx -0.16$, that is a physically senseless negative
value for the condensate density.

The impossibility of applying simple perturbation theory to liquids can be
easily understood remembering that for this theory to be applicable requires
the smallness of the ratio of the mean potential energy to the mean kinetic
energy. However, for liquids, this ratio is never small, but, on the
contrary, it is usually larger than one. For helium, as follows from
theoretical calculations \cite{Vakarchuk90,Lam88,Lam88b} and experiments 
\cite{Sears83}- \cite{Bogoyavlenskii90}, this ratio is about two.

\section{Definition of Superfluid Density}

One commonly believes that superfluidity appears simultaneously with Bose
condensation, although the relation between the superfluid and condensate
fractions has never been established for liquids. Thus, at zero temperature,
all volume of helium is superfluid, while the condensate fraction does not
exceed a value of about $10\%$, and no general relation between these
fractions is known.

The condensate density is defined as the difference 
\begin{equation}  \label{17.11}
\rho_0 =\rho-\tilde\rho
\end{equation}
between the total density $\rho$ and the density of noncondensed atoms, 
\begin{equation}  \label{17.12}
\tilde\rho\equiv \frac{1}{(2\pi)^3} \int n(k)\; d{\bf k} \; , \qquad
n(k)\equiv\; <a_k^\dagger\; a_k>\; ,
\end{equation}
in which $n(k)$ is the momentum distribution.

The superfluid density can be determined by analyzing the response of the
fluid to the motion imposed by boundary conditions \cite
{Lifshits78,Mahan81,Pollock87}. For this purpose, one needs to study what
happens when the system is subject to an external perturbation, such that
the liquid starts moving uniformly with velocity ${\bf v}$. This motion
could be achieved by pushing the liquid through a tube having a pressure
difference at its ends or enclosing the system between two rotating
cylinders of radii much larger than the distance between the cylinder walls.

For a system uniformly moving with velocity ${\bf v}$, the field operator in
the laboratory frame, $\psi_v$, is connected with the field operator $\psi$
in the frame, where the system is immovable, through the Galilean
transformation 
\begin{equation}  \label{17.13}
\psi_v({\bf r}) = \psi({\bf r})\exp\left ( i\; \frac{m_0}{\hbar}\; {\bf v}
\cdot{\bf r}\right ) \; .
\end{equation}
Then the operators of observables in the laboratory frame are obtained by
taking $\psi_v$ instead of $\psi$. For instance, the Hamiltonian (\ref{8.13}%
) becomes 
\begin{equation}  \label{17.14}
H_v = H + \int\psi^\dagger({\bf r})\left ( {\bf v}\cdot\hat{{\bf p}} + \frac{%
1}{2}\; m_0 v^2 \right ) \psi({\bf r}) \; d{\bf r} \; ,
\end{equation}
where $\hat{{\bf p}}\equiv -i\hbar{\bf \nabla}$. The number-of-atoms
operator does not change, 
\begin{equation}  \label{17.15}
\hat N_v = \hat N = \int \psi^\dagger({\bf r})\; \psi({\bf r})\; d{\bf r} \;
.
\end{equation}
And the momentum operator 
\[
\hat{{\bf P}} \equiv \int \psi^\dagger({\bf r})\; \hat{{\bf p}}\psi({\bf r})
\; d{\bf r} 
\]
transforms to 
\begin{equation}  \label{17.16}
\hat{{\bf P}}_v = \hat{{\bf P}} + m_0N{\bf v} \; .
\end{equation}
Observable quantities from the algebra of observables ${\cal A}$ are given
by the average 
\begin{equation}  \label{17.17}
<{\cal A}>_v \; \equiv {\rm Tr}\; \hat\rho_v{\cal A} \; ,
\end{equation}
with the statistical operator 
\begin{equation}  \label{17.18}
\hat\rho_v \equiv \frac{\exp\{-\beta(H_v-\mu\hat N)\}} {{\rm Tr}%
\;\exp\{-\beta(H_v-\mu\hat N)\}} \; ,
\end{equation}
where $\beta\equiv(k_BT)^{-1}$. For the momentum operator (\ref{17.16}), one
gets 
\begin{equation}  \label{17.19}
<\hat{{\bf P}}_v>_v\; = \; <\hat{{\bf P}}>_v \; + m_0 N{\bf v}\; .
\end{equation}

The part of the liquid, which nontrivially responds to the perturbative
motion, defines the superfluid component with the density 
\begin{equation}  \label{17.20}
\rho_s \equiv \frac{\rho}{m_0N}\; \lim_{{\bf v}\rightarrow 0}\; \frac{%
\partial}{\partial v^\alpha}\; <\hat P_v^\alpha>_v \; .
\end{equation}
Here, it is taken into account that, for an initially isotropic system, the
density (\ref{17.20}) should not depend on the direction of the probing
velocity, that is, $\rho_s$ does not depend on the index $\alpha$.

To calculate the superfluid density (\ref{17.20}), one has to analyze the
limit ${\bf v}\rightarrow 0$. In this limit, linearizing the statistical
operator (\ref{17.18}), one finds 
\begin{equation}  \label{17.21}
\hat\rho_v \simeq \hat\rho\left [ 1 +\beta{\bf v}\cdot \left ( <\hat{{\bf P}}%
>\; - \hat{{\bf P}}\right )\right ] \; ,
\end{equation}
which for the average (\ref{17.17}) yields 
\begin{equation}  \label{17.22}
<{\cal A}>_v\; \simeq \; <{\cal A}>\; + \beta{\bf v}\cdot \left ( <\hat{{\bf %
P}}><{\cal A}>\; - \; <\hat{{\bf P}}{\cal A}>\right )\; .
\end{equation}
Here, $<{\cal A}>\equiv{\rm Tr}\;\rho{\cal A}$ implies an average in the
frame at rest.

For an isotropic system, one has 
\begin{equation}  \label{17.23}
<\hat{{\bf P}}>\; = \hbar\sum_k {\bf k}\; n(k) = 0 \; .
\end{equation}
Because of this, the statistical operator (\ref{17.21}) simplifies as 
\begin{equation}  \label{17.24}
\hat\rho_v \simeq \hat\rho \left ( 1 - \beta{\bf v}\cdot\hat{{\bf P}}\right )
\end{equation}
and the average (\ref{17.22}) reduces to 
\begin{equation}  \label{17.25}
<{\cal A}>_v\; \simeq\; <{\cal A}>\; - \beta{\bf v}\cdot <\hat{{\bf P}}{\cal %
A}> \; .
\end{equation}
For example, 
\begin{equation}  \label{17.26}
<\hat{{\bf P}}>_v\; \simeq -\beta\; <({\bf v}\cdot\hat{{\bf P}} )\; \hat{%
{\bf P}}>\; .
\end{equation}
Using this, for the average (\ref{17.19}) one obtains 
\begin{equation}  \label{17.27}
<\hat{{\bf P}}_v>_v\; \simeq m_0N{\bf v} -\beta\; <({\bf v}\cdot\hat{{\bf P}}%
)\;\hat{{\bf P}}> \; .
\end{equation}
In the case of an isotropic system, one can employ the equality 
\[
<\hat P^\alpha\hat P^\beta>\; = \delta_{\alpha\beta}\; <(\hat P^\alpha)^2>\;
= \frac{1}{3}\; \delta_{\alpha\beta}\; <\hat{{\bf P}}^2> \; , 
\]
which gives 
\[
\frac{\partial}{\partial v_\alpha} \; <\hat P^\alpha_v>_v\; \simeq m_0N -\; 
\frac{\beta}{3}\; <\hat{{\bf P}}^2> \; . 
\]
Finally, the superfluid density (\ref{17.20}) takes the form 
\begin{equation}  \label{17.28}
\rho_s = \rho -\; \frac{\rho\beta}{3m_0 N}\; <\hat{{\bf P}}^2> \; .
\end{equation}
Since one also has the relation 
\begin{equation}  \label{17.29}
\rho_s =\rho-\rho_n \; ,
\end{equation}
where $\rho_n$ is the density of the normal component, the comparison of
Eqs. (\ref{17.28}) and (\ref{17.29}) yields 
\begin{equation}  \label{17.30}
\rho_n =\frac{\rho\beta}{3m_0N}\; <\hat{{\bf P}}^2> \; .
\end{equation}
This tells that the normal component is related to the dissipated energy of
motion, while the superfluid component corresponds to nondissipative motion.

The dissipative term $<\hat{{\bf P}}^{2}>$ can be written in several forms.
It can be expressed through the momentum-momentum correlation function as 
\begin{equation}
<\hat{{\bf P}}^{2}>\;=\int <\hat{{\bf p}}({\bf r})\;\hat{{\bf p}}({\bf r}%
^{\prime })>\;d{\bf r}\;d{\bf r}^{\prime }\;,  \label{17.31}
\end{equation}
where the momentum-density operator is 
\[
\hat{{\bf p}}({\bf r})\equiv \psi ^{\dagger }({\bf r})\;(-i\hbar {\bf \nabla 
})\;\psi ({\bf r})\;. 
\]
It can also be connected with the two-particle Green function 
\[
G_{2}(1234)\equiv -\;<\hat{T}\psi (1)\psi (2)\psi ^{\dagger }(3)\psi
^{\dagger }(4)>\;, 
\]
in which $\hat{T}$ is the time-ordering operator and $\psi (j)\equiv \psi (%
{\bf r}_{j},t_{j})$. One has 
\begin{equation}
<\hat{{\bf P}}^{2}>\;=\hbar ^{2}\int \lim_{3142}({\bf \nabla }_{1}\cdot {\bf %
\nabla }_{2})\;G_{2}(1234)\;d{\bf r}_{1}\;d{\bf r}_{2}\;,  \label{17.32}
\end{equation}
where the limit means 
\[
\lim_{3142}\equiv \lim_{{\bf r}_{3}\rightarrow {\bf r}_{1}}\;\lim_{{\bf r}%
_{4}\rightarrow {\bf r}_{2}}\;\lim_{t_{i}\rightarrow t} 
\]
under the condition $t_{3}>t_{1}>t_{4}>t_{2}$. Passing to the momentum
representation by means of the Fourier transform 
\[
\psi ({\bf r})=\frac{1}{\sqrt{V}}\;\sum_{k}\;a_{k}\;e^{i{\bf k}\cdot {\bf r}%
}\;, 
\]
one gets 
\begin{equation}
<\hat{{\bf P}}^{2}>\;=\hbar ^{2}\sum_{k\;k^{\prime }}({\bf k}\cdot {\bf k}%
^{\prime })<a_{k}^{\dagger }\;a_{k}\;a_{k^{\prime }}^{\dagger
}\;a_{k^{\prime }}>\;.  \label{17.33}
\end{equation}

To find explicit expressions for the condensate density (\ref{17.11}) and
the superfluid density (\ref{17.28}), one needs to specify the problem. It
is straightforward to show how to accomplish calculations for the ideal gas
with the Hamiltonian 
\begin{equation}  \label{17.34}
H =\sum_k (\hbar\omega_k -\mu)\; a_k^\dagger\; a_k \; ,
\end{equation}
where $\omega_k$ is a particle spectrum and $\mu$, the chemical potential.
Then one has 
\[
<a_k^\dagger\; a_k\; a_{k^{\prime}}^\dagger\; a_{k^{\prime}}>\; = n(k)\;
n(k^{\prime}) -\; \frac{\delta_{k\;k^{\prime}}}{\beta\hbar}\; \frac{\partial
n(k)}{\partial\omega_k} \; , 
\]
which can be directly checked by differentiating $\partial n(k)/
\partial\omega_k$. Assuming the thermodynamic limit with the standard
replacement 
\[
\sum_k \rightarrow \; \frac{V}{(2\pi)^3} \; \int d{\bf k} \; , 
\]
and using condition (\ref{17.23}), one finds 
\begin{equation}  \label{17.35}
<\hat{{\bf P}}^2> \; = -\; \frac{\hbar V}{(2\pi)^3\beta}\; \int k^2\; \frac{
\partial n(k)}{\partial\omega_k}\; d{\bf k} \; .
\end{equation}
Hence, the normal density (\ref{17.30}) becomes 
\begin{equation}  \label{17.36}
\rho_n = -\; \frac{\hbar}{3m_0(2\pi)^3}\; \int k^2\;\frac{\partial n(k)}{%
\partial\omega_k}\; d{\bf k} \; .
\end{equation}
With the ideal-gas Hamiltonian (\ref{17.34}), the momentum distribution is 
\begin{equation}  \label{17.37}
n_k =\frac{1}{\exp\{\beta(\hbar\omega_k -\mu)\}-1} \; .
\end{equation}
Specifying the spectrum $\hbar\omega_k=\hbar^2k^2/2m_0$, one finds the
density (\ref{17.12}) of noncondensed atoms, 
\[
\tilde\rho=\left ( \frac{2m_0}{\beta\hbar^2}\right )^{3/2} \frac{e^{\beta\mu}%
}{4\pi^2} \; \int_0^\infty \; \frac{x^{1/2}dx}{e^x -e^{\beta\mu}}\; , 
\]
and the density (\ref{17.36}) of the normal component, 
\[
\rho_n=\left ( \frac{2m_0}{\beta\hbar^2}\right )^{3/2} \frac{e^{\beta\mu}}{%
6\pi^2} \; \int_0^\infty \; \frac{x^{3/2}e^x\; dx}{(e^x -e^{\beta\mu})^2}\;
. 
\]
The integrals here are related through the equality 
\[
\int_0^\infty \; \frac{x^{3/2}e^x\; dx}{(e^x -e^{\beta\mu})^2}=
-\lim_{\lambda\rightarrow 1} \; \frac{\partial}{\partial\lambda}\;
\int_0^\infty \; \frac{x^{1/2}dx}{e^{\lambda x} -e^{\beta\mu}}\; . 
\]
Below the condensation temperature, one has $\mu=0$, and the integrals
simplify to the form 
\[
\int_0^\infty\; \frac{x^{p-1}dx}{e^{\lambda x}-1} = \frac{1}{\lambda^p}\;
\Gamma(p)\;\zeta(p) \; . 
\]
In this way, one comes to the conclusion that below the condensation
temperature 
\[
T_c =\frac{2\pi\hbar^2}{m_0 k_B} \left [ \frac{\rho}{\zeta(3/2)}\right %
]^{2/3} 
\]
the density of noncondensed atoms (\ref{17.12}) and the normal density (\ref
{17.30}) coincide, 
\begin{equation}  \label{17.38}
\tilde\rho =\rho_n =\rho\left ( \frac{T}{T_c}\right )^{3/2}\; .
\end{equation}
Consequently, the condensate density (\ref{17.11}) coincides with the
superfluid density (\ref{17.29}).

Such a coincidence is to be treated rather as an occasion than as a rule,
since the general forms of the densities (\ref{17.12}) and (\ref{17.30}) are
very different. The coincidence in Eq. (\ref{17.38}) happened because of the
particular case of an ideal gas with a parabolic spectrum. If, with the same
momentum distribution (\ref{17.37}), the spectrum $\omega_k$ is slightly
changed, the coincidence of $\tilde\rho$ and $\rho_n$ will not occur. As an
illustration, one may take the phonon spectrum $\omega_k=ck$. Then the
density of noncondensed particles becomes 
\begin{equation}  \label{17.39}
\tilde\rho =\frac{\zeta(3)}{\pi^2}\left ( \frac{k_BT}{\hbar c}\right )^3 \; .
\end{equation}
The calculation of the normal density reduces to the integral 
\[
\int_0^\infty\; \frac{x^{2n}e^x\; dx}{(e^x-1)^2} = 2^{2n-1}\pi^{2n}
|B_{2n}|\; , 
\]
where $B_n$ are the Bernoulli numbers. Equation (\ref{17.36}) results in 
\begin{equation}  \label{17.40}
\rho_n =\frac{2\pi^2(k_BT)^4}{45m_0\hbar^3c^5}\; .
\end{equation}
As is seen, expressions (\ref{17.39}) and (\ref{17.40}) are neither
coinciding with nor proportional to each other, but they even have different
temperature dependence.

For a nonideal system, both the particle spectrum $\omega _{k}$ and momentum
distribution $n(k)$ differ from those of the ideal gas, as a result of which
the condensate density $\rho _{0}$ is, in general, very different from the
superfluid density $\rho _{s}$. The momentum distribution for liquid helium
has little in common with that for an ideal gas. Instead of $n(k)$, one
often considers the combination $k^{2}n(k)$. The latter, for an ideal gas
below $T_{c}$, has the maximum $2m_{0}k_{B}T/\hbar ^{2}$ at $k=0$. But for
liquid helium, the function $k^{2}n(k)$ is zero at $k=0$ and possesses a
maximum at $k\approx 0.7\AA ^{-1}$, as follows from theoretical calculations 
\cite{Manousakis85}-\cite{Manousakis91,Lam88,Lam88b} and experiments \cite
{JacksonHW74}-\cite{Sokol89}. In this way, there is no general relation
between the condensate and superfluid densities. The coincidence of these
for the ideal gas with a parabolic spectrum is rather occasional. Moreover,
this coincidence is even confusing, since the Landau criterion of
superfluidity 
\[
\min_{k}\;\frac{\varepsilon (k)}{k}>0 
\]
cannot be satisfied for a parabolic spectrum. Hence, the ideal gas should
not posses the property of superfluidity at all.

\section{Spectrum of Collective Excitations}

The most convenient technique for the theoretical description of collective
excitations in quantum liquids is the method of Green functions. The
spectrum of collective excitations is defined by the poles of the
two-particle Green function or by the poles of the density response function 
$\chi(k,\omega)$. This is equivalent to saying that the collective spectrum $%
\varepsilon(k)$ is defined by the zeros of the inverted response function $%
\chi^{-1}(k,\omega)$, that is by the equation 
\begin{equation}  \label{17.41}
\chi^{-1}(k,\varepsilon(k)) = 0 \; .
\end{equation}
This method of describing collective excitations can also be employed for
nonuniform systems, such as gases of trapped atoms. Therefore, it is worth
mentioning here some relevant points of this approach to Bose systems.

The calculation of the density response function for strongly interacting
Bose liquids, such as helium, is a very nontrivial task \cite{Sridhar87}.
Actually, there exist no reliable theoretical methods of treating strongly
interacting quantum liquids, being based on microscopic theories. Because of
this, the consideration here will be limited by weakly nonideal systems, for
which the so-called random-phase approximation is valid. This approximation
corresponds to the usage of the Hartree form for self-energy. To be more
accurate, one has to employ the {\it correlated Hartree approximation} \cite
{Yukalov98,Yukalov90b} taking account of interatomic correlations, as a
result of which the bare interaction potential is replaced by the smoothed
potential (\ref{17.8}). This is especially important for atoms interacting
through nonintegrable potentials for which the Hartree self-energy diverges
because of the divergence of the Fourier transform of the interaction
potential, while the Fourier transform 
\begin{equation}
\tilde{\Phi}(k)=\int \overline{\Phi }(r)\;e^{-i{\bf k}\cdot {\bf r}}\;d{\bf r%
}  \label{17.42}
\end{equation}
of the smoothed potential (\ref{17.8}) perfectly exists.

Considering collective excitations for the same system, it is very
instructive to compare the spectrum obtained under different assumptions, in
order to understand what would be the difference between the collective
spectra for the cases: (i) when the system is in a coherent state and when
it is incoherent, and (ii) when gauge symmetry is broken and when it is
conserved. As follows from Chapter 13, collective excitations for a weakly
nonideal Bose system in a coherent state possess the same spectrum as that
for a system with broken symmetry \cite{Bogolubov67,Bogolubov70,Lifshits78}.

For an incoherent system, with conserved gauge symmetry, the single-particle
spectrum in the correlated Hartree approximation is 
\begin{equation}
\omega (k)=\frac{k^{2}}{2m_{0}}+\rho \tilde{\Phi}(0)-\mu \;.  \label{17.43}
\end{equation}
Here and in what follows, the system of units is used where $\hbar \equiv 1$%
. Recall that the single-particle spectrum is given by the poles of the
single-particle Green function. These poles, when gauge symmetry is
conserved, are different from those of the two-particle Green function,
giving the spectrum of collective excitations. This is contrary to the case
of broken gauge symmetry when the single-particle and collective spectra
coincide \cite{Bogolubov67,Lifshits78}. The single-particle Green function
for a Bose system with conserved gauge symmetry has the form 
\[
G(k,\omega )=\frac{1+n(k)}{\omega -\omega (k)+i0}\;-\;\frac{n(k)}{\omega
-\omega (k)-i0}\;, 
\]
in which the single-particle spectrum $\omega (k)$ is defined in Eq. (\ref
{17.43}), for the approximation considered, and 
\[
n(k)=\left[ \exp \{\beta \omega (k)\}-1\right] ^{-1} 
\]
is the momentum distribution. In this case, the density response function
becomes 
\begin{equation}
\chi (k,\omega )=\frac{\Pi (k,\omega )}{1-\Pi (k,\omega )\tilde{\Phi}(k)}\;,
\label{17.44}
\end{equation}
with the polarization function 
\begin{equation}
\Pi (k,\omega )=\frac{k^{2}}{(2\pi )^{3}m_{0}}\int \;\frac{n(k^{\prime })\;d%
{\bf k}^{\prime }}{(\omega -{\bf k}\cdot {\bf k}^{\prime
}/m_{0})^{2}-(k^{2}/2m_{0})^{2}}\;.  \label{17.45}
\end{equation}
The equation (\ref{17.41}) for the spectrum of collective excitations can be
written as 
\begin{equation}
1-\Pi (k,\varepsilon (k))\;\tilde{\Phi}(k)=0\;.  \label{17.46}
\end{equation}

The polarization function (\ref{17.45}) can be simplified noticing that the
momentum distribution $n(k)$ quickly diminishes as $k$ increases. Then one
can put ${\bf k}^{\prime }=0$ in the denominator of Eq. (\ref{17.45}), which
yields 
\begin{equation}
\Pi (k,\omega )=\frac{\rho k^{2}/m_{0}}{\omega ^{2}-(k^{2}/2m_{0})^{2}}\;.
\label{17.47}
\end{equation}
Substituting this in Eq. (\ref{17.46}) results in the Bogolubov spectrum 
\begin{equation}
\varepsilon _{B}(k)=\sqrt{c^{2}(k)\;k^{2}+\left( \frac{k^{2}}{2m_{0}}\right)
^{2}}\;,  \label{17.48}
\end{equation}
in which 
\begin{equation}
c(k)\equiv \sqrt{\frac{\rho }{m_{0}}\;\tilde{\Phi}(k)}\;.  \label{17.49}
\end{equation}
If one assumes here the delta-potential, as in Eq. (\ref{9.3}), one gets the
same Bogolubov spectrum (\ref{13.8}) as for a coherent system. The same
collective spectrum (\ref{17.48}) follows for a system with broken gauge
symmetry \cite{Bogolubov67,Bogolubov70,Lifshits78}.

The approach, based on density response functions, can also be applied to a
mixture of Bose liquids defined by the Hamiltonian (\ref{14.1}). It is
straightforward to demonstrate \cite{Yukalov80b} that the spectrum of
collective excitations branches, and, for a mixture with conserved gauge
symmetry, acquires the same form (\ref{14.27}) as for a coherent mixture of
section 14.2 or for a mixture with broken gauge symmetry \cite
{Nepomnyashchy74}. The condition of dynamical stability for a binary mixture
is 
\begin{equation}
\tilde{\Phi}_{11}(0)\tilde{\Phi}_{22}(0)>\tilde{\Phi}_{12}^{2}(0)\;.
\label{17.50}
\end{equation}
For the case of the delta-potentials (\ref{14.11}), one has to replace $%
\tilde{\Phi}_{ij}(k)$ by $A_{ij}$, which reduces the inequality (\ref{17.50}%
) to condition (\ref{14.34}).

Moreover, considering the mixture with relative motion of components, it is
possible to show that the spectra of collective excitations and,
respectively, the conditions of dynamic stability are the same for a
coherent mixture discussed in section 14.4, for a mixture with broken gauge
symmetry \cite{Nepomnyashchy74}, as well as for a normal mixture with
conserved gauge symmetry \cite{Yukalov80b}. Thus, the spectrum of collective
excitations does not depend on whether the system is coherent or normal,
whether gauge symmetry is broken or conserved.

For superfluid $^{4}$He, the first spectrum of collective excitations was
proposed by Landau \cite{Landau41,Landau47} in the course of analyzing
thermodynamic properties. Feynman \cite{Feynman54} suggested a microscopic
basis for Landau's phenomenological dispersion curve, connecting the
excitations spectrum with the static structure factor $S(k)$, which resulted
in the spectrum 
\[
\varepsilon _{F}(k)=\frac{k^{2}/2m_{0}}{S(k)}\;. 
\]
Neither Landau nor Feynman mentioned the broken gauge symmetry. Bogolubov 
\cite{Bogolubov67,Bogolubov47}, deriving the excitation spectrum, introduced
gauge symmetry breaking. But, as is demonstrated above, the same Bogolubov
spectrum can be derived without breaking gauge symmetry. All of them,
Landau, Feynman and Bogolubov, considered the phonon-roton curve of
excitations in helium as a {\it unified branch}, so that it is impossible to
speak strictly of phonons and rotons as of different types of elementary
excitations. But it is more correct to speak of the phonon and roton parts
of the same unique spectrum. Note that the Bogolubov spectrum (\ref{17.48})
can reproduce the phonon-roton spectrum of liquid helium for an appropriate
interaction potential defining the effective sound velocity (\ref{17.49}).
This spectrum is reproduced under simple conditions on the Fourier transform 
$\tilde{\Phi}(k)$: 
\[
\tilde{\Phi}(0)>0\;,\qquad \min_{k}\tilde{\Phi}(k)=\tilde{\Phi}%
(k_{r})<0\;,\qquad \lim_{k\rightarrow \infty }\tilde{\Phi}(k)=0\;, 
\]
where $k_{r}$ corresponds to the point of roton minimum. Such conditions are
easy to achieve even for rather simple potentials \cite{Brueckner57}.

The difficulty of calculating the spectrum of collective excitations for the
realistic strongly interacting liquids, such as helium, prompted some
authors to construct phenomenological or semiphenomenological models. We
shall not give here a complete survey of these models but will mention only
one of them, which recently provoked a vivid discussion. This is the model
advanced by Glyde and Griffin \cite{Glyde90}-\cite{Glyde92b}. The basic
assumption of this model is that there are in superfluid $^{4}$He two
principally different branches of excitations: one is the phonon branch due
to density excitations at low wave vectors, and another part is the
quadratic single-particle branch at higher wave vectors. These two branches
exist independently of each other, so that they remain above as well as
below the temperature of superfluid transition $T_{\lambda }$. But below $%
T_{\lambda }$, these branches become coupled via the appeared Bose
condensate accompanied by broken gauge symmetry. However, the existence of
two separate branches apparently contradicts the unified picture of Landau,
Feynman, and Bogolubov. By exact microscopic consideration, Nepomnyashchy 
\cite{Nepomnyashchy92} showed that model propagators, employed in the
discussed model \cite{Glyde90}-\cite{Glyde92b}, are not consistent with the
general structure of Green function equations, while the latter support the 
{\it unified nature of the phonon-roton spectrum}. The temperature
dependence of the excitation spectrum was studied experimentally \cite
{Montfrooij94,Svensson96}. These experiments demonstrated that there is no
indication of a well-defined single-particle branch, becoming the roton
mode, that would suddenly appear as one goes below $T_{\lambda }$. Contrary
to this, the superfluid transition is marked by a complete softening of the
roton mode and its rapid attenuation, when one approaches $T_{\lambda }$
from below. Above $T_{\lambda }$, the roton mode continues to an overdamped
diffusive mode of zero frequency. In this way, experiments \cite
{Montfrooij94,Svensson96}, as is concluded by their authors, imply ''{\it a
qualitative disagreement with the interpretation proposed by Glyde and
Griffin}''.

Thus, the phonon-roton spectrum of superfluid helium must be considered as a
unified branch. The question remains whether there could exist some remnants
of low-frequency quasiparticle excitations {\it in addition} to the
phonon-roton branch, with the energies below the broad multiphonon
component. There have been some theoretical arguments \cite
{Yukalov80b,Griffin86}-\cite{Yarunin98} concerning the possible existence of
an additional quasiparticle excitation branch. In a series of papers \cite
{Blagoveshchenskii93}-\cite{Bogoyavlenskii97}, the authors find experimental
indications that an excitation branch, additional to the phonon-roton
spectrum, could exist. However, the experiments have not yet been confirmed
by other groups. It is worth emphasizing that, even if some additional
excitation branch does exist, one has, first of all, to understand its
physical origin and, second, no interpretation should contradict the fact
that the phonon-roton curve is a unified branch \cite{Nepomnyashchy92}-\cite
{Svensson96}.

The phonon-roton spectrum of superfluid $^{4}$He has been carefully studied
in many experiments (see review \cite{Woods73}). Its commonly accepted form,
at saturated-vapor pressure and low temperature $T\leq 1.2$ K, is presented
in Fig. \ref{Fig40}. In the long-wave limit, one has the phonon spectrum 
\[
\varepsilon (k)\simeq c_{0}\;k\;,\qquad c_{0}=2.38\times 10^{4}\; {\rm cm/s}
\;. 
\]
In the vicinity of the roton minimum, the dispersion curve is 
\[
\varepsilon (k)\simeq \Delta _{r}+\frac{(k-k_{r})^{2}}{2m_{r}}\;, 
\]
with $\Delta _{r}=8.6$ K, $k_{r}=1.9\;\AA ^{-1}$, $m_{r}=0.16m_{0}$. But let
us stress it again that the phonon and roton parts of the spectrum are the
pieces of a unified branch.

The phonon-roton spectrum terminates at around $k=3.5\;\AA ^{-1}$, becoming
unstable with respect to the decay of excitations into several other
excitations with lower energies \cite{Pitaevskii59}-\cite{Nepomnyashchy74b}.

\section{Dynamic Structure Factor}

The spectrum of collective excitations can be experimentally measured by
means of neutron scattering described by the double differential cross
section 
\begin{equation}
\frac{d^{2}\sigma }{d\Omega \;d\omega }=b_{s}^{2}\;\frac{k_{f}}{k_{i}}%
\;S(k,\omega )\;,  \label{17.51}
\end{equation}
where $b_{s}$ is the scattering length of a neutron on a helium atom, $k_{i}$
and $k_{f}$ are the initial and final wavevectors of the scattering neutron,
and $k$ and $\omega $ are the momentum and energy transfer from the neutron
to the sample. The dynamics of the liquid are contained in the {\it dynamic
structure factor} 
\begin{equation}
S(k,\omega )=\frac{1}{2\pi \rho }\;\int \left[ R({\bf r},t,0,0)-\rho ^{2}%
\right] \;e^{-i({\bf k}\cdot {\bf r}-\omega t)}\;d{\bf r}\;dt\;,
\label{17.52}
\end{equation}
in which 
\begin{equation}
R({\bf r},t,{\bf r}^{\prime },t^{\prime })\equiv \;<\hat{\rho}({\bf r},t)\;%
\hat{\rho}({\bf r}^{\prime },t^{\prime })>  \label{17.53}
\end{equation}
is the density-density correlation function, with the density operator 
\[
\hat{\rho}({\bf r},t)\equiv \psi ^{\dagger }({\bf r},t)\;\psi ({\bf r},t)\;. 
\]
Using the Fourier integral 
\[
R({\bf r},t,{\bf r}^{\prime },t^{\prime })=\frac{1}{(2\pi )^{4}}\;\int
R(k,\omega )\;e^{i{\bf k}\cdot ({\bf r}-{\bf r}^{\prime })-i\omega
(t-t^{\prime })}\;d{\bf k}\;d\omega \;, 
\]
one gets 
\begin{equation}
S(k,\omega )=\frac{1}{2\pi \rho }\;\left[ R(k,\omega )-(2\pi )^{4}\;\rho
^{2}\;\delta ({\bf k})\;\delta (\omega )\right] \;.  \label{17.54}
\end{equation}
Employing the properties of Green functions \cite{Yukalov98}, one can find 
\begin{equation}
R(k,\omega )=-\;\frac{2{\rm Im}\;\chi (k,\omega )}{1+e^{-\beta \omega }}%
+(2\pi )^{4}\;\rho ^{2}\;\delta ({\bf k})\;\delta (\omega )\;.  \label{17.55}
\end{equation}
Then for the dynamic structure factor (\ref{17.54}) one has 
\begin{equation}
S(k,\omega )=-\;\frac{{\rm Im}\;\chi (k,\omega )}{\pi \rho (1+e^{-\beta
\omega })}\;.  \label{17.56}
\end{equation}
The response function $\chi (k,\omega )$, on the complex $\omega $-plane,
possesses the spectral representation 
\begin{equation}
\chi (k,\omega )=\frac{1}{2\pi }\;\int_{-\infty }^{+\infty }\;\frac{\kappa
(k,\omega ^{\prime })}{\omega -\omega ^{\prime }}\;d\omega ^{\prime }\;,
\label{17.57}
\end{equation}
in which the spectral function is 
\begin{equation}
\kappa (k,\omega )=i\;\left[ \chi (k,\omega +i0)-\chi (k,\omega -i0)\right]
\;.  \label{17.58}
\end{equation}
From the properties of Green functions \cite{Yukalov98} it follows that 
\begin{equation}
{\rm Im}\;\chi (k,\omega )=-\;\frac{1}{2}\;\kappa (k,\omega )\;{\rm coth}%
\left( \frac{\beta \omega }{2}\right) \;.  \label{17.59}
\end{equation}
Hence, Eq. (\ref{17.56}) can be written as 
\begin{equation}
S(k,\omega )=\frac{\kappa (k,\omega )}{2\pi \rho (1-e^{-\beta \omega })}\;.
\label{17.60}
\end{equation}
The latter, with the notation for the Bose function 
\[
b(\omega )\equiv \left( e^{\beta \omega }-1\right) ^{-1}\;, 
\]
takes the form 
\begin{equation}
S(k,\omega )=\frac{1+b(\omega )}{2\pi \rho }\;\kappa (k,\omega )\;.
\label{17.61}
\end{equation}

The dynamic structure factor satisfies the following sum rules 
\begin{equation}  \label{17.62}
\int_{-\infty}^{+\infty} S(k,\omega)\; d\omega = S(k) \; ,
\end{equation}
which defines the static structure factor $S(k)$, 
\begin{equation}  \label{17.63}
\int_{-\infty}^{+\infty}\omega S(k,\omega)\; d\omega = E_k \; ,
\end{equation}
which gives the kinetic energy 
\begin{equation}  \label{17.64}
E_k \equiv \frac{k^2}{2m_0} \; ,
\end{equation}
and 
\begin{equation}  \label{17.65}
\int_{-\infty}^{+\infty} \;\frac{S(k,\omega)}{\omega} \; d\omega = -\; \frac{%
{\rm Re}\;\chi(k,0)}{2\rho} \; .
\end{equation}

The relation of the dynamic structure factor to the density-density
correlation function and to the density response function means that these
are the density fluctuations which contribute to $S(k,\omega)$. In their
turn, the density fluctuations define the spectrum of collective
excitations, because of which the dynamic structure factor is directly
related to the latter. This relation can be clearly illustrated using the
random-phase approximation for the density response function (\ref{17.44})
and the form (\ref{17.47}) for the polarization function, which yields 
\begin{equation}  \label{17.66}
\chi(k,\omega) =\frac{2\rho E_k}{\omega^2-\varepsilon^2(k)} \; ,
\end{equation}
where $\varepsilon(k) =\varepsilon_B(k)$ is the spectrum of collective
excitations in the Bogolubov approximation (\ref{17.48}). With Eq. (\ref
{17.66}), the spectral function (\ref{17.58}) becomes 
\begin{equation}  \label{17.67}
\kappa(k,\omega) =\frac{2\pi\rho E_k}{\varepsilon(k)} \; \left [
\delta(\omega-\varepsilon(k)) - \delta(\omega +\varepsilon(k)) \right ] \; .
\end{equation}
Then the dynamic structure factor (\ref{17.61}) is 
\begin{equation}  \label{17.68}
S(k,\omega) = [ 1+b(\omega)] \; \frac{E_k}{\varepsilon(k)} \; \left [
\delta(\omega -\varepsilon(k)) - \delta(\omega +\varepsilon(k))\right ] \; .
\end{equation}
For the static structure factor (\ref{17.62}) one gets 
\begin{equation}  \label{17.69}
S(k) =\frac{E_k}{\varepsilon(k)} \; {\rm coth}\; \frac{\beta\varepsilon(k)}{2%
} \; .
\end{equation}
The sum rule (\ref{17.63}) is identically valid, and from Eq. (\ref{17.65})
one has 
\begin{equation}  \label{17.70}
\int_{-\infty}^{+\infty} \; \frac{S(k,\omega)}{\omega} \; d\omega = \frac{E_k%
}{\varepsilon^2(k)} \; ,
\end{equation}
in agreement with the form (\ref{17.66}).

Expression (\ref{17.68}) shows that the dynamic structure factor has a sharp
peak at the frequency $\omega$ coinciding with the spectrum of collective
excitations $\varepsilon(k)$. The delta-function shape of this peak is the
result of the simplicity of the approximation used. In reality, the observed
peaks are, of course, finite and can be fitted to the measured data by means
of the Lorentzian or Gaussian forms.

\section{Measurement of Condensate Fraction}

The dynamic structure factor, as is shown above, gives information on the
spectrum of collective excitations in liquid helium. Hohenberg and Platzman 
\cite{Hohenberg66} suggested that this factor can also be used for
extracting the value of the condensate fraction 
\begin{equation}
n_{0}\equiv \frac{\rho _{0}}{\rho }\;.  \label{17.71}
\end{equation}
For this purpose, one has to invoke deep-inelastic neutron scattering with
very high transferred momenta $k$, such that the scattering could be treated
as occurring on single atoms and the scattering atoms could be assumed to be
in a free particle state. This implies that the recoil energy $k^{2}/2m_{0}$
must be much larger than the mean potential energy $E_{pot}$ of an atom, 
\begin{equation}
k\gg \sqrt{2m_{0}E_{pot}}\;.  \label{17.72}
\end{equation}
For superfluid helium, with $m_{0}=6.64\times 10^{-24}$~g, this gives $k\gg
10$~\AA $^{-1}$, that is one should have $k\approx 100$~\AA $^{-1}$. Then,
for the dynamic structure factor, the impulse approximation is valid
yielding 
\begin{equation}
S_{IA}(k,\omega )=\frac{1}{\rho }\;\int n(p)\;\delta (\omega
-E_{k+p}+E_{p})\;\frac{d{\bf p}}{(2\pi )^{3}}\;,  \label{17.73}
\end{equation}
where $E_{k}$ is defined in Eq.~(\ref{17.64}). Substituting the momentum
distribution 
\begin{equation}
n(k)=(2\pi )^{3}\;n_{0}\rho \;\delta ({\bf k})+\tilde{n}(k)  \label{17.74}
\end{equation}
in the impulse approximation (\ref{17.73}), one has 
\begin{equation}
S_{IA}(k,\omega )=S_{0}(k,\omega )+S_{n}(k,\omega )\;,  \label{17.75}
\end{equation}
with the terms 
\[
S_{0}(k,\omega )=n_{0}\delta (\omega -E_{k})\;, 
\]
\[
S_{n}(k,\omega )=\frac{1}{\rho }\;\int \tilde{n}(p)\;\delta (\omega
-E_{k+p}+E_{p})\;\frac{d{\bf p}}{(2\pi )^{3}}\;. 
\]
Hence, the existence of a condensate should result in the appearance of a
sharp peak in $S_{0}(k,\omega )$ above the broad distribution due to $%
S_{n}(k,\omega )$.

For the deep-inelastic scattering, it is convenient to use the West \cite
{West75} scaling variable 
\begin{equation}
Y\equiv \frac{m_{0}}{k}\;(\omega -E_{k})  \label{17.76}
\end{equation}
and to define the so-called Compton profile 
\begin{equation}
J(k,Y)\equiv \frac{k}{m_{0}}\;S(k,\omega )\;,  \label{17.77}
\end{equation}
whose name comes from the initial usage of such variables in electron
scattering. The convenience of using the profile (\ref{17.77}) is due to the
fact that at high momenta it tends to a value 
\begin{equation}
J(Y)\simeq J(k,Y)\qquad (k\rightarrow \infty )\;,  \label{17.78}
\end{equation}
which does not depend on $k$. Thus, for the impulse approximation (\ref
{17.73}), one gets 
\begin{equation}
J_{IA}(Y)\equiv \frac{k}{m_{0}}\;S_{IA}(k,\omega )=\frac{1}{(2\pi )^{2}\rho }%
\;\int_{|Y|}^{\infty }\;p\;n(p)\;dp\;.  \label{17.79}
\end{equation}
Inverting the last equation gives the momentum distribution 
\begin{equation}
n(k)=-(2\pi )^{2}\;\frac{\rho }{k}\;\frac{\partial }{\partial k}%
\;J_{IA}(k)\;.  \label{17.80}
\end{equation}
If this distribution would have the form (\ref{17.74}), one could directly
measure the condensate fraction $n_{0}$.

However, there exist several principal difficulties prohibiting the
extraction of the momentum distribution from the observed scattering. First,
any experimental observation is affected by the statistical uncertainty of
the measurements. These uncertainties will translate into uncertainties in
the inferred $n(k)$. The most striking feature of the inferred momentum
distribution is the increase in the statistical noise near $k=0$, due to the
division by $k$ in Eq. (\ref{17.80}). Even very large differences in $n(k)$
at small $k$ only cause small changes in the Compton profile $J(Y)$. Thus,
the statistical noise present in $J(Y)$ allows a whole family of $n(k)$ that
are consistent with the observed data \cite{Sokol89}. The predicted small $k$
singular behavior makes little contribution to the observed scattering, and
with the experimental techniques now available, will be difficult, if not
impossible, to observe. Due to the finite statistical errors inherent in any
experiment, the experimental results can not definitely prove the existence
of a condensate, which formally corresponds to a $\delta $-function. Some
other singular, or even not singular, behavior, but not a condensate, could
be responsible for the increase in the scattering at small $k$ observed in
the superfluid \cite{Sokol89}. Hence, the experimental results can not rule
out a ground state which does not contain a condensate or which corresponds
to something like a smeared condensate \cite{Shenoy77}. The measured
scattering is consistent with many different forms for $n(k)$, including
models that do not include a condensate at all \cite{JacksonHW74,Sosnick89}.

Another weak point in the attempts to measure the condensate fraction in
superfluid helium is the usage of the impulse approximation (\ref{17.73}),
which assumes that helium atoms behave as free particles. The latter
requires that the transferred momenta satisfy inequality (\ref{17.72}),
being about $100\;\AA ^{-1}$ for liquid helium. However, the majority of
neutron-scattering experiments have been performed at momentum transfers not
higher than $23\;\AA ^{-1}$. Some experiments \cite{Ikeda87} used the
transferred momenta as high as $150\;\AA ^{-1}$, but the accuracy of these
measurements was so low that it did not allow one to decide anything about
the value of $n_{0}$.

The fact that helium atoms inside a liquid are not free but strongly
interact with their surrounding leads to what one calls the final-state
effects \cite{Silver88}-\cite{Carraro90} and the initial-state effects \cite
{Mayers89}-\cite{Mayers90}. The former can be taken into account by defining
a convolution 
\begin{equation}
J(Y)=\int_{-\infty }^{+\infty }F(Y-Y^{\prime })\;J_{IA}(Y^{\prime
})\;dY^{\prime }  \label{17.81}
\end{equation}
of the impulse-approximation result with a final-state broadening $F(Y)$
that is to be calculated from a microscopic model \cite{Silver88,Silver89}.
Taking account of the initial-state effects requires to change the
definition of the scattering variable (\ref{17.76}) itself \cite{Stringari87}%
.

One more problem which is to be taken into account is that what one actually
measures is not the profile (\ref{17.81}), but the effects of instrumental
resolution must be involved in order to determine the true scattering. In
general, the instrumental broadening is a complicated function depending on
the energy and momentum transfer and the instrument geometry, and a simple
closed-form expression for the resolution function is not possible. In the
case of helium, an effective resolution function $I(Y)$ can be calculated by
a Monte Carlo simulation of the spectrometer. In terms of this instrumental
resolution function, the observed broadened Compton profile is given by the
convolution 
\begin{equation}  \label{17.82}
J_{obs}(Y) = \int_{-\infty}^{+\infty} I(Y-Y^{\prime})\; J(Y^{\prime})\;
dY^{\prime}\; ,
\end{equation}
where $J(Y)$ is defined by Eq. (\ref{17.81}).

In interpreting experimental scattering data, one usually does the following 
\cite{Sosnick90,Herwig90,Sosnick91}. Rather than attempt to deconvolute the
instrumental resolution and the final-state broadening, one assumes a {\it %
model profile} $J_{mod}(Y)$, which is substituted in Eq. (\ref{17.81})
instead of the profile (\ref{17.79}). After this, one fits the convolutions
with $J_{mod}(Y)$ to the observed scattering profile $J_{obs}(Y)$. The most
often employed model profile \cite{Herwig90,Sosnick91} is a sum of Gaussians 
\begin{equation}
J_{mod}(Y)=\frac{A_{1}}{\sqrt{2\pi }\;\sigma }\;\exp \left\{ -\;\frac{%
(Y-Y_{0})^{2}}{2\sigma _{1}^{2}}\right\} +\frac{A_{2}}{\sqrt{2\pi }\;\sigma
_{2}}\;\exp \left\{ -\;\frac{(Y-Y_{0})^{2}}{2\sigma _{2}^{2}}\right\} \;,
\label{17.83}
\end{equation}
whose amplitudes, widths, and common center may be varied. This form is,
certainly, not unique, and many other forms could be used to fit the data.
These two Gaussians model the two terms in the dynamic structure factor (\ref
{17.75}). The term that is narrower is assumed to model the condensate peak $%
S_{0}(k,\omega )$, while the wider Gaussian is supposed to model $%
S_{n}(k,\omega )$. Thus for superfluid helium at $T=0.35$ K, one finds \cite
{Sosnick91} $\sigma _{1}=0.95\;\AA ^{-1}$ and $\sigma _{2}=0.29\;\AA ^{-1}$,
so that the latter width should be related to the condensate. But it is
worth noting that even for normal helium the observed scattering is not well
characterized by a single Gaussian, and a sum of two Gaussians much better
describes the observed scattering. For example, for normal helium at $T=3.5$
K, one has \cite{Sosnick91} $\sigma _{1}=1\;\AA ^{-1}$ and $\sigma
_{2}=0.45\;\AA ^{-1}$. Therefore, the two-Gaussian model may show not the
appearance of a condensate but just non-Gaussian behavior of the momentum
distribution \cite{Herwig90,Sosnick91}.

In this way, the original goal for much of the work with liquid helium, a
direct observation of the condensate fraction, has not come to pass. In view
of the current understanding of the final-state effects in helium, it is
unlikely that this goal will ever be reached in deep inelastic neutron
scattering experiments \cite{Sosnick90,Sokol89,Sosnick91}. While the current
experimental results do not definitely prove the existence of a condensate,
they do provide indirect evidence for its existence, which agrees with many
theoretical calculations predicting $n_{0}\approx 10\%$ at zero temperature.

Several other ways have been suggested for indirectly extracting information
on the value of the condensate fraction; the interpretation of such methods
being based on model assumptions. Sears \cite{Sears83} tried to determine $%
n_{0}$ by assuming a relation between the value of the mean kinetic energy 
\[
<E_{k}>\;=\int \;\frac{k^{2}}{2m_{0}}\;n(k)\;\frac{d{\bf k}}{(2\pi )^{3}} 
\]
at $T=T_{\lambda }$ and that value at $T<T_{\lambda }$. The mean kinetic
energy could be determined by using the impulse approximation for the
dynamic structure factor, 
\begin{equation}
<E_{k}>\;=\lim_{k\rightarrow \infty }\;\frac{3\pi }{4E_{k}}\;\int_{-\infty
}^{+\infty }(\omega -E_{k})^{2}S_{IA}(k,\omega )\;d\omega \;.  \label{17.84}
\end{equation}
Campbell \cite{Campbell83} suggested to consider a relation between the
condensate fraction and the surface tension of superfluid helium. Wyatt \cite
{Wyatt98} studied quantum evaporation from the free surface of liquid $^{4}$%
He. The mentioned ways of determining the condensate fraction, being based
on several model assumptions, provide the {\it upper limit} for $n_{0}$.

An interesting proposal was made by Cummings, Hyland, and Rowlands \cite
{Cummings70}-\cite{Cummings81} who advanced the relation 
\begin{equation}
\rho ^{2}[g(r)-1]=\rho _{n}^{2}[g_{n}(r)-1]\;,  \label{17.85}
\end{equation}
assumed to be valid for $r\geq 4.5\;\AA ^{-1}$ and connecting the pair
correlation function 
\begin{equation}
g(|{\bf r}-{\bf r}^{\prime }|)\equiv \frac{1}{\rho ^{2}}\;<\psi ^{\dagger }(%
{\bf r})\;\psi ^{\dagger }({\bf r}^{\prime })\;\psi ({\bf r}^{\prime
})\;\psi ({\bf r})>\;,  \label{17.86}
\end{equation}
measured at $T<T_{\lambda }$, with the pair correlation function $g_{n}(r)$
identified as the function either just above $T_{\lambda }$ or that function
extrapolated to the temperature under consideration. The pair correlation
function (\ref{17.86}) and the density-density correlation function (\ref
{17.53}) are connected as 
\begin{equation}
R({\bf r},t,{\bf r}^{\prime },t^{\prime })=\rho ^{2}\;g(|{\bf r}-{\bf r}
^{\prime }|)+\rho \;\delta ({\bf r}-{\bf r}^{\prime })\;.  \label{17.87}
\end{equation}
From Eqs. (\ref{17.54}) and (\ref{17.62}), it follows that 
\begin{equation}
g(r)=1+\frac{1}{(2\pi )^{3}\rho }\;\int [S(k)-1]\;e^{i{\bf k}\cdot {\bf r}
}\;d{\bf k}\;.  \label{17.88}
\end{equation}
Therefore, the pair correlation function can be calculated by using Eq. (\ref
{17.88}) with the measured static structure factor $S(k)$. Then, by
substituting $\rho _{n}=\rho -\rho _{0}$ into the relation (\ref{17.85}),
one has 
\begin{equation}
n_{0}=1-\left[ \frac{g(r)-1}{g_{n}(r)-1}\right] ^{1/2}\;.  \label{17.89}
\end{equation}
This method of calculating the condensate fraction was employed together
with the data for the pair correlation function obtained through neutron
scattering \cite{Sears79,Svensson81} and $x$-ray scattering techniques \cite
{Wirth83,Wirth87}. The values of $n_{0}$, found by applying Eq. (\ref{17.89}%
), are in good agreement with those obtained by other methods. However, the
derivation of the relation (\ref{17.85}) was criticized by several authors 
\cite{Griffin80}, \cite{Chester80}-\cite{Griffin87}. The main argument
against this relation is that the latter does not appropriately take into
account the anomalous averages existing in a system with broken gauge
symmetry. But if gauge symmetry is conserved, the relation (\ref{17.85}) can
be approximately valid \cite{Yukalov81} in the region $4\;\AA
^{-1}<r<12\;\AA ^{-1}$.

An accurate analysis of different experimental methods of measuring $n_{0}$
was done by Wirth and Hallock \cite{Wirth87}. They fitted each of the sets
of experimental data to the function 
\begin{equation}
n_{0}(T)=n_{0}(0)\left[ 1-\left( \frac{T}{T_{\lambda }}\right) ^{\alpha }%
\right] \;.  \label{17.90}
\end{equation}
While there is little theoretical justification for the use of this form for
liquid helium, it provides a uniform methodology for obtaining values $%
n_{0}(0)$. Summarizing the results of various experiments, one has $%
n_{0}(0)\approx 0.10$ and $5\leq \alpha \leq 10$.

\vskip5mm

{\Large {\bf Concluding Remarks}}

\vskip 3mm

Bose-Einstein condensation of trapped atoms is now a very vast and quickly
developing branch of physics. Because it is so vast, it is impossible to
touch, on a reasonable level of explanation, all related directions in one
review. This especially concerns theoretical aspects. Therefore, we
preferred to concentrate on the principal points which the theory of
nonuniform Bose systems is based on. We have tried to clearly elucidate
these main points. The choice of the most important problems is, of course,
subjective, and many interesting questions concerning Bose atoms were left
aside. The theoretical description of the degenerate trapped Fermi atoms 
\cite{Bruun99,Bruun99b} has not been touched at all, as well as the
description of trapped Bose-Fermi mixtures \cite{Miyakawa99}.

The majority of theoretical considerations here have been based on the
Gross-Pitaevskii equation. Temperature effects were only slightly touched.
This is because of the following reasons. First of all, it was necessary to
concentrate on the principal features of Bose-Einstein condensate at zero or
low temperatures, and a detailed discussion of its thermal properties would
essentially enlarge the review. Another reason is that there are not yet
enough reliable experiments on trapped atoms with Bose-Einstein condensates
at finite temperatures which theory could be compared with. Thermal
properties of trapped atoms are to be studied more accurately, both
theoretically as well as experimentally.

\vskip1cm

{\bf Acknowledgment}

\vskip 3mm

We are very grateful to our colleagues for the kind permission to use
figures from their papers. Several theoretical results included in this
review have been obtained in collaboration with E.P.~Yukalova whose help is
very much appreciated. We acknowledge financial support from the S\~{a}o
Paulo State Research Foundation FAPESP. Ph.W.~C. wishes to thank the
Deutscher Akademischer Austauschdienst DAAD for financial support.

\bibliographystyle{LaserPhys}
\bibliography{slava,vander}

\backmatter
\appendix

\newpage

\begin{center}
{\LARGE Figure Captions}
\end{center}

\begin{enumerate}
\item[{\bf Fig. \ref{FigBoseCondensation}.}]  
\end{enumerate}

Condensed fraction and heat capacity at the phase transition for a
homogeneous gas (dotted line) and for a harmonically trapped gas (solid
line).

\begin{enumerate}
\item[{\bf Fig. \ref{FigPeakDensity}.}]  
\end{enumerate}

Peak density at the phase transition for a harmonically trapped ideal
Bose-gas of $10^{6}$ rubidium atoms. The trap secular frequency is set to $%
\omega _{trap}=2\pi 16$~Hz.

\begin{enumerate}
\item[{\bf Fig. \ref{FigHeatCapacity}.}]  
\end{enumerate}

Heat capacity at the phase transition for $N=100$\ (continuous line), $%
N=1000 $\ (dashed line) and $N=10000$\ (dotted line).

\begin{enumerate}
\item[{\bf Fig. \ref{FigAbsorptionPictures}.}]  
\end{enumerate}

Time-of-flight absorption pictures above (a), slightly below (b), and well
below the phase transition (c) (figures taken from \cite{Han98}).

\begin{enumerate}
\item[{\bf Fig. \ref{FigHydrogenSpectrum}.}]  
\end{enumerate}

Two-photon absorption spectrum of hydrogen. The narrow Doppler-free peak at
negative detunings and the broad Doppler-sensitive peak at positive
detunings acquire characteristic shoulders when a BEC is present (courtesy
of \cite{Fried98}).

\begin{enumerate}
\item[{\bf Fig. \ref{FigReleaseEnergy}.}]  
\end{enumerate}

Measurement of the scaled release energy per particle versus reduced
temperature at the phase transition. Straight line is ideal Boltzmann-gas,
dashed line finite number ideal Bose-gas \cite{Grossmann95} and solid curved
line fit to the data (courtesy of \cite{Ensher96}).

\begin{enumerate}
\item[{\bf Fig. \ref{FigAspectRatio}.}]  
\end{enumerate}

Temporal evolution of the aspect ratio of suddenly released BECs. The
cloverleaf trap had the trapping frequencies $\omega _{r}=2\pi \cdot 248$~Hz
and $\omega _{z}=2\pi \cdot 16$~Hz (courtesy of \cite{Ketterle98}).

\begin{enumerate}
\item[{\bf Fig. \ref{FigCompressionOscillations}.}]  
\end{enumerate}

Compression oscillations in experiment (dots) and theory (solid line) along
the radial and axial directions (courtesy of \cite{Matthews98}).

\begin{enumerate}
\item[{\bf Fig. \ref{FigRabiTwist}.}]  
\end{enumerate}

Measured (upper curve) and calculated (lower curve) Rabi oscillations of the
space-integrated fractional population of the lower hyperfine state
(courtesy of \cite{Williams99}).

\begin{enumerate}
\item[{\bf Fig. \ref{FigImpurityScattering}.}]  
\end{enumerate}

Impurity scattering within a BEC. Elastic collisions between the condensate
and impurity atoms traveling at $6$~cm$/$~s (towards the left in images)
distributed the momentum of the collision partners over a sphere showing up
as a halo in $50$~ms time-of-flight absorption images (a). In Fig.~(b) a
Stern-Gerlach type magnetic field gradient has been applied to separate the $%
m_{F}=0$ atoms from the $m_{F}=-1$ condensate. The fringes are an imaging
artifact (courtesy of \cite{Chikkatur00}).

\begin{enumerate}
\item[{\bf Fig. \ref{FigVortex}.}]  
\end{enumerate}

Density distribution (a) of the vortex state (the visible atoms are in the
upper hyperfine state), (b) after a $\pi /2$\ pulse, and (c) after a $\pi $\
pulse (the visible atoms are in the lower hyperfine state). The images (d)
and (e) visualize the phase slip around the vortex (courtesy of \cite
{Matthews99b}).

\begin{enumerate}
\item[{\bf Fig. \ref{FigVortexLattice}.}]  
\end{enumerate}

Array of $7$, $8$, and $11$ vortices in a Bose-Einstein condensate stirred
by a laser beam. The absorption image was taken after a $27$~ms period of
free expansion (courtesy of \cite{Madison00}).

\begin{enumerate}
\item[{\bf Fig. \ref{FigSoliton}.}]  
\end{enumerate}

Dark solitons in a Bose-Einstein condensate. The images (A to E) show
experimental measurements, and the images (F to J) are calculated density
distributions for various times after a phase imprint of $1.5\pi $ on the
top half of the condensate. A positive density disturbance moved rapidly in
the $+x$ direction, and a dark soliton moved oppositely and significantly
slower than the speed of sound (reprinted with permission from \cite
{Denschlag00}).

\begin{enumerate}
\item[{\bf Fig. \ref{FigStimulationCurves}.}]  
\end{enumerate}

Bosonic stimulation. The curves show the growth of the condensate towards
thermal equilibrium after a sudden initial desequilibration for various
initial numbers of condensed atoms (courtesy of \cite{Miesner98}).

\begin{enumerate}
\item[{\bf Fig. \ref{FigInterferenceScheme}.}]  
\end{enumerate}

Scheme of the setup for interference observation. A cigar-shaped condensate
is built in a cloverleaf trap, it is split into two parts with a
blue-detuned far-off resonance laser beam, suddenly released from the trap
and partially illuminated by a laser light sheet. The interference patterns
are recorded by absorption imaging.

\begin{enumerate}
\item[{\bf Fig. \ref{FigInterferenceFringes}.}]  
\end{enumerate}

Interference patterns of two released condensates recorded with the setup
sketched in Fig.~\ref{FigInterferenceScheme} for three different values of
the height of the potential barrier ({\em i.e.} intensity of the laser light
sheet that separates the trapped condensates). The three pictures on the
right hand side are calculated patterns \cite{Wallis98} (courtesy of \cite
{Ketterle98}).

\begin{enumerate}
\item[{\bf Fig. \ref{FigOutputCoupling}.}]  
\end{enumerate}

Output coupling of parts of a BEC by irradiation of radiofrequency pulses
(courtesy of \cite{Mewes97}).

\begin{enumerate}
\item[{\bf Fig. \ref{FigBraggDiffractionScheme}.}]  
\end{enumerate}

Bragg scattering for matter-waves. The figure on the left shows the
geometric arrangement used in the experiments \cite
{Kozuma99,Stenger99,Stamper-Kurn99b}. Short pulses of Raman beams enclosing
the angle $\vartheta $ and detuned by $\Delta \omega $\ from one another are
shone into the BEC. The figure in the right shows the parabolic dispersion
relation, which strictly holds only in the limit of negligible mean-field
interaction.

\begin{enumerate}
\item[{\bf Fig. \ref{FigBraggSpectroscopy}.}]  
\end{enumerate}

Bragg spectroscopy of recoil-induced resonances. Fig.~(a) shows the shift
(solid line) and halfwidth (gray area) of the RIR in the case of
particle-like excitations, $\varepsilon _{part}(p)=h\cdot 100$~kHz, as a
function of density. Fig.~(b) shows the RIR shift $\Delta \varepsilon $\ and
Fig.~(c) shows the RIR strength $S(p)$, {\em i.e.} the fraction of atoms
deflected into the first Bragg order, for phonon-like excitations, $%
\varepsilon _{phon}(p)=h\times 1.54$~kHz.

\begin{enumerate}
\item[{\bf Fig. \ref{FigFourWaveMixing}.}]  
\end{enumerate}

Four-wave mixing can be illustrated in the laboratory frame (a), in the
moving frame defined by $p_{1}=-p_{3}$\ (b), and in the moving defined by $%
p_{1}=-p_{2}$\ (c) and accordingly be interpreted in different ways (see
text).

\begin{enumerate}
\item[{\bf Fig. \ref{FigFourWaves}.}]  
\end{enumerate}

False color absorption picture of \ the atomic density distribution after
4WM after $6$~ms time of flight. The newly created wavepacket $\psi _{4}$\
is smaller than the others (reprinted with permission from \cite{Deng99}).

\begin{enumerate}
\item[{\bf Fig. \ref{FigRaleighWaves}.}]  
\end{enumerate}

Superradiant Rayleigh scattering (reprinted with permission from \cite
{Inouye99}). The time-of-flight images show the momentum distribution of the
condensate after irradiation of a single laser pulse polarized
perpendicularly to the long axis and having the durations (a) $35$~$\mu $s,
(b) $75$~$\mu $s and (c) $100$~$\mu $s. For the longer pulse durations,
repeated scattering processes give rise to additional peaks.

\begin{enumerate}
\item[{\bf Fig. \ref{FigFeshbachResonance}.}]  
\end{enumerate}

Feshbach resonance in collisions of ground-state $^{85}$Rb atoms \cite
{Courteille98b}. The atoms collide in the $f=2+f=2$\ channel (scattering
wavefunction $u_{coll}$). A vibrational bound state of the $f=3+f=3$\
channel has almost the same energy (wavefunction $u_{res}$). As the energies
are tuned to resonance, the wavefunction $u_{res}$\ is resonantly enhanced.

\begin{enumerate}
\item[{\bf Fig. \ref{FigFeshbachScatteringLength}.}]  
\end{enumerate}

Magnetic field dependence of the scattering length close to the strong
Feshbach resonance near $156$~G\ in $^{85}$Rb atoms. The shaded area
emphasizes the range of positive values of the scattering length.

\begin{enumerate}
\item[{\bf Fig. \ref{FigTwoPhotonPhotoassociation}.}]  
\end{enumerate}

Free-bound-bound two-photon photoassociation in $^{87}$Rb. While two $%
^{2}S_{1/2},f=1,m_{f}=-1$\ ground state atoms in the hyperfine state are
colliding, they may undergo a photoassociative Raman transition to the bound
vibrational state $v=-2,l=0,F=2,m_{F}=-2$\ located$\ 636.0094$~MHz\ below
the ionization threshold. The intermediate excited state is $v,J=0$\ at $%
12555$~cm$^{-1}$\ of the$\ 0_{g}^{-}$\ potential connected to the $%
^{2}S_{1/2}-^{2}P_{1/2}$\ asymptote. The levels are chosen to optimize the
Franck-Condon overlap.

\begin{enumerate}
\item[Fig. {\bf \ref{Fig23}.}]  
\end{enumerate}

The ground-state energy for the one-dimensional nonlinear Schr\"{o}dinger
equation. The self-similar approximant $E_{\ast }(g)$ (solid line) is given
by Eq.~(\ref{11.25}), the crossover approximant $E_{2}^{\ast }(g)$ (dashed
line) is defined in Eq.~(\ref{11.21}), and the Thomas-Fermi approximant $%
E_{TF}(g)$ (dashed line with diamonds) is the energy (\ref{11.28}).

\begin{enumerate}
\item[{\bf Fig. \ref{Fig24}.}]  
\end{enumerate}

The density (\ref{11.29}) for the corresponding wave functions in the
self-similar approximation (\ref{11.24}) (solid line), Gaussian
approximation (\ref{11.26}) (dashed line), and Thomas-Fermi approximation (%
\ref{11.27}) (dashed line with diamonds) for different coupling parameters:
(a) $g=0.2$; (b) $g=1$; (c) $g=5$; (d) $g=20$; (e) $g=50$; (f) $g=100$.

\begin{enumerate}
\item[{\bf Fig. \ref{Fig25}.}]  
\end{enumerate}

The residual $R(x)$ defined in Eq.~(\ref{11.30}) for the self-similar
solution (\ref{11.24}) (solid line), Gaussian solution (\ref{11.26}) (dashed
line), and Thomas-Fermi solution (\ref{11.27}) (dashed line with diamonds)
for several coupling parameters: (a) $g=5$; (b) $g=50$; (c) $g=100$.

\begin{enumerate}
\item[{\bf Fig. \ref{Fig26}.}]  
\end{enumerate}

The ground-state energy of atoms confined in a spherically symmetric trap:
The self-similar approximant (\ref{11.39}) (solid line); second-order
crossover approximant (\ref{11.32}) (dashed line); and the Thomas-Fermi
energy (\ref{9.34}) (short-dashed line).

\begin{enumerate}
\item[{\bf Fig. \ref{Fig27}.}]  
\end{enumerate}

Percentage errors of the first crossover approximants for the ground-state
energy of a spherical trap: $E_{1}^{\ast }(s)$ (solid line); $E_{2}^{\ast
}(s)$ (dashed line); $E_{3}^{\ast }(s)$ (short-dashed line).

\begin{enumerate}
\item[{\bf Fig. \ref{Fig28}.}]  
\end{enumerate}

Percentage errors of the higher crossover approximants for the ground-state
energy of a spherical trap: $E_{2}^{\ast }(s)$ (solid line); $E_{3}^{\ast
}(s)$ (dashed line); $E_{4}^{\ast }(s)$ (short-dashed line); $E_{5}^{\ast
}(s)$ (dotted line).

\begin{enumerate}
\item[{\bf Fig. \ref{Fig29}.}]  
\end{enumerate}

Percentage errors of $E_{1}^{\ast }$ (solid line), $E_{2}^{\ast }$ (dashed
line) and $E_{3}^{\ast }$ (short-dashed line) as functions of the coupling $%
g $ for several energy levels and trap shapes: (a) $\nu =0.1,\;n=m=k=0$
(ground-state); (b) $\nu =0.1,\;n=k=0,\;m=1$ (vortex state); (c) $\nu
=0.1,\;n=3,\;m=2,\;k=1$; (d) $\nu =100,\;n=m=k=0$; (e) $\nu
=100,\;n=k=0,\;m=2$.

\begin{enumerate}
\item[{\bf Fig. \ref{Fig30}.}]  
\end{enumerate}

The ground-state energy of atoms confined in a cylindrical trap with $\nu
=10 $: The optimized approximant (\ref{10.54}) (solid line) and the
Thomas-Fermi energy (\ref{9.34}) (dashed line).

\begin{enumerate}
\item[{\bf Fig. \ref{Fig31}.}]  
\end{enumerate}

The vortex energies as functions of $\nu g$: the basic-vortex energy $\Omega
_{010}$ (solid line) and the energy $\Omega _{020}$ of the vortex with the
winding number $m=2$ (dashed line).

\begin{enumerate}
\item[{\bf Fig. \ref{Fig32}.}]  
\end{enumerate}

The self-similar crossover approximants $f_{k}^{\ast }(r)$ for the vortex as
compared to exact numerical data marked by diamonds: $f_{1}^{\ast }(r)$ is
shown by the solid line; $f_{2}^{\ast }(r)$, by the long-dashed line; $%
f_{3}^{\ast }(r)$, by the short-dashed line; and $f_{4}^{\ast }(r)$ is
presented by the dotted line.

\begin{enumerate}
\item[{\bf Fig. \ref{Fig33}.}]  
\end{enumerate}

The fractional populations $n_{0}(t)$ (dashed line) and $n_{j}(t)$ (solid
line) as functions of dimensionless time, measured in units of $\alpha ^{-1}$%
. The transition amplitude is fixed, $b=0.4999$, and the detuning is varied:
(a) $\delta =0$; (b) $\delta =0.0001$; (c) $\delta =0.0001001$; (d) $\delta
=0.00011$.

\begin{enumerate}
\item[{\bf Fig. \ref{Fig34}.}]  
\end{enumerate}

The time dependence of the fractional populations $n_{0}(t)$ (dashed line)
and $n_{j}(t)$ (solid line) under the fixed detuning $\delta =0$ and varied
transition amplitude: (a) $b=0.45$; (b) $b=0.4999$; (c) $b=0.5$; (d) $%
b=0.5001$; (e) $b=0.6$; (f) $b=1$.

\begin{enumerate}
\item[{\bf Fig. \ref{Fig35}.}]  
\end{enumerate}

Excitation of the radial dipole mode with the quantum numbers $%
n=1,\;m=0,\;k=0$ with the parameters $g=100,\;\nu =10,\;b=0.4,\;\delta =0.01$%
. The ground-state density $\rho _{0}$ (solid line) and the density $\rho
_{100}$ (dashed line) as functions of the radial variable $r$ at the point $%
z=0$ for different times measured in units of $\alpha ^{-1}$: (a) $t=0$; (b) 
$t=2$; (c) $t=4$.

\begin{enumerate}
\item[{\bf Fig. \ref{Fig36}.}]  
\end{enumerate}

Excitation of the vortex mode with $n=0,\;m=1,\;k=0$ with the same
parameters as in Fig.~38. The ground-state density $\rho _{0}$ (solid line)
and the density $\rho _{010}$ (dashed line) as functions of the radial
variable $r$ at the point $z=0$ for different times: (a) $t=2$; (b) $t=4$.

\begin{enumerate}
\item[{\bf Fig. \ref{Fig37}.}]  
\end{enumerate}

Excitation of the axial dipole mode, with $n=0,\;m=0,\;k=1$ with the same
parameters as in Fig.~38. The ground-state density $\rho _{0}$ (solid line)
and the density $\rho _{001}$ (dashed line) as functions of the axial
variable $z$ at the point $r=0$ for different times: (a) $t=0$; (b) $t=2$;
(c) $t=4$.

\begin{enumerate}
\item[{\bf Fig. \ref{Fig38}.}]  
\end{enumerate}

Phase portrait for the period of time $0\leq t\leq 50$ for atoms starting
from the trap center $x_{0}=y_{0}=z_{0}=0$ with velocities $\dot{x}_{0},\; 
\dot{y}_{0},\;\dot{z}_{0}$ varied in the interval $[-0.1,0.1]$. The trap
parameters are $R=10,\;L=10$, and $\lambda =20$. Note that the picture
practically does not change upon independently varying the trap radius and
length between $10$ and $100$. The gravity parameters are $\delta
_{x}=0.01,\;\delta _{y}=0,\;\delta _{z}=-0.01$. Shown are: (a) trajectories;
(b) velocities.

\begin{enumerate}
\item[{\bf Fig. \ref{Fig39}.}]  
\end{enumerate}

Trajectories and velocities during the period of time $0\leq t\leq 50$ for
atoms with the same initial conditions as in Fig.~41, but for the trap
parameters $R=1,\;L=1$, and $\lambda =20$, and for the gravity parameters $%
\delta _{x}=0.05,\;\delta _{y}=0,\;\delta _{z}=-0.05$. Here: (a)
trajectories; (b) velocities.

\begin{enumerate}
\item[{\bf Fig. \ref{Fig40}.}]  
\end{enumerate}

Spectrum of collective excitations in superfluid $^{4}$He at saturated-vapor
pressure and low temperature. The energy $\varepsilon (k)$ is measured in K
and the wave vector $k$ in $\AA ^{-1}$.

\newpage

\begin{center}
{\LARGE Table Captions}
\end{center}

\begin{enumerate}
\item[{\bf Table \ref{TabCriticalTemperature}.}]  
\end{enumerate}

Critical temperature, condensed fraction, heat capacity and its
discontinuity at the phase transition for various trapping potentials. $V$\
denotes a three-dimensional and $S$\ a two-dimensional volume.

\begin{enumerate}
\item[{\bf Table \ref{TabBoseSpecies}.}]  
\end{enumerate}

Nuclear spin, scattering lengths and transition parameters for various
isotopes. The fifth, sixth and seventh column give the linewidth and the
transition frequencies of the $D1$\ and $D2$\ lines, where applicable. The
last column gives the ground state hyperfine splitting, where applicable.

\begin{enumerate}
\item[{\bf Table \ref{TabExcitationRegimes}.}]  
\end{enumerate}

Characteristic length scales for elementary excitations.

\newpage

\begin{center}
{\bf Table 1}
\end{center}

\vskip2cm

\begin{tabular}{||l|l|l|l|l||}
\hline\hline
$U({\bf r})$ & $\eta $ & $k_{B}T_{c}^{0}$ & $C(T_{c}^{-})/Nk_{B}$ & $\Delta
C(T_{c}^{0})/Nk_{B}$ \\ \hline\hline
3D box & $\frac{3}{2}$ & $\left[ \frac{Nh^{3}}{(2\pi m)^{3/2}}\frac{1}{%
Vg_{3/2}(1)}\right] ^{2/3}$ & $1.92$ & $0$ \\ \hline
$\left( \frac{z}{a}\right) ^{2}$ & $2$ & $\left[ \frac{Nh^{3}}{(2\pi m)^{3/2}%
}\frac{1}{\pi ^{2}aSg_{2}(1)}\right] ^{1/2}$ & $4.38$ & $0$ \\ \hline
$\left\{ 
\begin{array}{cc}
\frac{z}{a}, & z>0 \\ 
\infty , & z<0
\end{array}
\right\} $ & $\frac{5}{2}$ & $\left[ \frac{Nh^{3}}{(2\pi m)^{3/2}}\frac{1}{%
aS1.4}\right] ^{2/5}$ & $6.88$ & $3.35$ \\ \hline
$\left( \frac{x}{a}\right) ^{2}+\left( \frac{y}{b}\right) ^{2}+\left( \frac{z%
}{c}\right) ^{2}$ & $3$ & $\left[ \frac{Nh^{3}}{(2\pi m)^{3/2}}\frac{1}{\pi
^{3/2}abcg_{3}(1)}\right] ^{1/3}=\hbar \omega _{trap}\left( \frac{N}{g_{3}(1)%
}\right) ^{1/3}$ & $10.82$ & $6.57$ \\ \hline\hline
\end{tabular}

\newpage

\begin{center}
{\bf Table 2}
\end{center}

\vskip2cm

\begin{tabular}{||c|c|c|c|c|c|c|c||}
\hline\hline
Element & $I$ & $a_{mixed}$ & $a_{triplett}$ & $\gamma _{D2}/2\pi $ & $D1$ & 
$D2$ & $\nu _{HFS}[S_{1/2}]$ \\ \hline
&  & [$a_{B}$] & [$a_{B}$] & [~MHz] & $[$~cm$^{-1}]$ & $[$~cm$^{-1}]$ & $[$%
~MHz$]$ \\ \hline\hline
\multicolumn{1}{||c|}{$^{1}$H} & $1/2$ &  & $1.23$ & $99.58$ & $82264.$ & $%
82264.$ &  \\ \cline{1-4}\cline{8-8}
\multicolumn{1}{||c|}{$^{2}$H} & $1$ &  & $-6.8$ &  &  &  &  \\ \hline
$^{6}$Li & $1$ &  & $-2160$ & $5.92$ & $14901.$ & $14901.$ & $228.2$ \\ 
\cline{1-4}\cline{8-8}
\multicolumn{1}{||c|}{$^{7}$Li} & $3/2$ & $10$ & $-27.3$ &  &  &  & $803.5$
\\ \hline
$^{23}$Na & $3/2$ & $52$ & $85$ & $10.01$ & $16956.$ & $16973.$ & $1771.6$
\\ \cline{1-7}\cline{5-8}
\multicolumn{1}{||c|}{$^{39}$K} & $3/2$ & $118$ & $81.1$ &  &  &  & $461.7$
\\ \cline{1-4}\cline{8-8}
\multicolumn{1}{||c|}{$^{40}$K} & $4$ & $158$ & $1.7$ & $6.09$ & $12985.$ & $%
13043.$ & $-1285.8$ \\ \cline{1-4}\cline{8-8}
\multicolumn{1}{||c|}{$^{41}$K} & $3/2$ & $225$ & $286$ &  &  &  & $254.0$
\\ \hline
\multicolumn{1}{||c|}{$^{85}$Rb} & $5/2$ & $-450$ & $-363$ & $5.98$ & $%
12579. $ & $12816.$ & $3035.7$ \\ \cline{1-4}\cline{8-8}
\multicolumn{1}{||c|}{$^{87}$Rb} & $3/2$ & $105$ & $109.3$ &  &  &  & $%
6834.7 $ \\ \hline
\multicolumn{1}{||c|}{$^{133}$Cs} & $7/2$ & $-240$ & $-350$ & $5.18$ & $%
11182.$ & $11737.$ & $9192.6$ \\ \cline{1-4}\cline{8-8}
\multicolumn{1}{||c|}{$^{135}$Cs} & $7/2$ & $163$ & $138$ &  &  &  &  \\ 
\hline\hline
\end{tabular}

\newpage

\begin{center}
{\bf Table 3}
\end{center}

\vskip2cm

\begin{tabular}{||l|l|l||}
\hline\hline
{\bf regime} & $k^{-1}$ & {\bf method} \\ \hline\hline
hydrodynamic & $\gg l_{mfp}$ & large BECs, high temperatures \\ \hline
collisionless & $\ll l_{mfp}$ & trap modulation \\ \hline
collective discrete modes & $\gg a_{trap}$ & trap modulation, standing\
soundwave \\ \hline
pulsed localized modes & $\ll a_{trap}$ & dipole force laser beam,
propagating soundwave \\ \hline
phonon-like & $\gg \xi $ & Bragg scattering \\ \hline
free particle-like & $\ll \xi $ & Bragg scattering \\ \hline\hline
\end{tabular}

\newpage

\setcounter{equation}{0} 
\begin{equation}
\text{Figure}  \label{FigBoseCondensation}
\end{equation}
\begin{equation}
\text{Figure}  \label{FigPeakDensity}
\end{equation}
\begin{equation}
\text{Figure}  \label{FigHeatCapacity}
\end{equation}
\begin{equation}
\text{Figure}  \label{FigAbsorptionPictures}
\end{equation}
\begin{equation}
\text{Figure}  \label{FigHydrogenSpectrum}
\end{equation}
\begin{equation}
\text{Figure}  \label{FigReleaseEnergy}
\end{equation}
\begin{equation}
\text{Figure}  \label{FigAspectRatio}
\end{equation}
\begin{equation}
\text{Figure}  \label{FigCompressionOscillations}
\end{equation}
\begin{equation}
\text{Figure}  \label{FigRabiTwist}
\end{equation}
\begin{equation}
\text{Figure}  \label{FigImpurityScattering}
\end{equation}
\begin{equation}
\text{Figure}  \label{FigVortex}
\end{equation}
\begin{equation}
\text{Figure}  \label{FigVortexLattice}
\end{equation}
\begin{equation}
\text{Figure}  \label{FigSoliton}
\end{equation}
\begin{equation}
\text{Figure}  \label{FigStimulationCurves}
\end{equation}
\begin{equation}
\text{Figure}  \label{FigInterferenceScheme}
\end{equation}
\begin{equation}
\text{Figure}  \label{FigInterferenceFringes}
\end{equation}
\begin{equation}
\text{Figure}  \label{FigOutputCoupling}
\end{equation}
\begin{equation}
\text{Figure}  \label{FigBraggDiffractionScheme}
\end{equation}
\begin{equation}
\text{Figure}  \label{FigBraggSpectroscopy}
\end{equation}
\begin{equation}
\text{Figure}  \label{FigFourWaveMixing}
\end{equation}
\begin{equation}
\text{Figure}  \label{FigFourWaves}
\end{equation}
\begin{equation}
\text{Figure}  \label{FigRaleighWaves}
\end{equation}
\begin{equation}
\text{Figure}  \label{FigFeshbachResonance}
\end{equation}
\begin{equation}
\text{Figure}  \label{FigFeshbachScatteringLength}
\end{equation}
\begin{equation}
\text{Figure}  \label{FigTwoPhotonPhotoassociation}
\end{equation}
\begin{equation}
\text{Figure}  \label{Fig23}
\end{equation}
\begin{equation}
\text{Figure}  \label{Fig24}
\end{equation}
\begin{equation}
\text{Figure}  \label{Fig25}
\end{equation}
\begin{equation}
\text{Figure}  \label{Fig26}
\end{equation}
\begin{equation}
\text{Figure}  \label{Fig27}
\end{equation}
\begin{equation}
\text{Figure}  \label{Fig28}
\end{equation}
\begin{equation}
\text{Figure}  \label{Fig29}
\end{equation}
\begin{equation}
\text{Figure}  \label{Fig30}
\end{equation}
\begin{equation}
\text{Figure}  \label{Fig31}
\end{equation}
\begin{equation}
\text{Figure}  \label{Fig32}
\end{equation}
\begin{equation}
\text{Figure}  \label{Fig33}
\end{equation}
\begin{equation}
\text{Figure}  \label{Fig34}
\end{equation}
\begin{equation}
\text{Figure}  \label{Fig35}
\end{equation}
\begin{equation}
\text{Figure}  \label{Fig36}
\end{equation}
\begin{equation}
\text{Figure}  \label{Fig37}
\end{equation}
\begin{equation}
\text{Figure}  \label{Fig38}
\end{equation}
\begin{equation}
\text{Figure}  \label{Fig39}
\end{equation}
\begin{equation}
\text{Figure}  \label{Fig40}
\end{equation}
\setcounter{equation}{0} 
\begin{equation}
\text{Table}  \label{TabCriticalTemperature}
\end{equation}
\begin{equation}
\text{Table}  \label{TabBoseSpecies}
\end{equation}
\begin{equation}
\text{Table}  \label{TabExcitationRegimes}
\end{equation}

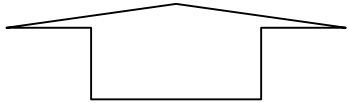
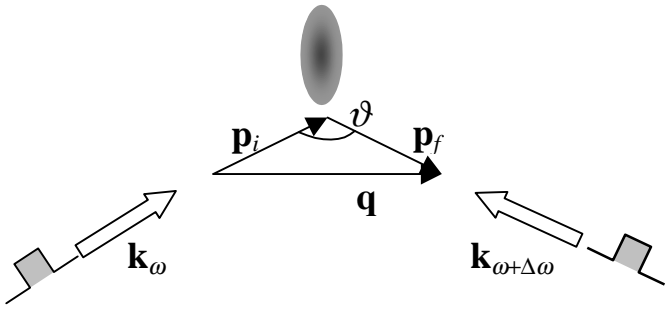
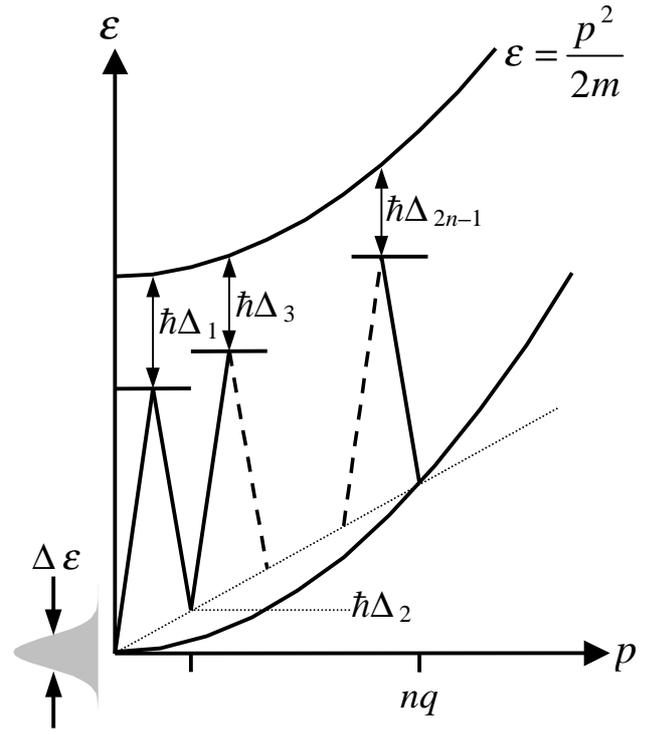
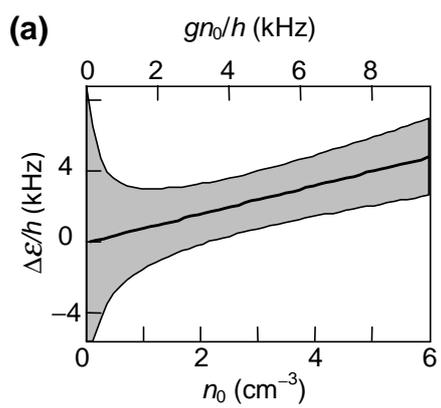
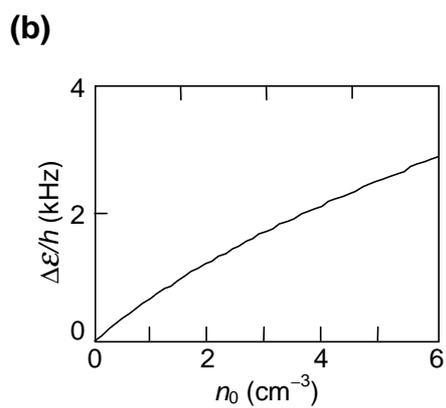
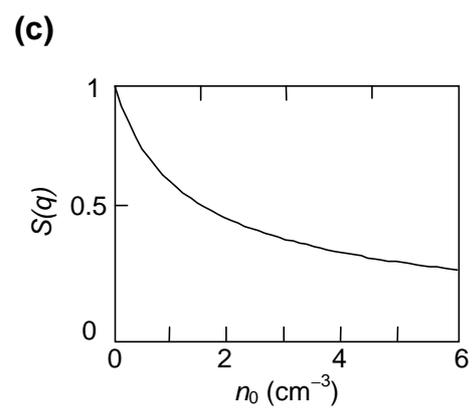
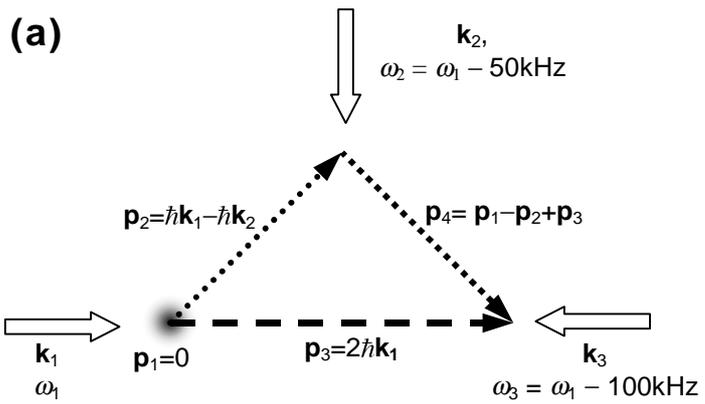
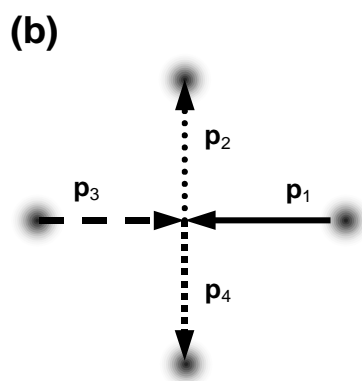
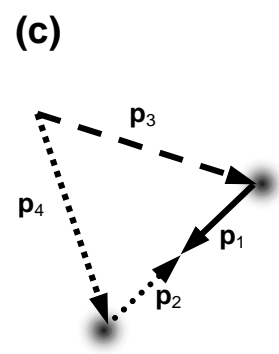
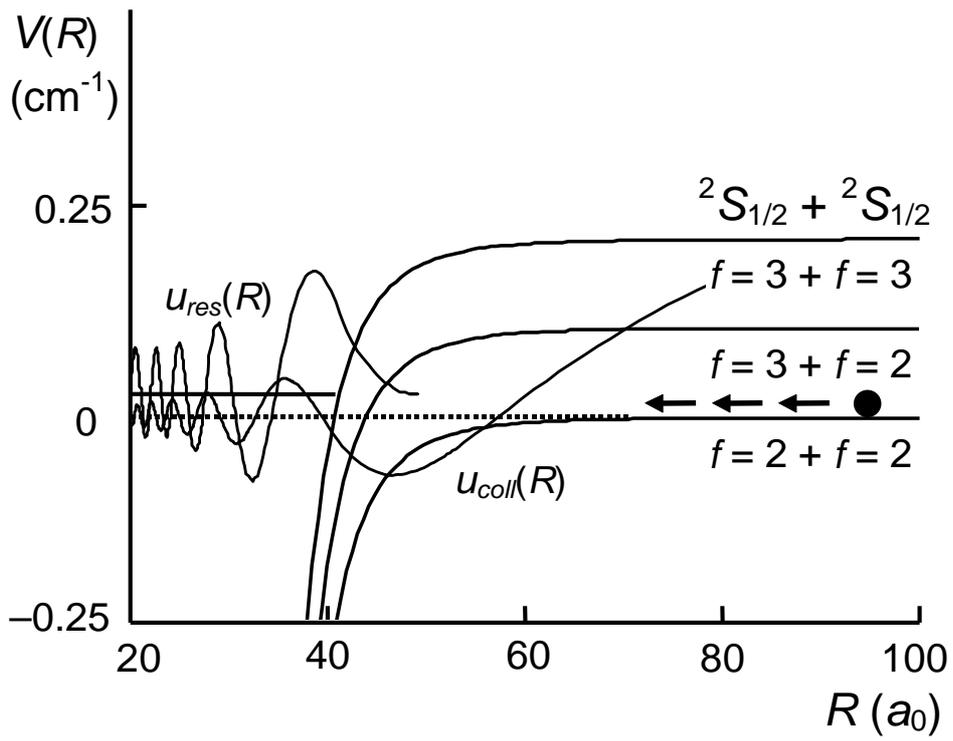
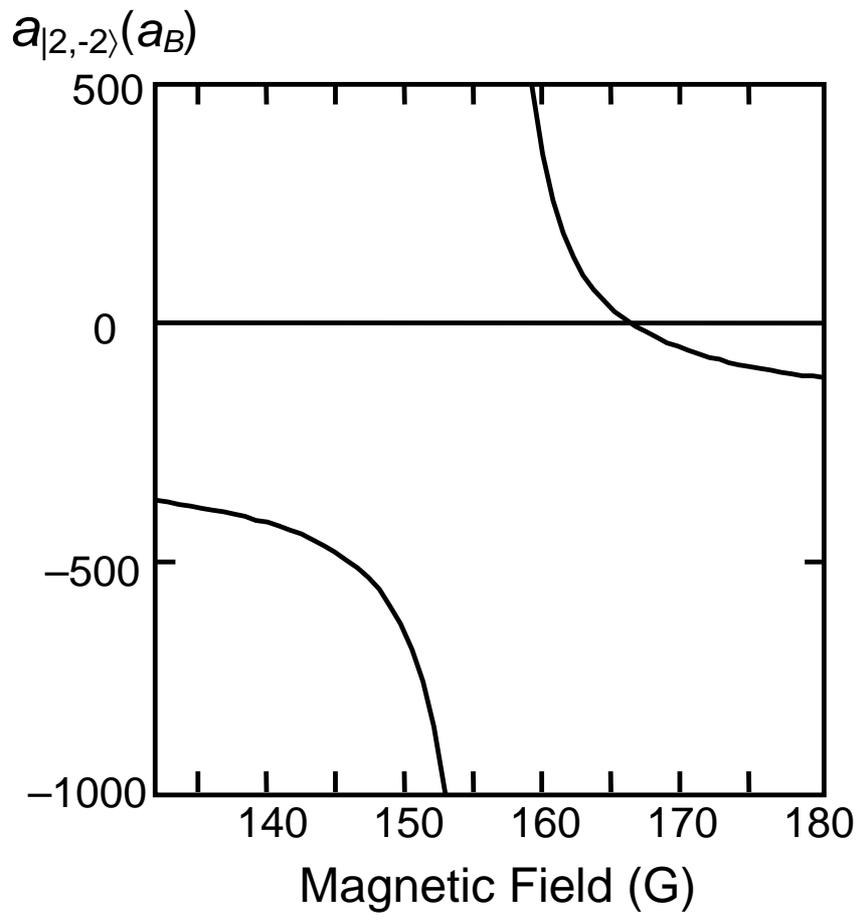
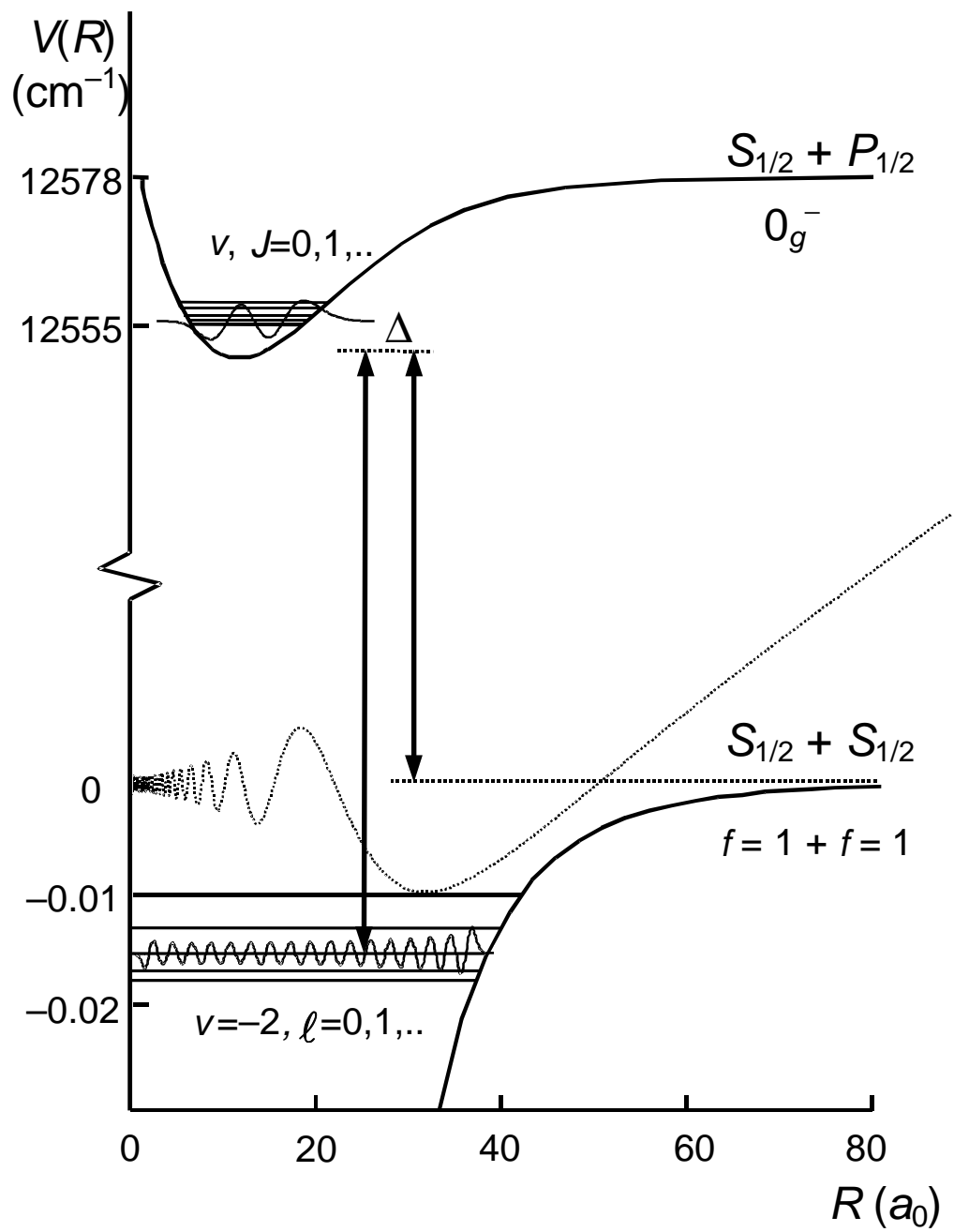
\end{document}